\documentclass[10pt,twocolumn,letterpaper]{article}

\usepackage{cvpr}              

\usepackage{amsmath,amsfonts,amsthm,amssymb}
\usepackage{mathtools}
\usepackage{bm}
\usepackage{nicefrac}
\usepackage{microtype}
\usepackage{lipsum}

\usepackage{color,xcolor}
\usepackage{epsfig}
\usepackage{graphicx}

\usepackage{adjustbox}
\usepackage{array}
\usepackage{booktabs}
\usepackage{colortbl}
\usepackage{wrapfig}
\usepackage{hhline}
\usepackage{multirow}
\usepackage{subcaption}
\usepackage[size=small]{caption}
\usepackage{float}

\usepackage{changepage}
\usepackage{extramarks}
\usepackage{fancyhdr}
\usepackage{lastpage}
\usepackage{setspace}
\usepackage{soul}
\usepackage{xspace}

\usepackage{url}

\usepackage{algorithm}
\usepackage{algorithmicx}
\usepackage{algpseudocode}
\usepackage{enumerate}
\usepackage{enumitem}  
\usepackage{makecell}
\usepackage{pifont}
\usepackage{titlecaps}
\usepackage[accsupp]{axessibility}
\usepackage{framed}

\usepackage[dvipsnames]{xcolor}
\usepackage{pifont}
\newcommand{\cmark}{\ding{51}}
\newcommand{\xmark}{\ding{55}}
\newcommand{\yes}{\textcolor{Green}{\cmark}}
\newcommand{\no}{\textcolor{red}{\xmark}}

\newcommand{\name}{DTC\xspace}

\renewcommand{\paragraph}[1]{\vspace{0.1cm}\noindent\textbf{#1}}

\definecolor{MyDarkBlue}{rgb}{0,0.08,1}
\definecolor{MyAqua}{rgb}{0,0.7,0.7}
\definecolor{MyDarkGreen}{rgb}{0.02,0.6,0.02}
\definecolor{MyDarkRed}{rgb}{0.8,0.02,0.02}
\definecolor{MyDarkOrange}{rgb}{0.40,0.2,0.02}
\definecolor{MyPurple}{RGB}{111,0,255}
\definecolor{MyRed}{rgb}{1.0,0.0,0.0}
\definecolor{MyGold}{rgb}{0.75,0.6,0.12}
\definecolor{MyDarkgray}{rgb}{0.66, 0.66, 0.66}

\definecolor{kwBlue}{HTML}{1f77b4}
\definecolor{kwGreen}{HTML}{2ca02c}
\definecolor{kwOrange}{HTML}{ff7f0e}

\definecolor{Cardinal}{rgb}{0.549,0.082,0.082}

\definecolor{cvprblue}{rgb}{0.21,0.49,0.74}
\usepackage[pagebackref,breaklinks,colorlinks,citecolor=cvprblue]{hyperref}

\def\paperID{11123} 
\def\confName{CVPR}
\def\confYear{2025}

\title{Digital Twin Catalog: A Large-Scale Photorealistic 3D Object Digital Twin Dataset}

\author{
Zhao Dong$^{1}$ \;
Ka Chen$^{1,*}$ \;
Zhaoyang Lv$^{1,*}$ \;
Hong-Xing Yu$^{2,*}$ \;
Yunzhi Zhang$^{2,*}$ \;
Cheng Zhang$^{1,*}$ \\
Yufeng Zhu$^{1,*}$ \;
Stephen Tian$^{2}$ \;
Zhengqin Li$^{1}$ \;
Geordie Moffatt$^{1}$\;
Sean Christofferson$^{1}$ \;
James Fort$^{1}$\\
Xiaqing Pan$^{1}$\;
Mingfei Yan$^{1}$ \;
Jiajun Wu$^{2}$ \;
Carl Yuheng Ren$^{1}$ \;
Richard Newcombe$^{1}$ 
\vspace{0.3cm} \\
$^{1}$Meta Reality Labs Research \quad $^{2}$Stanford University \\
}
\begin{document}

\vspace{-0.3cm}
\twocolumn[{%
\renewcommand\twocolumn[1][]{#1}%
\maketitle
\begin{center}
  \centering
  \captionsetup{type=figure}
  \includegraphics[width=\linewidth]{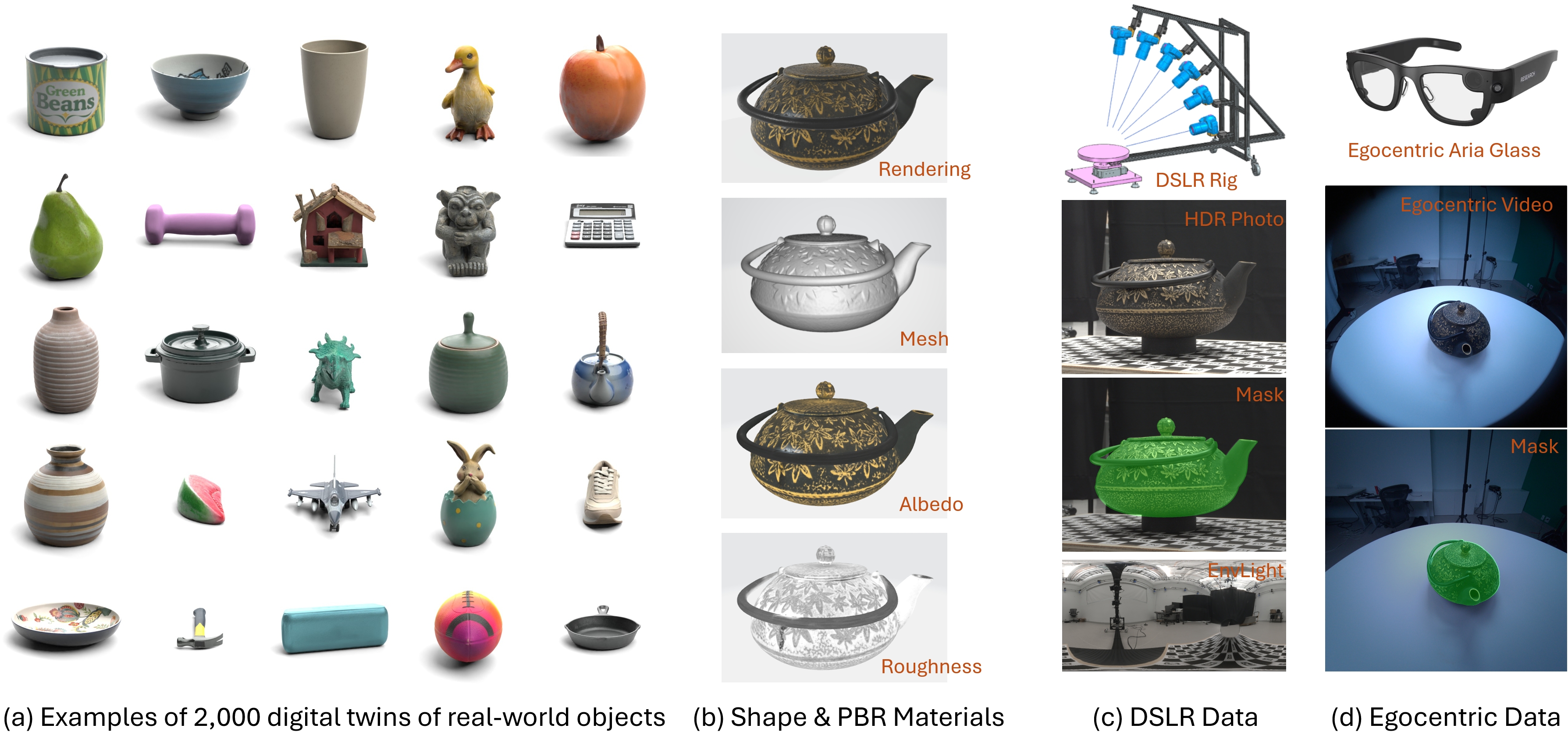}
   \caption{The Digital Twin Catalog (DTC) dataset comprises 2,000 digital twins of physical-world objects (a), characterized by millimeter-level geometric accuracy and photorealistic PBR materials (b). DTC includes evaluation data captured using both DSLR cameras and egocentric Aria glasses, featuring captured images with precise foreground object masks and environment lighting for relighting evaluation.}
   \label{fig:DTC_Teaser}
\end{center}
\vspace{-0.3em}
}]

\begingroup
\renewcommand\thefootnote{}\footnotetext{
$^*$Authors contributed equally and are listed in alphabetical order.
}
\endgroup

\begin{abstract}
We introduce the Digital Twin Catalog (DTC), a new large-scale photorealistic 3D object digital twin dataset. A digital twin of a 3D object is a highly detailed, virtually indistinguishable representation of a physical object, accurately capturing its shape, appearance, physical properties, and other attributes. Recent advances in neural-based 3D reconstruction and inverse rendering have significantly improved the quality of 3D object reconstruction. Despite these advancements, there remains a lack of a large-scale, digital twin-quality real-world dataset and benchmark that can quantitatively assess and compare the performance of different reconstruction methods, as well as improve reconstruction quality through training or fine-tuning. Moreover, to democratize 3D digital twin creation, it is essential to integrate creation techniques with next-generation egocentric computing platforms, such as AR glasses. Currently, there is no dataset available to evaluate 3D object reconstruction using egocentric captured images. To address these gaps, the DTC dataset features 2,000 scanned digital twin-quality 3D objects, along with image sequences captured under different lighting conditions using DSLR cameras and egocentric AR glasses. This dataset establishes the first comprehensive real-world evaluation benchmark for 3D digital twin creation tasks, offering a robust foundation for comparing and improving existing reconstruction methods. The DTC dataset is already released at \href{https://www.projectaria.com/datasets/dtc/}{https://www.projectaria.com/datasets/dtc/}, and we will also make the baseline evaluations open-source.
\end{abstract}

\begin{figure*}[t!]
    \centering
    \includegraphics[width=\linewidth]{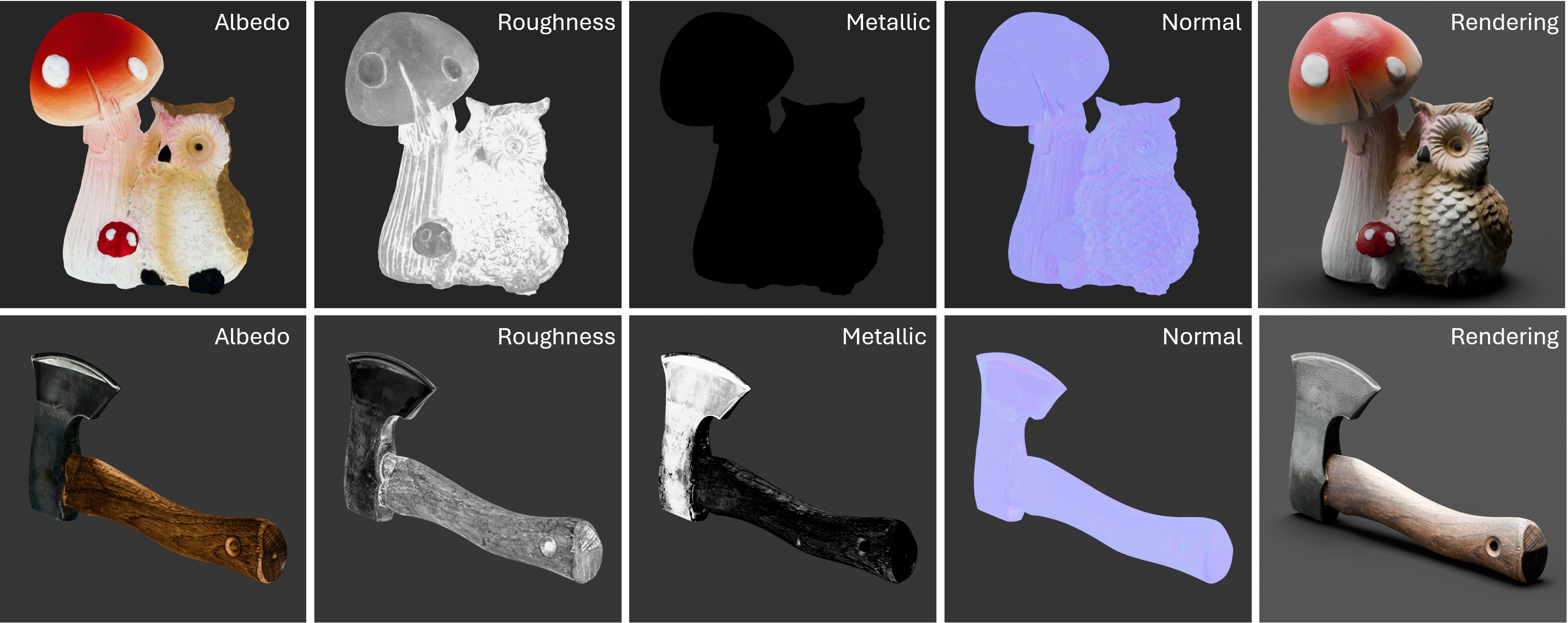}
    \caption{Example DTC models with photorealistic PBR materials.}
    \label{fig:DTC_PBR_Models}
\end{figure*}

\section{Introduction}
\label{sec:intro}

\begin{table*}[t]
\centering
\newcommand{\synthetic}{\textcolor{red}{synthetic}}
\newcommand{\studio}{\textcolor{red}{studio}}
\newcommand{\wild}{\textcolor{Green}{in-the-wild}}
\caption{Comparison with existing object-centric inverse rendering datasets.
*Objaverse~\citep{deitke2023objaverse} consists of both synthetic objects and real scans, only part of which contain PBR materials.
}
\footnotesize
\resizebox{\linewidth}{!}{

\begin{tabular}{lccccccccc}
\toprule
Dataset & \# Objects & Real & Scene Type & Multi-view & Shape  &PBR Mat. & Relit Image  & Lighting & Egocentric Cap.\\
\midrule
ShapeNet-Intrinsics~\citep{shi2017learning} & 31K & \no & \synthetic&\yes & \yes & \no & \yes & \yes & \no\\
NeRD Synthetic~\citep{boss2021nerd} & 3 & \no & \synthetic&\yes& \yes & \yes & \yes & \yes & \no  \\
ABO~\citep{collins2022abo} & 8K & \no & \synthetic&\yes& \yes & \yes & \yes &\yes & \no \\
\midrule
MIT Intrinsics~\citep{grosse2009mitintrinsics}              & 20  & \yes & \studio&\yes & \no & \no & \no & \no & \no \\
DTU-MVS~\citep{jensen2014dtu}    & 80 & \yes & \studio&\yes & \yes & \no &\no & \no  & \no \\
Objaverse~\citep{deitke2023objaverse} & 818K & (\yes)* & \studio & \yes & \yes & (\yes)* & \no & \no & \no \\
DiLiGenT-MV~\citep{li2020multi}  & 5 & \yes & \studio&\yes & \yes & \no & \no & \yes & \no \\
ReNe~\citep{toschi2023relight} & 20  & \yes & \studio&\yes & \no & \no & \no &\yes & \no  \\
OpenIllumination~\citep{liu2023openillumination} &64&\yes&\studio&\yes &\no&\yes&
\yes&\yes& \no\\
GSO~\citep{downs2022google} &1030&\yes&\studio&\yes&\yes &\no&\no&\no& \no\\
Lombardi~\etal~\citep{lombardi2012reflectance} &6&\yes& \wild&\no &\yes& \no& \yes&\yes & \no\\
NeRD Real~\citep{boss2021nerd} & 4 & \yes & \wild&\yes & \no & \yes&\yes & \no & \no  \\
NeROIC~\citep{kuang2022neroic}  & 3 & \yes & \wild&\yes & \no & \yes & \yes & \no & \no  \\
Oxholm~\etal~\citep{oxholm2014multiview}&4&\yes&\wild&\yes &\yes&\no&\yes&\yes& \no\\
OmniObject3D~\cite{wu2023omniobject3d} &6k&\yes&\wild&\yes &\yes&\no&\no&\no& \no\\
Stanford-ORB~\cite{kuang2023stanfordorb} & 14 & \yes & \wild&\yes & \yes & \yes & \yes & \yes & \no  \\
\name (ours) & 2k & \yes & \wild&\yes & \yes & \yes & \yes & \yes & \yes  \\
\bottomrule
\end{tabular}
}
\label{tab:datasets}
\end{table*}

A digital twin of a 3D object is a highly detailed, virtually indistinguishable representation of a physical object, capturing its shape, appearance, physical properties and other attributes with precision. Such a digital twin enables visualization, analysis, and interaction as if it were the real object, supporting simulation, automation, and real-world problem-solving across a wide range of applications in AR/VR~\cite{aloqaily2022integrating, hamidouche2024immersive}, spatial/contextual AI~\cite{alexopoulos2020digital}, and robotics~\cite{gao2021ObjectFolder, li24simpler}. As fundamental properties of an object, its shape and appearance form the basis for recognizing and interpreting the 3D object, enabling identification, manipulation, and realistic rendering. Recovering these attributes has long been a foundational topic in computer vision and graphics, inspiring extensive research in 3D reconstruction and inverse rendering. Recent breakthroughs in neural-based representation and reconstruction techniques, such as NeRF~\cite{mildenhall2020nerf} and 3D Gaussian splatting (3DGS)~\cite{kerbl3Dgaussians}, have significantly elevated the quality of novel view synthesis (NVS) to photorealistic levels. Many subsequent works~\cite{sun2023neural, Jin2023TensoIR} integrate neural reconstruction with physically-based inverse rendering, enabling relightable appearances. Furthermore, leveraging priors from large reconstruction models (LRMs)~\cite{hong2024lrm}, high-quality shape and appearance reconstruction can now be achieved with as few as one to four views~\cite{Li_2024_ICLR, wei2024meshlrm, zhang2024gs}. 

Despite the rapid advancements in 3D object reconstruction, one question remains: \textit{does the reconstruction quality truly meet the standard of a digital twin, where virtual representations are \textbf{indistinguishable} from reality?} This digital twin standard demands both \textbf{highly accurate shape matching} and \textbf{photorealistic appearance across different lighting}, which present significant acquisition challenges for real-world objects. Existing object-centric datasets for 3D reconstruction or inverse rendering have focused on either dataset size~\cite{deitke2023objaverse} or quality of specific aspects~\cite{collins2022abo, jensen2014dtu, shi2017learning, downs2022google, wu2023omniobject3d, liu2023openillumination}, often sacrificing comprehensive fidelity and limiting their application scope. This trade-off has led to a lack of datasets that fully satisfy the digital twin criteria, hindering current 3D reconstruction methods from achieving digital twin fidelity. To bridge this gap, we developed the Digital Twin Catalog (DTC) dataset, comprising 2,000 scanned 3D object models (Fig.~\ref{fig:DTC_Teaser}(a)), each with millimeter geometry accuracy and photorealistic PBR materials (Fig.~\ref{fig:DTC_Teaser}(b), Fig.~\ref{fig:DTC_PBR_Models}).

In addition to 3D digital twin models, the DTC dataset includes evaluation data designed to support 3D object reconstruction research. This evaluation data features multi-view image sequences with precise foreground object masks and environment lighting information for relighting evaluation. Traditionally, high-quality HDR images captured with modern DSLR cameras have been the standard for 3D reconstruction research. Looking ahead, we encourage the integration of 3D reconstruction research with next-generation human-centric computing platforms, such as egocentric AR glasses, aiming to democratize 3D reconstruction techniques and empower everyone to effortlessly create 3D digital twins. To this end, alongside DSLR-captured evaluation data (Fig.~\ref{fig:DTC_Teaser}(c)), the DTC dataset also provides egocentric evaluation data captured using Project Aria glasses (\href{https://www.projectaria.com}{https://www.projectaria.com}) (Fig.~\ref{fig:DTC_Teaser}(d)).

The DTC dataset offers extensive opportunities for advancing research in object digital twin creation. We provide a benchmark for state-of-the-art 3D object reconstruction and inverse rendering methods
These benchmarks evaluate performance across novel view synthesis (NVS), shape reconstruction, and relightable appearance reconstruction. We further provide the evaluation of novel view synthesis methods using the egocentric aligned DTC data. Additionally, we explore the dataset's potential in downstream robotics applications by assessing its effectiveness in training robotic policies for pushing and grasping tasks in simulation. These benchmarks and applications provide valuable insights, highlight existing challenges, and uncover promising directions for future research in 3D digital twin creation.

\section{Related Work}
\label{sec:related_work}
We provide a comparison of our DTC dataset to existing object-centric datasets in Table \ref{tab:datasets}. We provide the largest 3D dataset with PBR materials and real world multi-view recordings with digital twin counterparts. We further provide digital twin aligned egocentric recordings, the first of their kind in the egocentric domain. We will discuss the related datasets and methods they can empower as follows. 

\paragraph{3D Digital Twin Datasets}
Existing 3D digital twin datasets with PBR materials often serve as ground truth for evaluating inverse rendering results. Early efforts \cite{grosse2009mitintrinsics,bell2014intrinsic} provide small-scale intrinsic image of real objects and do not provide shape or PBR material information.
Synthetic datasets~\cite{shi2017learning, li2018learning, boss2021nerd, wu2021derender, wimbauer2022derender, collins2022abo} are widely used for evaluation but do not represent the complexity in a real world environment. 
For datasets that contain real objects,~\cite{jensen2014dtu, deitke2023objaverse, li2020multi, toschi2023relight, lombardi2012reflectance,oxholm2014multiview}, the reconstruction quality can vary, which leads to imprecise evaluations. Table~\ref{tab:datasets} provides a comparison to the previous work in this domain. Compared to Objaverse\cite{deitke2023objaverse,deitke2024objaverse}, which is a collection of existing 3D models with only a small subset containing PBR materials with varying quality, we offer a high quality collection of 3D object data that is also aligned with real world recordings. Compared to OmniObject3D \cite{wu2023omniobject3d}, the DTC models provide higher-quality shape and additional PBR materials that are necessary for high quality inverse rendering. We offer the largest quantity of 3D object models compared to all counterparts in various tasks. For real-world evaluation, Stanford-ORB \cite{kuang2023stanfordorb} was the prior largest inverse rendering benchmark with in-the-wild lighting. In contrast, we provide more object diversity and higher quality for each object model. The Aria Digital Twin dataset \cite{pan2023adt} was the first dataset to provide digital twin aligned environments for the scenes and recorded using egocentric device. However, their scene environments are limited and the contained object ground truths inside do not have high quality geometries with PBR materials. 

\paragraph{Object Reconstruction \& Inverse Rendering.} 
Using object-centric multi-view images as input, early object reconstruction methods focused on estimating individual object properties, such as shape from shading~\cite{bichsel1992simple, zhang1999shape, barron2014shape}, material acquisition~\cite{li2018materials, li2018learning, oxholm2015shape, oxholm2014multiview, yamashita2023nlmvs}, and lighting estimation~\cite{weber2018learning, yu2023accidental}. Some approaches also aimed to recover reflectance and illumination assuming known object shapes~\cite{lombardi2012reflectance, lombardi2015reflectance}. Inverse rendering, which seeks to invert the rendering equation~\cite{kajiya1986rendering}, estimates an image's intrinsic components—geometry, materials, and lighting. The advent of differentiable renderers~\cite{luan2021unified, chen2021dib, cai2022physics} enabled full-fledged inverse rendering methods to simultaneously recover all these properties for object reconstruction~\cite{marschner1998inverse}.

Neural volumetric representations such as Neural Radiance Fields (NeRFs)~\cite{mildenhall2020nerf} and the like~\cite{Bi2020DeepECCV, boss2021nerd, boss2021neural, boss2022samurai, zhang2021nerfactor, rudnev2022nerf, yu2023learning} encode geometry and appearance as volumetric densities and radiance with a Multi-Layer Perceptron (MLP) network, and render images using the volume rendering equation~\cite{max1995optical}.
3D-GS \cite{kerbl3Dgaussians} introduces 3D Gaussian primitives and rasterization and its following-up variants \cite{huang20242Dgs} demonstrates high quality geometry prediction as well.

Other surface-based representations~\cite{wang2021neus, physg2021, zhang2022iron, Munkberg2022nvdiffrec, zhang2022modeling, hasselgren2022nvdiffrecmc, wu2023nefii, sun2023neural} extract surfaces as the zero level set, for instance, of a signed distance function (SDF) or an occupancy field~\cite{Niemeyer2020CVPR}, allowing them to efficiently model the appearance on the surface with an explicit material model, such as bidirectional reflectance distribution functions (BRDFs).
This also enables modeling more complex global illumination effects, such as self-shadows.
Most of these methods focus on per-scene optimization and require dense multiple views as input.
Recently, researchers have incorporated learning-based models, distilling priors from large training datasets for fast inference on limited test views~\cite{shi2017learning, janner2017intrinsic, li2018learning, bi2020deep, Boss2020brdf, wu2021derender, wimbauer2022derender, zhang2023rose}.

%
In this work, we provide the DTC dataset with real-world object recordings that can serve as the benchmark to evaluate object-centric inverse rendering tasks. We evaluate representative baselines from existing work.

\section{Digital Twin Catalog: A Large-Scale Photorealistic 3D Object Digital Twin Dataset}
\label{sec:dataset}

\subsection{Dataset Composition}
\label{subsec:composition}
Our DTC dataset contains: (1) \textbf{2,000} scanned 3D object models, featuring millimeter geometric accuracy relative to their physical counterparts, along with a full set of photorealistic PBR material maps (Fig.~\ref{fig:DTC_Teaser}), (2) \textbf{100} DSLR-captured evaluation data of \textbf{50} objects under different lighting conditions, and (3) \textbf{200} egocentric Aria-captured evaluation data of \textbf{100} objects with both \textbf{active} and \textbf{casual} observation modes.
\begin{figure}
    \centering
    \includegraphics[width=0.99\linewidth]{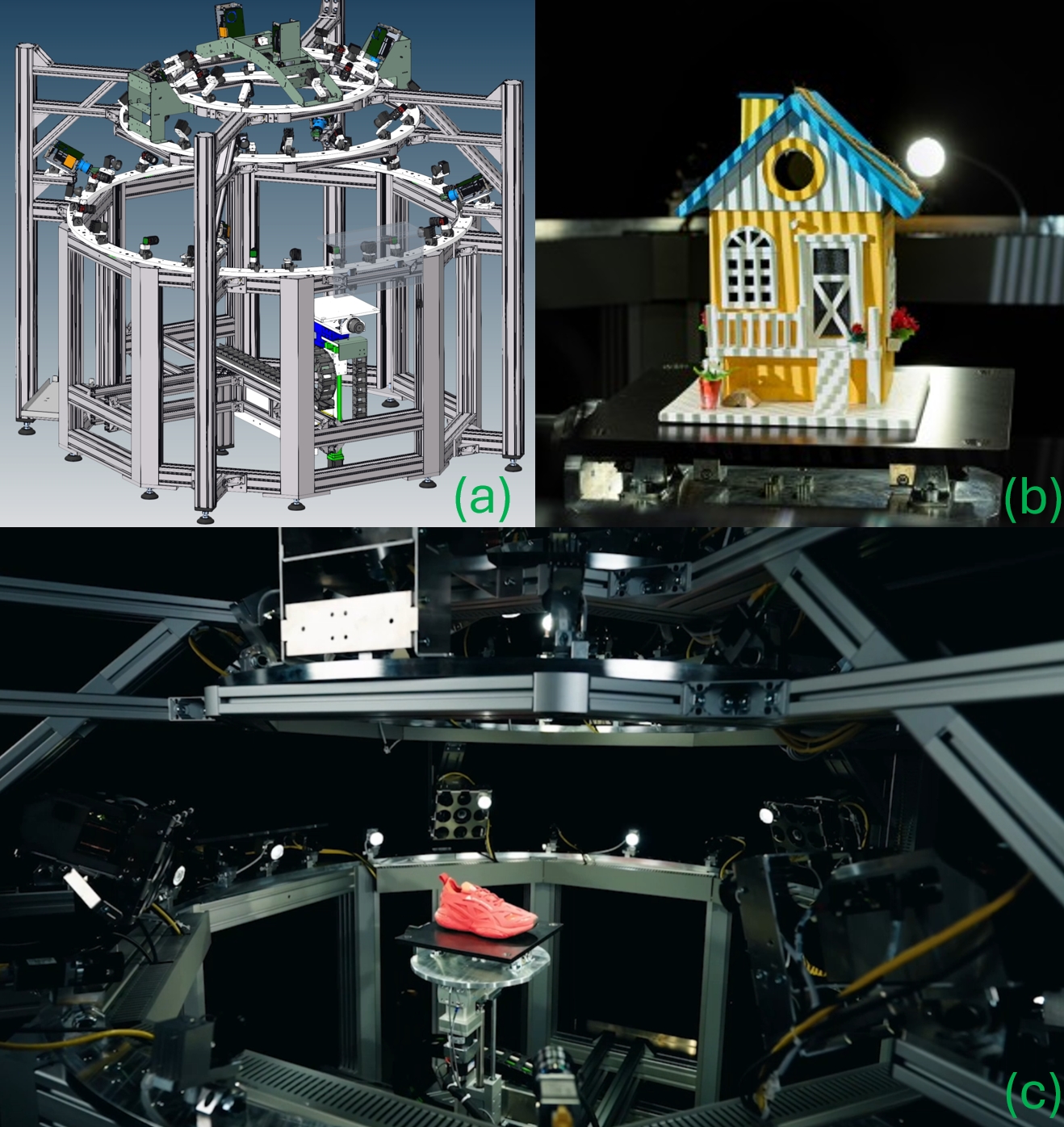}
    \caption{3D object scanner by Covision Media\textsuperscript{\textregistered}.}
    \label{fig:Covision_scanner}
\end{figure}
\vspace{-0.25em}

\subsection{Creation of 3D Object Models}
\label{subsec:3d_models}
Utilizing the state-of-the-art industrial 3D object scanner~\cite{Covision_3D_Scanner}, we selected 2,000 physical-world objects spanning 40 LVIS~\cite{gupta2019lvis} categories, carefully chosen to ensure both category diversity and compatibility with the scanner's capabilities. As illustrated in Fig.~\ref{fig:Covision_scanner}(a) (c), the scanner features a fixed lighting-camera setup within an upper-hemisphere dome, equipped with 8 structured lights for geometry scanning, and 29 spotlights and 29 cameras for material acquisition. During the scanning process, the object is placed on a central holder, and its pose can be adjusted with multi-round scanning to achieve a complete 360-degree scan. For our dataset, each object typically undergoes three pose changes, with a total scanning time of approximately 20 minutes per object. 
\begin{figure}
    \centering
    \includegraphics[width=\linewidth]{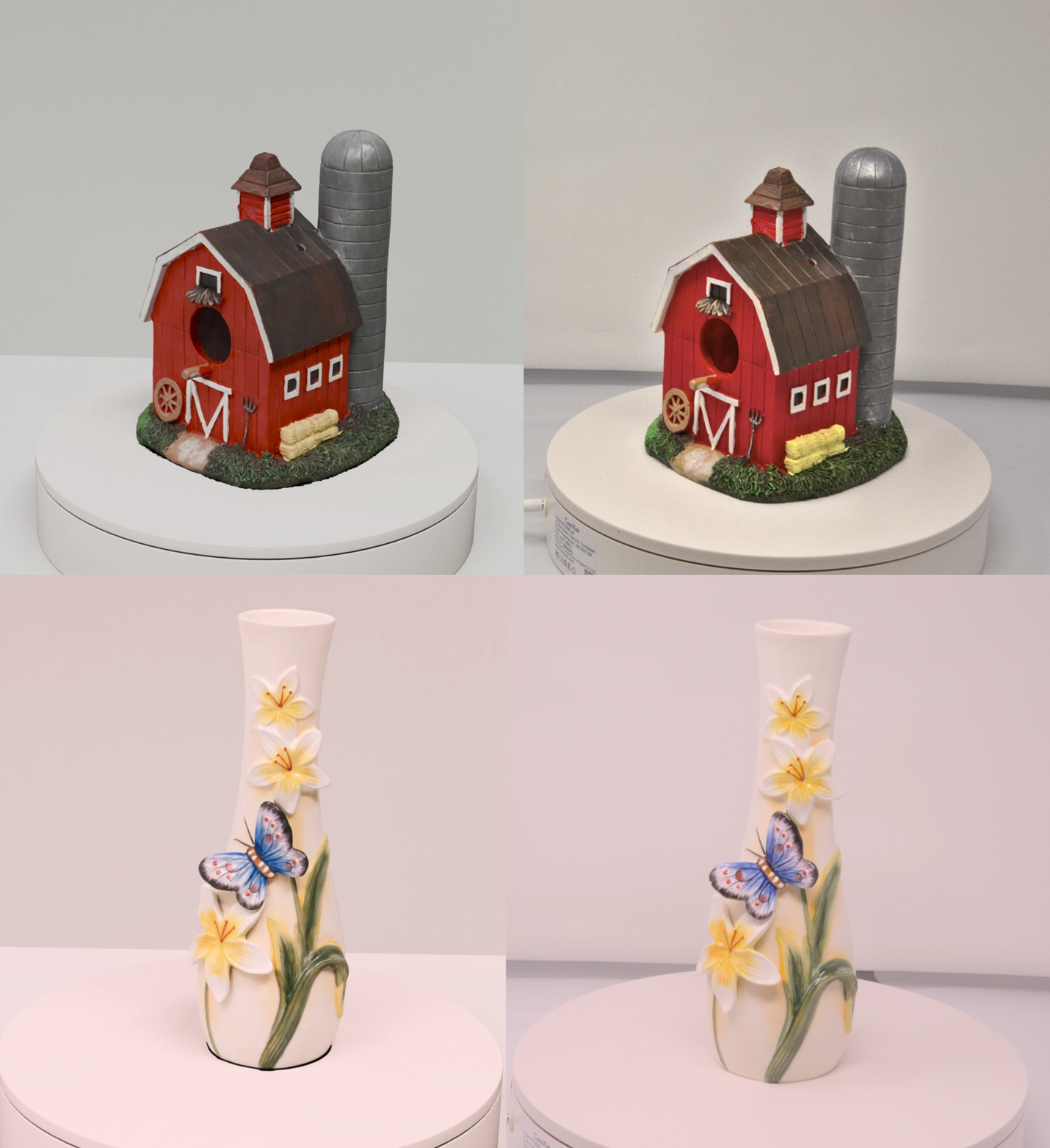}
    \caption{Rendered DTC models (left) v.s. Photo (Right).}
    \label{fig:DTC_SBS}
\end{figure}

After scanning, a proprietary post-processing pipeline reconstructs both the geometry and the PBR material maps. For 4K-resolution PBR material maps, the post-processing requires approximately 4 hours per object. In terms of quality, the structured-light-based shape reconstruction (Fig.~\ref{fig:Covision_scanner} (b)) in the post-processing achieves millimeter-level geometric accuracy. However, the material optimization process performs best for diffuse objects and often struggles with glossy or shiny surfaces. To address this limitation, we hired technical artists to develop a workflow to refine materials for glossy and shiny objects, ensuring that the material quality meets the standards of a digital twin.

\paragraph{3D Model Accuracy.} To validate the material and geometry accuracy of the 3D models, we compared a rendered image of our scanned and processed model with a photograph of the same object taken inside a light box. A virtual light box was meticulously modeled to replicate the light intensity and color temperature of the real light box. The scanned object was then placed in the virtual light box to generate the rendered image. The side-by-side comparison demonstrates a remarkable match between the rendered and real images (Fig.~\ref{fig:DTC_SBS}). 

\begin{figure}
    \centering
    \includegraphics[width=\linewidth]{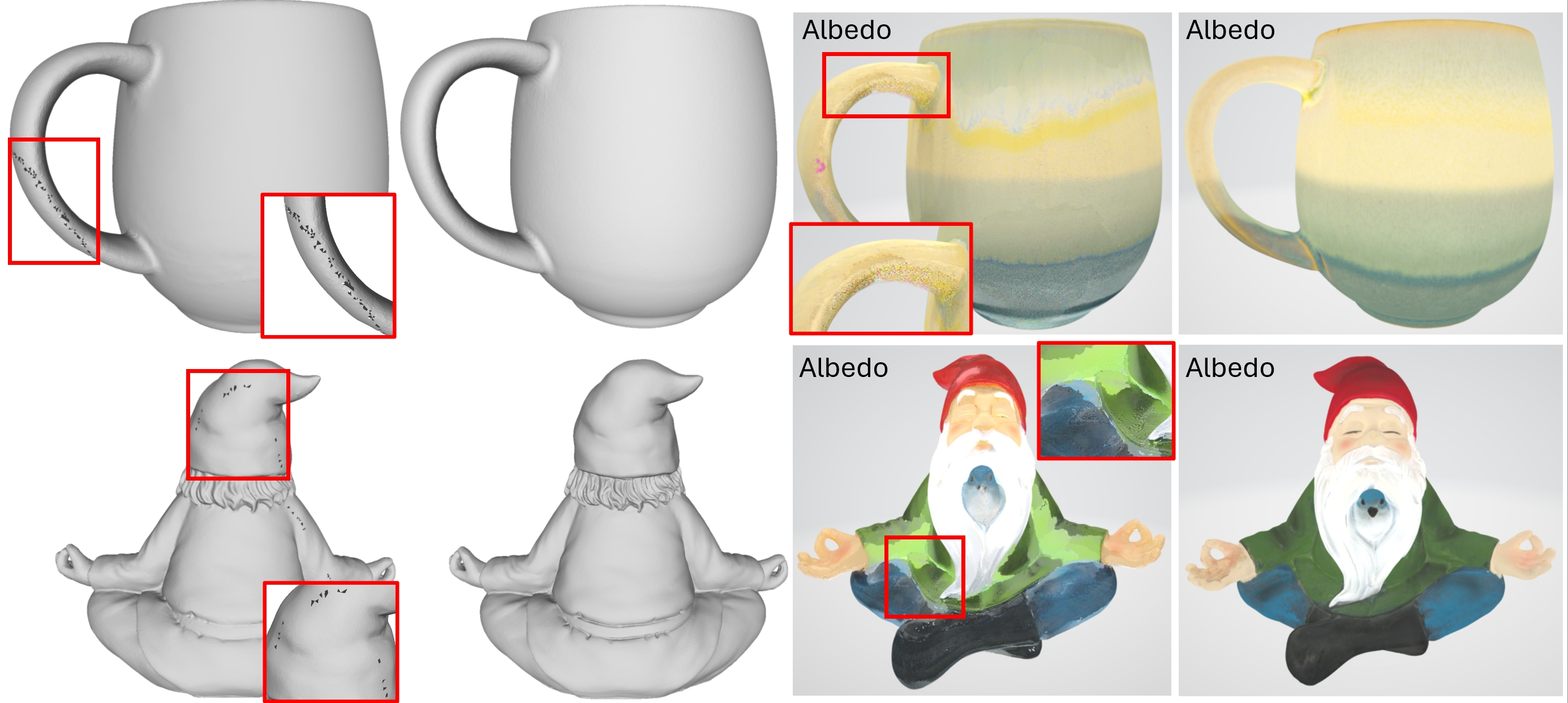}
    \caption{Shape and material (albedo) quality comparison between Stanford-ORB~\cite{kuang2023stanfordorb} (left) and our DTC (right).}
    \label{fig:orb_comparison}
\end{figure}

\paragraph{Comparison Against Stanford-ORB.} We also scanned the objects used in Stanford-ORB~\cite{kuang2023stanfordorb} to compare the shape and appearance quality. As illustrated in Fig.~\ref{fig:orb_comparison}, the Stanford-ORB models exhibit shape artifacts and noisy, lower-quality materials compared to our models.
\vspace{-0.5em}

\subsection{DSLR Evaluation Data}
\label{subsec:dslr}
Within DTC, we include a DSLR-captured evaluation dataset of 50 objects from Sec.~\ref{subsec:3d_models} captured under two different lighting conditions, resulting in 100 distinct image sequences. For every sequence, we provide (a) approximately 120 HDR and LDR images from different viewing directions, (b) one object pose and (c) per-image camera pose. The two lighting conditions are represented using two environment maps.

\begin{figure}
    \centering
    \includegraphics[width=\linewidth]{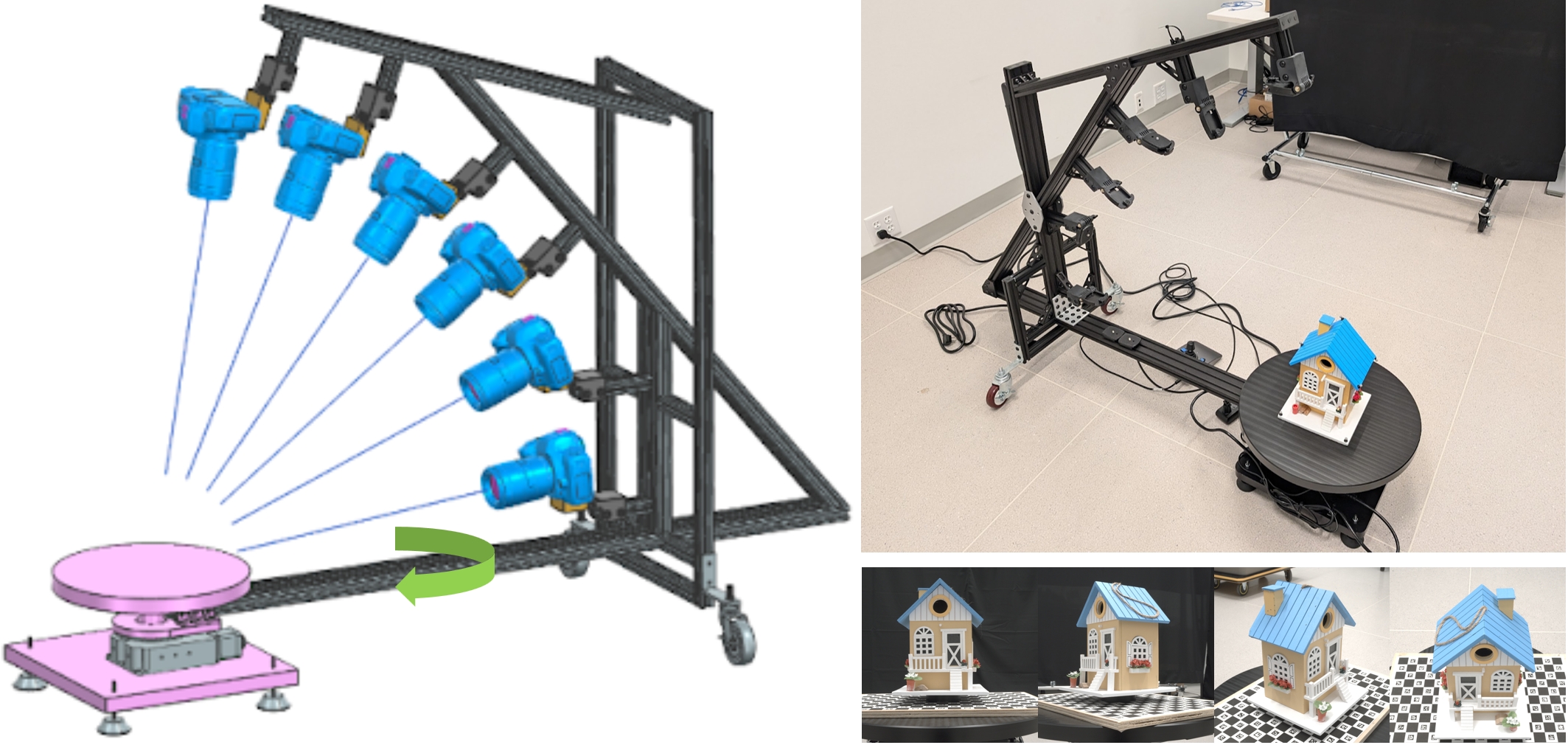}
    \caption{DSLR rig for capturing evaluation data.}
    \label{fig:dslr_rig}
\end{figure}

\paragraph{Data Capture.}
To ensure the DSLR evaluation data quality, we designed and built a DSLR camera rig to automate the capture process (Fig.~\ref{fig:dslr_rig}). The rig is designed to rotate the cameras around the centralized object, assuming the environment lighting remains unchanged during the capture. It features a motorized rotary stage with a centrally mounted stationary platform. Attached to the rotary stage is an extrusion frame that forms the gantry arm, supported by a set of castors to bear its weight and enable smooth rotation around the central axis. The extrusion frame is equipped with adjustable camera mounts, allowing DSLR cameras to be positioned flexibly to optimize the capture setup. For our capture process, we utilized three DSLR cameras to perform a 360-degree rotation around the object, capturing images at 9-degree intervals, resulting in 120 photos per object. To ensure precise camera pose estimation, a ChArUco board was placed beneath the object during the capture. Example images from this setup are shown in Fig.~\ref{fig:dslr_rig}.

\vspace{-0.5em}

\paragraph{Environment Maps.} Following a similar approach in Stanford-ORB~\cite{kuang2023stanfordorb}, we capture the two environment maps using chrome ball images obtained with the capture rig described earlier. With precise camera poses provided by a ChArUco board placed beneath the chrome ball, we first fit a synthetic 3D sphere to the chrome ball by optimizing its 3D position using a geometry-friendly differentiable renderer~\cite{Laine2020diffrast, psdr_jit}. Subsequently, using a differentiable Monte Carlo-based renderer~\cite{Mitsuba3}, we refine the environment map to match the reflection on the chrome ball, employing the single-view light estimation method proposed in~\cite{yu2023accidental}. The coordinate system of the environment map is determined by the ChArUco detection.

\paragraph{Pose Registration for Camera and Object.} We first obtain the initial camera poses using the ChArUco board. In most cases, for images captured from the top and middle views, the pose estimates are typically accurate. However, for bottom-view images, inaccuracies arise due to pattern distortion at grazing angles. To mitigate this issue, we refine the camera poses by fitting the rendering of a virtual ChArUco board to the real captured photo using a differentiable renderer~\cite{Laine2020diffrast, Mitsuba3, psdr_jit}. Once the camera poses have been accurately refined, we optimize the object pose by minimizing the mask loss between the rendered mask and the reference one generated by~\cite{sun2023neural}. By providing separate poses for camera and object anchored by ChArUco, the cameras and object are automatically aligned with the optimized environment maps.

\subsection{Egocentric Evaluation Data}
\label{subsec:aria_eval_data}

\begin{figure*}
    \centering
    \includegraphics[width=0.86\linewidth]{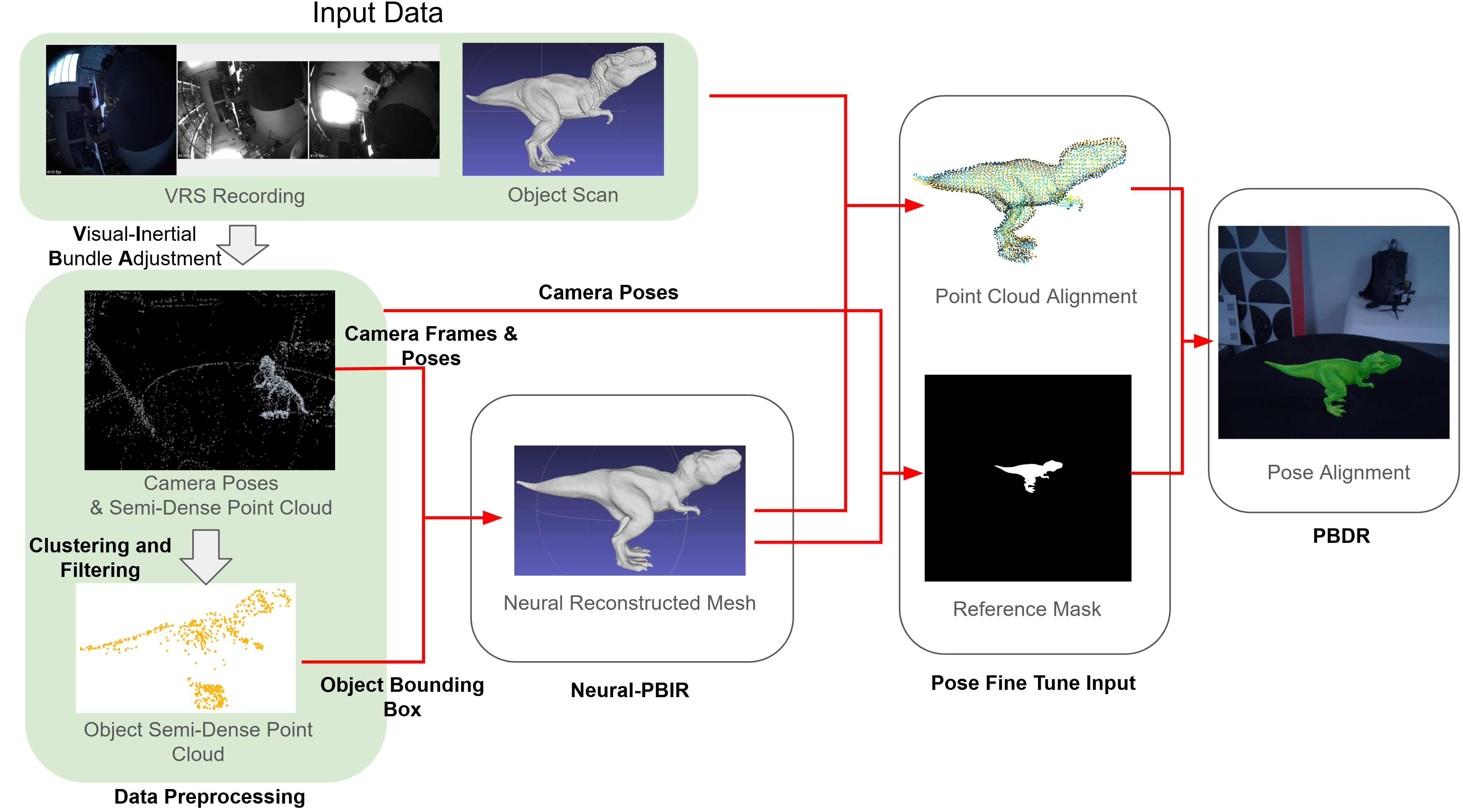}
    \caption{The workflow to align egocentric video and 3D objects. We acquire the object's semi-dense point cloud from the egocentric recording and the neural reconstructed mesh using Neural-PBIR \cite{sun2023neural}. Finally, we align the 3D object with the rendered mask from the neural reconstructed mesh using physics-based differentiable rendering (PBDR).}
    \label{fig:egocentric_alignment}
\end{figure*}
\vspace{-0.5em}

We include an evaluation dataset of real world recordings captured by an egocentric device paired with 100 objects from the \name dataset. We capture the egocentric recordings using the open-sourced Project Aria device~\cite{engel2023aria} and acquire the additional 3D ground truth using its machine perception tool, which includes online device calibration, device trajectory and semi-dense point clouds for each recording. Each recording contains a single 3D object and precisely aligned object poses in the Aria trajectory coordinate frame. Both the object and the trajectory are in metric scale. We can derive additional ground truth for each video from the aligned 3D object properties.

\paragraph{Data Capture.} To feature real world recordings observed from human perspectives that can be representative of 3D object reconstruction for AR/MR devices or robots, we provide two types of recording trajectories for selected objects, termed \emph{active} and \emph{passive} respectively, collected by human wearers. The active recording features a complete 360 view of the objects, which is similar to existing object 3D novel view synthesis dataset. The \emph{passive} recording features casual looks from human wearer, which only contain partial views of the object from certain viewing angles along the trajectory. In dataset creation, we collected the active and passive recordings in the same environment and generate their 3D information in the shared 3D space. This helps reduces potential failures when aligning the object to passive 3D recordings, which are shorter and contains less 3D information. To reduce the effect of noises and motion blur, which are common in egocentric videos in indoor low light environments, we light the capture environment to 3K+ lux illumination and used a fixed low-exposure and gain profile to collect each recording with the appropriate brightness. 

\paragraph{Alignment between Egocentric Video and Object.} We provide an illustration of the object alignment to the egocentric video in Fig.~\ref{fig:egocentric_alignment}. Given the images, camera poses and semi-dense point cloud acquired from Project Aria tools, we employ a neural-based mesh reconstruction method~\cite{sun2023neural} to create reference meshes for generating high-quality reference masks and used that to align with the corresponding 3D object mesh. This alignment step serves as an initialization for a more precise pose refinement, which leverages differentiable rendering. For certain objects with symmetric geometry, we observe this process can introduce ambiguities in point cloud registration and subsequent failures. To address such cases, we provide a GUI to manually align and correct the object alignment. Finally, akin to the DSLR camera pose registration phase, we optimize a mask loss over object poses to achieve fine-tuned pose registration.

\section{Benchmarking and Applications}
\label{sec:benchmark_SOTA}

\begin{table*}[t]
\centering
\scriptsize
\caption{Benchmark comparison of existing methods on inverse rendering for DSLR. Depth SI-MSE and Shape Chamfer distance $\times 10^{-3}$.
}
\resizebox{\linewidth}{!}{
\begin{tabular}{lccccccccccc}
\toprule
 \multirow{2}{*}{}  
 & \multicolumn{3}{c}{Geometry}
 & \multicolumn{4}{c}{Novel Scene Relighting}            & \multicolumn{4}{c}{Novel View Synthesis} 
 \\
 \cmidrule(l){2-4} \cmidrule(l){5-8} \cmidrule(l){9-12}
& Depth$\downarrow$ & Normal$\downarrow$ & Shape$\downarrow$ & PSNR-H$\uparrow$ & PSNR-L$\uparrow$ & SSIM$\uparrow$ & LPIPS$\downarrow$ & PSNR-H$\uparrow$ & PSNR-L$\uparrow$ & SSIM$\uparrow$ & LPIPS$\downarrow$\\\midrule
Neural-PIL~\cite{boss2021neural} & $5.71$ & $0.25$ & $25.02$ & \multicolumn{4}{c}{N/A} & $28.42$ & $35.76$ & $0.882$ & $0.096$\\
PhySG~\cite{physg2021} & $0.31$ & $0.16$ & $11.31$ & $27.28$ & $32.86$ & $0.959$ & $0.049$ & $28.54$ & $34.46$ & $0.964$ & $0.045$\\
NVDiffRec~\cite{Munkberg2022nvdiffrec} & $\mathbf{0.02}$ & $0.07$ & $1.64$ & $26.99$ & $33.27$ & $0.951$ & $\mathbf{0.037}$ & $28.95$ & $34.92$ & $0.967$ & $\mathbf{0.029}$\\
NeRD~\cite{boss2021nerd} & $4.55$ & $0.45$ & $108.20$ & $26.10$ & $32.60$ & $0.948$ & $0.061$ & $26.80$ & $33.40$ & $0.882$ & $0.102$\\
InvRender~\cite{wu2023nefii} & $0.22$ & $\mathbf{0.03}$ & $\mathbf{0.75}$ & $\mathbf{29.52}$ & $\mathbf{35.98}$ & $\mathbf{0.961}$ & $\mathbf{0.037}$ & $\mathbf{31.64}$ & $37.82$ & $0.970$ & $0.033$\\
NVDiffRecMC~\cite{hasselgren2022nvdiffrecmc} & $\mathbf{0.02}$ & $0.06$ & $1.34$ & $27.78$ & $34.55$ & $0.952$ & $0.042$ & $31.27$ & $\mathbf{38.17}$ & $\mathbf{0.972}$ & $0.032$\\

\bottomrule
\end{tabular}
}
\label{tab:benchmark}
\end{table*}
\vspace{-0.5em}

We first use our DLSR and egocentric dataset as a benchmark for existing state-of-the-art methods. For inverse rendering, we design metrics to evaluate the shape and material quality of the recovered 3D object digital twin from three perspectives. For egocentric recording, we evaluate the novel-view synthesis as the initial evaluation. We include additional tasks, e.g. sparse view reconstruction in the supplementary materials for both DSLR and egocentric recordings. Finally we can demonstrate our high quality 3D digital twin models can be beneficial to robotics domain using an application in robotics manipulation. 

\subsection{Application to Inverse Rendering for DSLR}
\label{subsec:inverse_rendering_dslr}

The DSLR dataset in \name provides accurate ground truth, including poses, lighting and 3D models, for inverse rendering tasks and serves as an evaluation suite to benchmark the performance of inverse rendering methods. We select six prior methods for this task and evaluate their performance using the ground truth provided by our dataset. In the following sections, we describe the data splitting strategy, evaluation metrics, and baselines. 

\paragraph{Data Splitting.} For benchmarking purposes, we select 15 objects from the DSLR dataset captured under two distinct lighting environments, resulting in a total of 30 image sequences. The selected objects encompass a diverse range of geometric and material properties to ensure a comprehensive evaluation. For each scene, 8 views are selected for testing, while the remaining views are reserved for training.

\paragraph{Evaluation Metrics.}
The metrics measure the accuracy of three aspects of baseline performance: geometry estimation , relighting, and novel view synthesis. 
For geometry estimation, we evaluate the accuracy of predicted depth and normal maps under held-out test views, as well as 3D meshes extracted from baseline methods, compared with the ground truth from our dataset. 
Relighting metrics evaluate the material decomposition quality of baselines by measuring the accuracy of predicted images under held-out lighting conditions. 
For view synthesis, we compare the predicted images from viewpoints unseen during training to ground truth captures. We refer to \citet{kuang2023stanfordorb} for metric details. 

\paragraph{Baselines.}
We include the following baselines: NVDiffRec~\citep{Munkberg2022nvdiffrec} and NVDiffRecMC~\citep{hasselgren2022nvdiffrecmc}, with a hybrid shape representation DMTet~\citep{shen2021deep}; InvRender~\citep{zhang2022modeling} and PhySG~\citep{physg2021}, which adopt signed distance functions (SDFs) to represent object geometry~\citep{yariv2020multiview} and utilize implicit neural fields for material decomposition; Neural-PIL~\citep{boss2021neural} and NeRD~\citep{boss2021nerd}, which use NeRFs~\citep{mildenhall2020nerf} as scene representations. 

\subsection{Application to Egocentric Reconstruction}
\label{subsec:egocentric_nvs}

\begin{table}[t]
\centering
\small
\caption{Benchmark on the egocentric aligned recordings.}
\resizebox{\linewidth}{!}{
\begin{tabular}{cccccc}
\toprule
      & PSNR $\uparrow$ & LPIPS $\downarrow$ & SSIM $\uparrow$ & Depth $\downarrow$ & Normal $\downarrow$  \\ \midrule
3D-GS \citep{kerbl3Dgaussians} & 28.81 & 0.020 & 0.9888 & 0.1768 & 0.3301\\
2D-GS \cite{huang20242Dgs} & 28.75 & 0.020 & 0.9886 & 0.1755 & 0.2112 \\
\bottomrule
\end{tabular}
}
\label{tab:aria_nvs}
\end{table}
\vspace{-0.5em}

Our digital twin models, aligned with real world video using the method described in Sec.~\ref{subsec:aria_eval_data}, can help obtain accurate ground truth for object-centric images that were previously difficult to acquire. We provide the first evaluation of object-centric novel view synthesis recorded from an egocentric device. We use the projected object shape given the 3D pose of the object and cameras in scene coordinates to acquire the image masks, depth and normal for each object. We selected 15 recordings from the egocentric recording sessions as the evaluation and used the \emph{active} recordings to benchmark novel view reconstruction. For each recording, we hold out every 8th image view as a testing view. 

\paragraph{Evaluations.} We build our baselines based on the gsplats~\cite{ye2024gsplat} implementation of the 3D Gaussian Splatting (GS) \citep{kerbl3Dgaussians} and 2D GS \citep{huang20242Dgs}, and handle the effect of lens shading from the Project Aria lens \cite{gu2024egolifter}. We calculate PSNR, depth and normal based on the observed objects with masks and provide SSIM and LPIPS score on images by masking out the non-object areas as black. Table~\ref{tab:aria_nvs} shows the benchmark results of the baselines. We use the same depth and normal metric in DSLR evaluation. We provide additional qualitative evaluations and analyses on egocentric data towards sparser view settings in the supplementary materials.

\subsection{Application to Robotic Manipulation}
High-quality object models have been leveraged in prior work to train real-world robotic agents in scenes represented explicitly~\cite{robocasa2024} or implicitly~\cite{torne2024rialto}. These object models have also been shown to facilitate object-centric pose and lighting parameter estimation, enabling model-based planning~\cite{tian2023multiobject}. 
In this section, we empirically evaluate the effectiveness of using \name dataset objects in training robotic policies. Specifically, we consider learning robotic pushing and grasping skills in simulation.

First, we sample a subset of $24$ \texttt{cup} category objects from the \name dataset and $24$ \texttt{cup} objects from Objaverse-XL~\citep{deitke2024objaverse}~\footnote{The version of Objaverse-XL used in this work excludes all 3D models sourced from Sketchfab. Further, no Polycam assets were obtained from the Polycam source site}.
Since not all Objaverse-XL objects come with textures, we randomize the colors of those objects uniformly in RGB space. To compute collision meshes for physical simulation, we perform convex decomposition on each object with CoACD~\citep{wei2022coacd}.
We import these objects along with a UR5e robot equipped with a Robotiq 2F-85 (pushing) or Robotiq 2F-140 (grasping) gripper into the PyBullet simulator~\citep{coumans2019} and collect data for each robotic task as described below. After training policies on data from each object set, we evaluate policy performance on a relatively high-quality unseen test object from the StanfordORB dataset~\citep{kuang2023stanfordorb}.

\paragraph{Pushing.} For the pushing task, we collect $5000$ trajectories of pseudo-random robotic interaction data for each object set. For each trajectory, a single object from the considered object set and its initial position are randomly selected. Then we train a goal-conditioned neural network policy  
$\pi(a | o, o_g)$ where $o$ is an image observation of the current scene and $o_g$ is a goal image indicating the desired final object and robot position, $o, o_g \in \mathbb{R}^{256 \times 256 \times 3}$, and the action $a \in \mathbb{R}^2$ represents a change in the robot's end-effector position in the $x$ and $y$ axes. The $z$ axis end-effector height is held fixed. We sample goals for training via hindsight relabeling~\citep{andrychowicz2017hindsight,ding2019goal,ghosh2019learning}.
We then perform evaluation on $100$ randomly sampled test goals manipulating an unseen test cup from StanfordORB~\cite{kuang2023stanfordorb}.

\paragraph{Grasping.} For grasping, we collect $5000$ successful grasp examples for each object set by first placing a single object into the scene, randomizing the object identity and initial position. We then randomly sample candidate grasp poses in a radius around the object's position and simulating their outcomes, rejecting unsuccessful grasps. 
We train a grasping policy $\pi(a | o)$ where $o \in \mathbb{R}^{256 \times 256 \times 3}$ is an image observation of the scene and $a \in \mathbb{R}^4$ represents the $x, y, z$ position and $\theta$ yaw rotation of the robot end-effector at which to attempt the grasp. 
Again we use $100$ test \texttt{cup} object poses. 

\begin{figure}
    \centering
    \includegraphics[width=\linewidth]{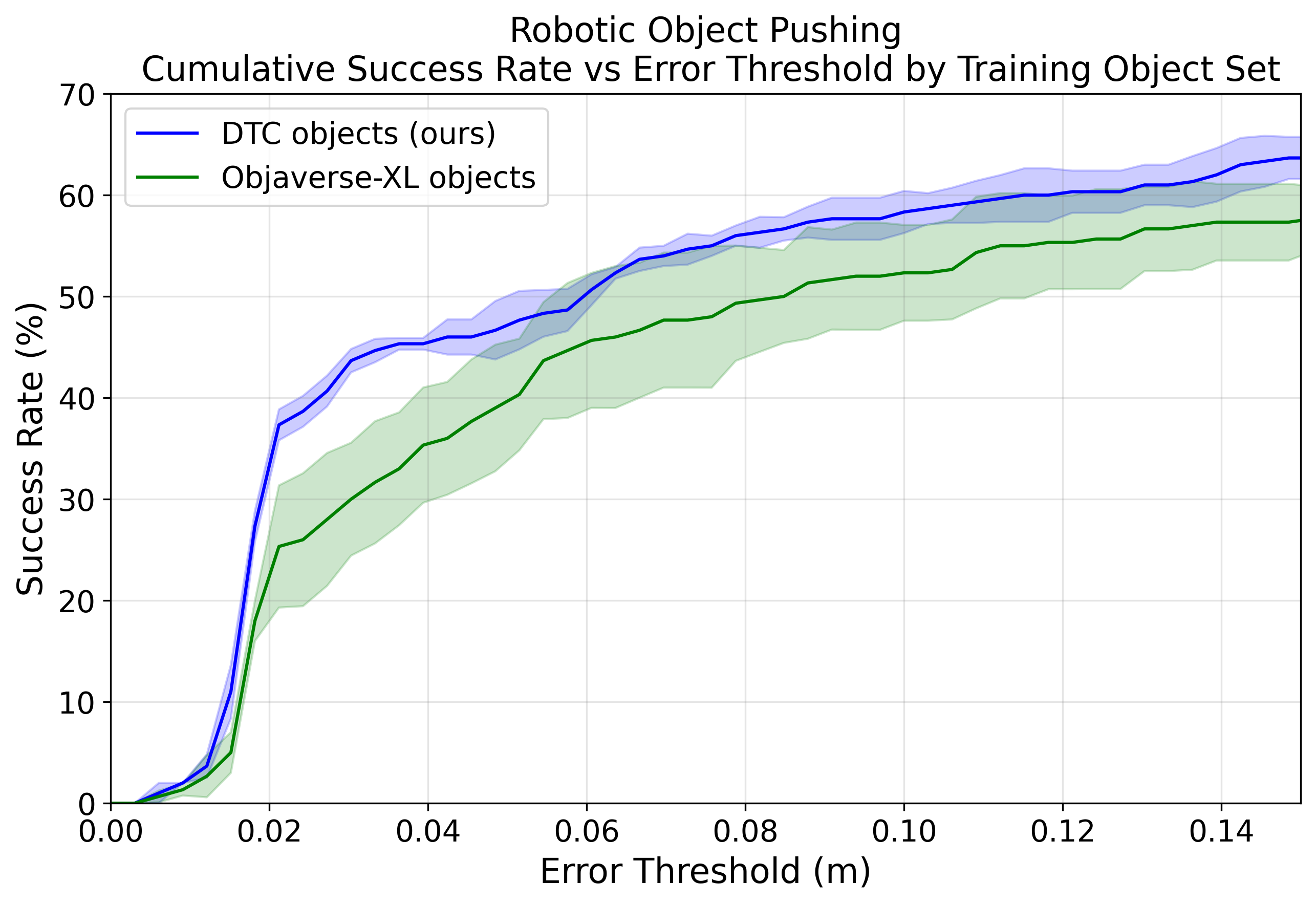}
    \caption{Success rates on robotic pushing task when training with our DTC objects and sampled Objaverse-XL objects. Particularly at lower error thresholds, policies trained on DTC objects outperform those trained using Objaverse-XL objects. Shaded bars represent sample standard deviations over policy training random seeds.}
    \label{fig:detailed_robotic_pushing_results}
    \vspace{-1.5em}
\end{figure}

\paragraph{Results.} We report the results in Table~\ref{tab:robotics_results} and Fig.~\ref{fig:detailed_robotic_pushing_results}. We find that across both pushing and grasping tasks, policies trained on \name dataset objects outperform those trained on Objaverse-XL objects when evaluated on the unseen test object. For pushing, we report performance by defining binary success thresholds based on the final Euclidean distance of the object position to the goal position. Training on \name objects appears to be especially helpful at enabling policies to make finer adjustments, improving pushing success rates at stricter thresholds. 
Additional experimental details can be found in the supplementary.

\begin{table}[]
\small
    \centering
    \begin{tabular}{lcc}
    \toprule
         Task & \name (ours) & Objaverse-XL~\cite{deitke2024objaverse} \\
         \midrule
         Pushing @ $2$cm & $36.3\% \pm 1.5\%$ & $25.3\% \pm 6.0\%$ \\
         Pushing @ $3$cm & $43.7\% \pm 1.2\%$ & $29.7\% \pm 6.0\%$ \\
         Pushing @ $5$cm & $47.0\% \pm 2.6\%$ & $40.3\% \pm 5.5\%$ \\
         Grasping & $42.7\% \pm 4.7\%$ & $38.6\% \pm 11.0\%$ \\ 
    \bottomrule
    \end{tabular}
    \caption{Success rate of policies trained on data collected using objects from our \name dataset and sampled from Objaverse-XL when evaluated on an unseen test object. Errors indicate sample standard deviation over  three policy training seeds.}
    \label{tab:robotics_results}
\end{table}

\section{Conclusion}

We presented a new large scale photorealistic 3D digital twin dataset with the real world recordings that contain its real world counterpart. We provide extensive evaluations of baselines on our DTC dataset serving as new benchmark for inverse rendering and novel view synthesis task. We also demonstrated that high quality digital twin models can be beneficial to applications in robotics domain. We believe our efforts can empower the research community to build and leverage digital twin models for future applications. 

\paragraph{Limitations.} Achieving high quality digital twin models currently requires deliberate hardware setup and human efforts in refinement. Solving this challenge without sacrificing quality can significantly further enhance the volume of digital twin models. Our hardware is also limited to objects within a certain size and can not yet recover objects that are deformable, highly specular, or transparent. 

\paragraph{Future work.} The existing digital twin model creation in \name dataset involves lengthy post-processing and may require subjective human refinement, hindering the automation of model generation. However, recent advancements in physics-based differentiable rendering hold promise for enabling faster and more accurate creation of digital twins, especially for material reconstruction. Furthermore, building large-scale digital twins for applications will necessitate efforts to enhance the diversity in object appearance (e.g., transparent objects) and to capture additional attributes, such as physical properties and functionalities.

For robotics applications, while deploying manipulation policies learned in simulation to the real world remains generally challenging, we hope that the high-quality digital twin data provided by DTC can serve as a stepping stone towards effective sim-to-real transfer.

\paragraph{Acknowledgments.}
This work is in part supported by NSF CCRI \#2120095, RI \#2211258, and RI \#2338203. YZ is in part supported by the Stanford Interdisciplinary Graduate Fellowship.

{
    \small
    \bibliographystyle{ieee_fullname}
    \bibliography{main}
}

\clearpage
\appendix

\clearpage
\setcounter{page}{1}
\maketitlesupplementary

We provide the following content in the supplementary materials: 
\begin{itemize}
    \item We provide a comprehensive dataset statistics of DTC dataset in Sec.~\ref{sec:DTC_statistics}, including the full category of DTC objects, the list of objects used in DLSR and egocentric recording respectively. We also provided visualizations for examples of DTC objects, DSLR and egocentric data. 
    \item We include complementary details of benchmark in Sec.~\ref{sec:complementary_benchmarks}. First, as indicated by the main paper, we include a sparse view setting of benchmark. We further provide the baseline comparisons and analysis for DSLR benchmark and egocentric benchmark. We use the same baselines and dataset split as described in the main paper. 
    \item We provide details of the simulation and experiments for our robotics experiments in Sec.~\ref{appendix:robotic_experiments_details}.
\end{itemize}

\section{DTC Dataset Statistics \& Details}
\label{sec:DTC_statistics}
\begin{figure*}[t!]
    \centering
    \includegraphics[width=\linewidth]{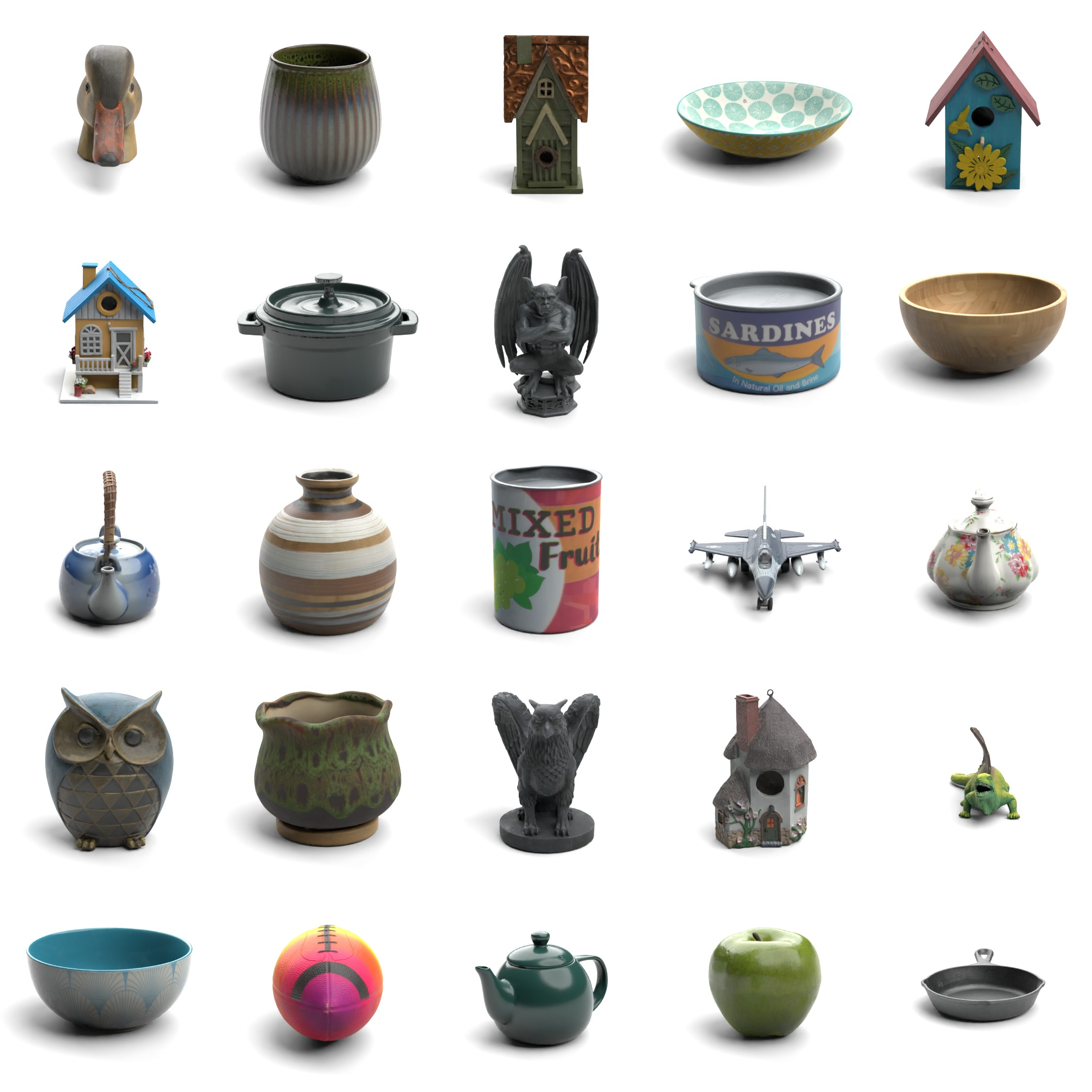}
    \caption{Examples of DTC 3D models. The PBR Materials of each object are presented in the following figures from Fig.~\ref{fig:more_PBR_maps_row1} to Fig.~\ref{fig:more_PBR_maps_row5}.}
    \label{fig:DTC_More_Models}
\end{figure*}

\begin{table}[htbp]
\centering
\resizebox{\columnwidth}{!}{%
\begin{tabular}{|l|c|c|c|c|}
\hline
\textbf{Category}        & \textbf{\# Models} & \textbf{avg \# vert / std} & \textbf{min / max} & \textbf{LVIS Category} \\ \hline
airplane & 15 & 129,153 / 15,535 & 80,719 / 144,587 & airplane
\\ \hline
axe & 3 & 133,588 / 1,922 & 132,096 / 136,301 & ax
\\ \hline
basketball & 55 & 132,260 / 3,013 & 128,652 / 140,985 & basketball
\\ \hline
birdhouse & 102 & 147,577 / 11,057 & 98,405 / 182,594 & birdhouse
\\ \hline
bowl & 106 & 127,456 / 1,019 & 125,883 / 130,964 & bowl
\\ \hline
building blocks & 129 & 96,592 / 46,377 & 28,181 / 301,475 & toy
\\ \hline
calculator & 38 & 131,244 / 3,638 & 126,379 / 139,970 & calculator
\\ \hline
candle & 31 & 108,221 / 58,406 & 15,897 / 309,601 & candle
\\ \hline
candle holder & 1 & 132,285 / 0 & 132,285 / 132,285 & candle\_holder
\\ \hline
cast iron & 31 & 165,954 / 127,424 & 128,074 / 711,536 & pan
\\ \hline
cup & 58 & 137,154 / 43,605 & 126,346 / 417,343 & cup
\\ \hline
cutting board & 16 & 133,796 / 11,244 & 127,543 / 173,141 & chopping\_board
\\ \hline
dino & 103 & 82,403 / 29,477 & 13,796 / 151,148 & animal
\\ \hline
dish & 51 & 129,732 / 4,130 & 127,014 / 147,451 & dish
\\ \hline
dumbbell & 39 & 156,915 / 165,650 & 126,593 / 1,177,907 & dumbbell
\\ \hline
fake food can & 79 & 117,829 / 22,772 & 78,646 / 270,271 & can
\\ \hline
fakefruit & 96 & 124,037 / 9,724 & 100,833 / 151,870 & fruit
\\ \hline
figurine & 77 & 57,152 / 47,753 & 15,482 / 140,812 & figurine
\\ \hline
football & 48 & 132,323 / 2,953 & 126,719 / 138,459 & football
\\ \hline
gargoyle & 50 & 137,921 / 5,019 & 130,206 / 151,325 & gargoyle
\\ \hline
gravestone & 24 & 85,245 / 55,760 & 10,750 / 150,966 & gravestone
\\ \hline
hammer & 33 & 133,413 / 36,272 & 46,808 / 320,845 & hammer
\\ \hline
hardcover\_book & 17 & 174,884 / 102,168 & 130,267 / 500,735 & hardback\_book
\\ \hline
key & 2 & 60,456 / 41,100 & 19,357 / 101,556 & key
\\ \hline
keyboard & 25 & 146,684 / 12,414 & 129,360 / 174,488 & computer\_keyboard
\\ \hline
knife & 10 & 126,359 / 8,118 & 102,569 / 133,250 & knife
\\ \hline
mallard (fake duck) & 48 & 99,728 / 45,926 & 9,000 / 131,058 & mallard
\\ \hline
marker & 54 & 93,394 / 39,460 & 34,796 / 306,272 & marker
\\ \hline
miscellaneous & 44 & 149,799 / 168,694 & 27,772 / 1,208,696 & NA
\\ \hline
mouse & 52 & 144,184 / 46,099 & 121,879 / 304,533 & mouse
\\ \hline
pistol & 2 & 133,587 / 91 & 133,496 / 133,678 & pistol
\\ \hline
pottery & 46 & 143,284 / 162,783 & 12,354 / 1,206,223 & pottery
\\ \hline
remote & 2 & 42,993 / 17,068 & 25,925 / 60,061 & remote\_control
\\ \hline
shampoo & 45 & 143,639 / 75,261 & 126,285 / 604,491 & shampoo
\\ \hline
shaver & 20 & 142,231 / 57,736 & 38,246 / 306,105 & shaver
\\ \hline
shoes & 121 & 139,363 / 4,171 & 130,604 / 148,465 & shoe
\\ \hline
speaker & 40 & 201,013 / 152,410 & 127,154 / 852,714 & speaker
\\ \hline
spoon & 34 & 107,649 / 56,459 & 29,712 / 302,920 & spoon
\\ \hline
teapot & 99 & 145,149 / 100,792 & 85,289 / 926,273 & teapot
\\ \hline
vase & 101 & 142,927 / 142,946 & 61,488 / 1,568,710 & vase
\\ \hline
volleyball & 52 & 133,921 / 4,056 & 129,653 / 144,770 & volleyball
\\ \hline
\end{tabular}
}
\caption{Categories of DTC Scanned Objects. We include the number of models per object (\#Models), the average number of vertices per object categories (avg \# vert) and its standard deviation (std), the minimum and maximum number of vertices within the object category (min/max), and the label name in LVIS category taxonomy.}
\label{tab:object_categories}
\end{table}

In this section, we present the category list of DTC 3D objects (Sec.~\ref{subsec:scanned_object_categories}) and showcase additional examples of DTC digital-twin-quality 3D objects (Fig.~\ref{fig:DTC_More_Models}). We also provide the list of selected objects used for creating the DSLR evaluation dataset (Sec.~\ref{subsec:model_list_DSLR}) and the egocentric evaluation dataset (Sec.~\ref{subsec:model_list_egocentric}). Furthermore, we include additional examples of real-world captures and recordings to better illustrate the DSLR (Sec.~\ref{subsec:example_dslr_data}) and Egocentric (Sec.~\ref{subsec:example_ego_data}) evaluation datasets. 

\subsection{Categories of DTC Scanned Objects}
\label{subsec:scanned_object_categories}
In DTC dataset, we selected and scanned 2,000 physical-world objects across 40 carefully curated categories from the taxonomy of LVIS~\cite{gupta2019lvis}. These categories were chosen to ensure a diverse representation of common daily objects while remaining compatible with the scanner’s capabilities. Table~\ref{tab:object_categories} provides an overview of these categories, including the number of models, average vertex count, minimum/maximum vertex numbers for each category and corresponding category label in LVIS taxonomy.

\begin{figure*}[t]
    \centering
    \includegraphics[width=0.19\linewidth]{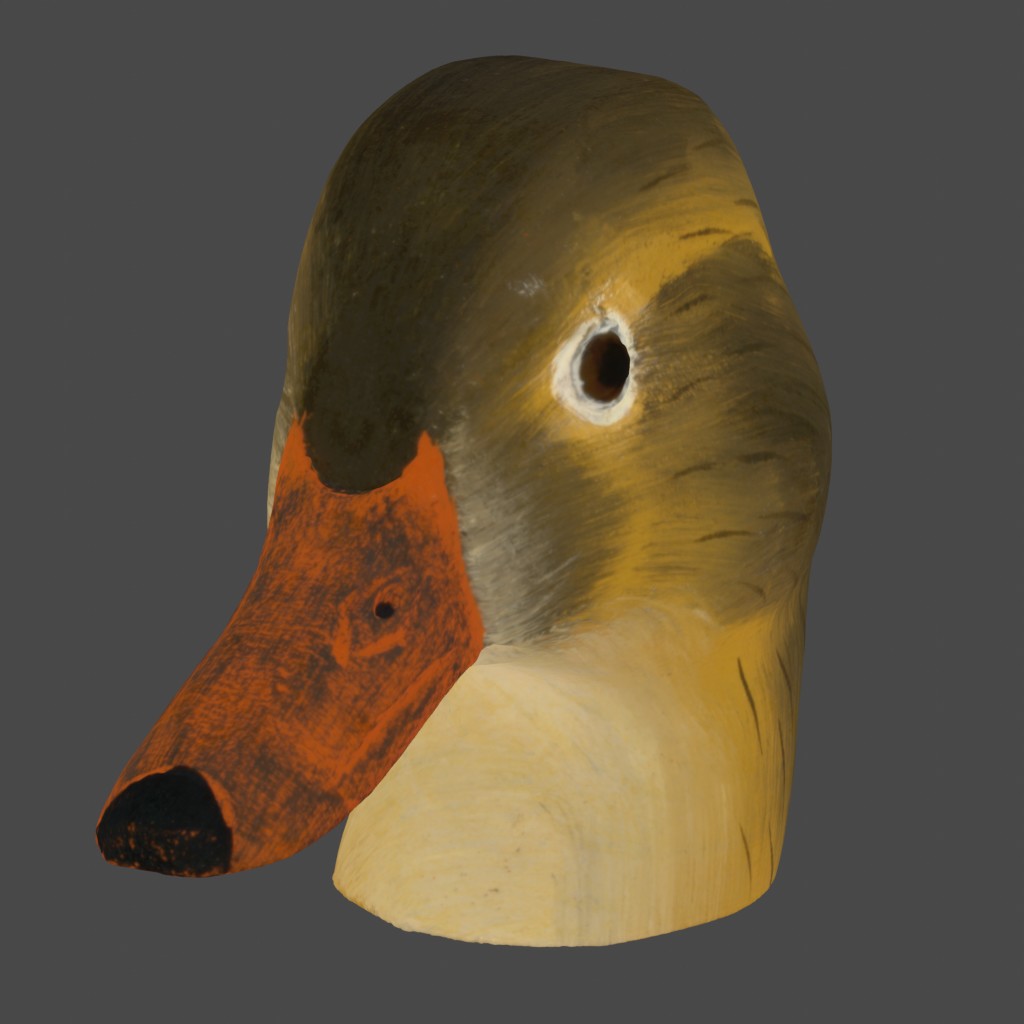}
    \includegraphics[width=0.19\linewidth]{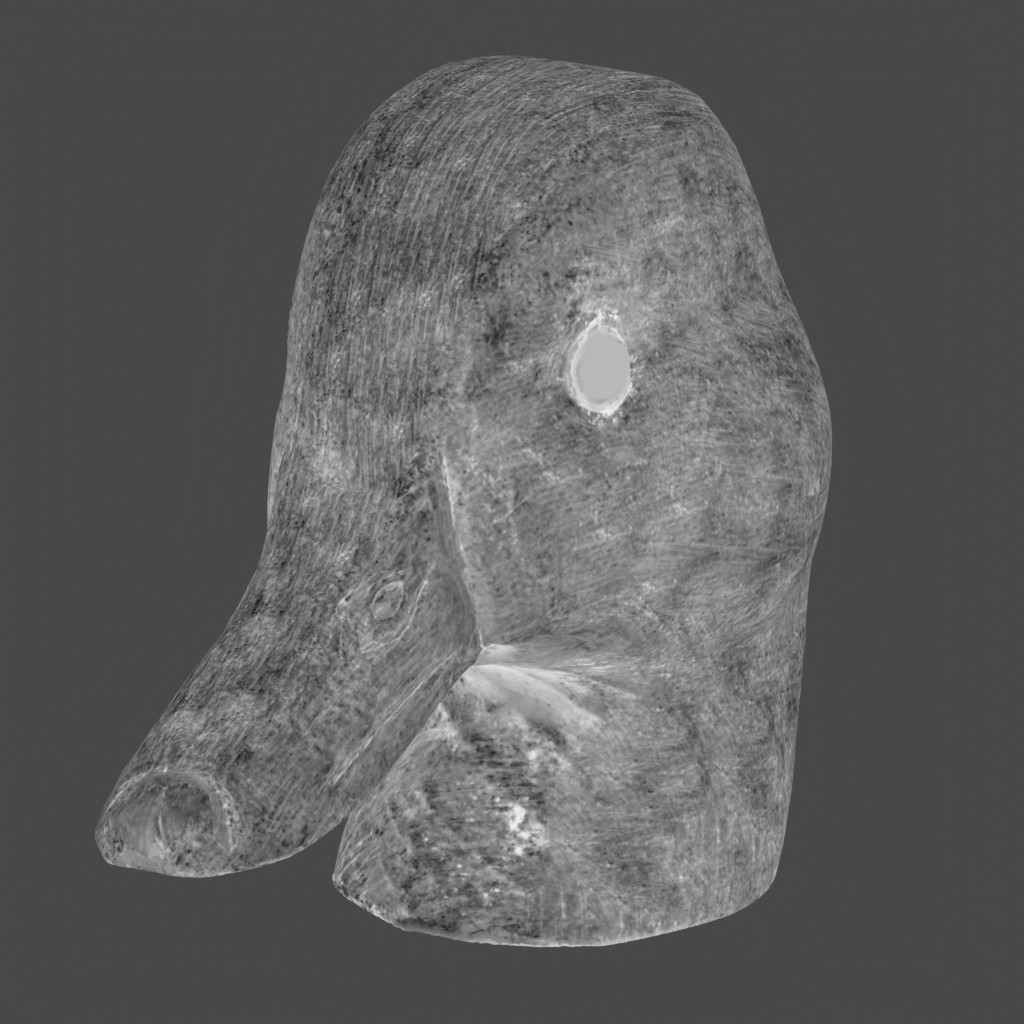}
    \includegraphics[width=0.19\linewidth]{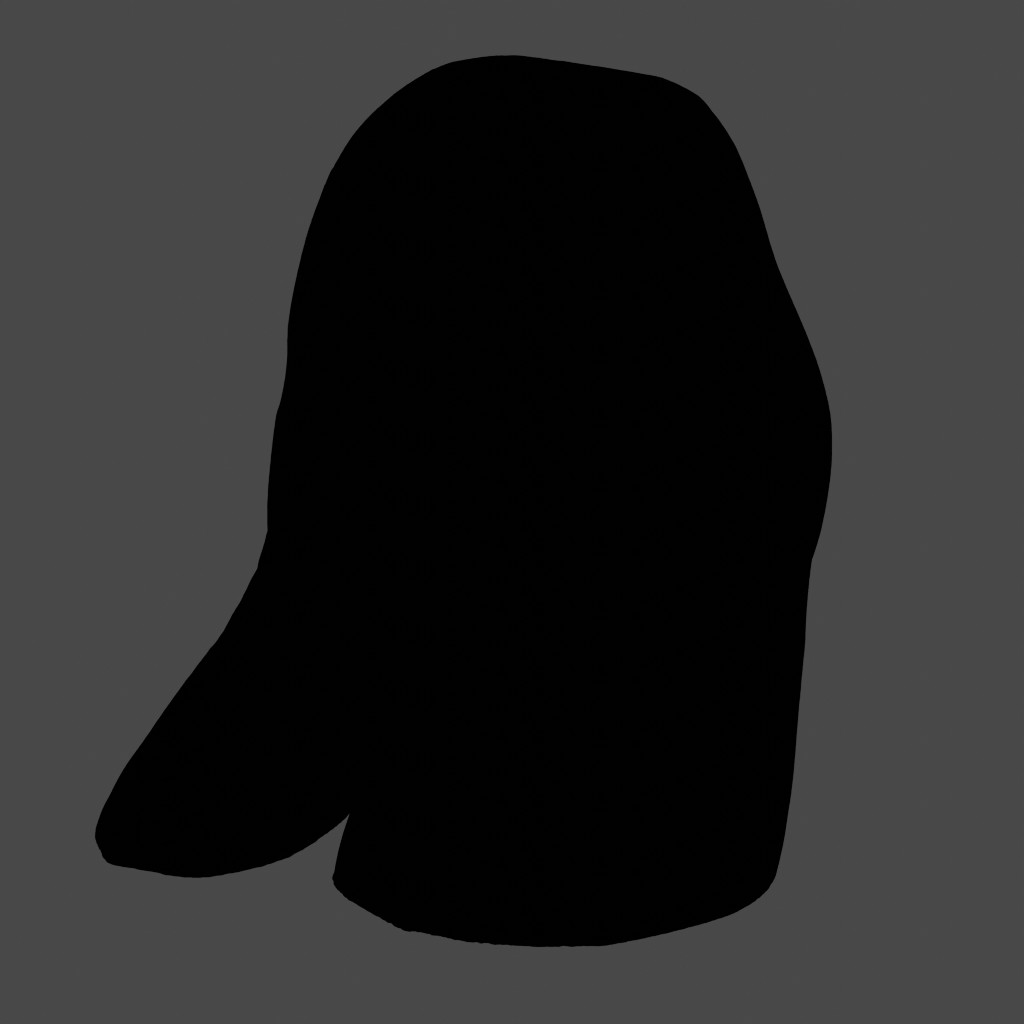}
    \includegraphics[width=0.19\linewidth]{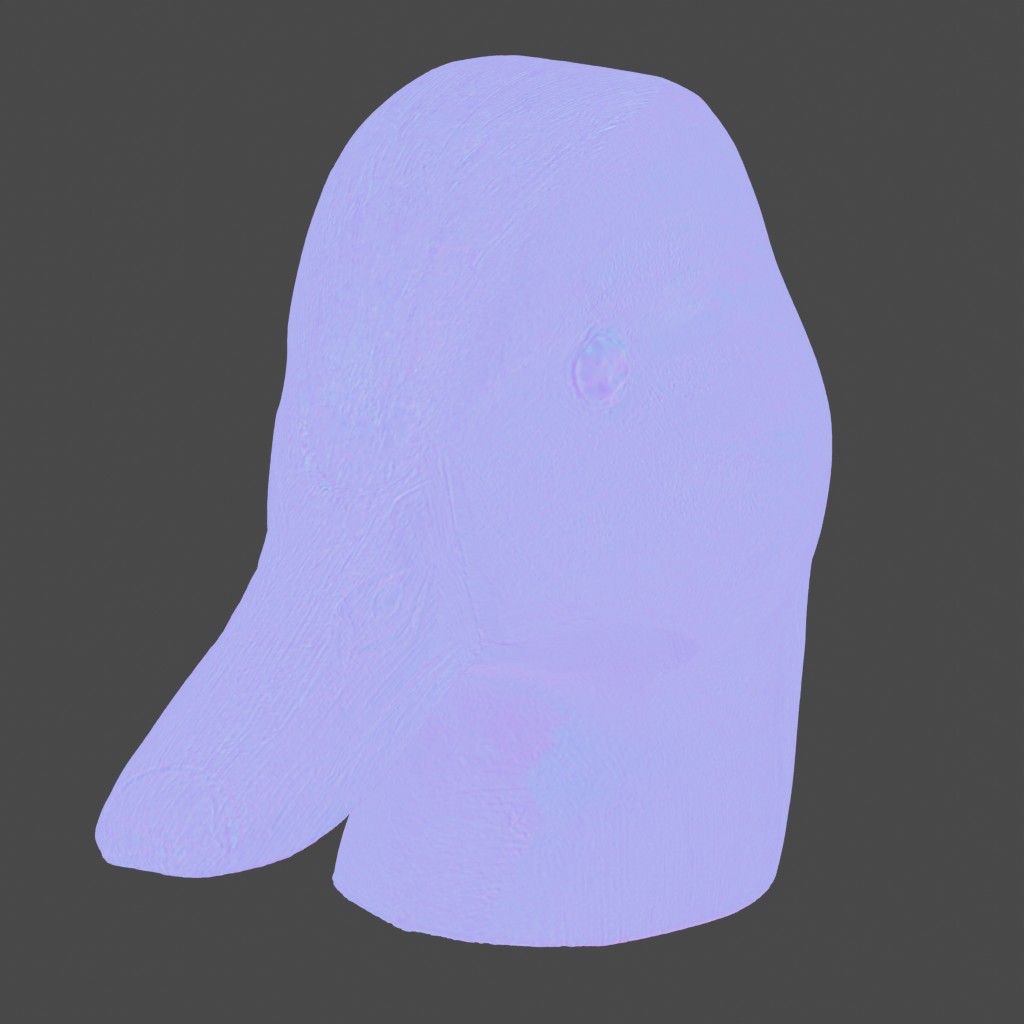}
    \includegraphics[width=0.19\linewidth]{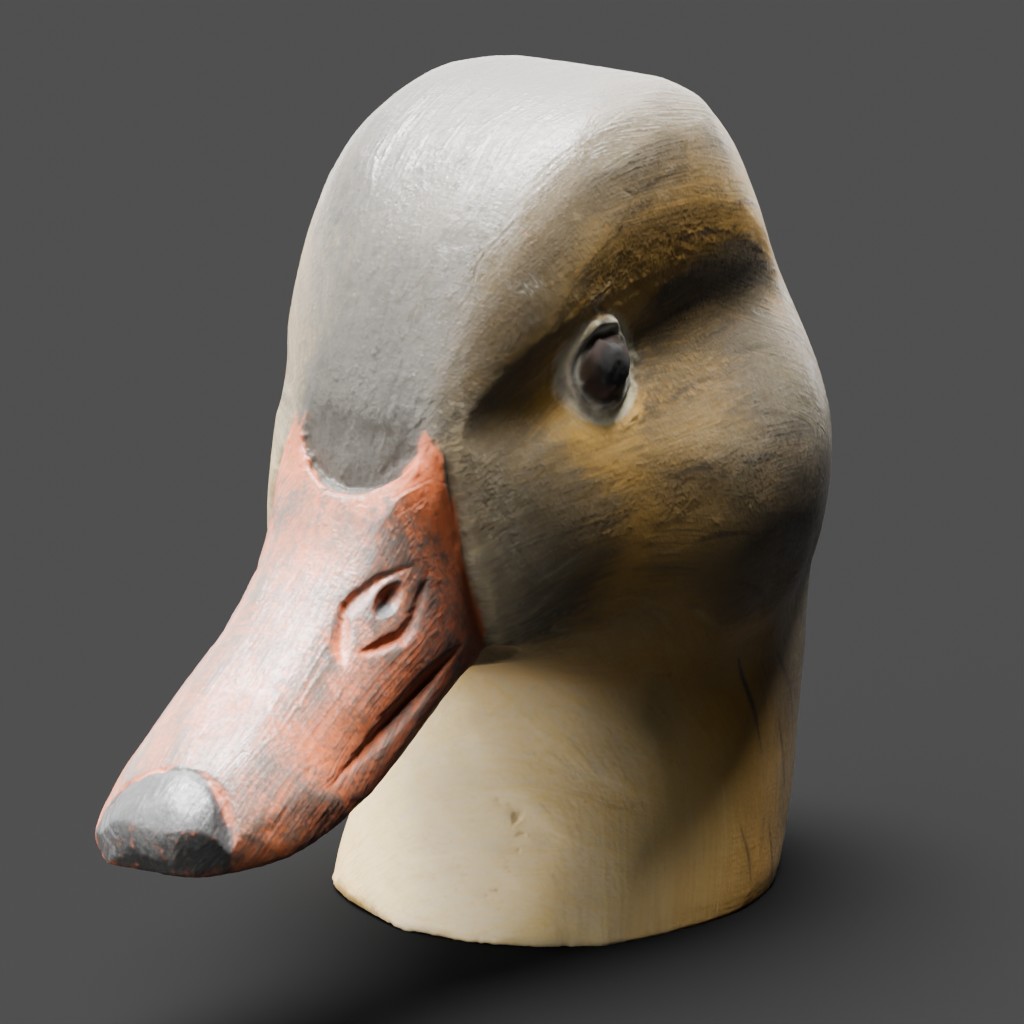} \\
    \includegraphics[width=0.19\linewidth]{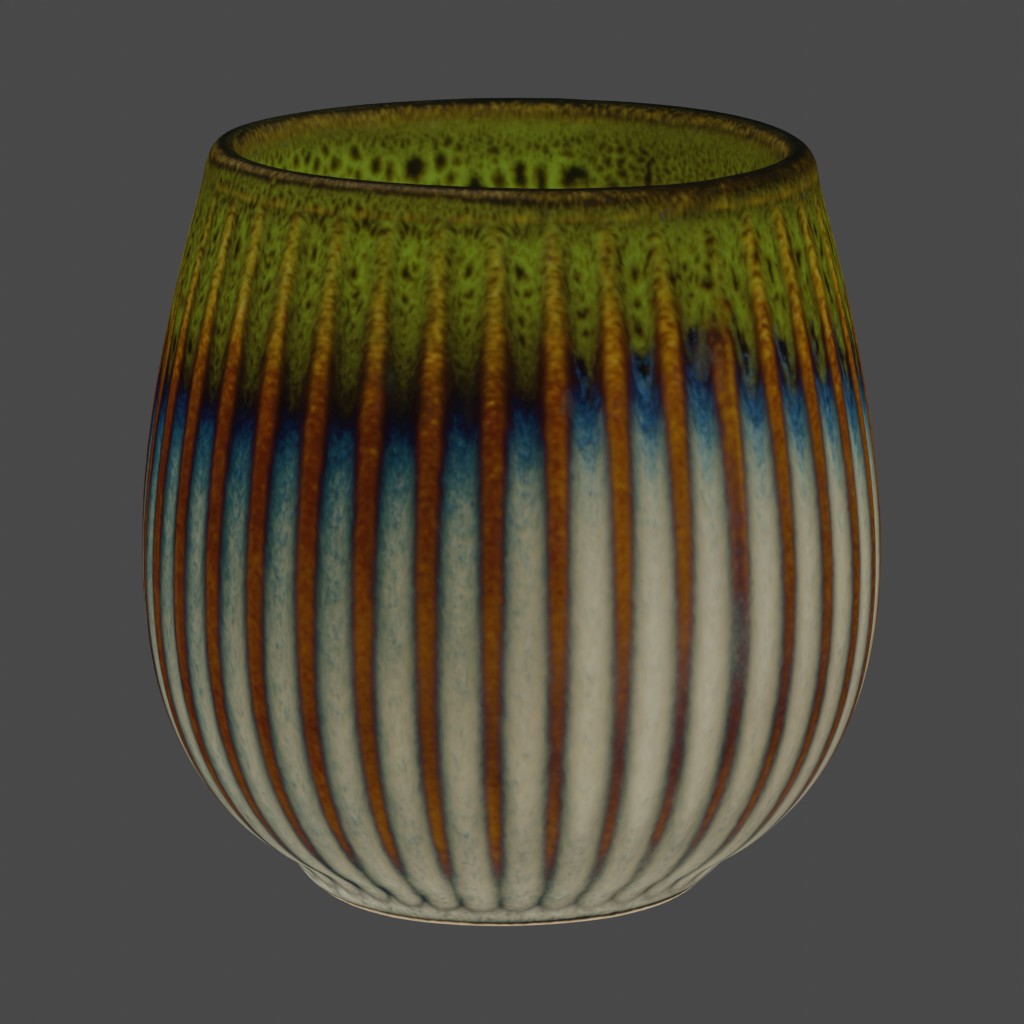}
    \includegraphics[width=0.19\linewidth]{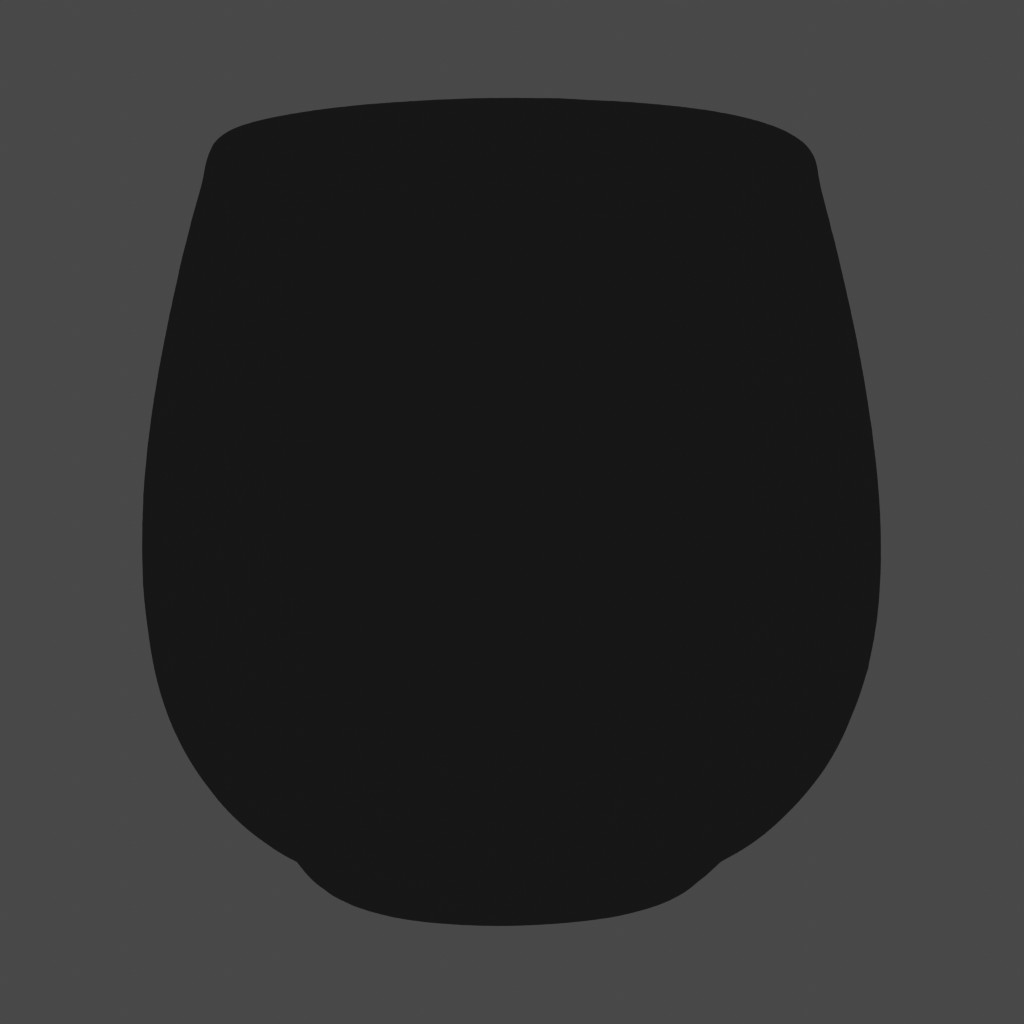}
    \includegraphics[width=0.19\linewidth]{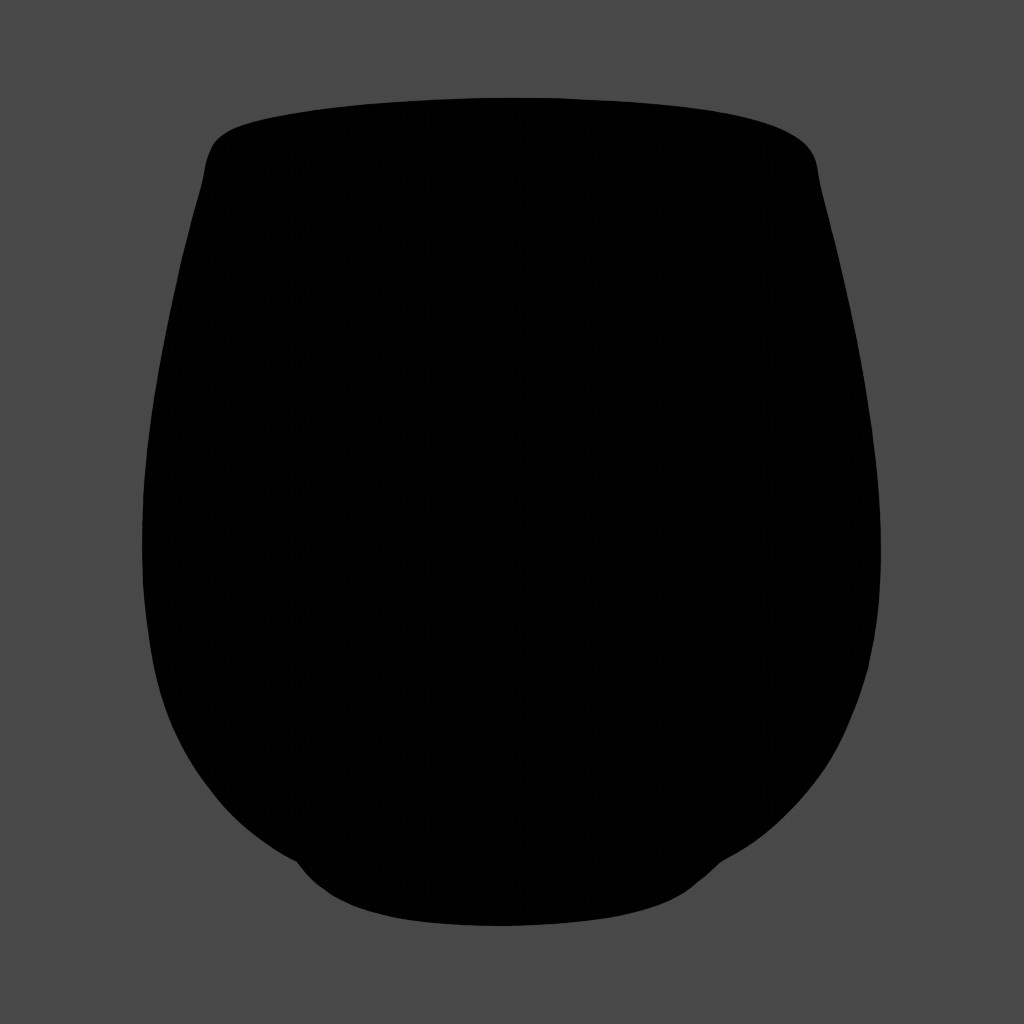}
    \includegraphics[width=0.19\linewidth]{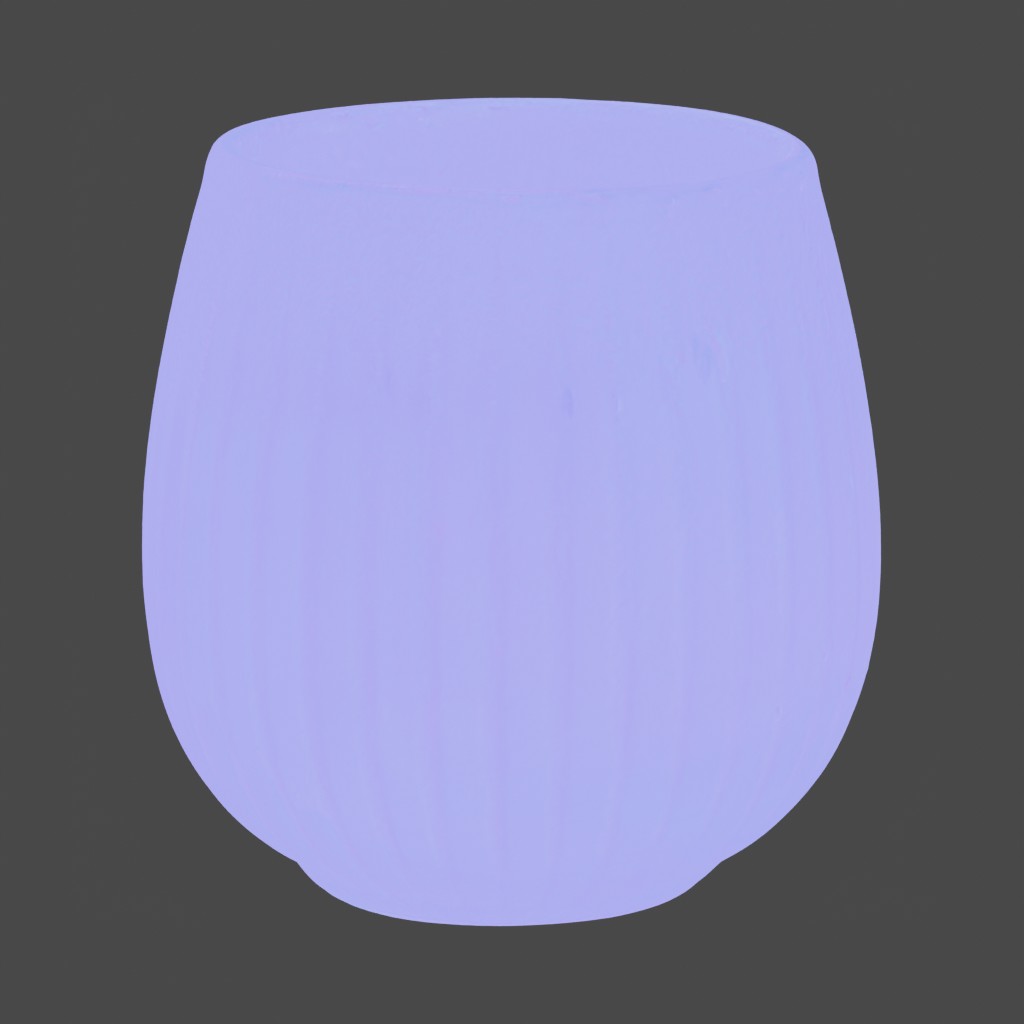}
    \includegraphics[width=0.19\linewidth]{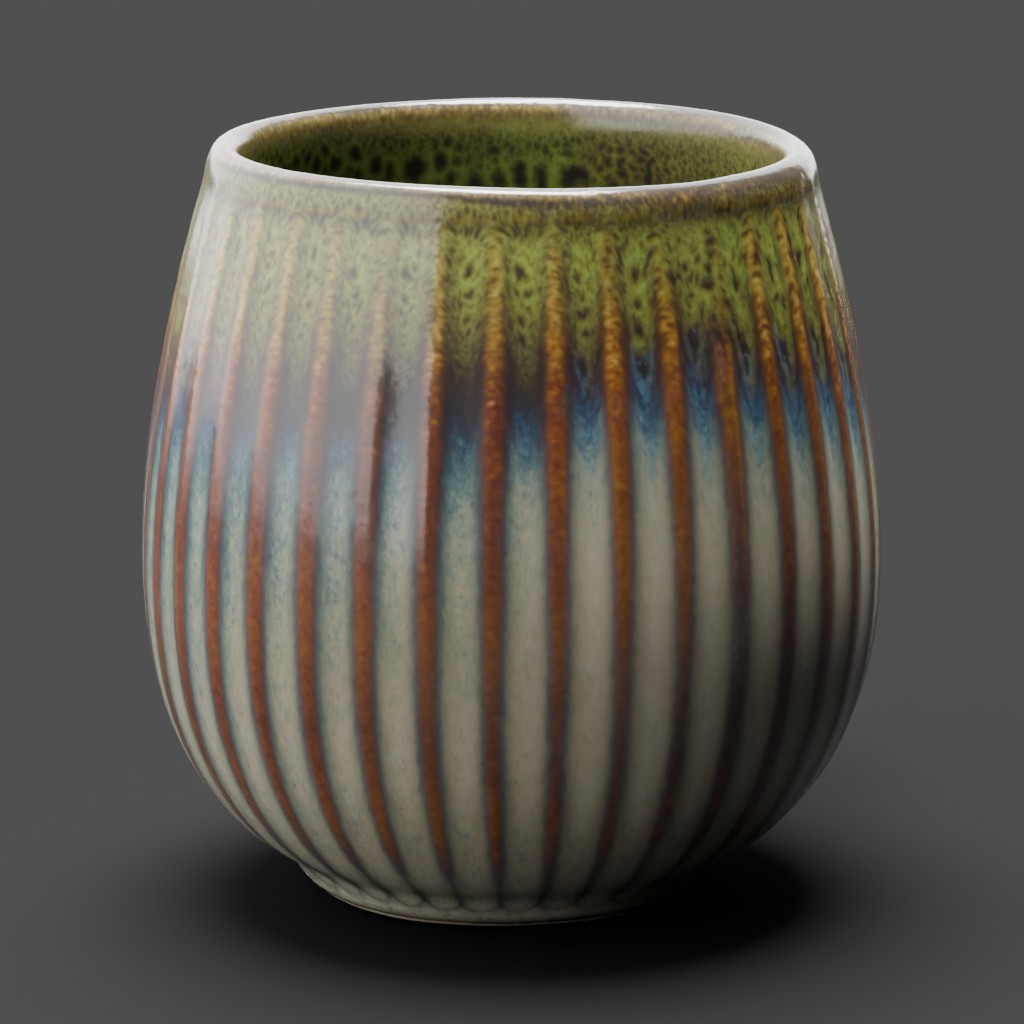} \\
    \includegraphics[width=0.19\linewidth]{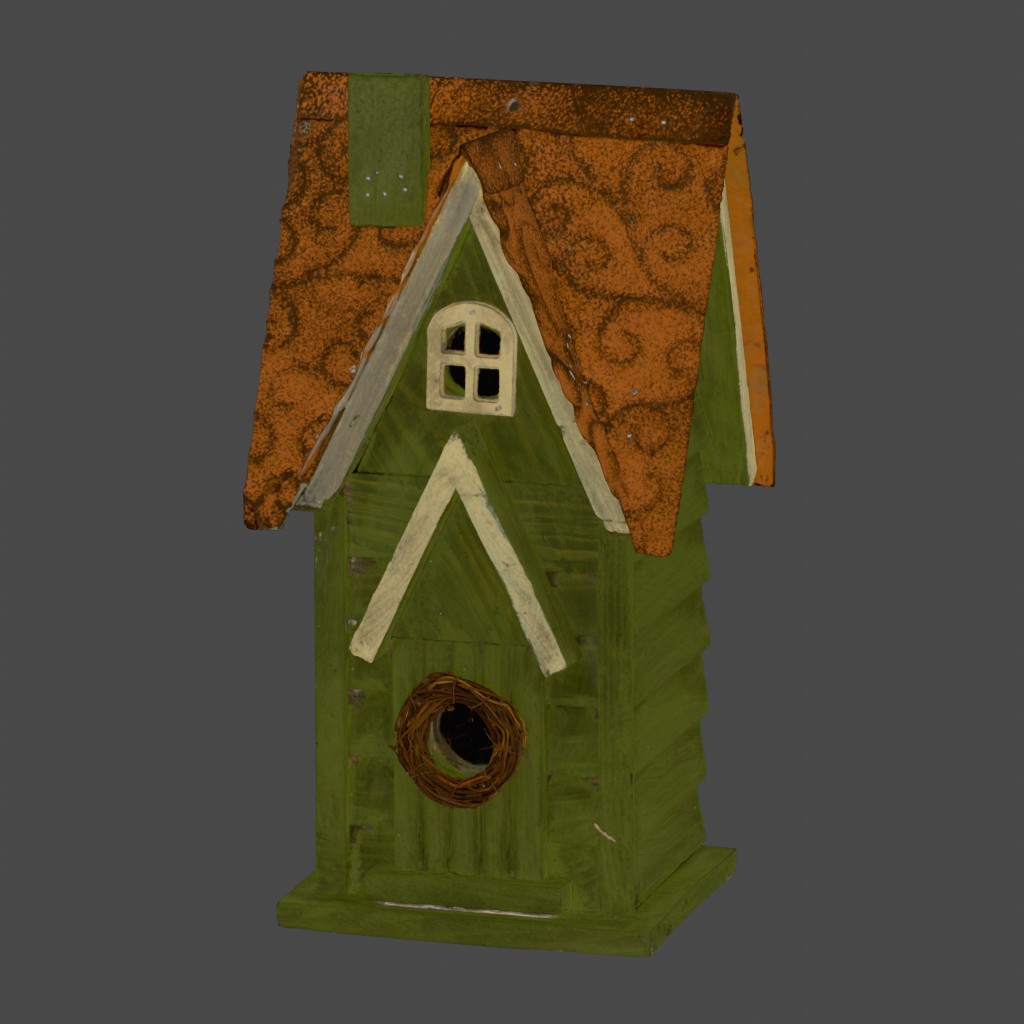}
    \includegraphics[width=0.19\linewidth]{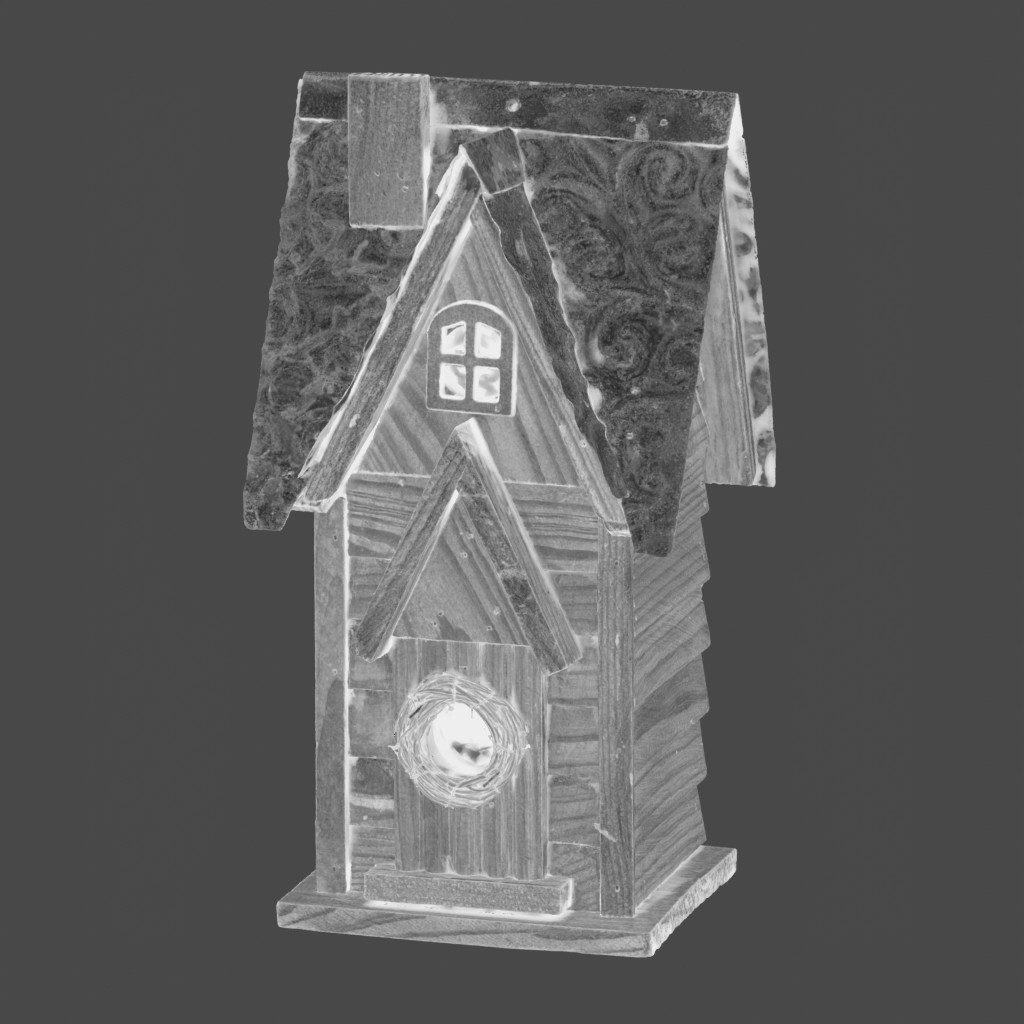}
    \includegraphics[width=0.19\linewidth]{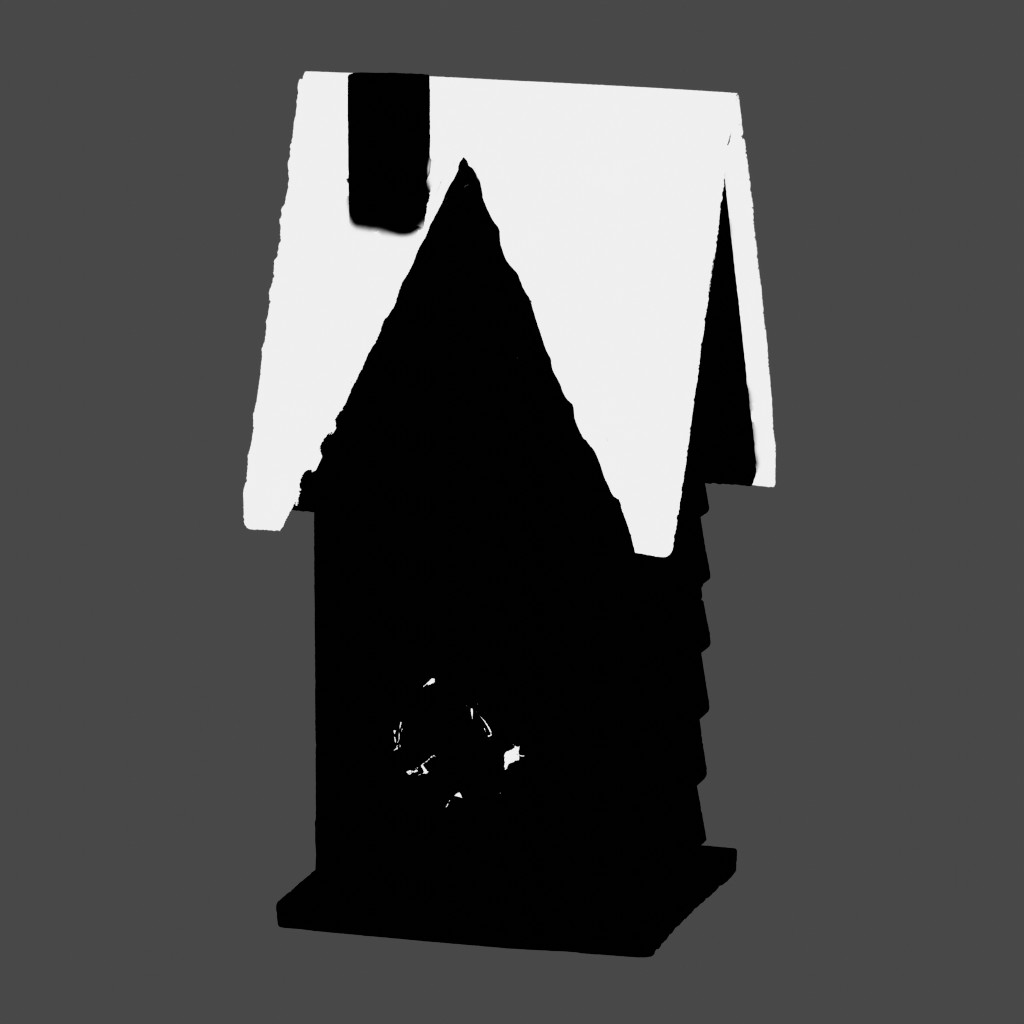}
    \includegraphics[width=0.19\linewidth]{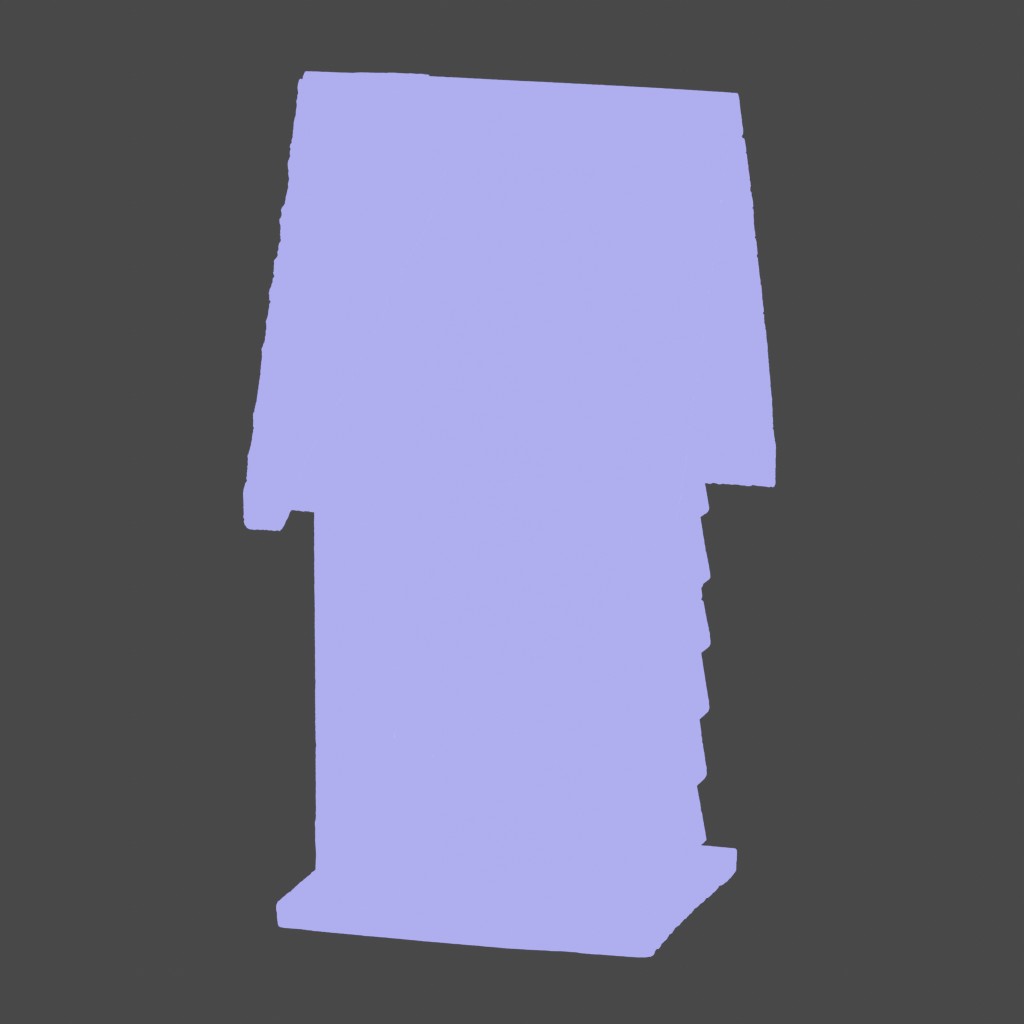}
    \includegraphics[width=0.19\linewidth]{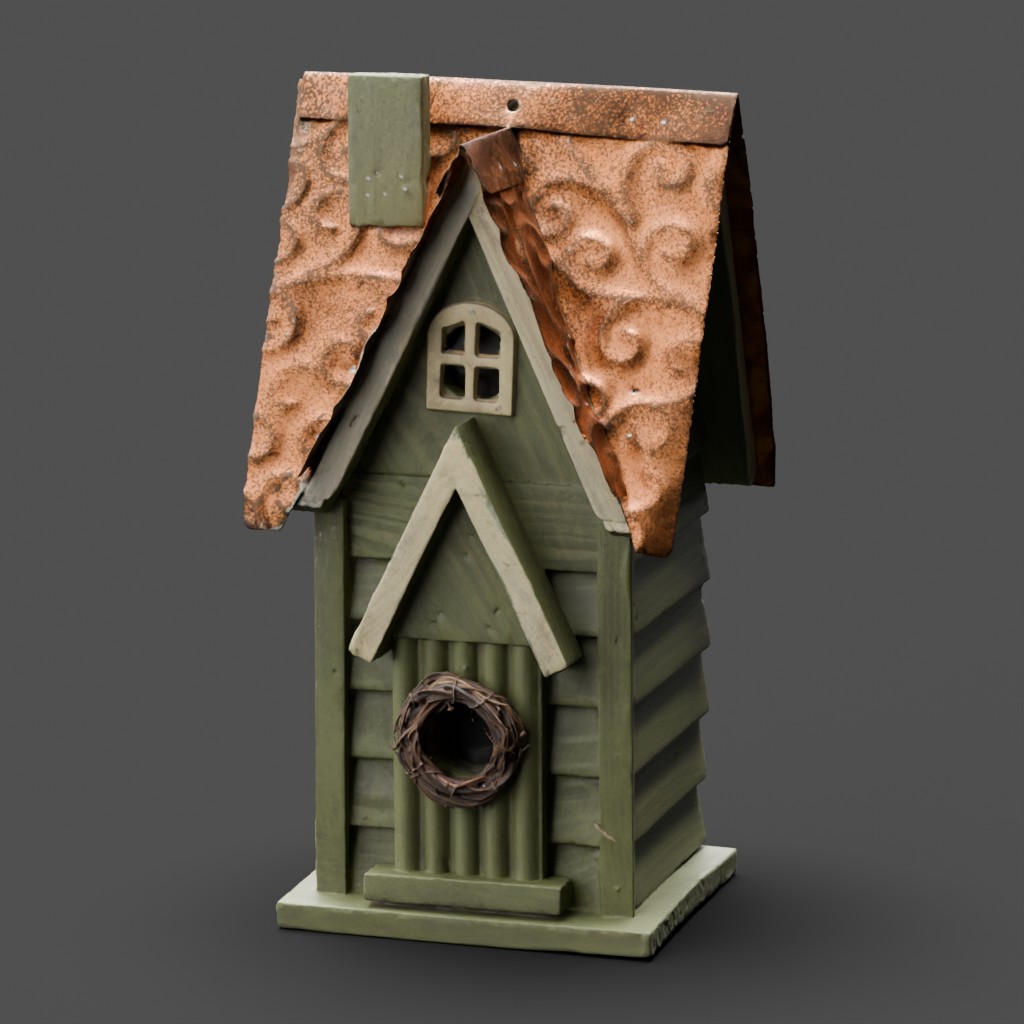} \\
    \includegraphics[width=0.19\linewidth]{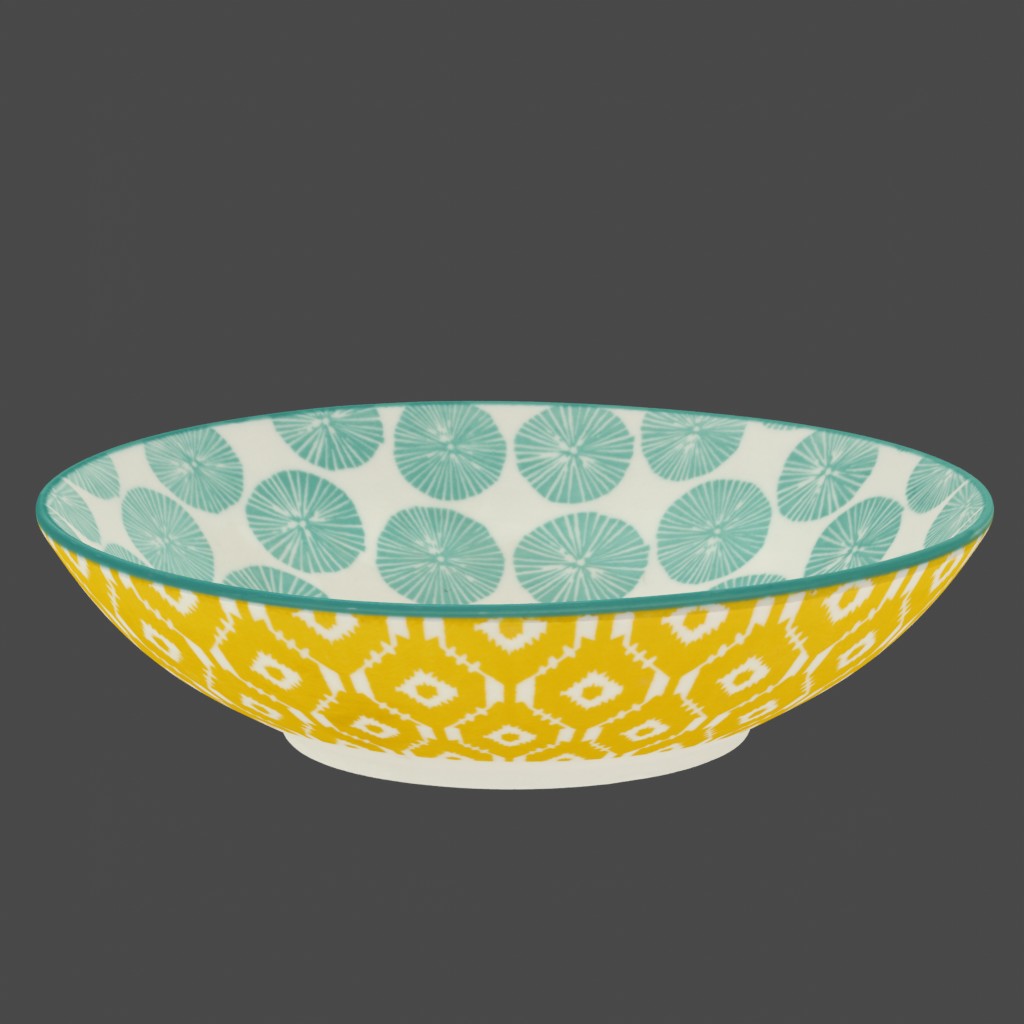}
    \includegraphics[width=0.19\linewidth]{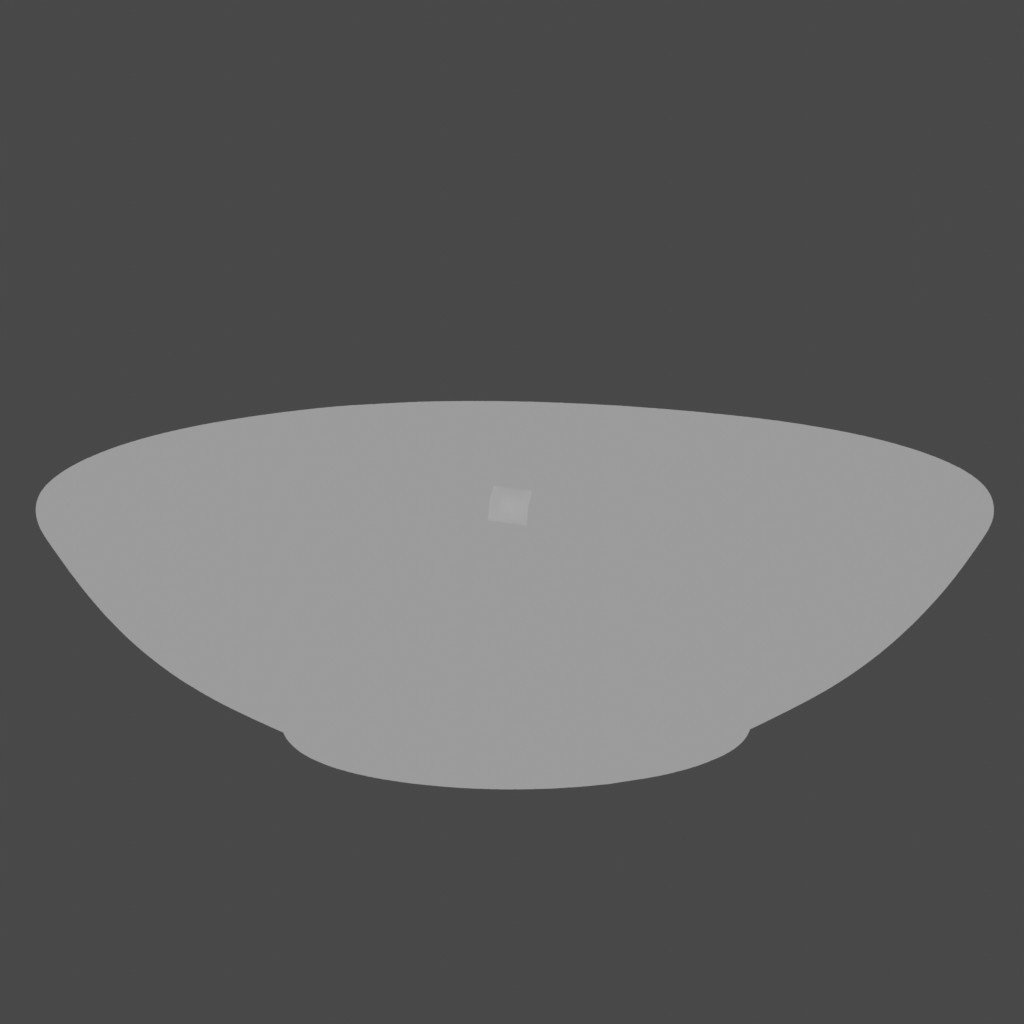}
    \includegraphics[width=0.19\linewidth]{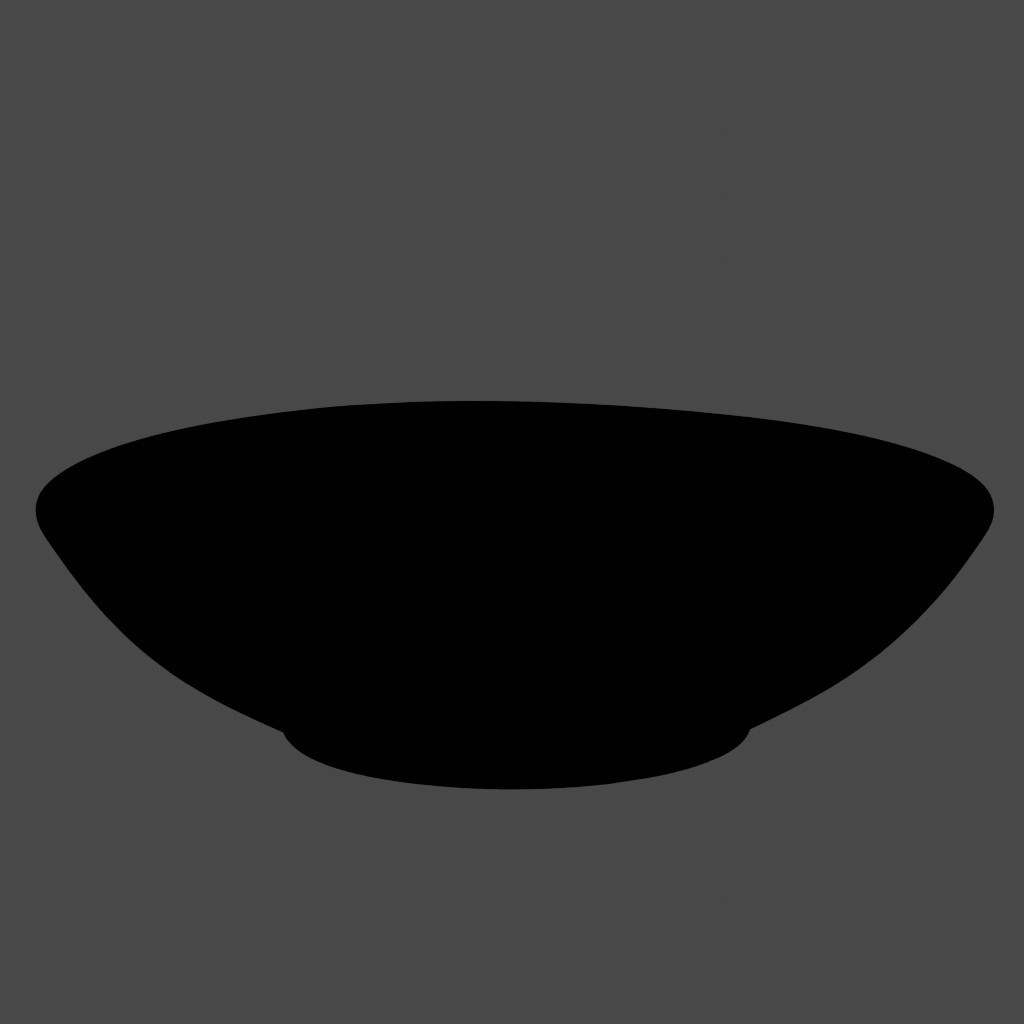}
    \includegraphics[width=0.19\linewidth]{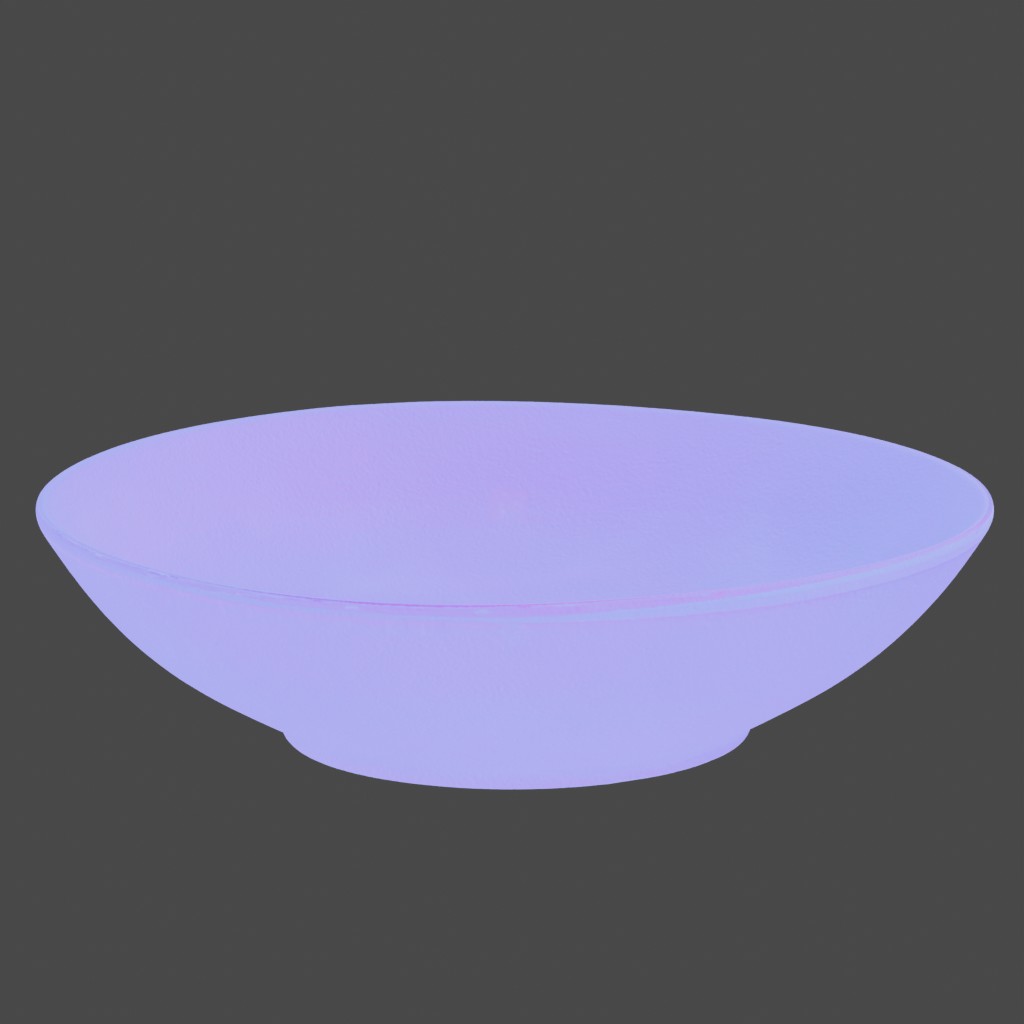}
    \includegraphics[width=0.19\linewidth]{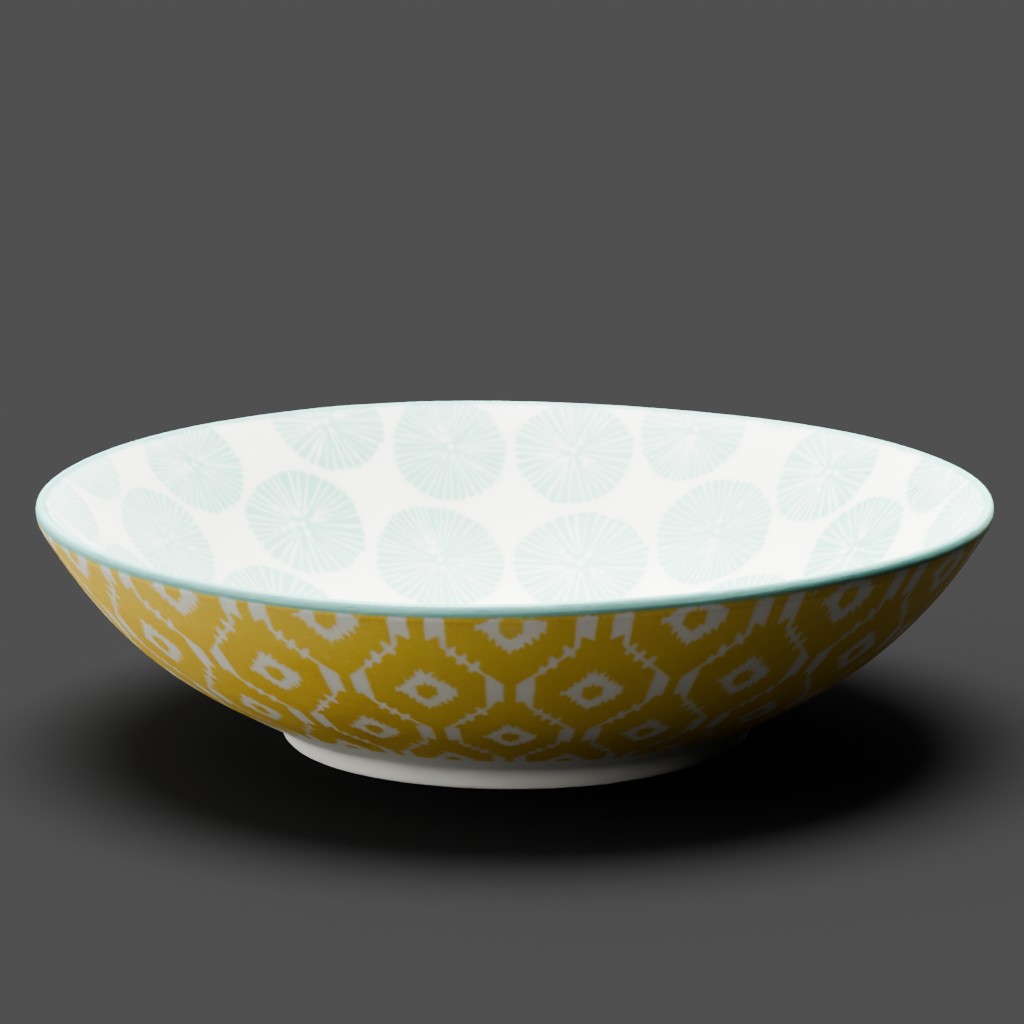} \\
    \includegraphics[width=0.19\linewidth]{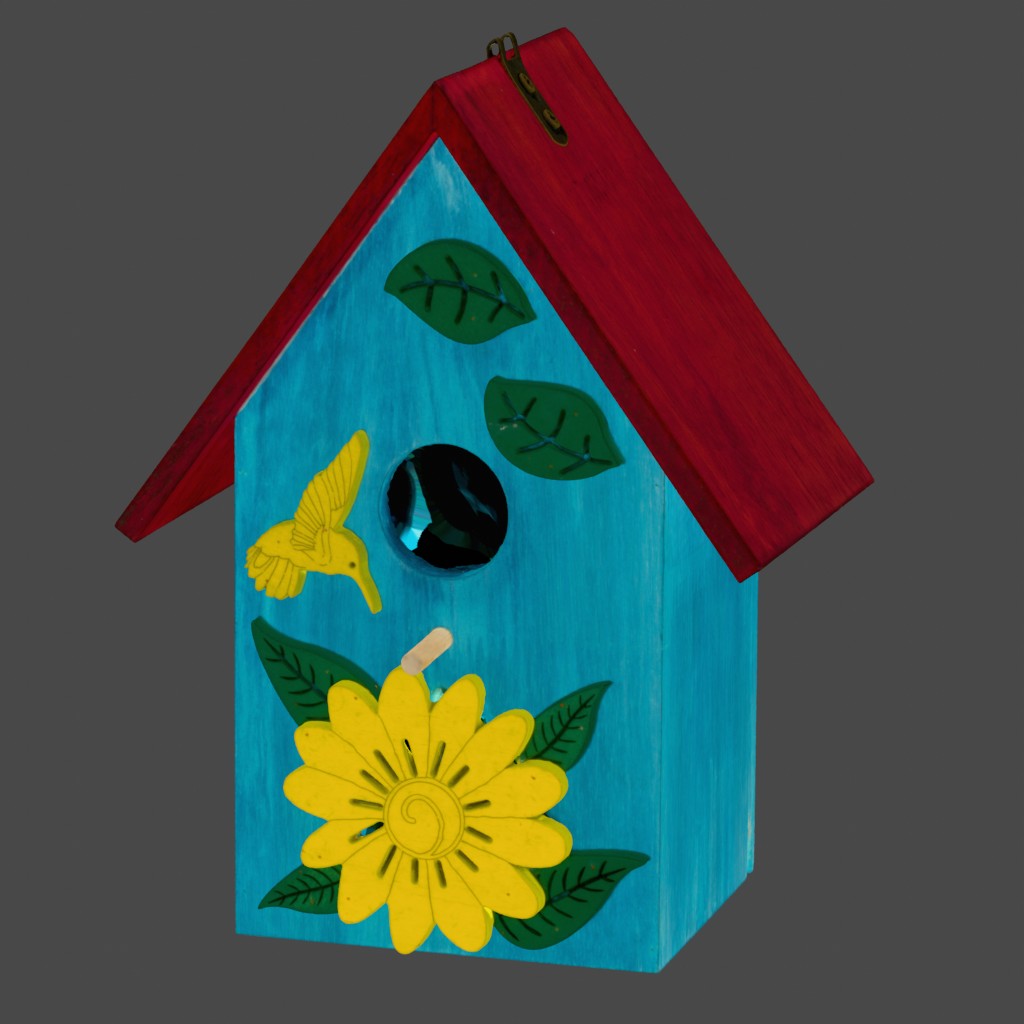}
    \includegraphics[width=0.19\linewidth]{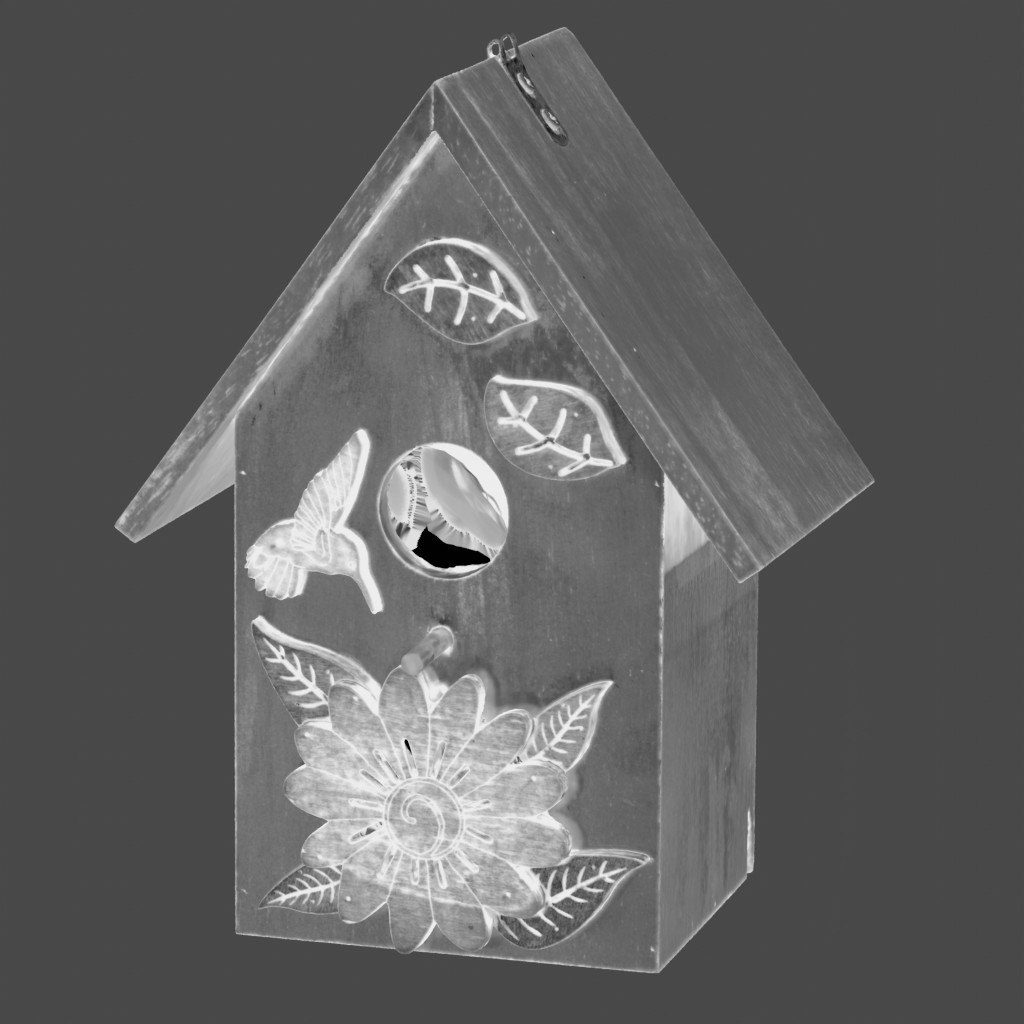}
    \includegraphics[width=0.19\linewidth]{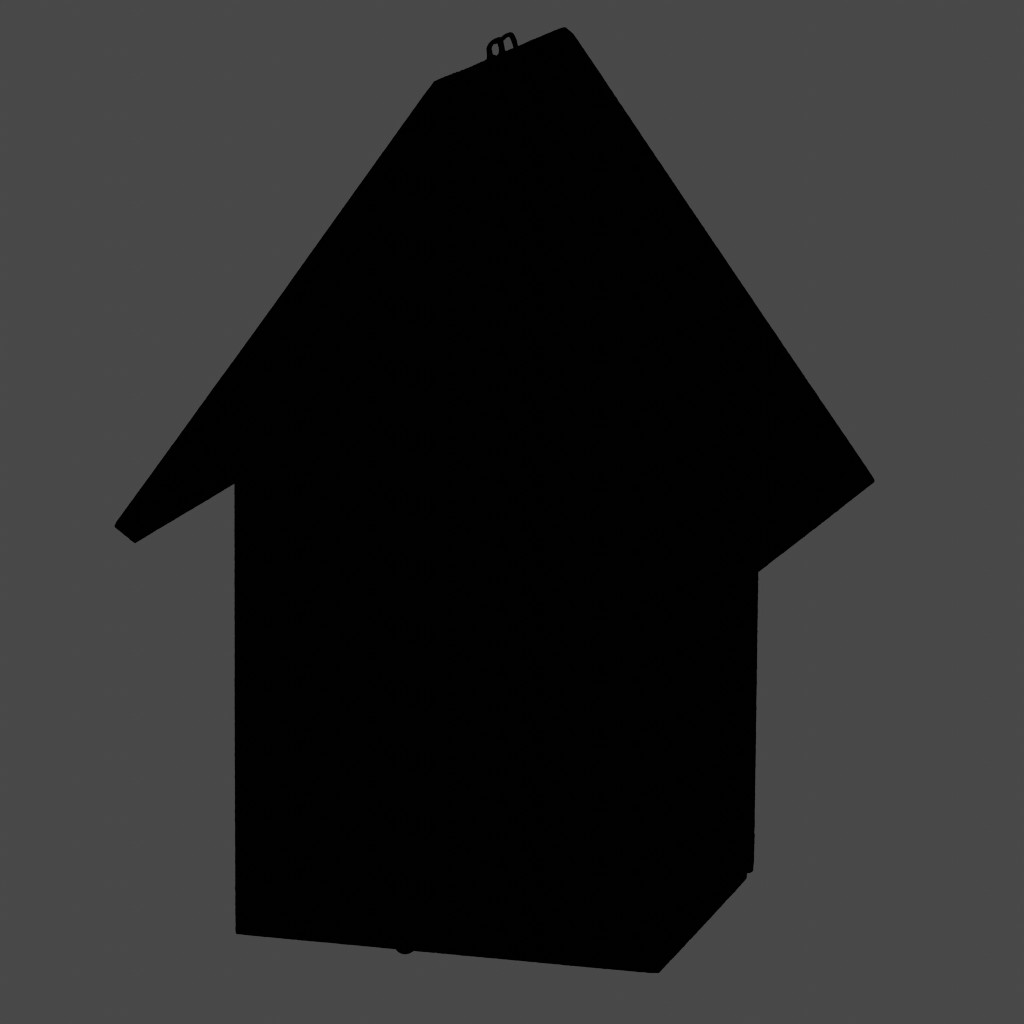}
    \includegraphics[width=0.19\linewidth]{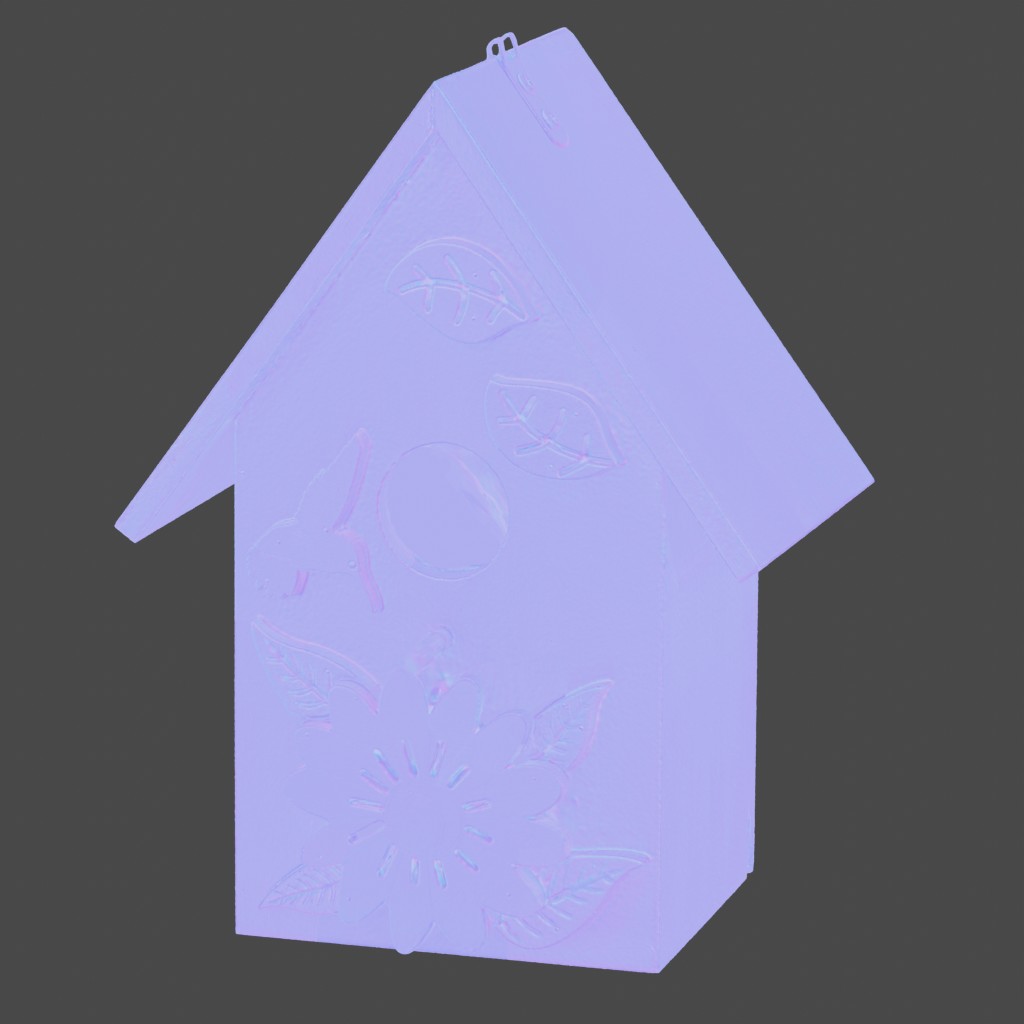}
    \includegraphics[width=0.19\linewidth]{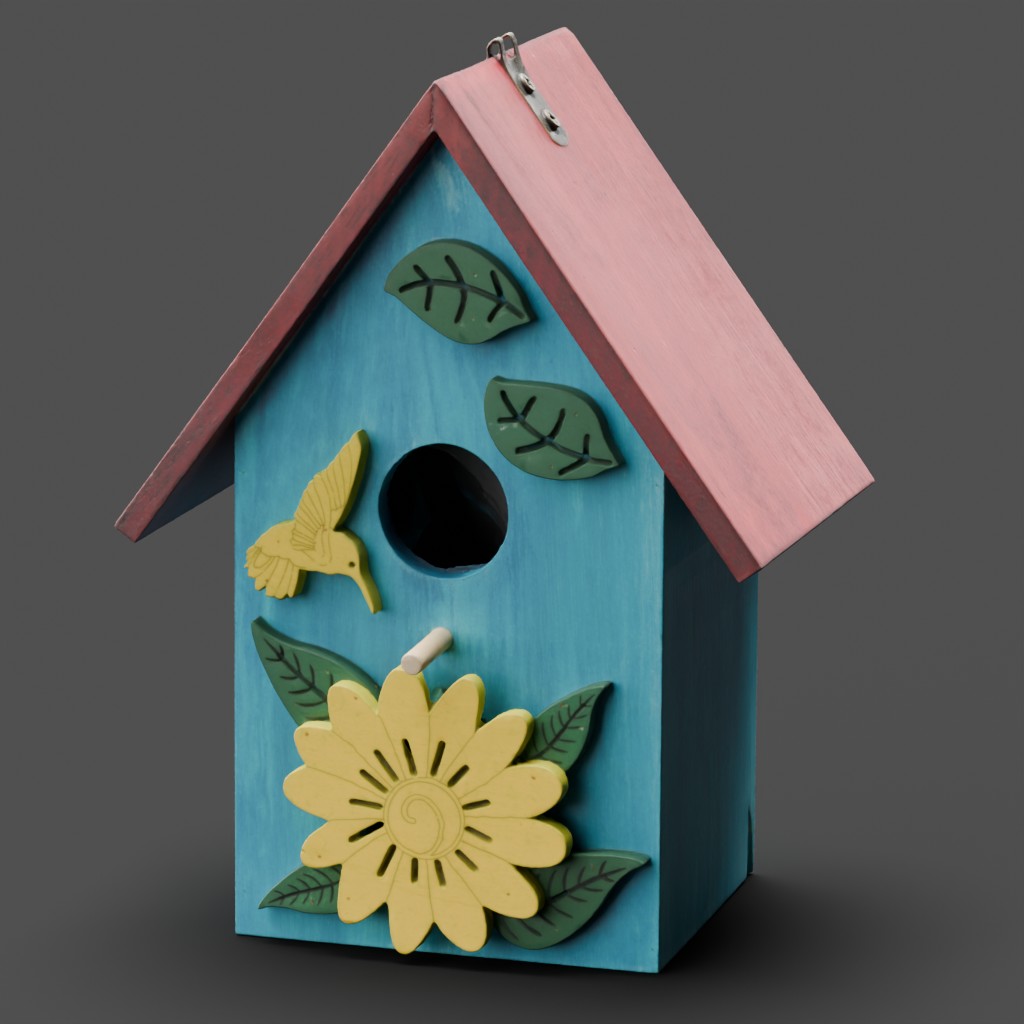}
    \caption{PBR Materials of the example DTC objects (the list of objects in Fig.\ref{fig:DTC_More_Models} \textbf{Row 1}), From left to right: albedo map, roughness map, metallic map, normal map, and PBR rendering.}
    \label{fig:more_PBR_maps_row1}
\end{figure*}

\begin{figure*}[t]
    \centering
    \includegraphics[width=0.19\linewidth]{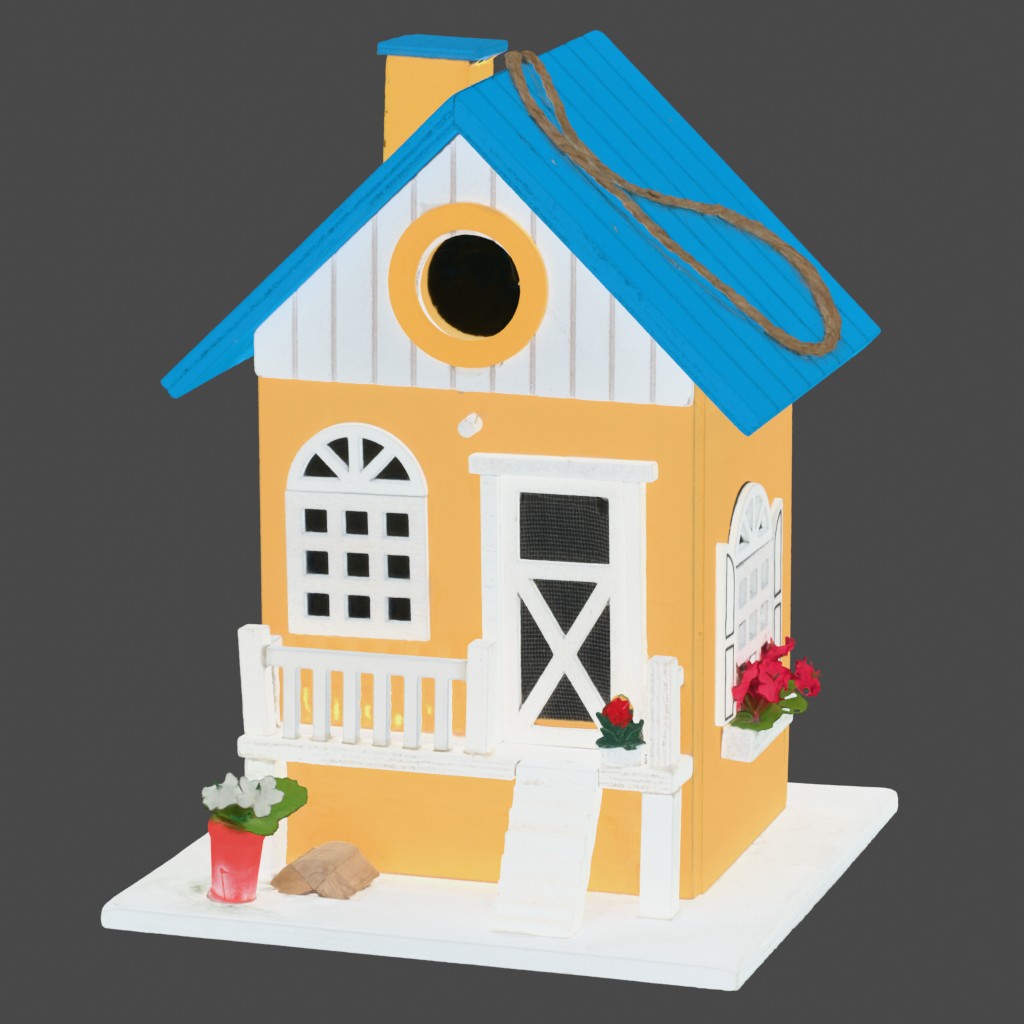}
    \includegraphics[width=0.19\linewidth]{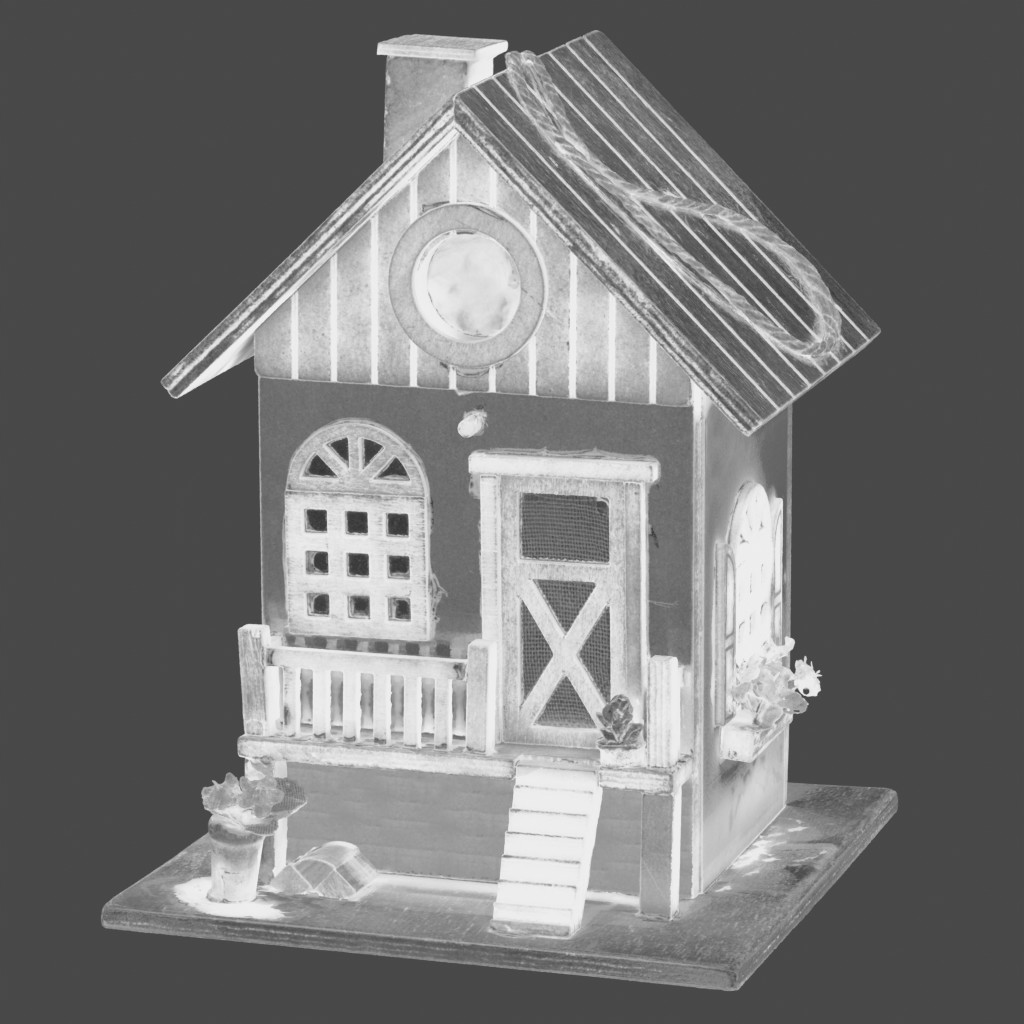}
    \includegraphics[width=0.19\linewidth]{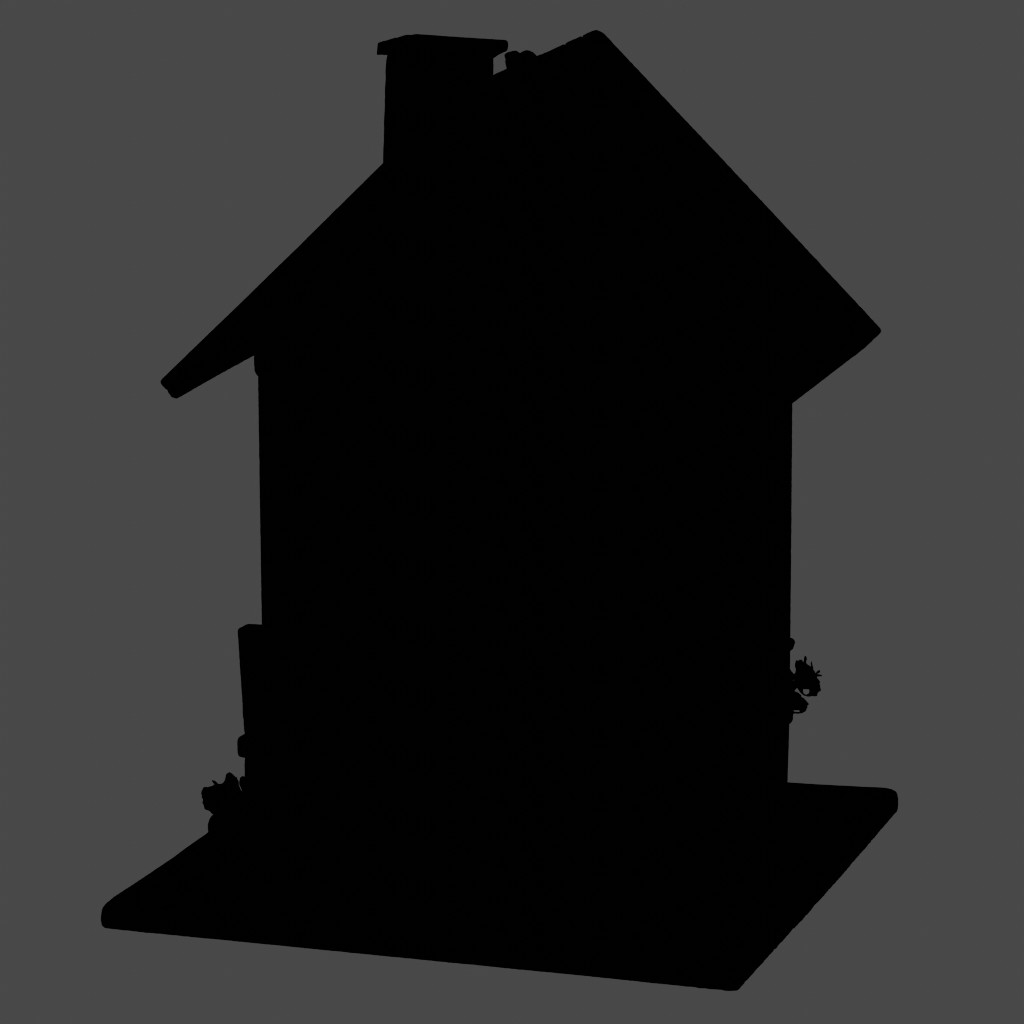}
    \includegraphics[width=0.19\linewidth]{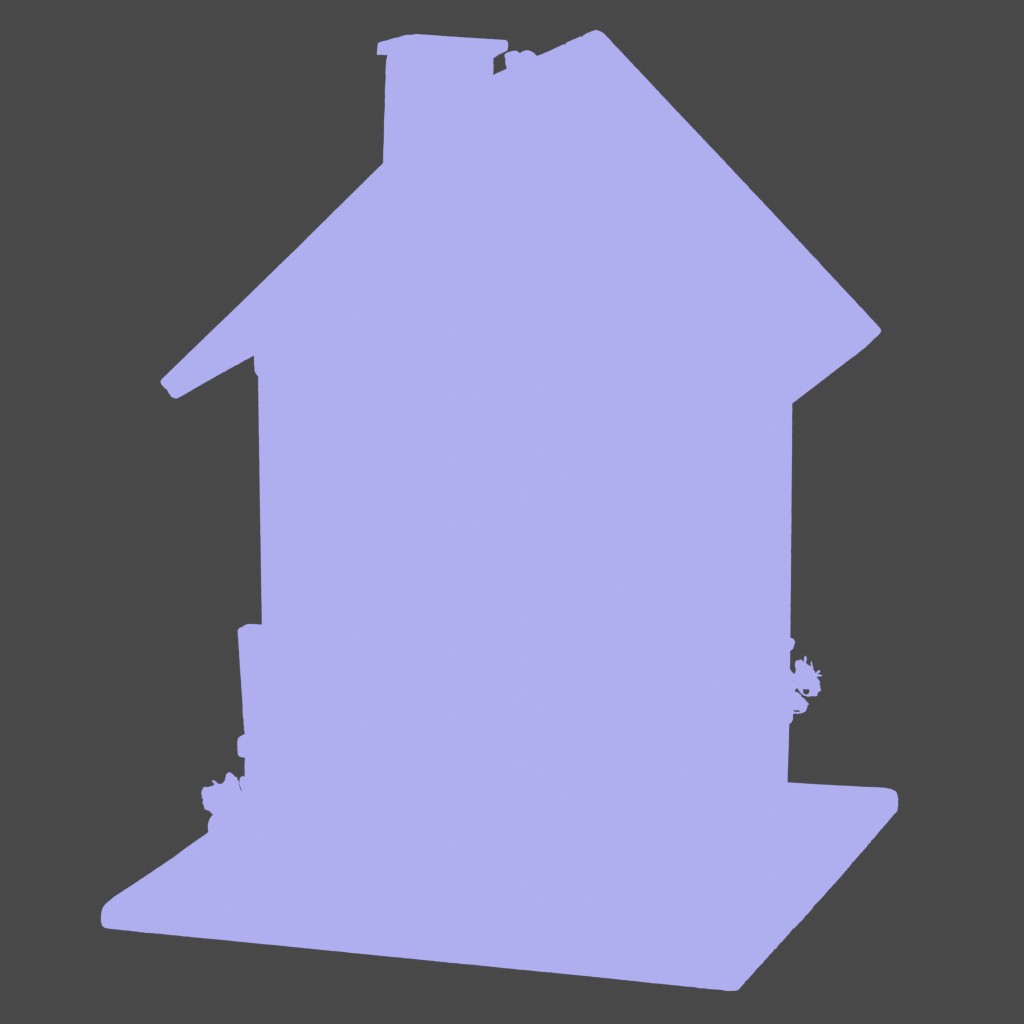}
    \includegraphics[width=0.19\linewidth]{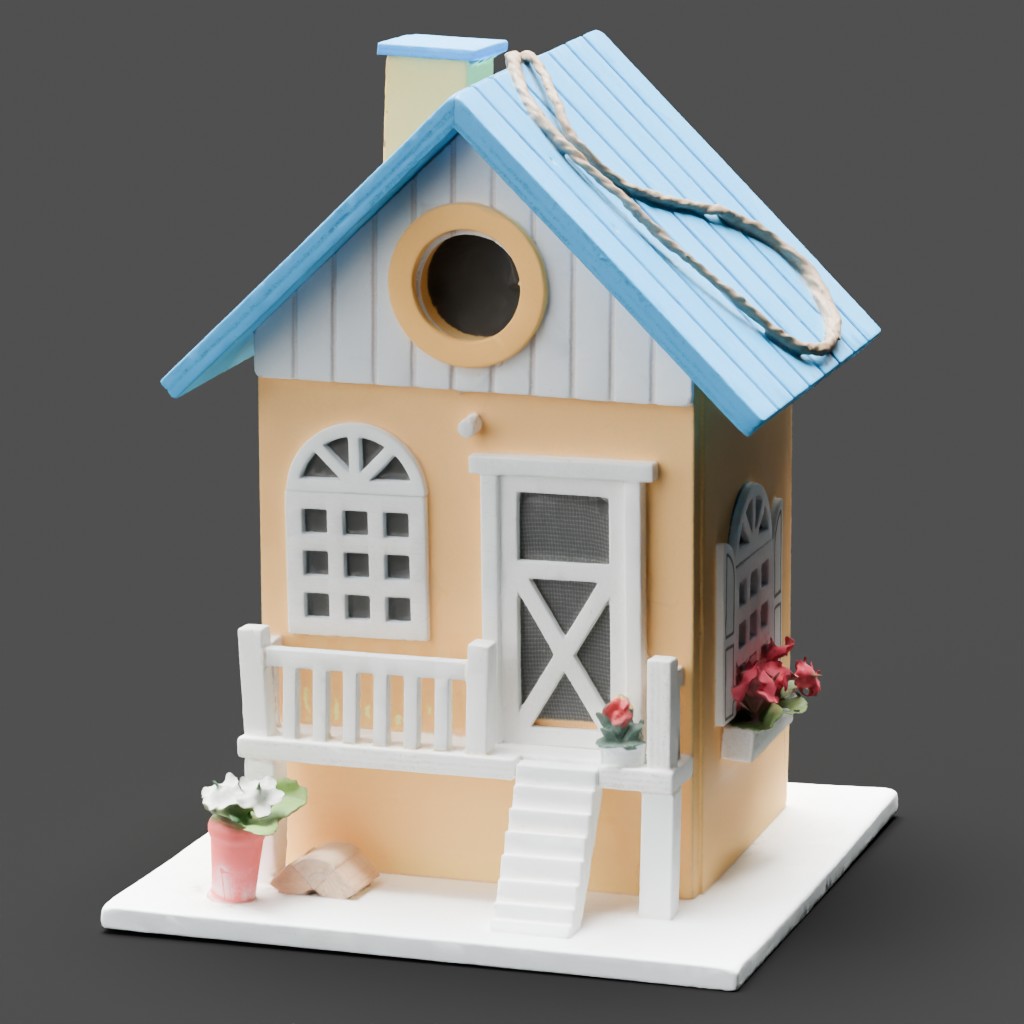} \\
    \includegraphics[width=0.19\linewidth]{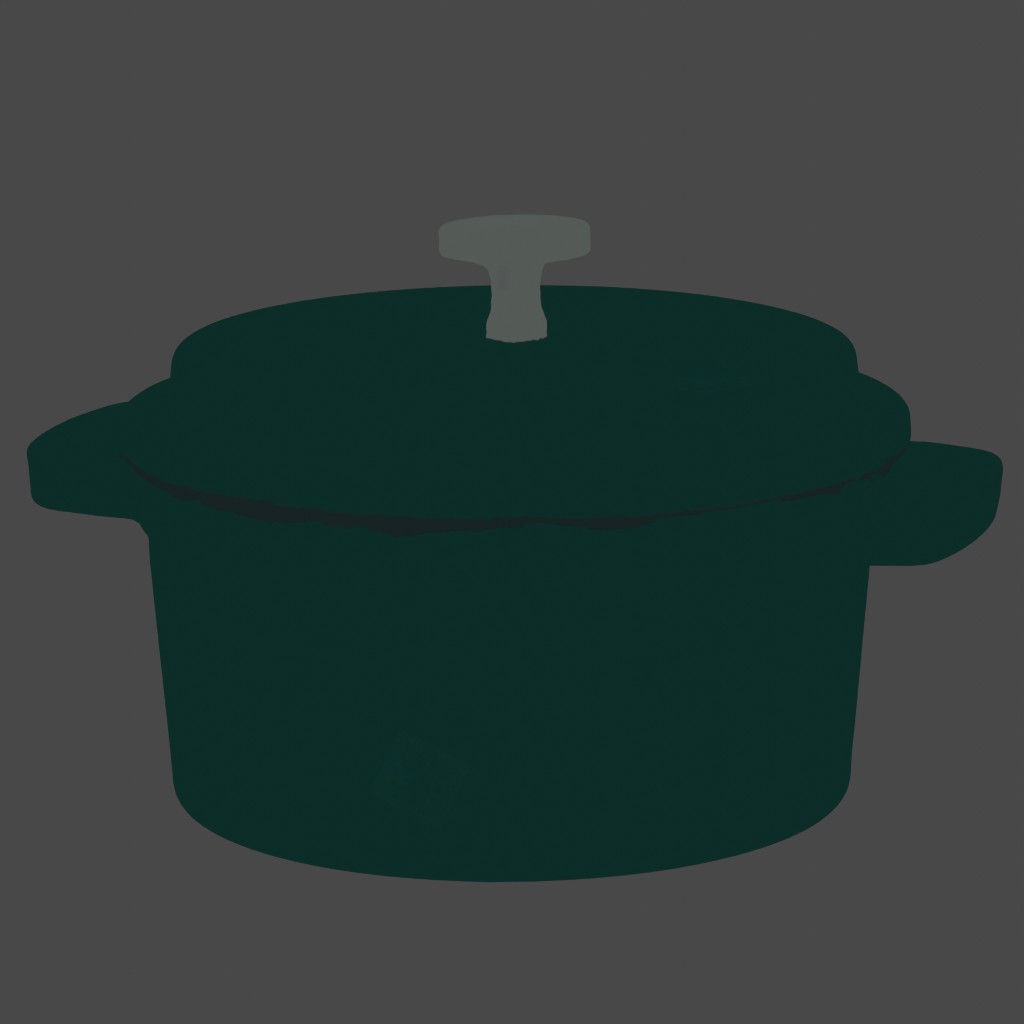}
    \includegraphics[width=0.19\linewidth]{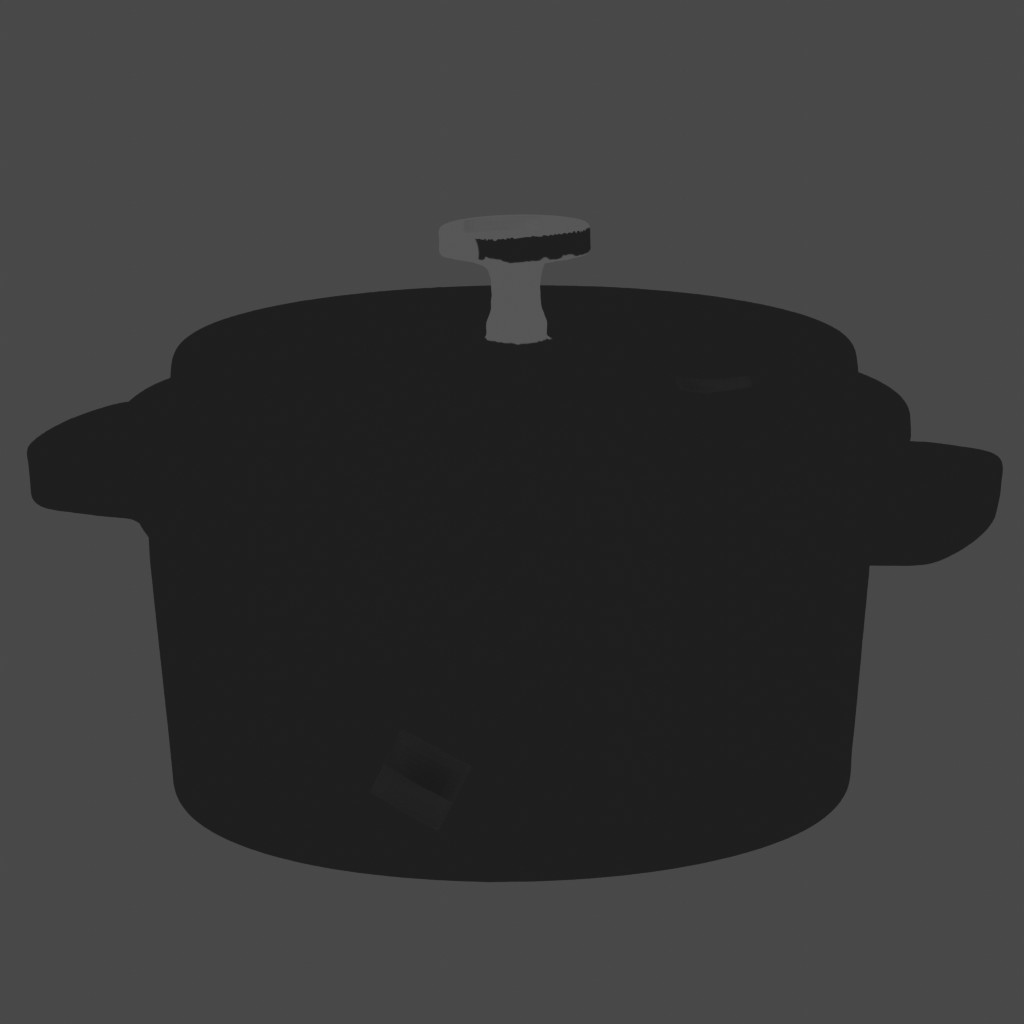}
    \includegraphics[width=0.19\linewidth]{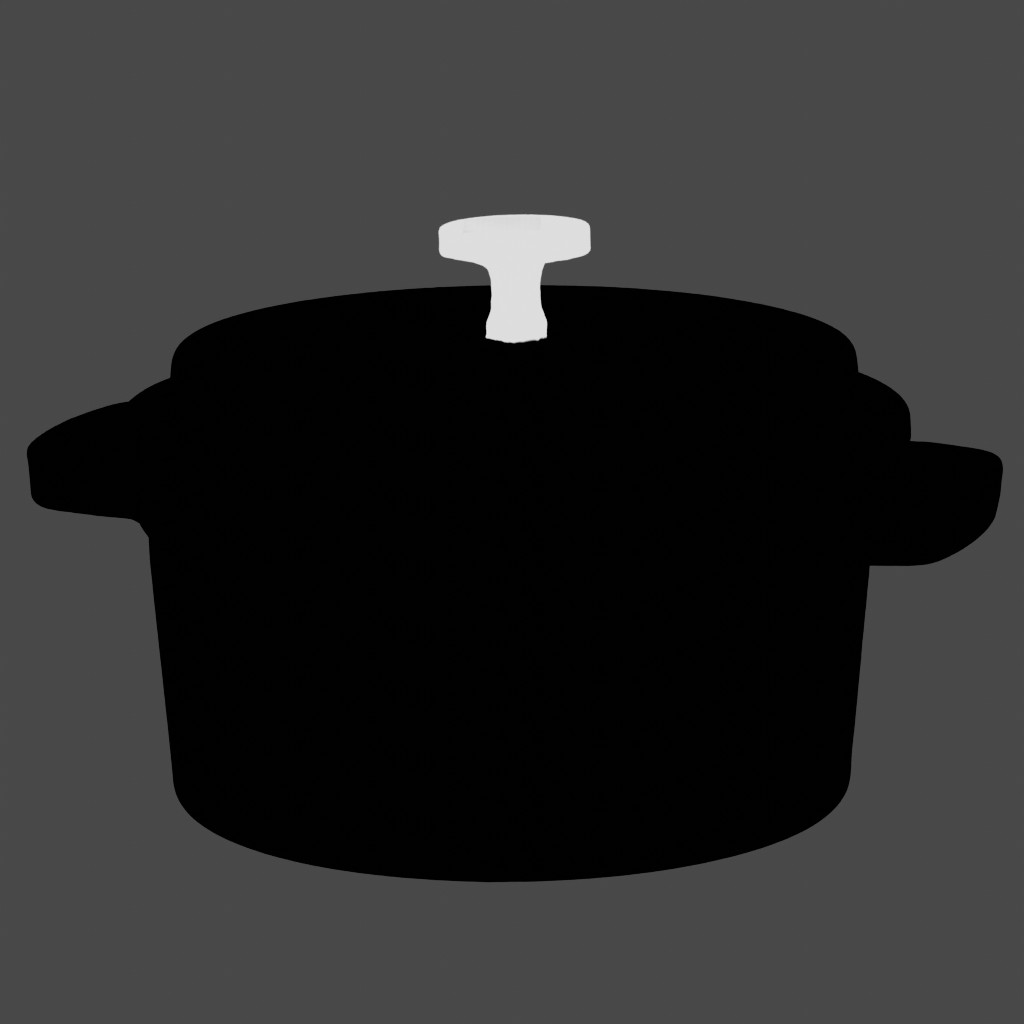}
    \includegraphics[width=0.19\linewidth]{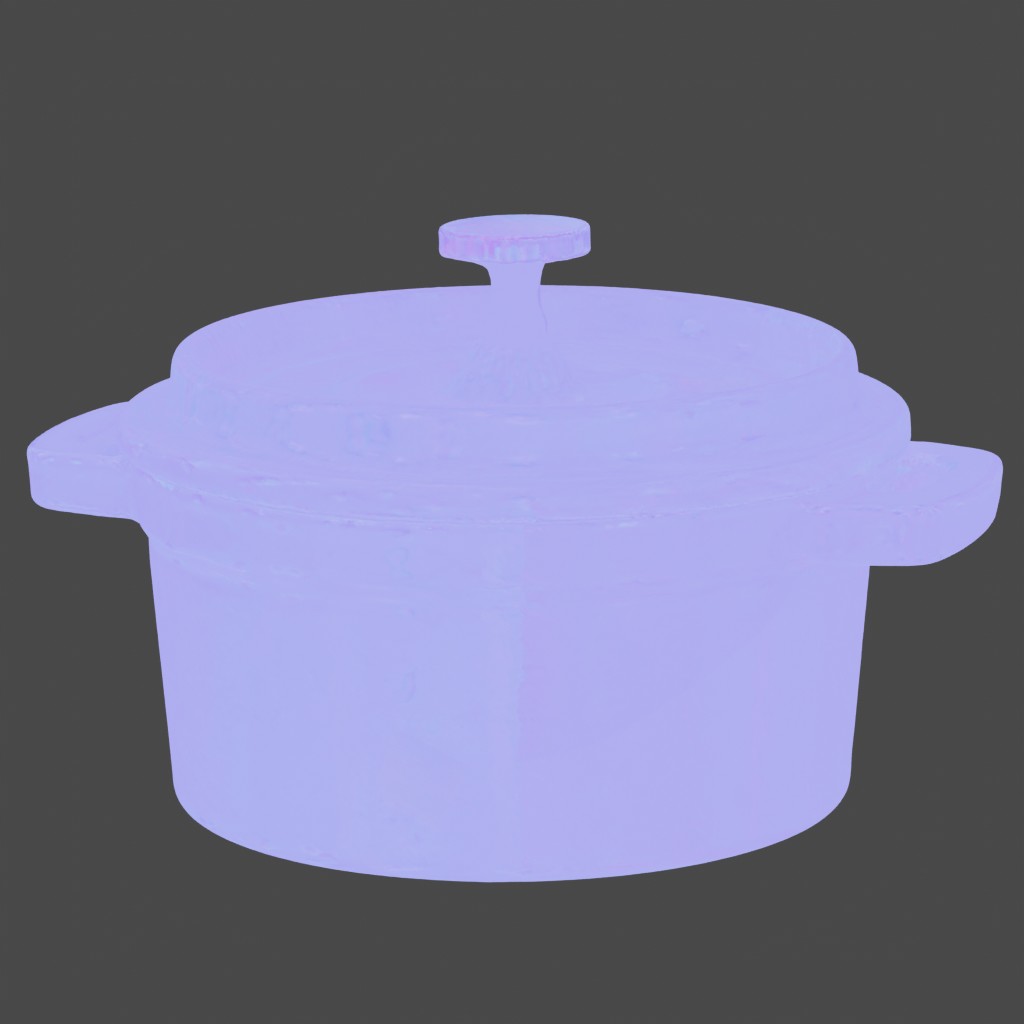}
    \includegraphics[width=0.19\linewidth]{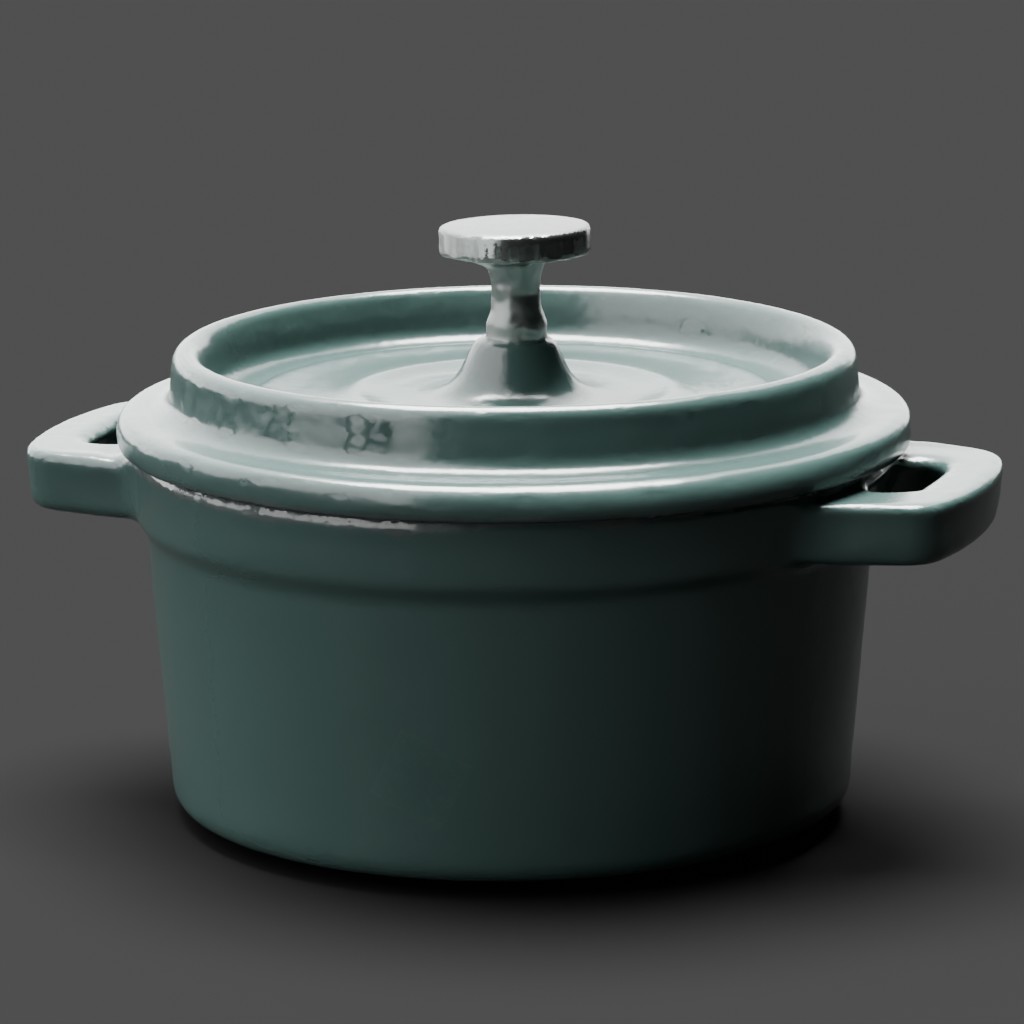} \\
    \includegraphics[width=0.19\linewidth]{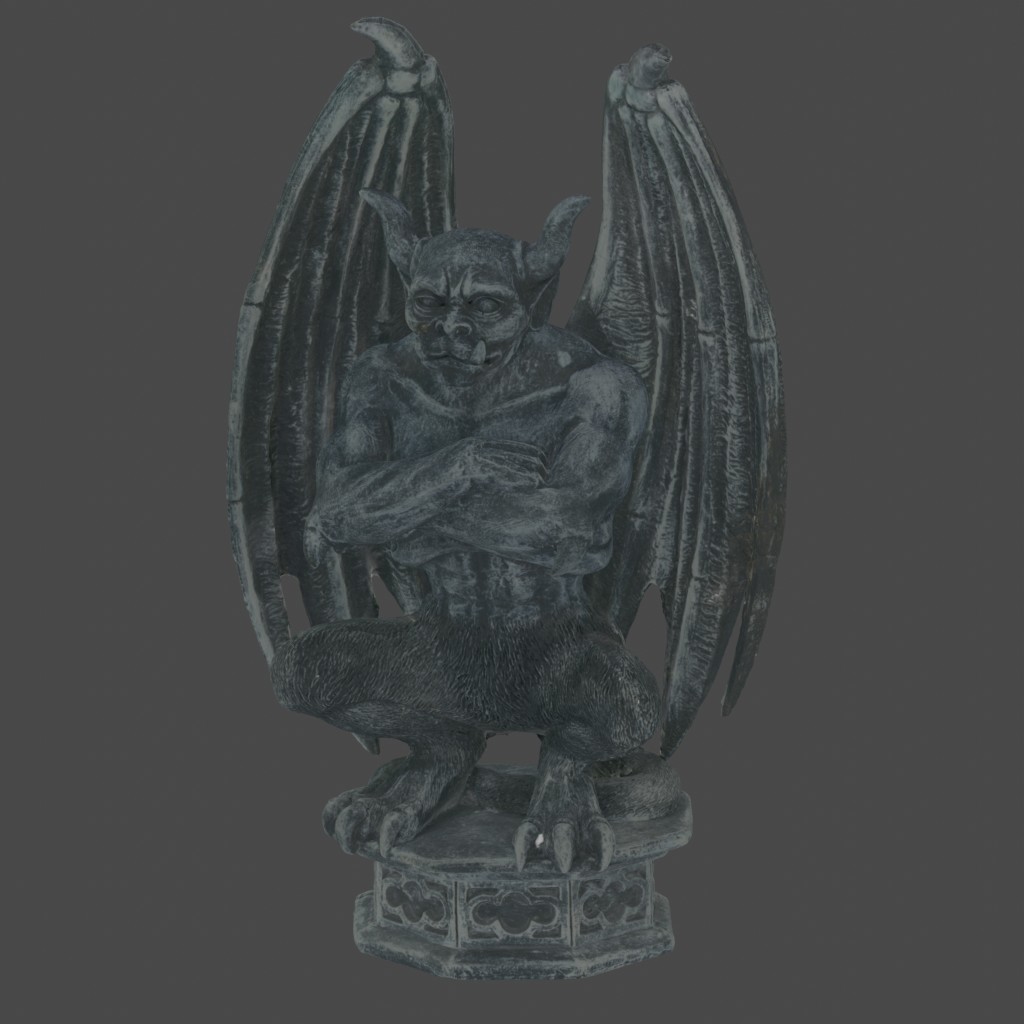}
    \includegraphics[width=0.19\linewidth]{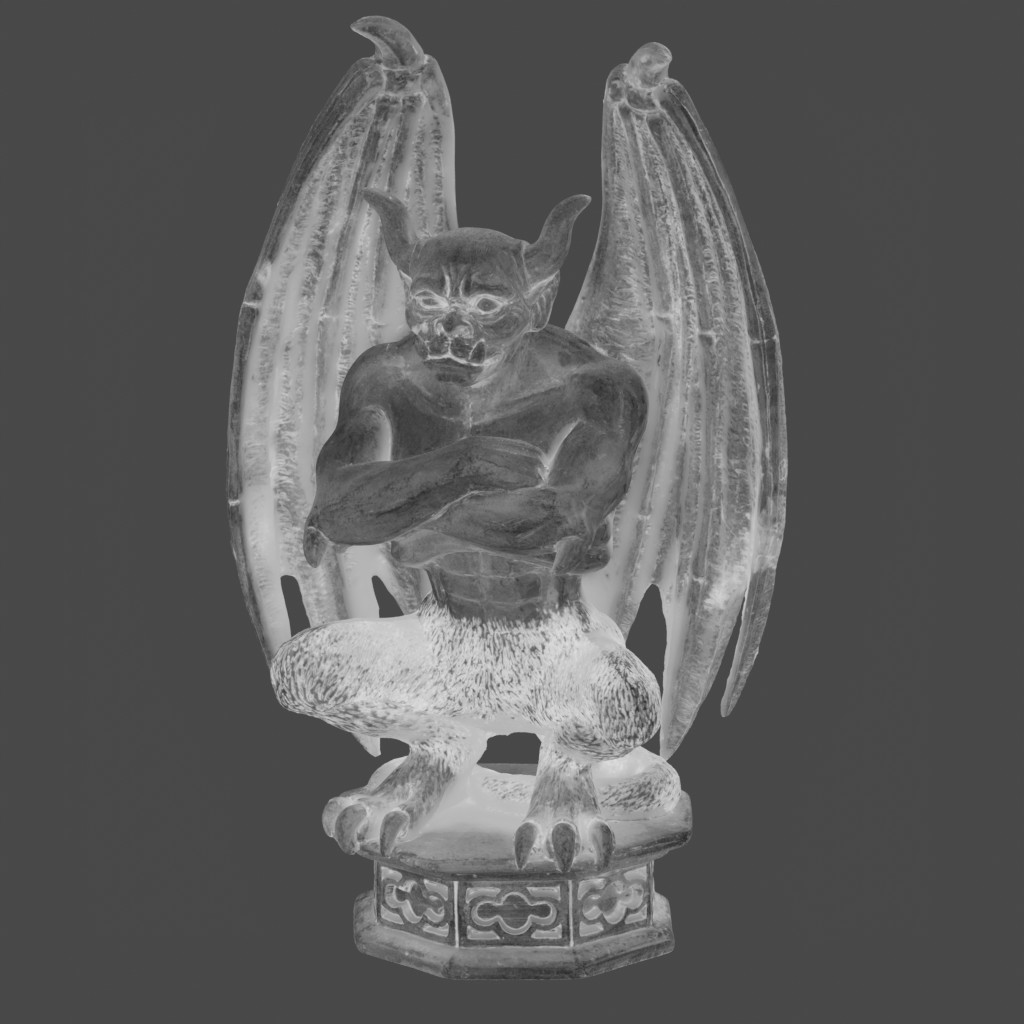}
    \includegraphics[width=0.19\linewidth]{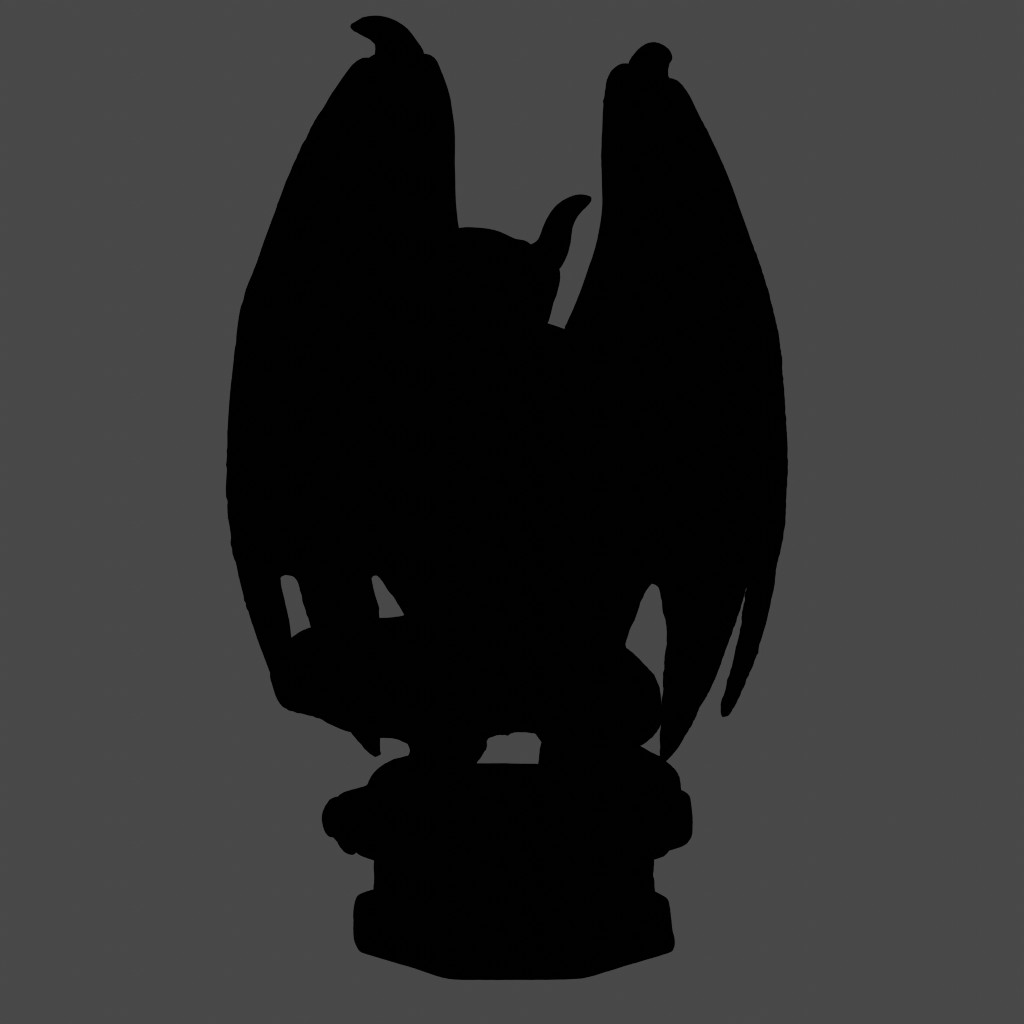}
    \includegraphics[width=0.19\linewidth]{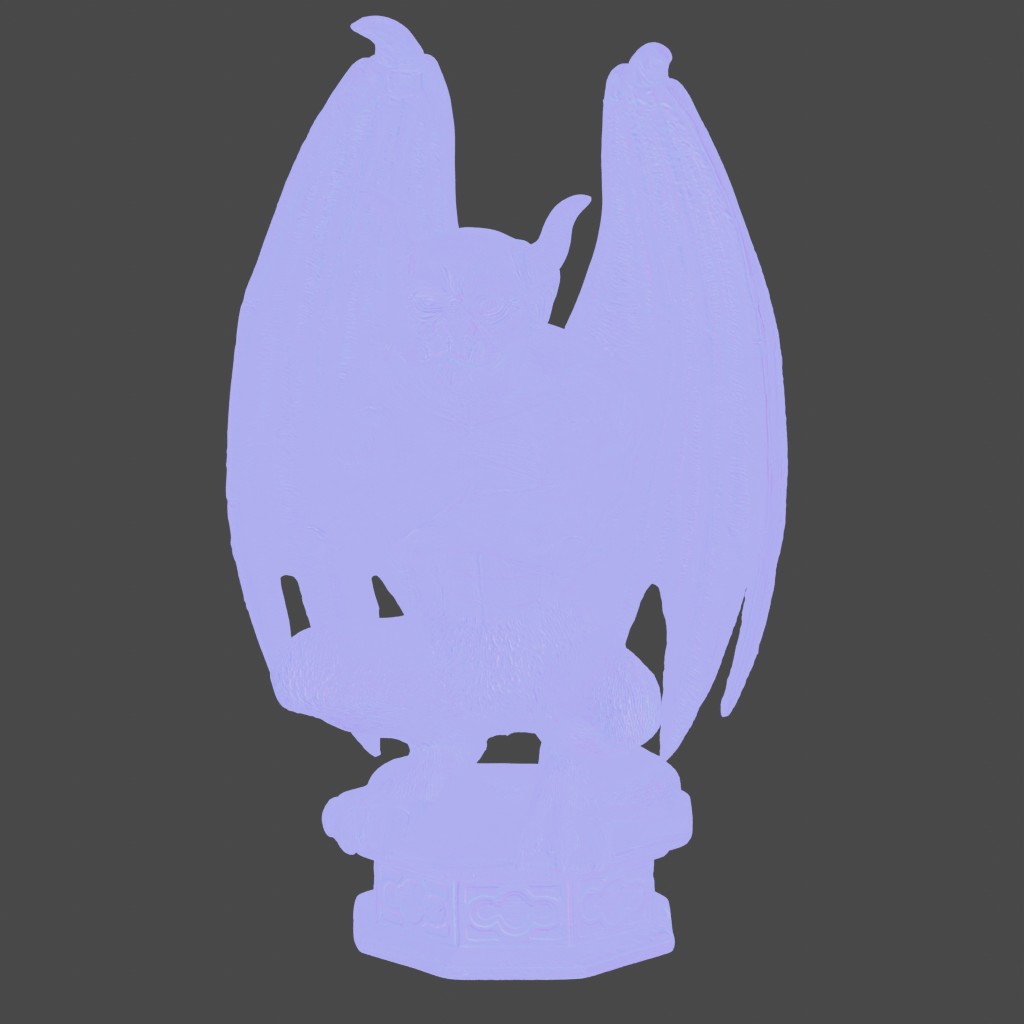}
    \includegraphics[width=0.19\linewidth]{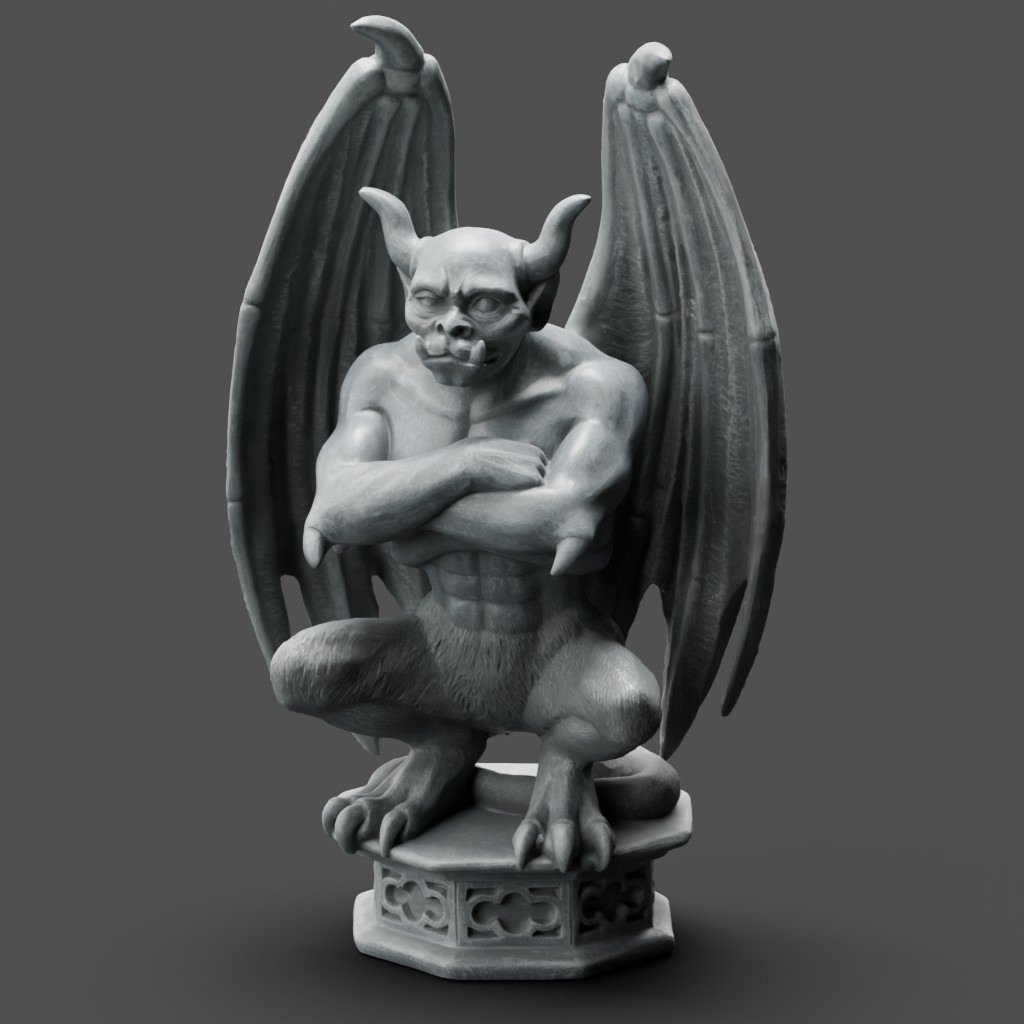} \\
    \includegraphics[width=0.19\linewidth]{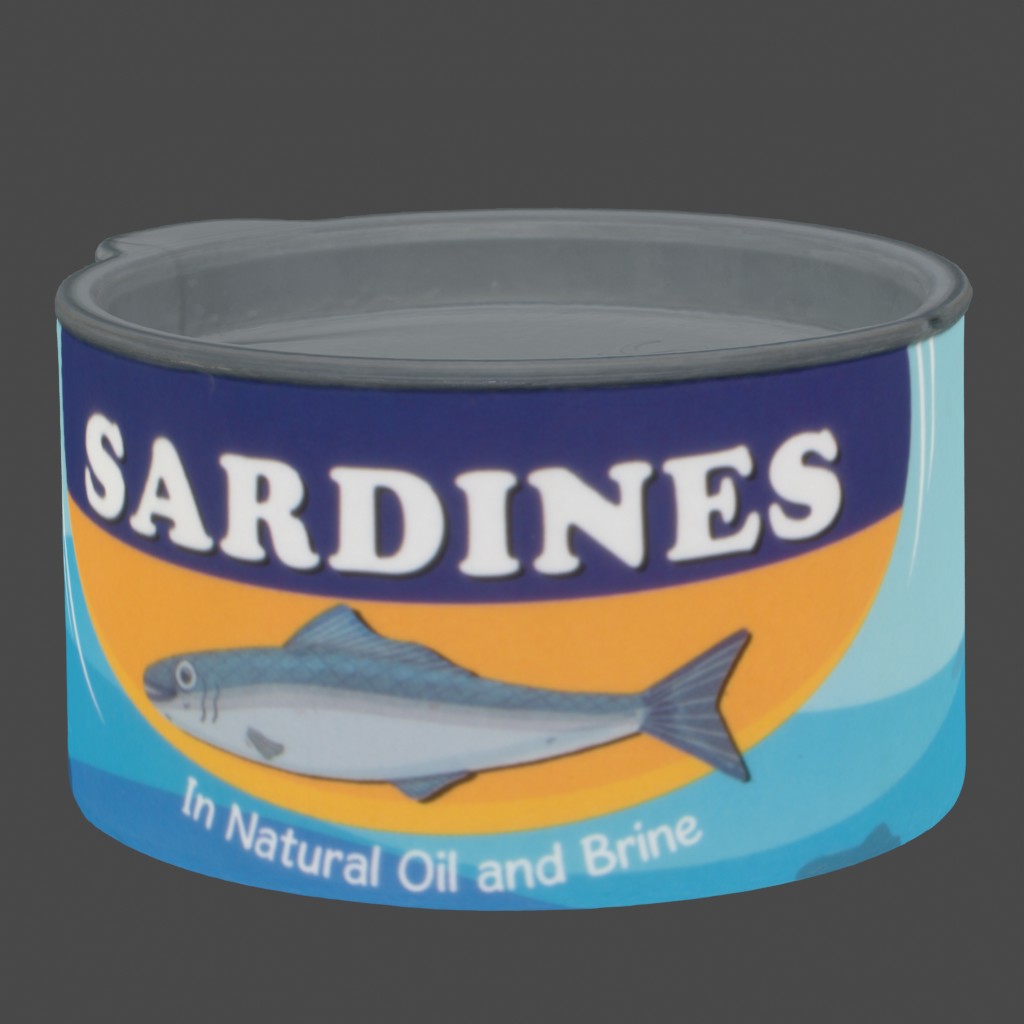}
    \includegraphics[width=0.19\linewidth]{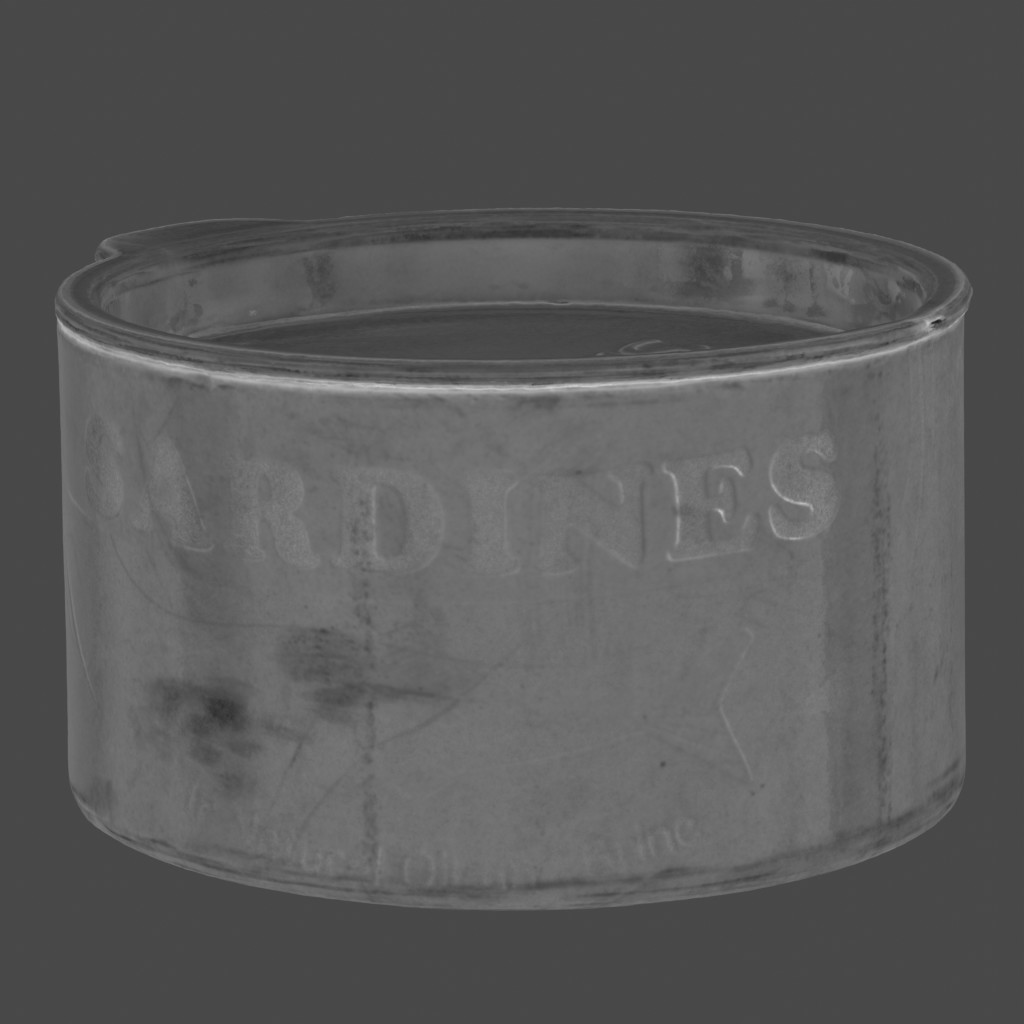}
    \includegraphics[width=0.19\linewidth]{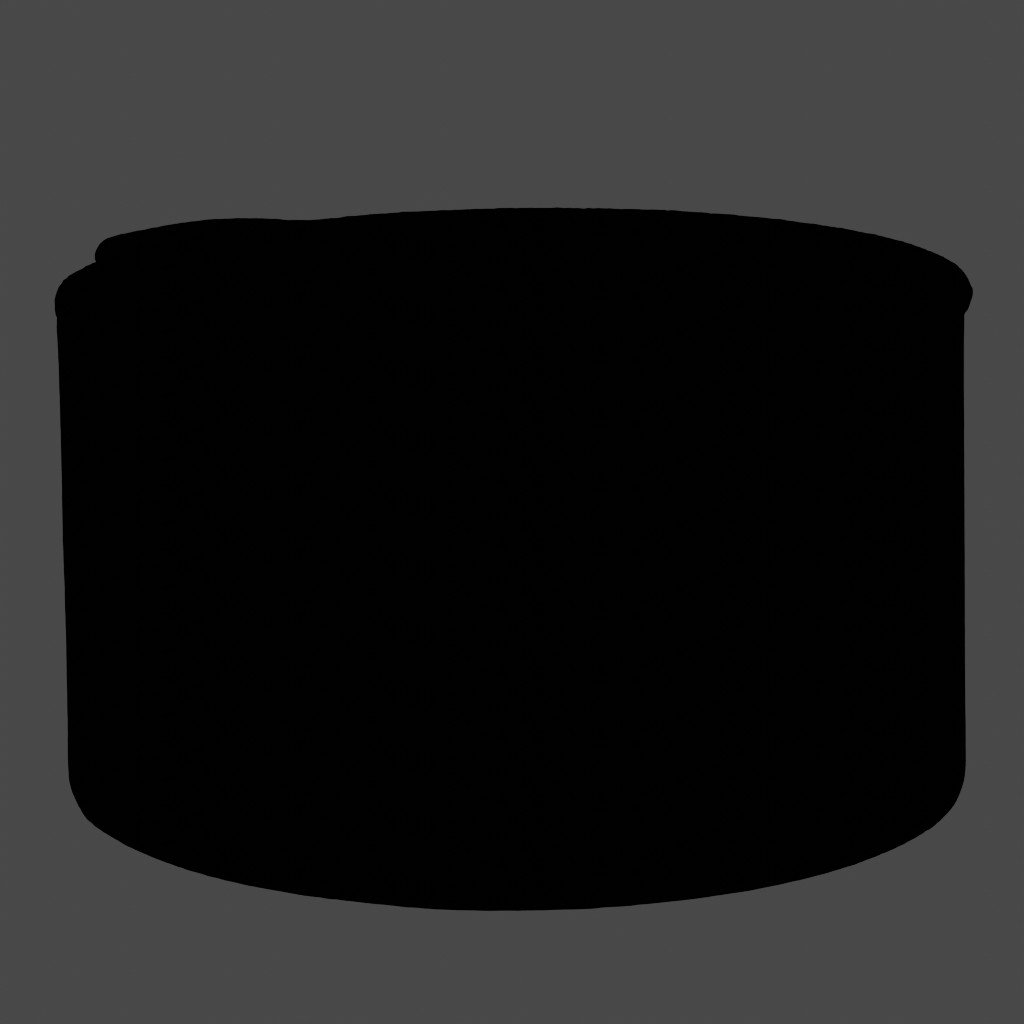}
    \includegraphics[width=0.19\linewidth]{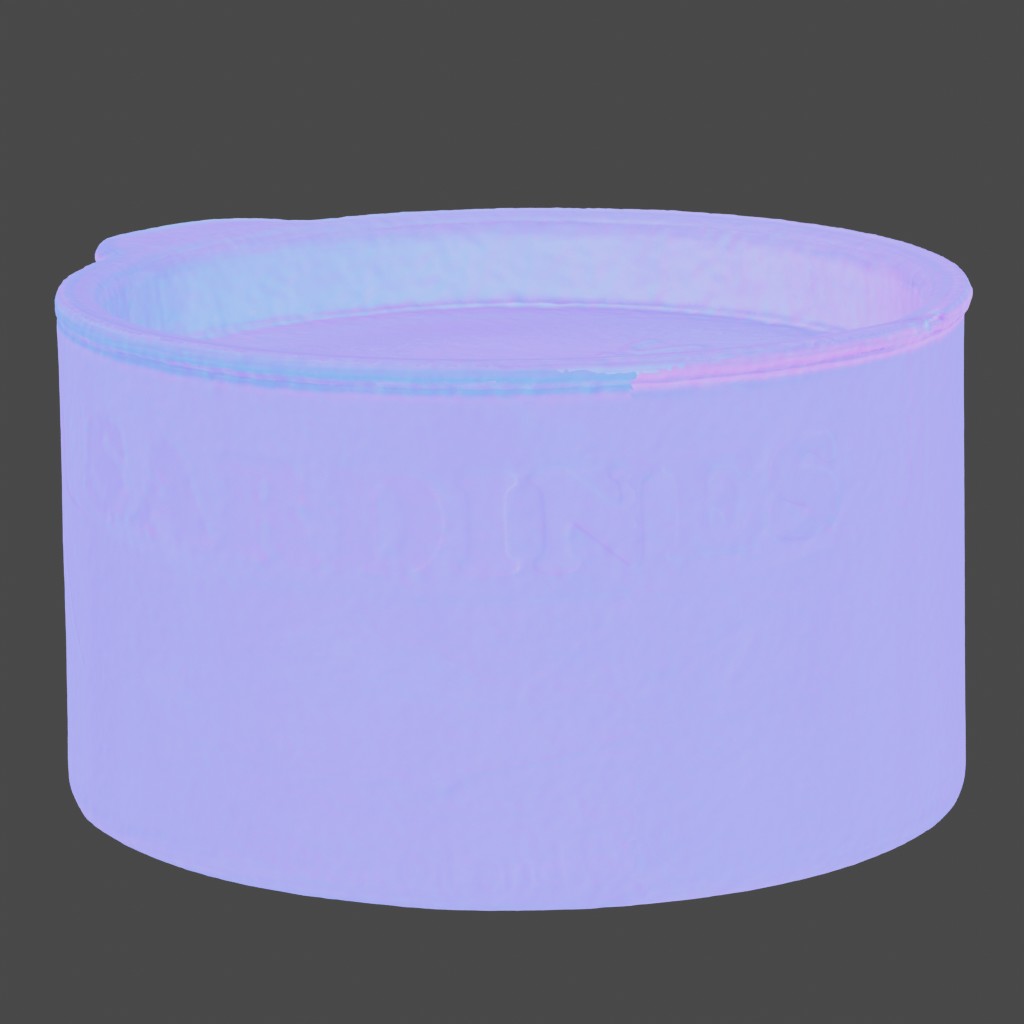}
    \includegraphics[width=0.19\linewidth]{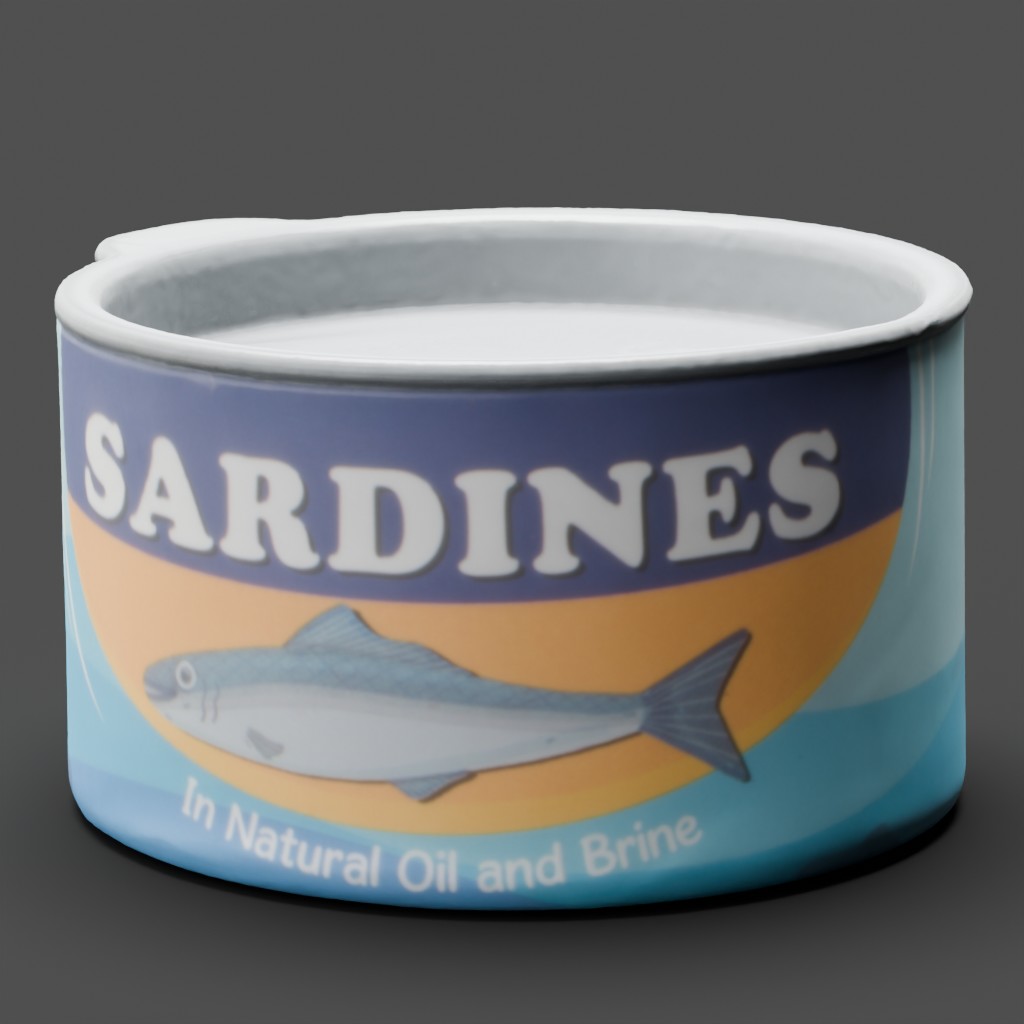} \\
    \includegraphics[width=0.19\linewidth]{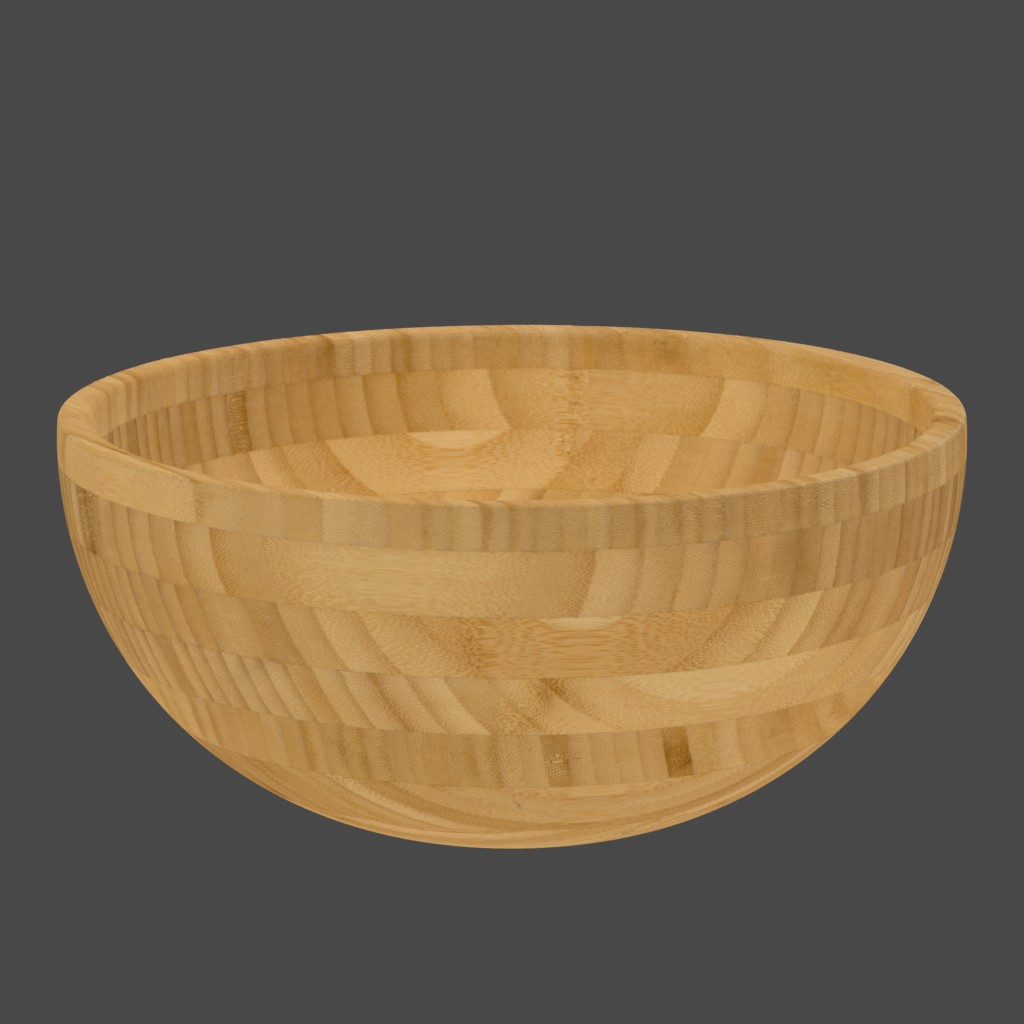}
    \includegraphics[width=0.19\linewidth]{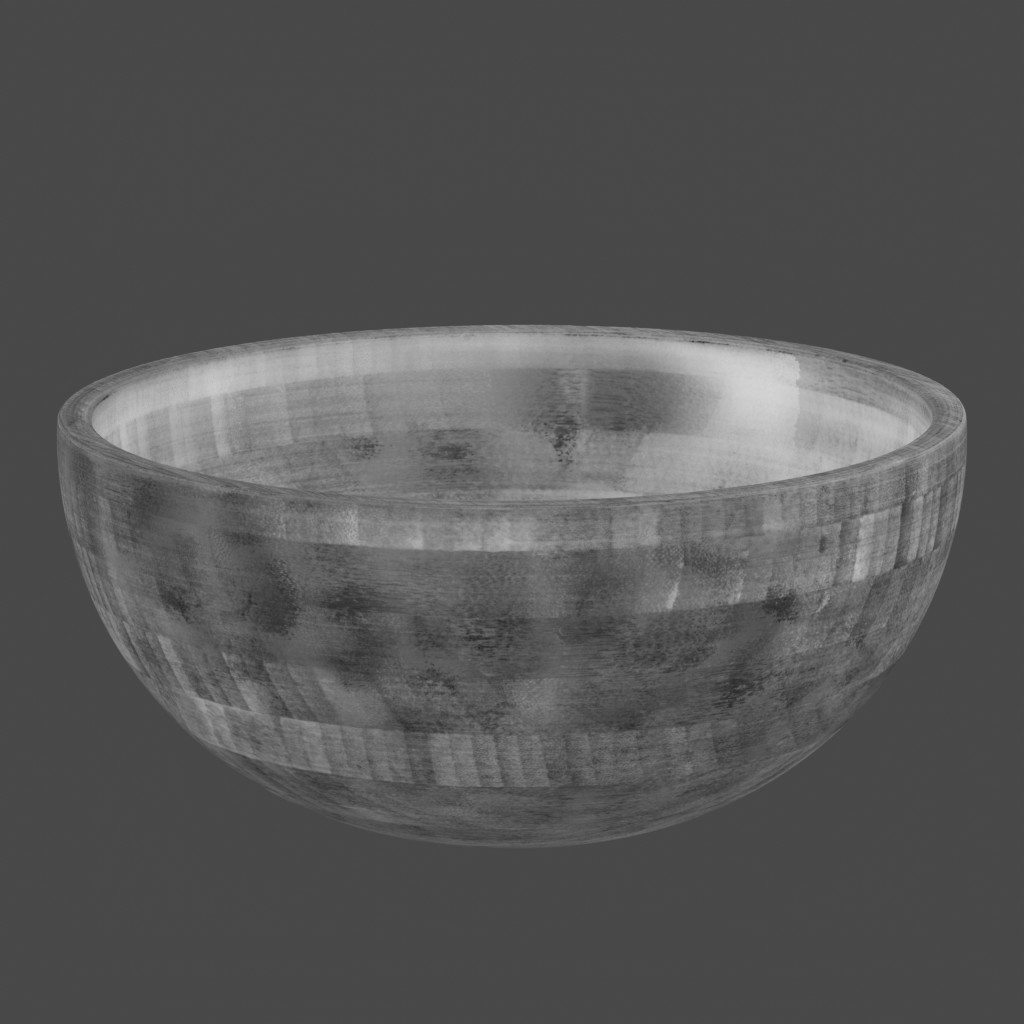}
    \includegraphics[width=0.19\linewidth]{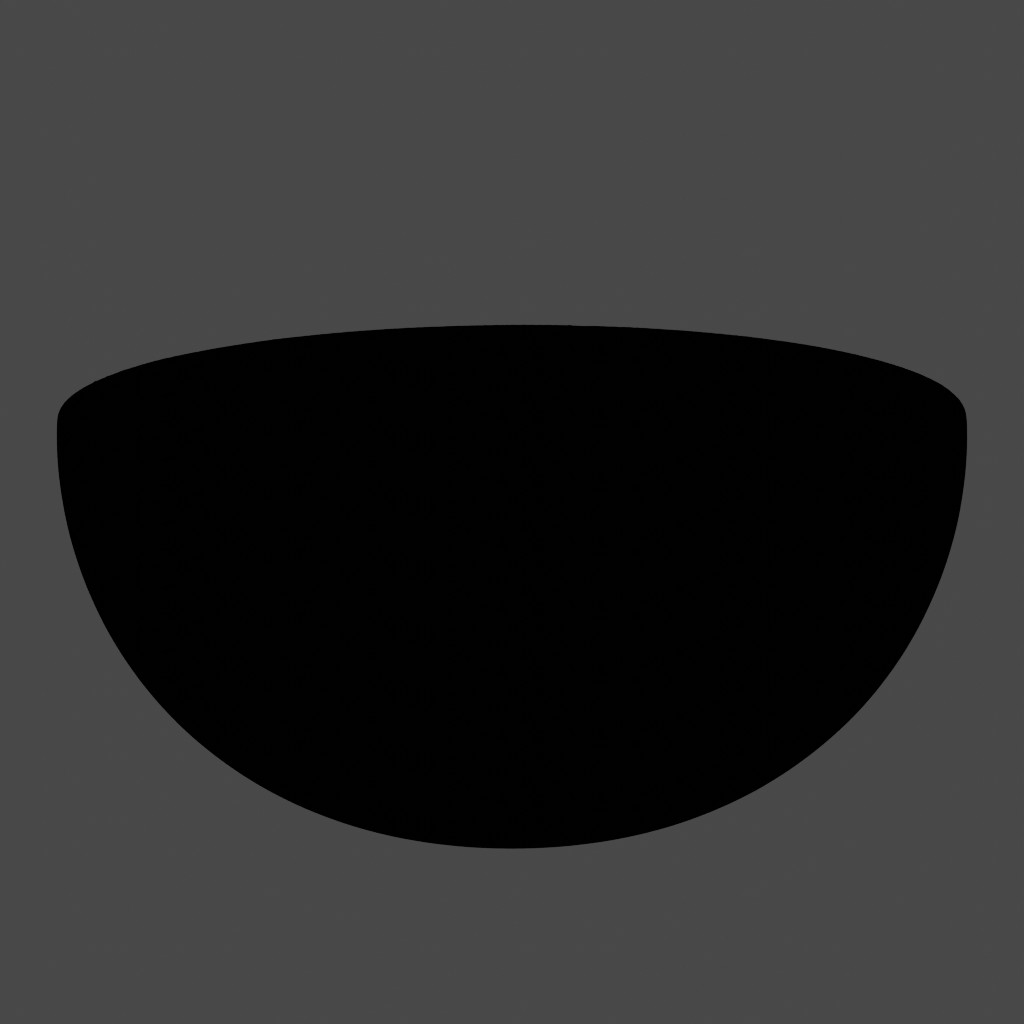}
    \includegraphics[width=0.19\linewidth]{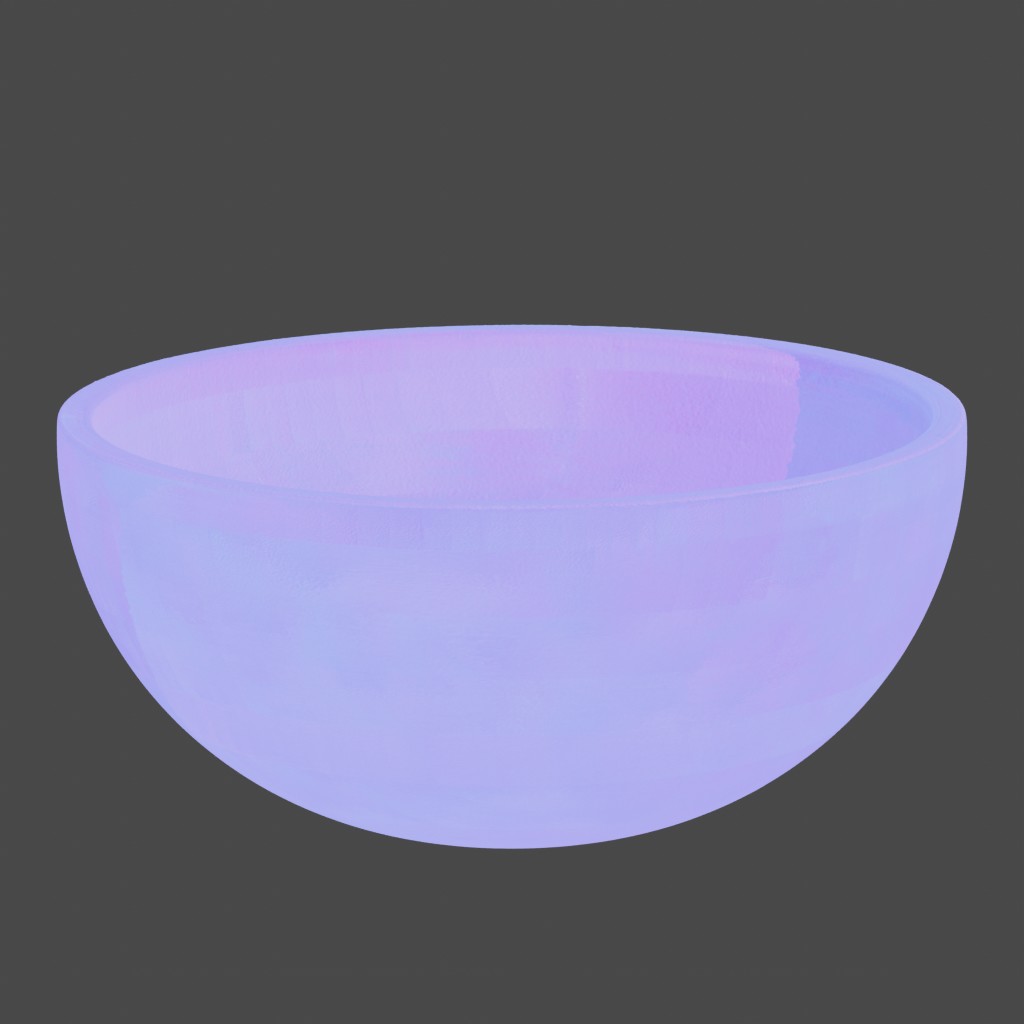}
    \includegraphics[width=0.19\linewidth]{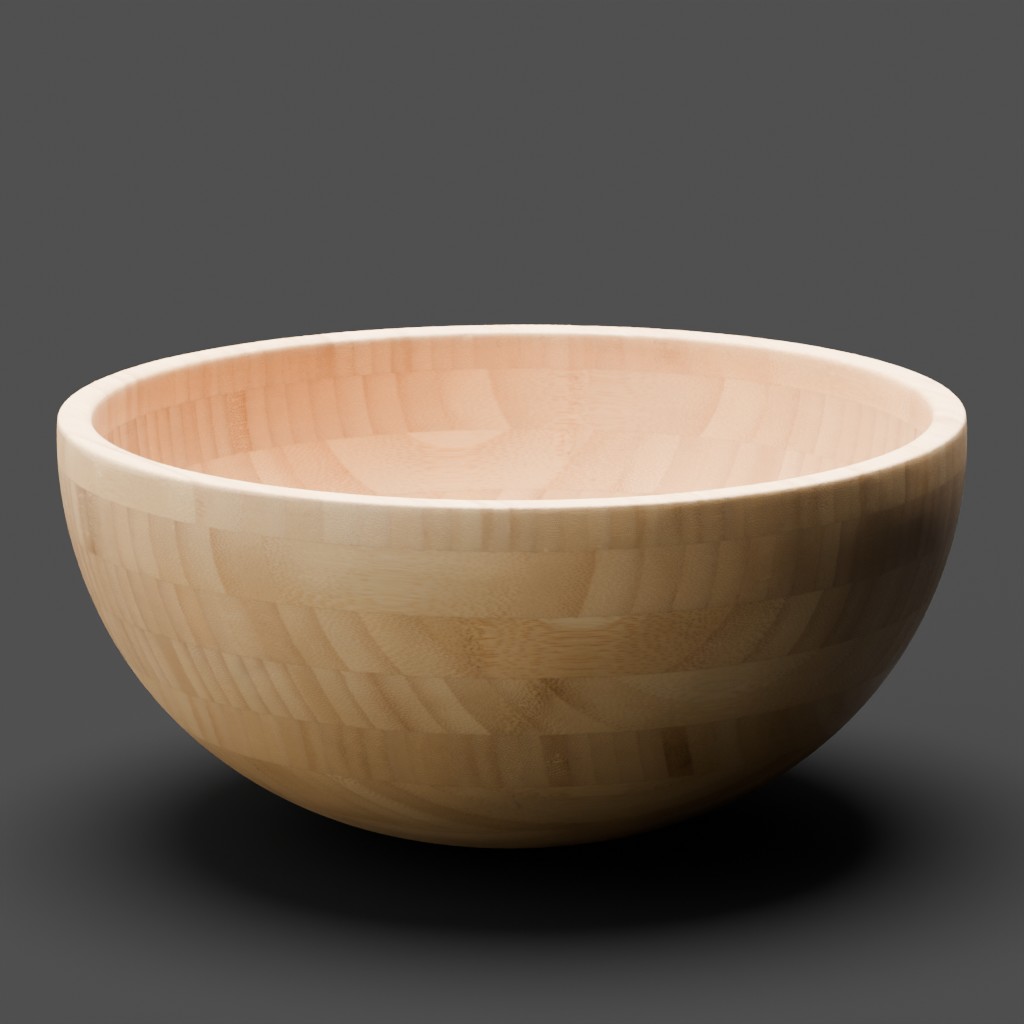}
    \caption{PBR Materials of the example DTC objects (the list of objects in Fig.\ref{fig:DTC_More_Models} \textbf{Row 2}). From left to right: albedo map, roughness map, metallic map, normal map, and PBR rendering.}
    \label{fig:more_PBR_maps_row2}
\end{figure*}

\begin{figure*}[t]
    \centering
    \includegraphics[width=0.19\linewidth]{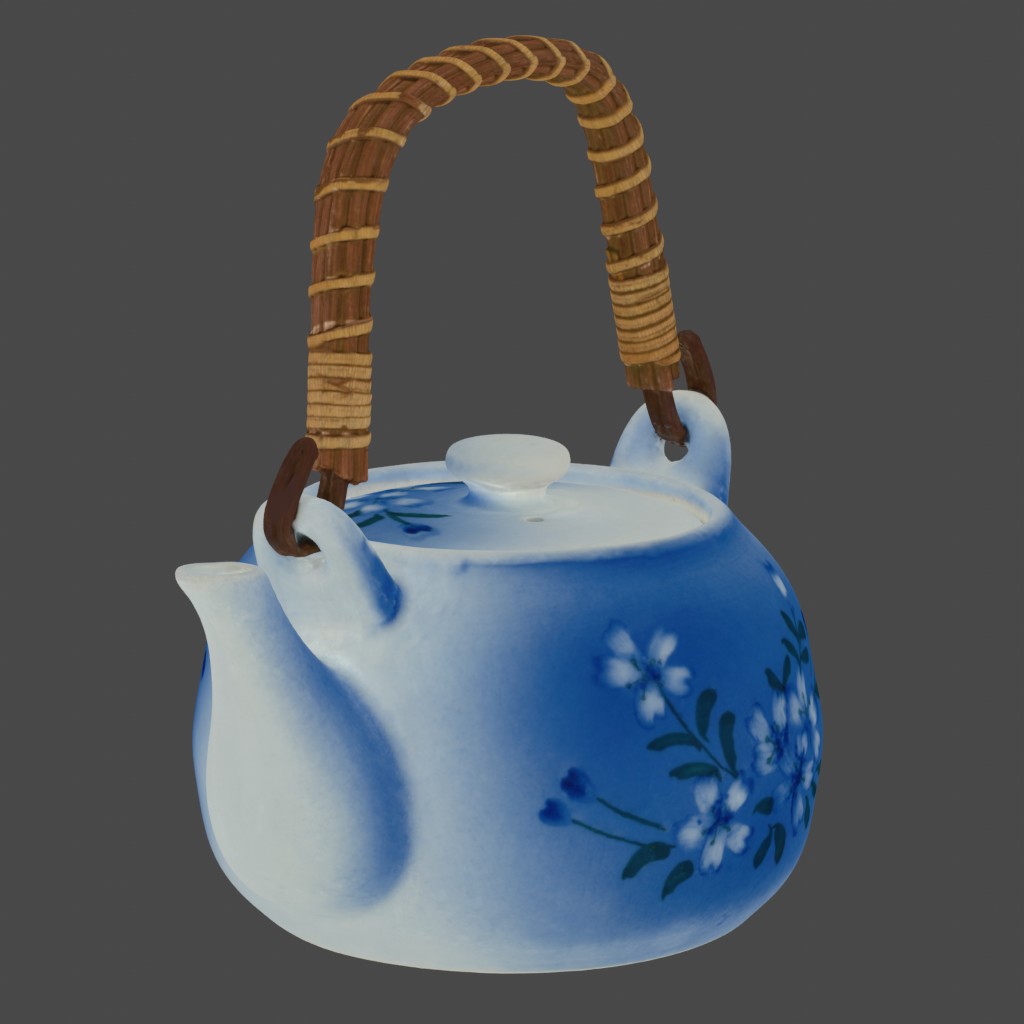}
    \includegraphics[width=0.19\linewidth]{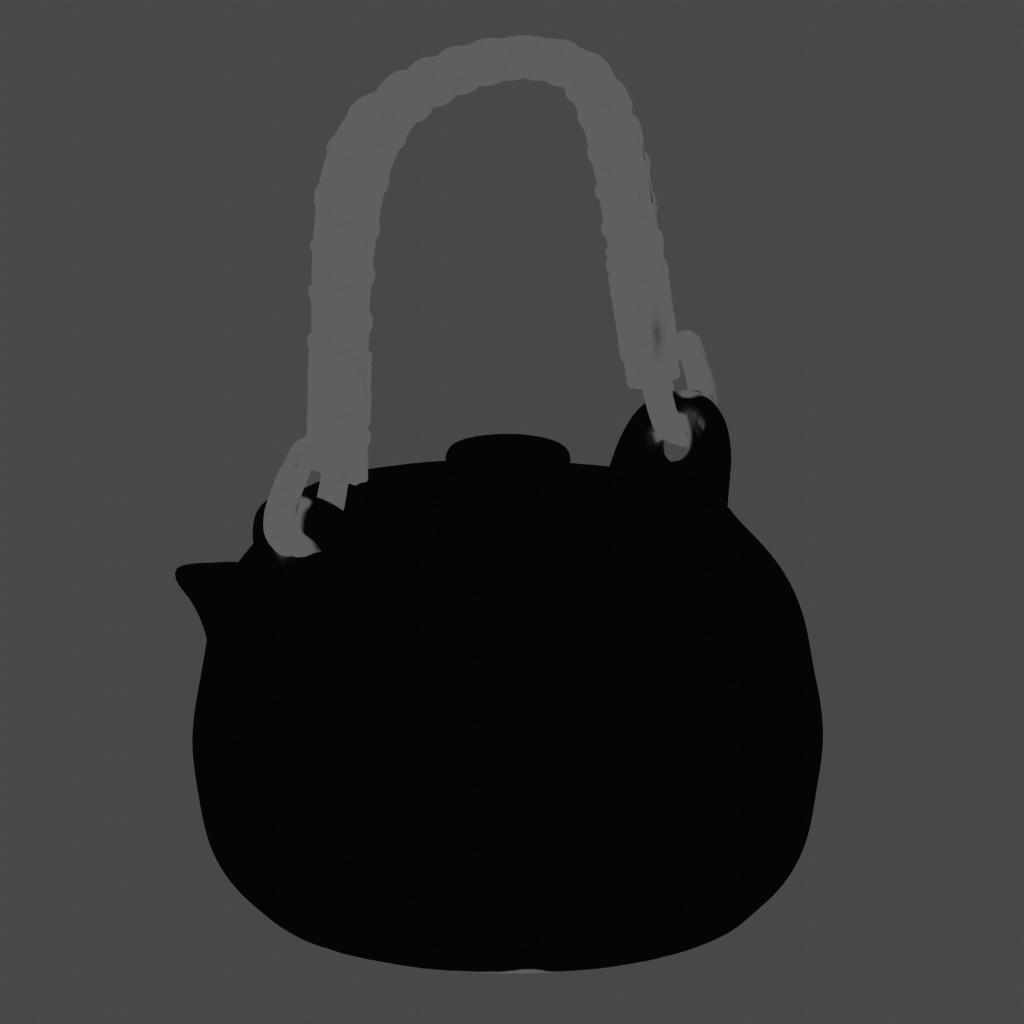}
    \includegraphics[width=0.19\linewidth]{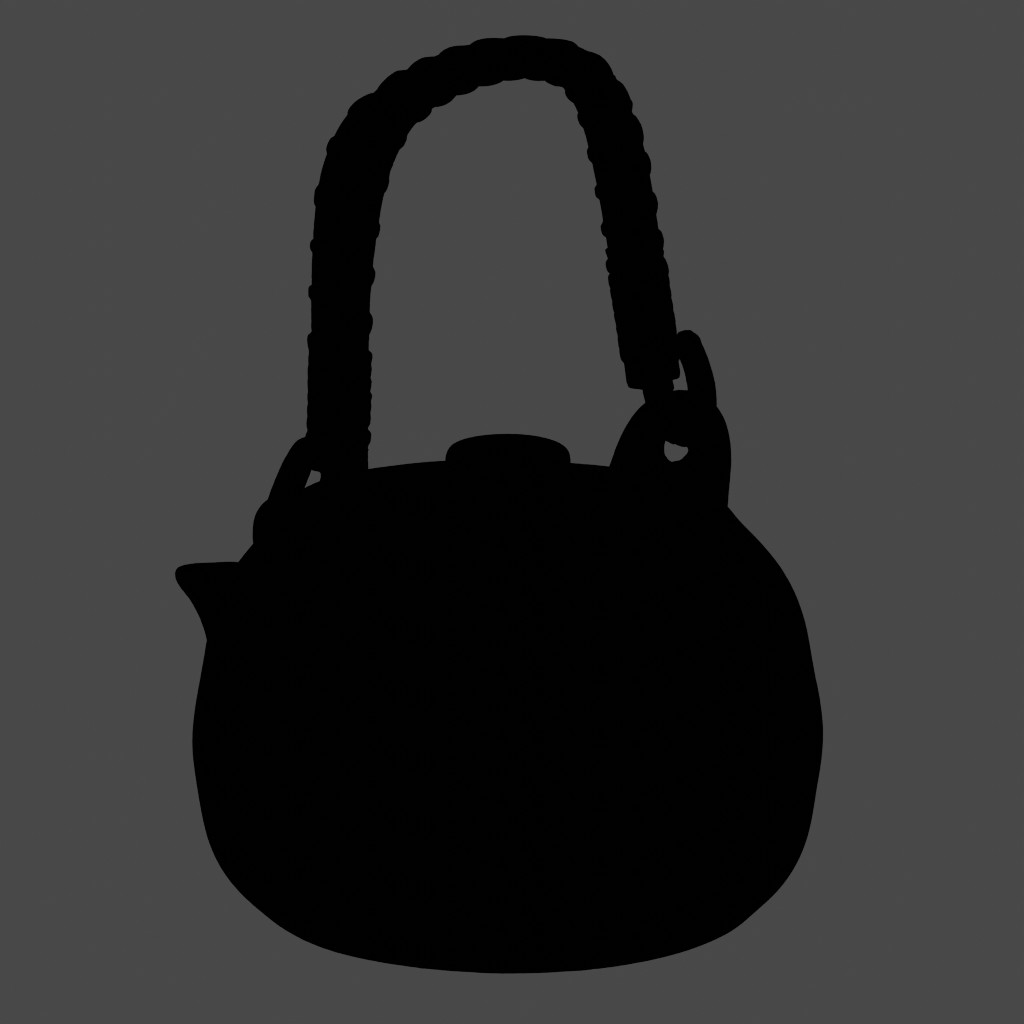}
    \includegraphics[width=0.19\linewidth]{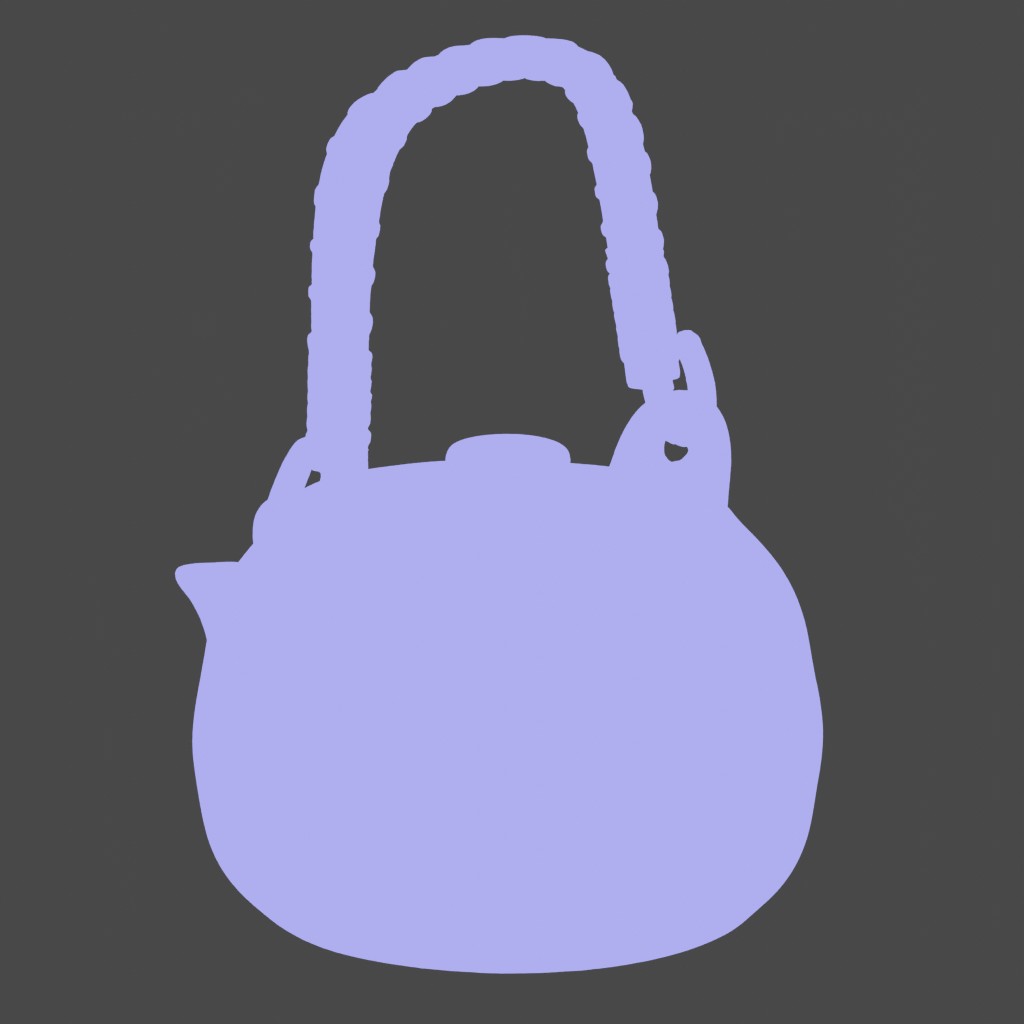}
    \includegraphics[width=0.19\linewidth]{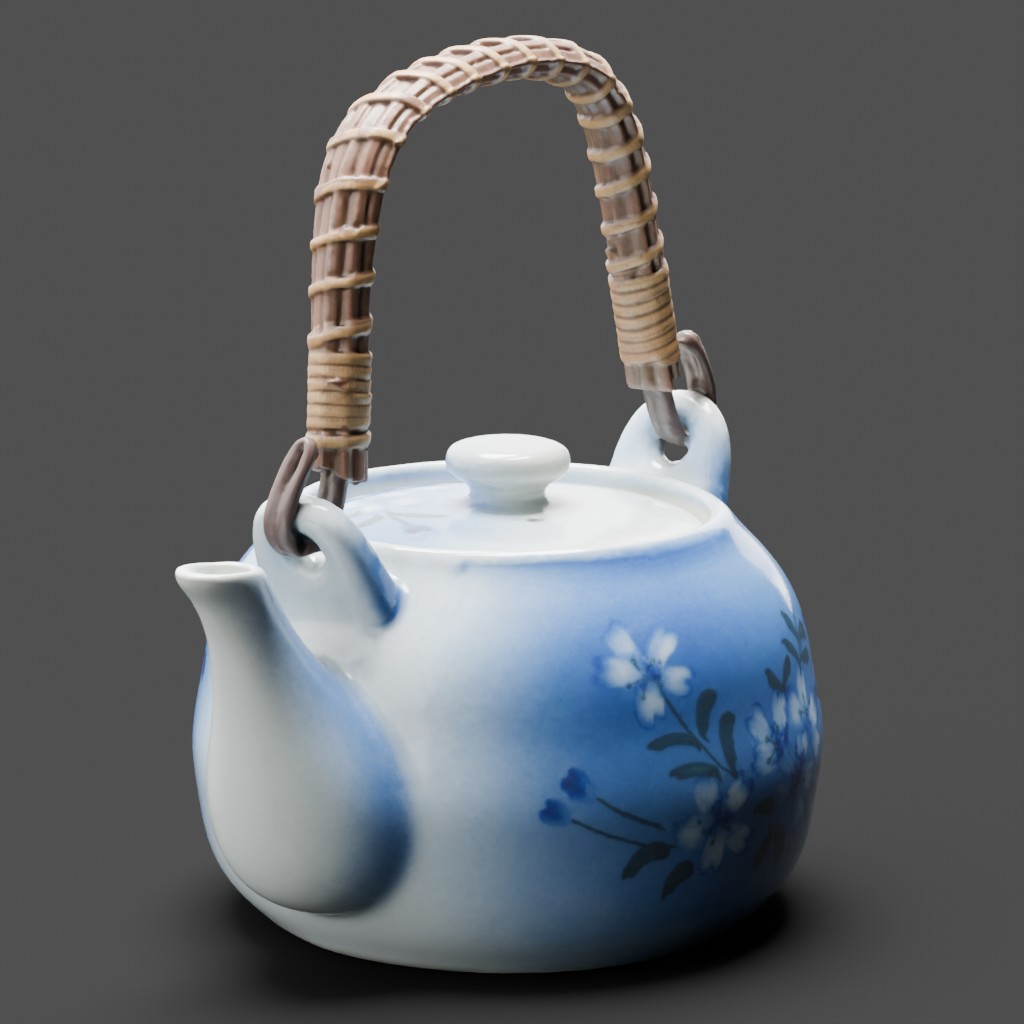} \\
    \includegraphics[width=0.19\linewidth]{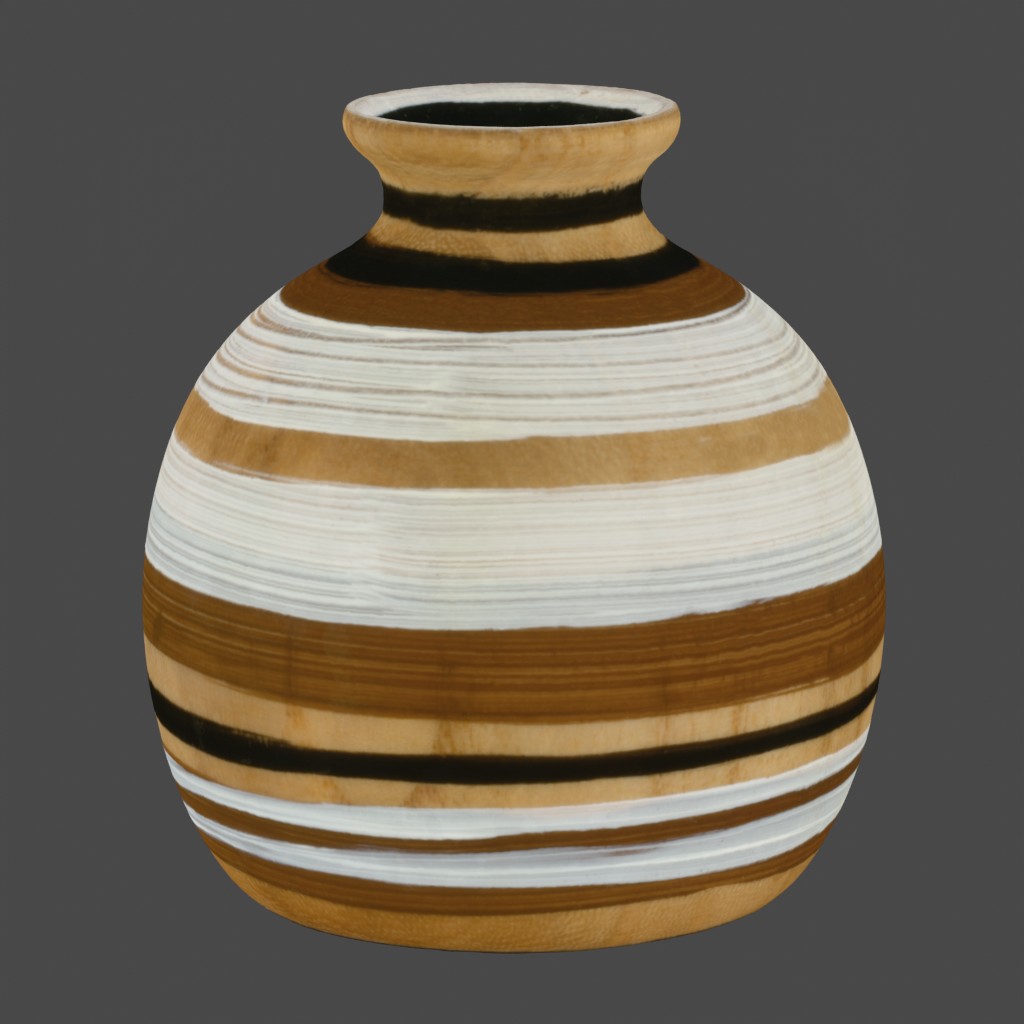}
    \includegraphics[width=0.19\linewidth]{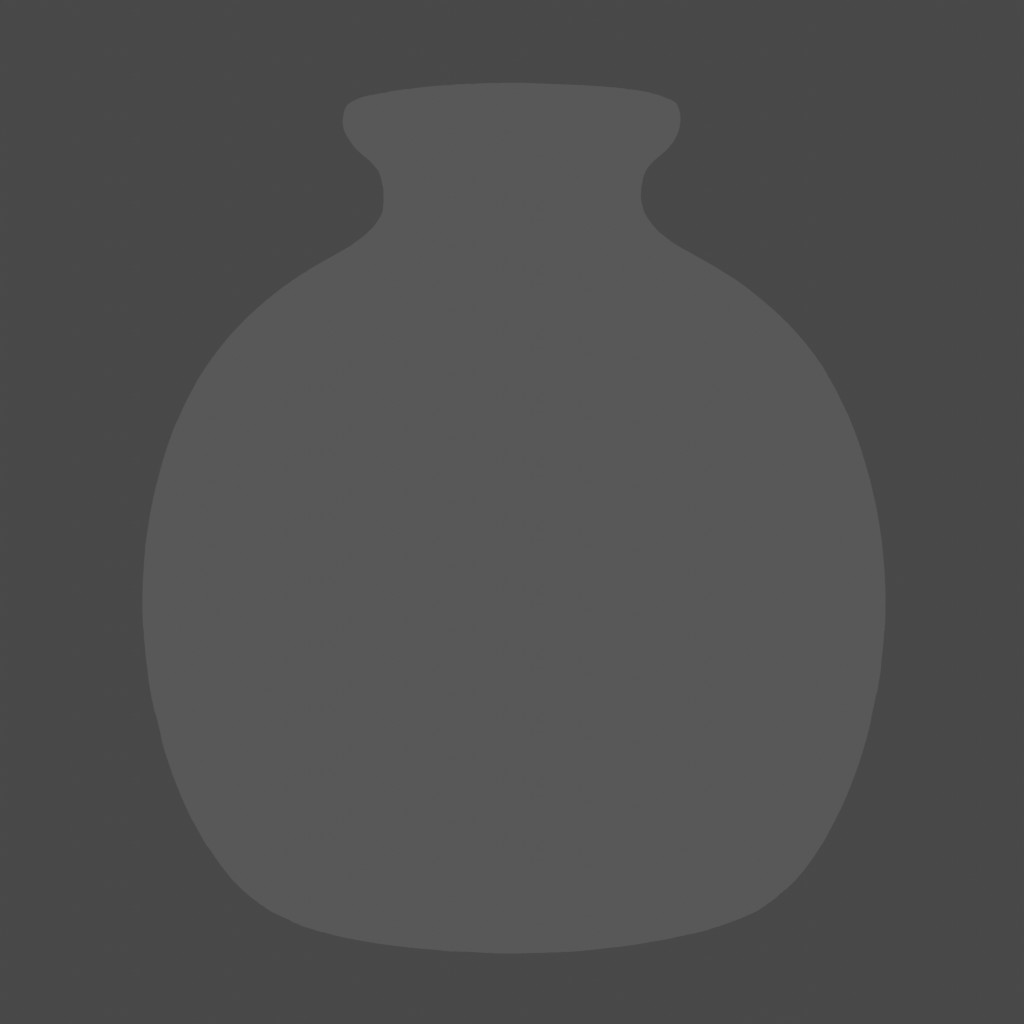}
    \includegraphics[width=0.19\linewidth]{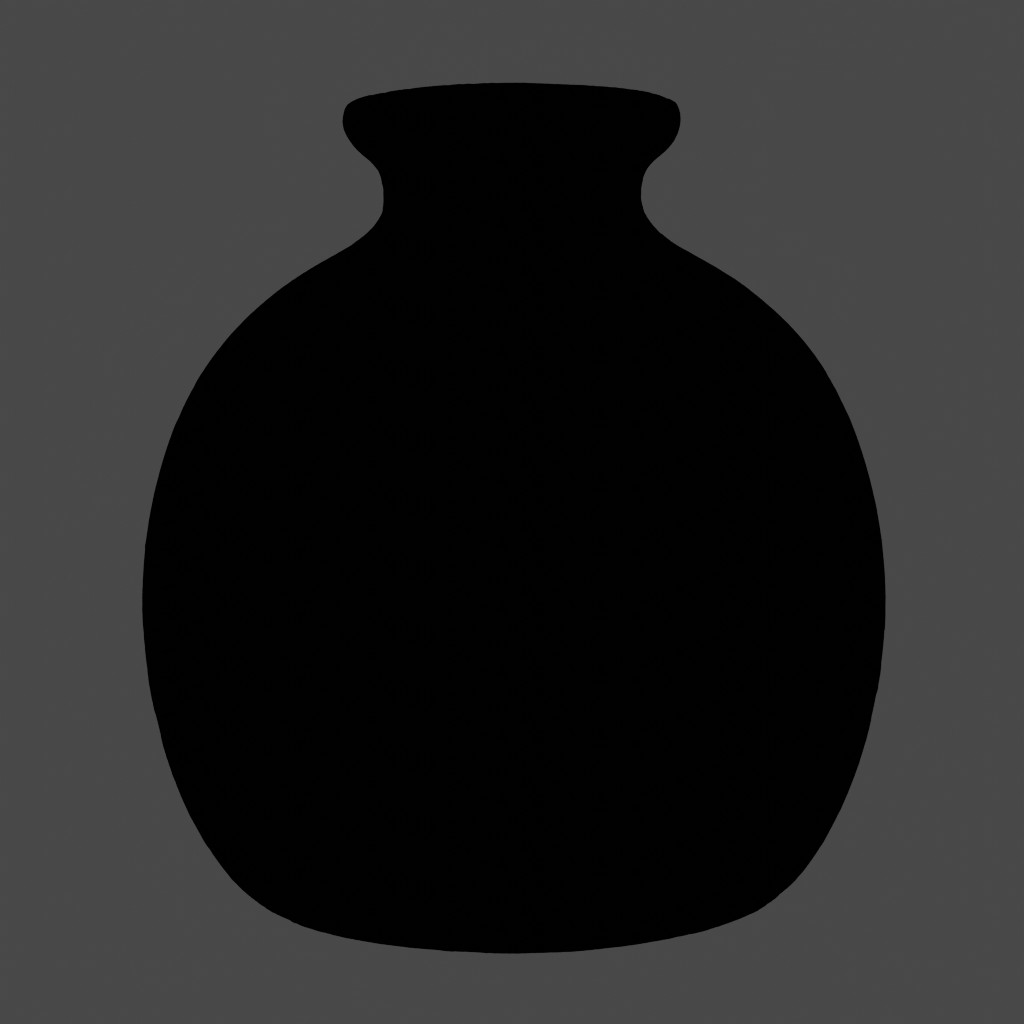}
    \includegraphics[width=0.19\linewidth]{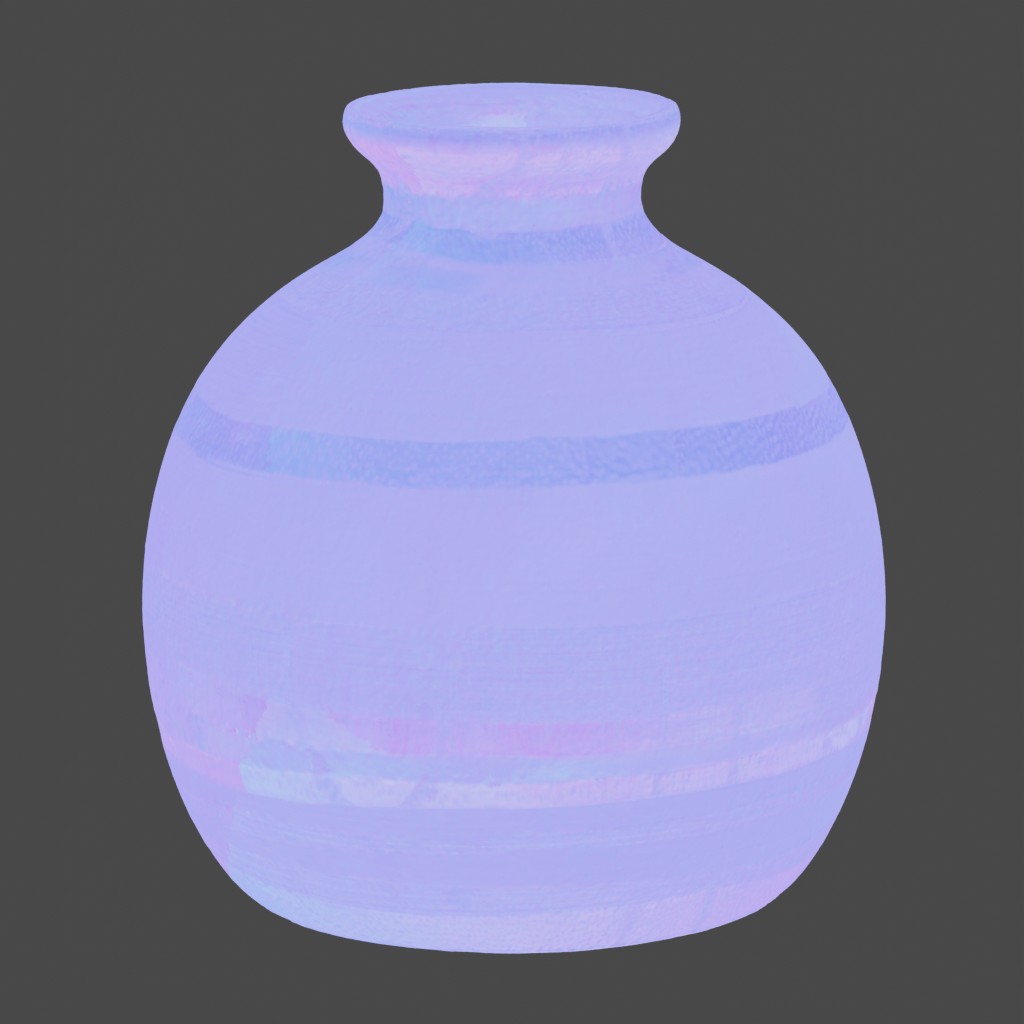}
    \includegraphics[width=0.19\linewidth]{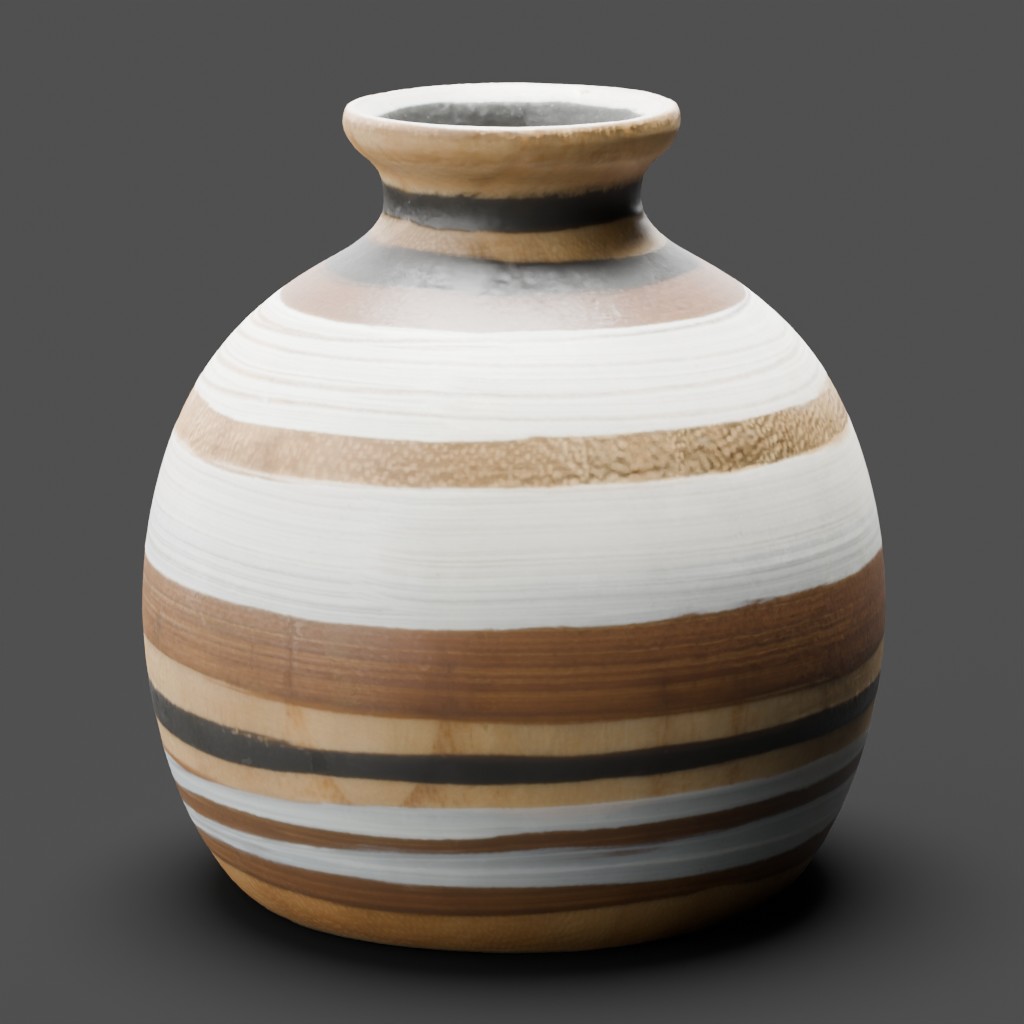} \\
    \includegraphics[width=0.19\linewidth]{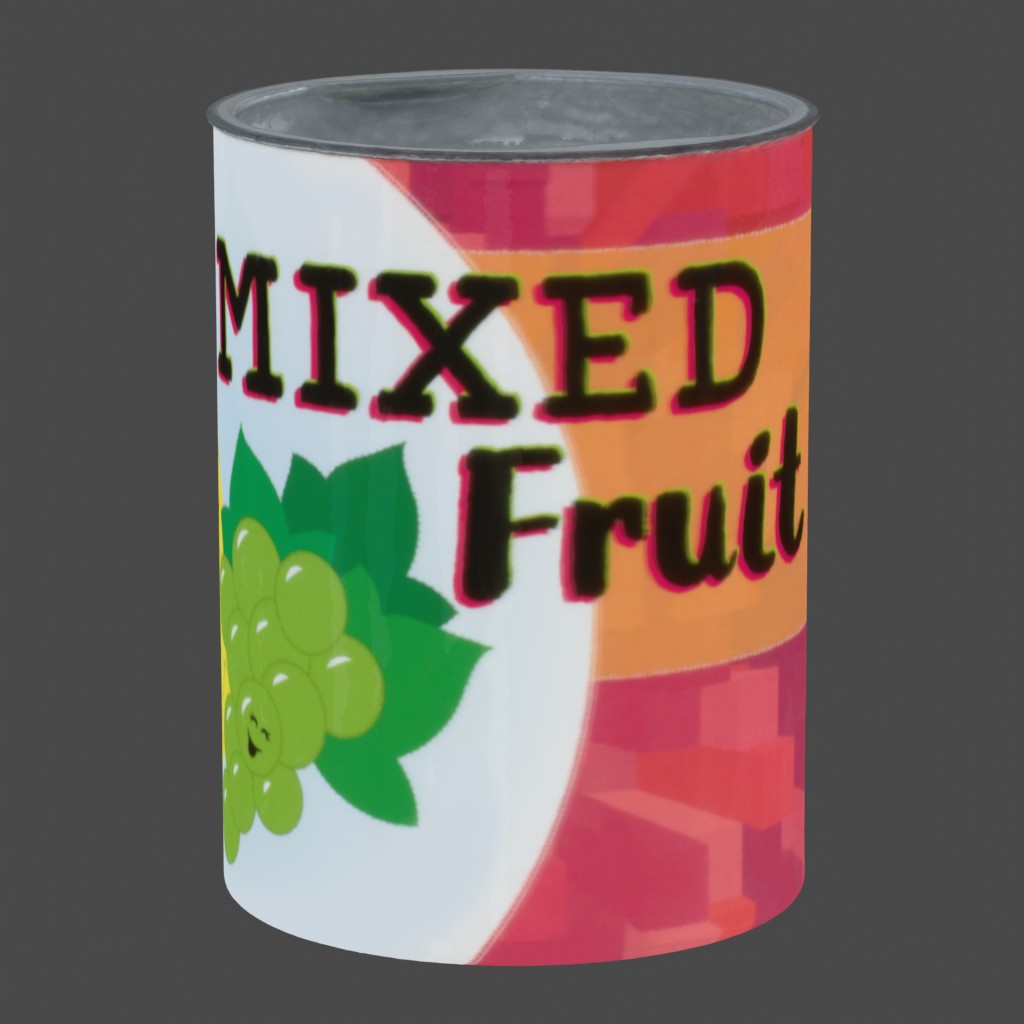}
    \includegraphics[width=0.19\linewidth]{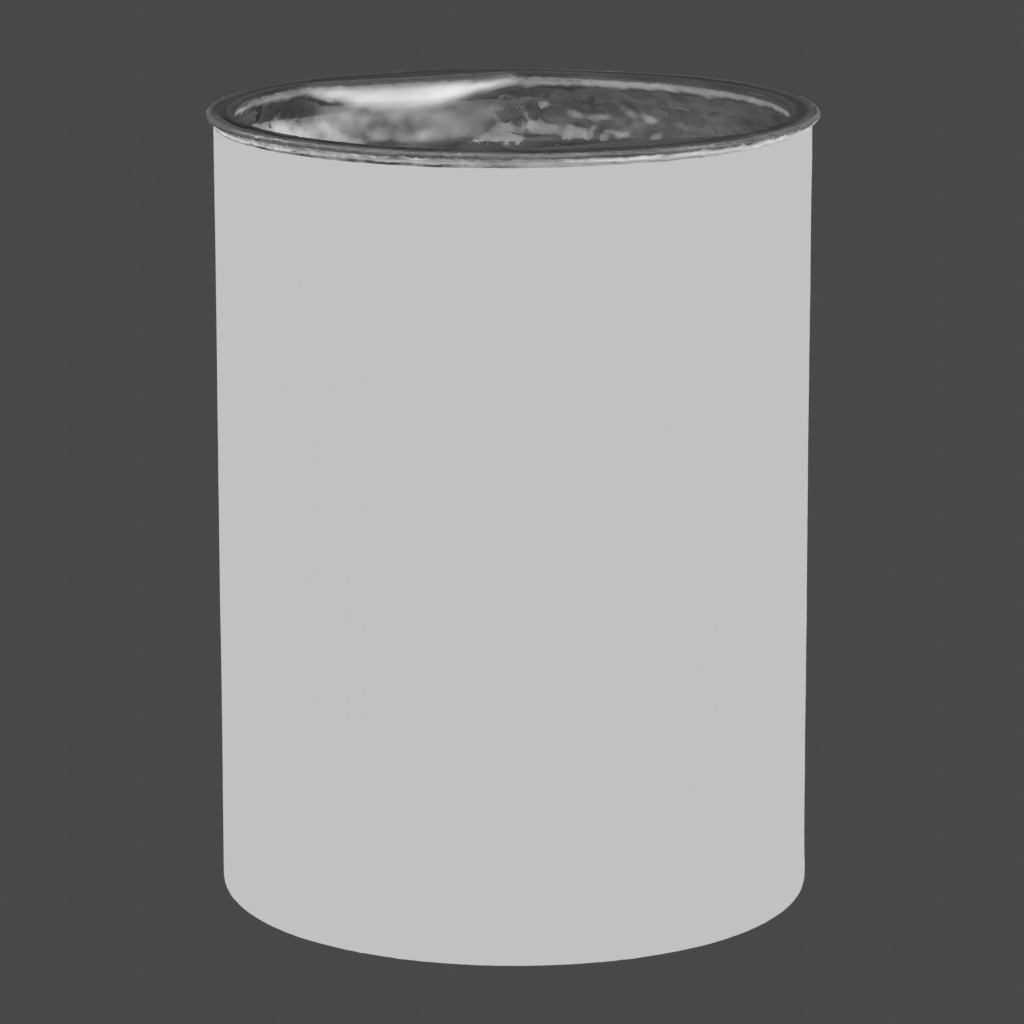}
    \includegraphics[width=0.19\linewidth]{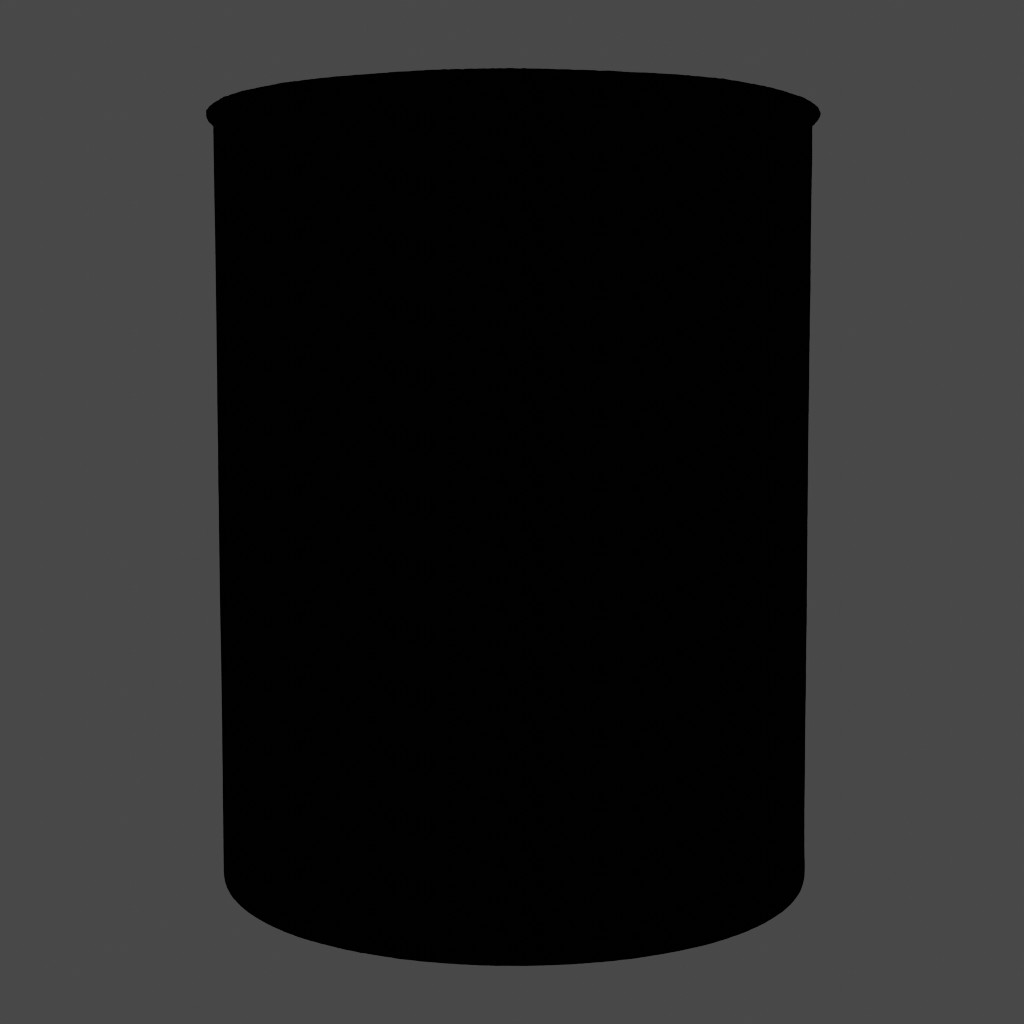}
    \includegraphics[width=0.19\linewidth]{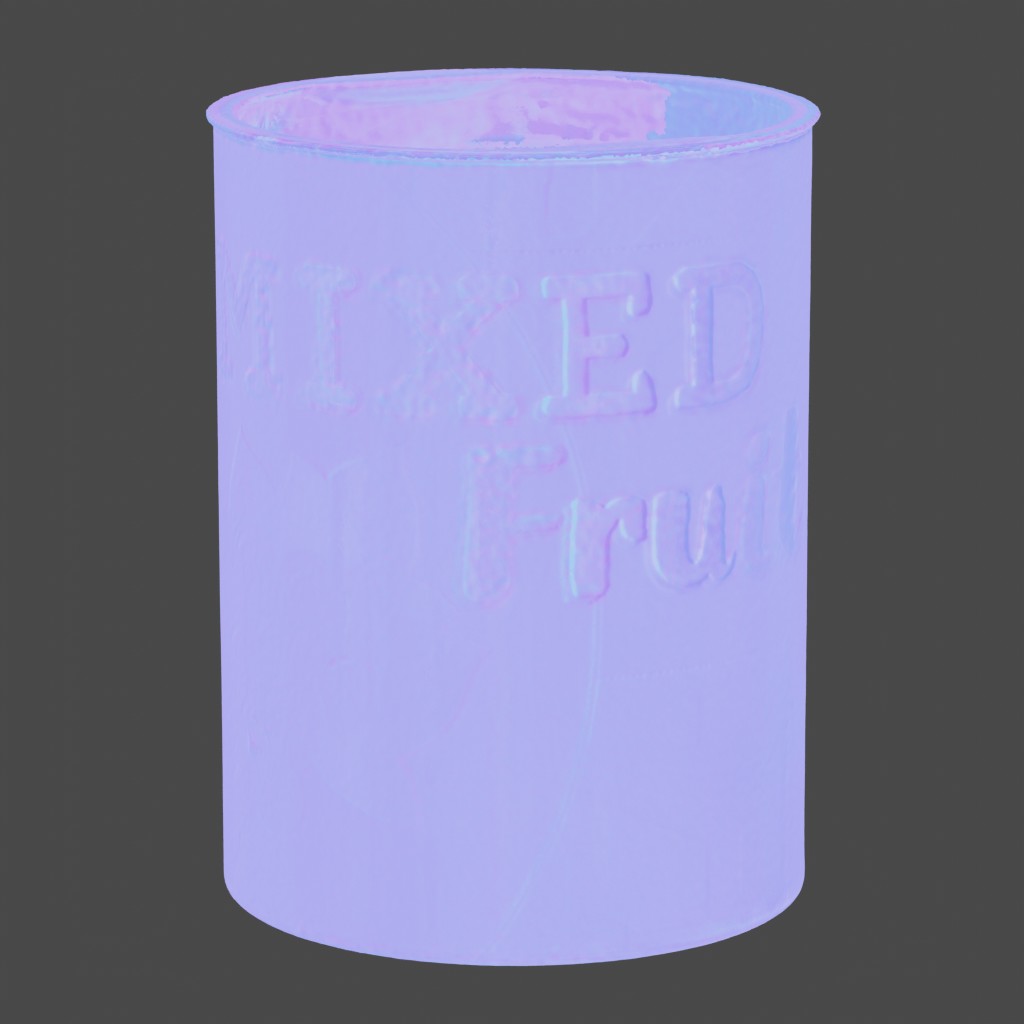}
    \includegraphics[width=0.19\linewidth]{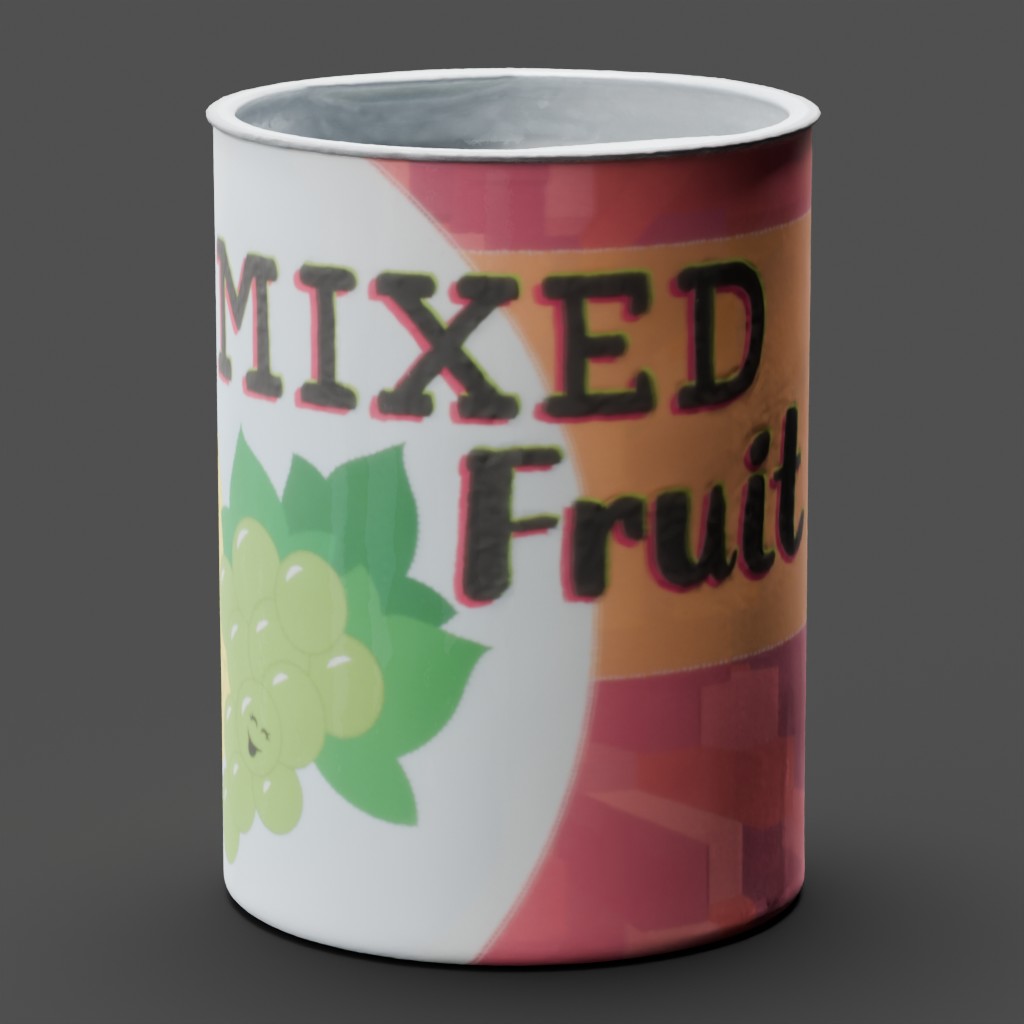} \\
    \includegraphics[width=0.19\linewidth]{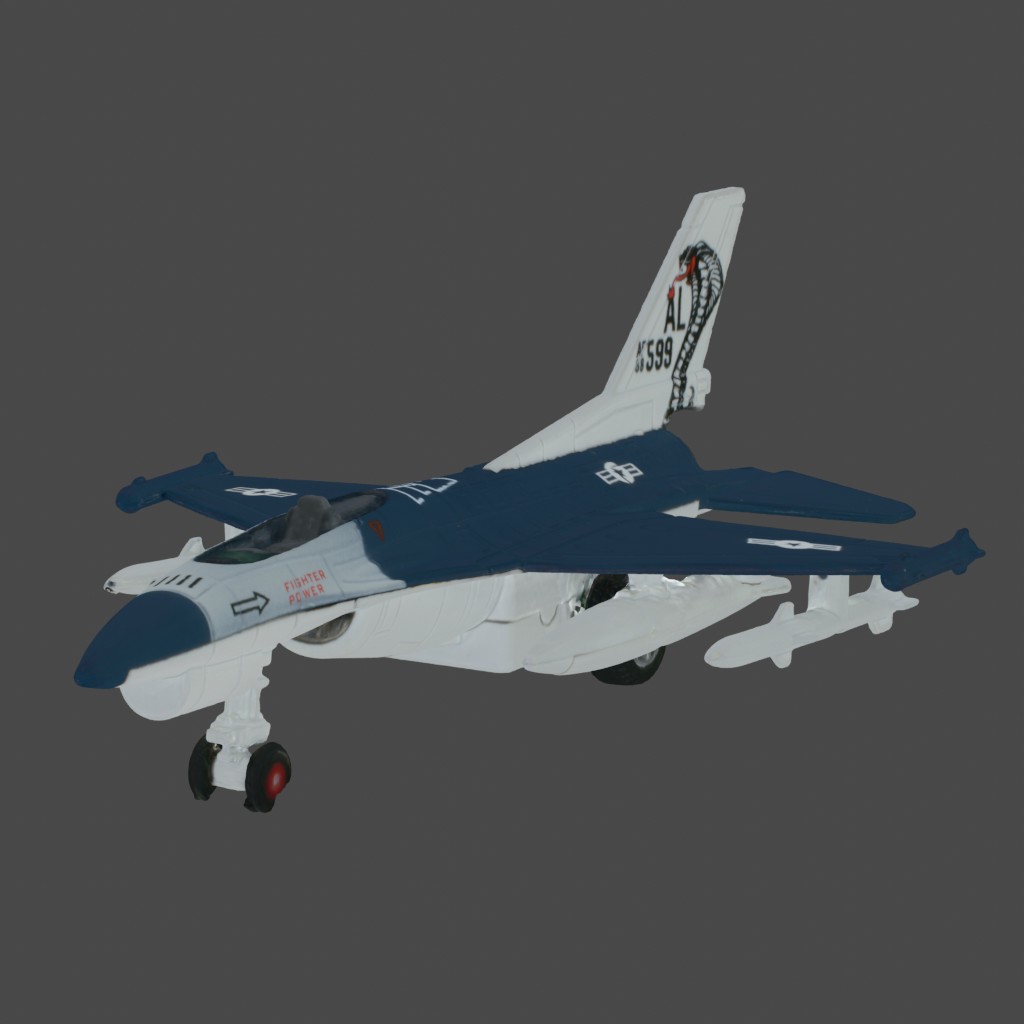}
    \includegraphics[width=0.19\linewidth]{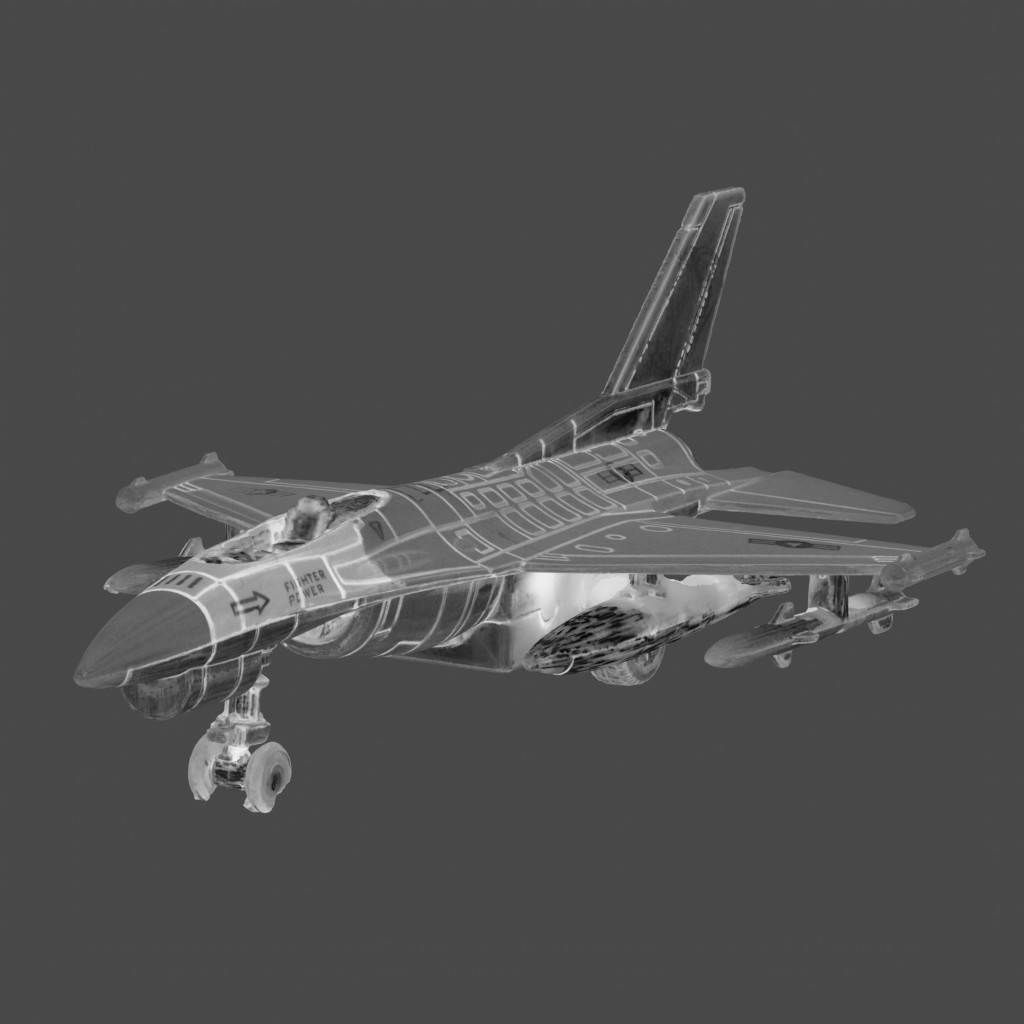}
    \includegraphics[width=0.19\linewidth]{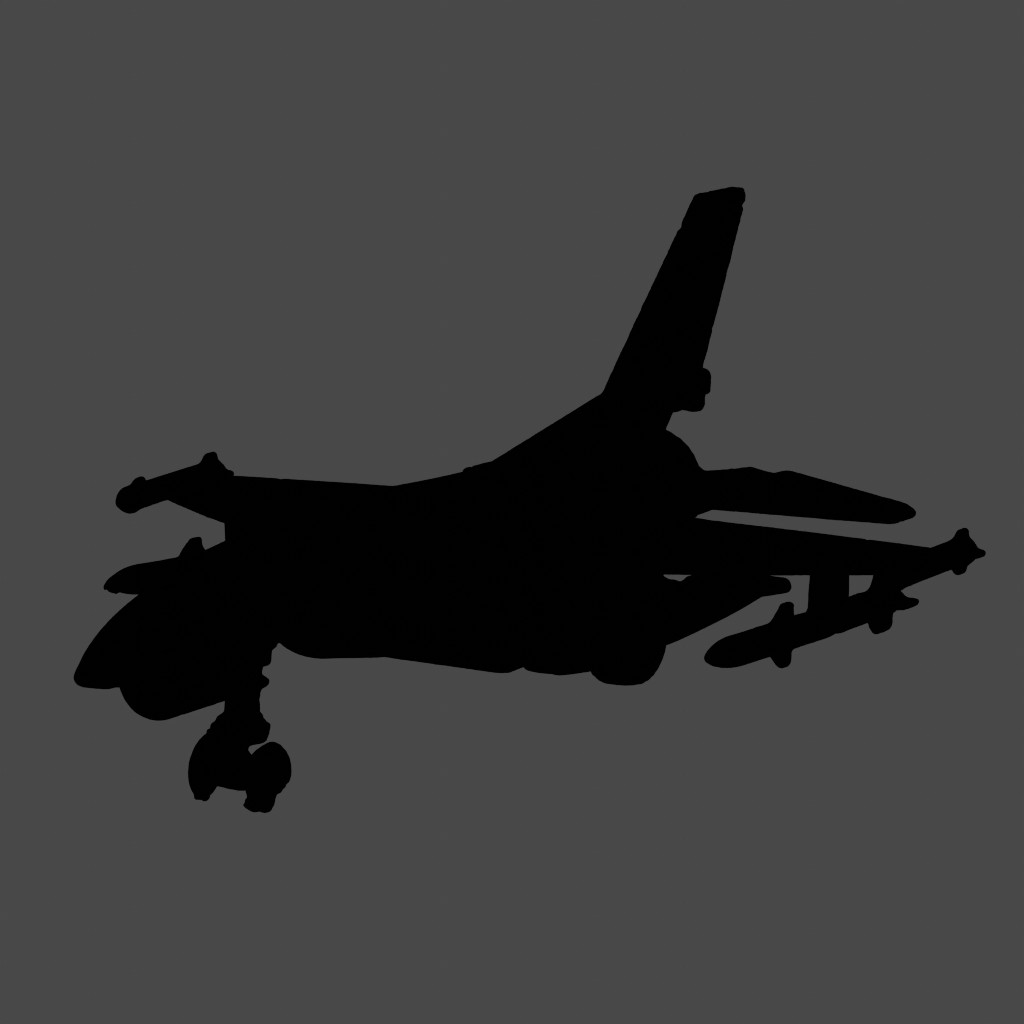}
    \includegraphics[width=0.19\linewidth]{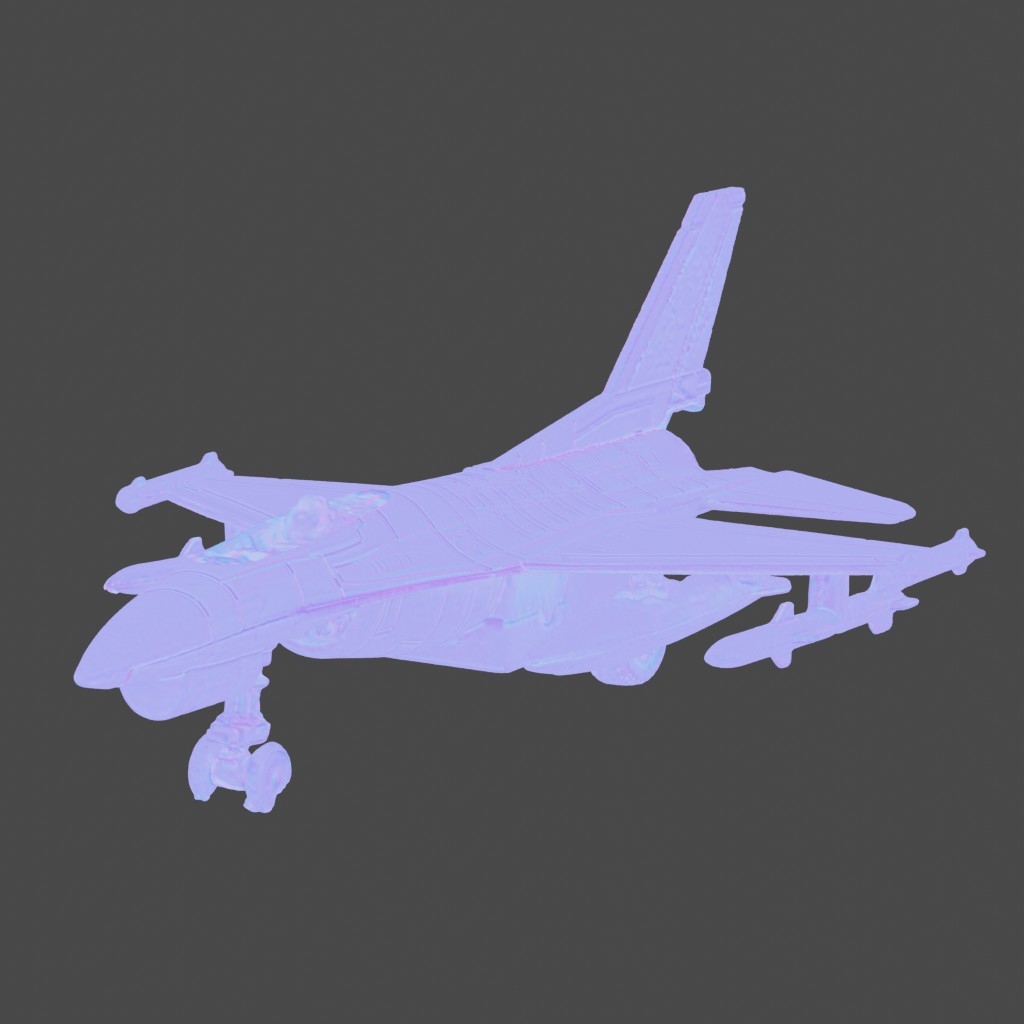}
    \includegraphics[width=0.19\linewidth]{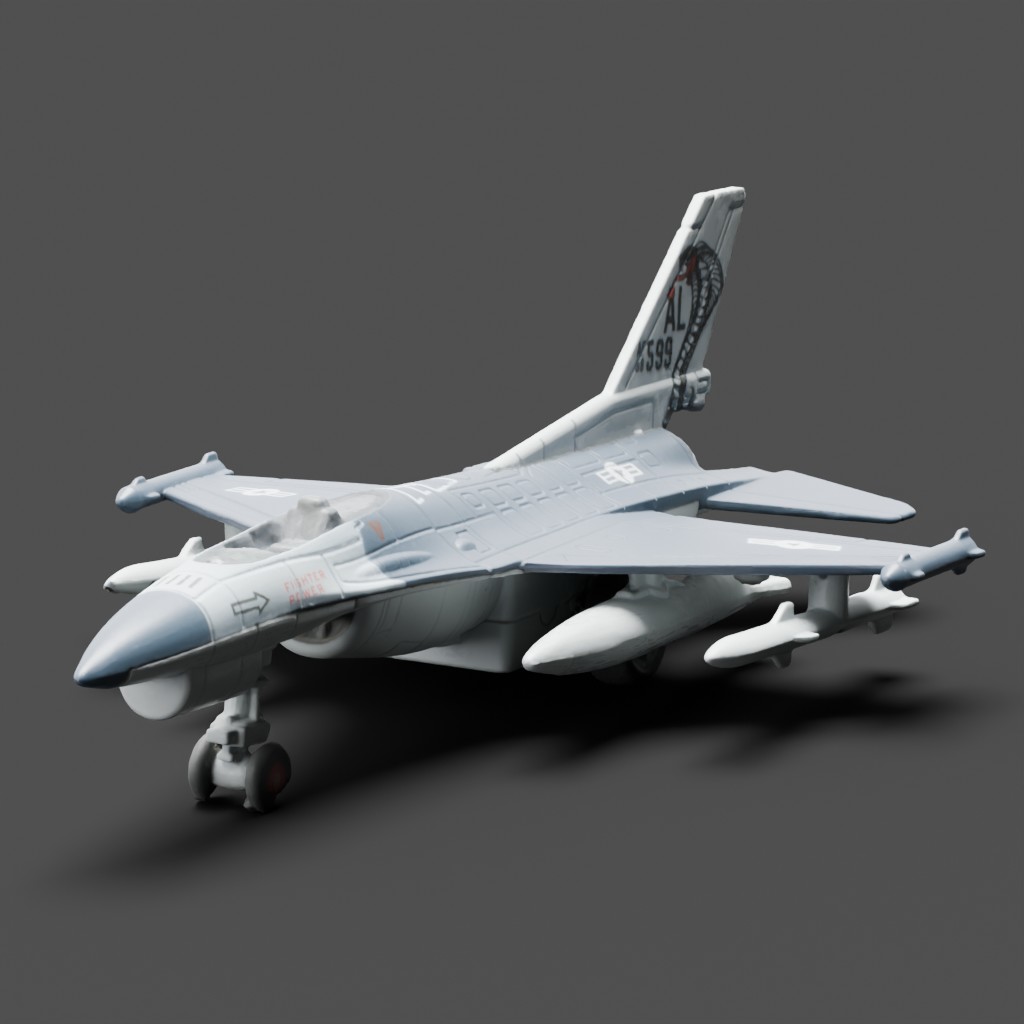} \\
    \includegraphics[width=0.19\linewidth]{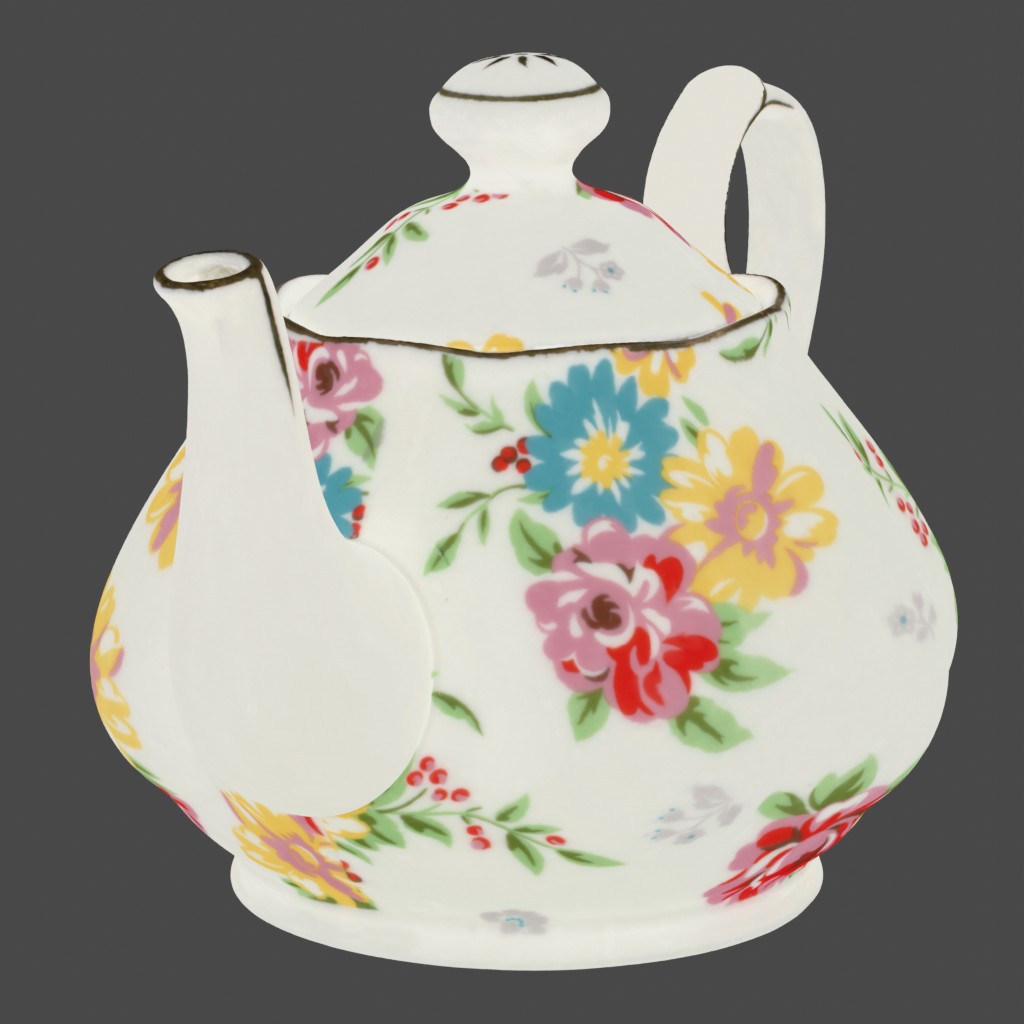}
    \includegraphics[width=0.19\linewidth]{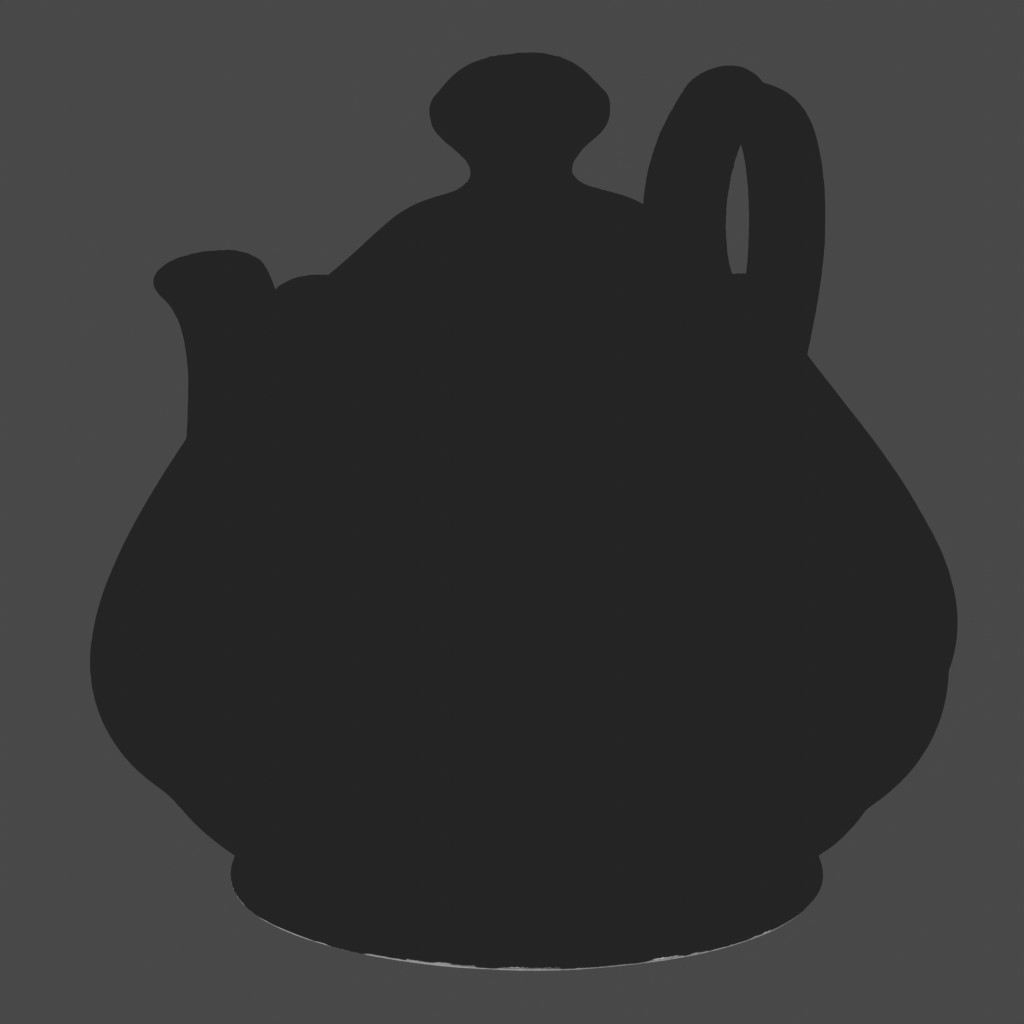}
    \includegraphics[width=0.19\linewidth]{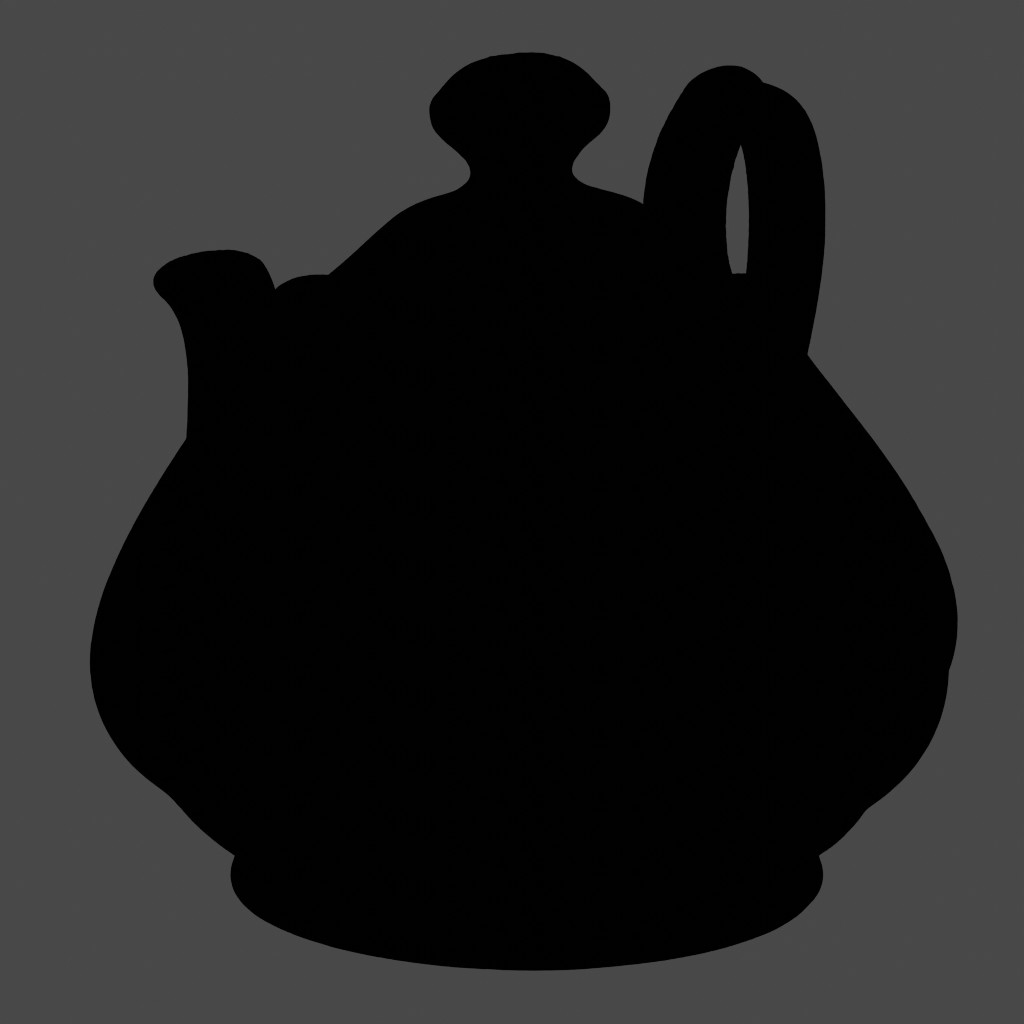}
    \includegraphics[width=0.19\linewidth]{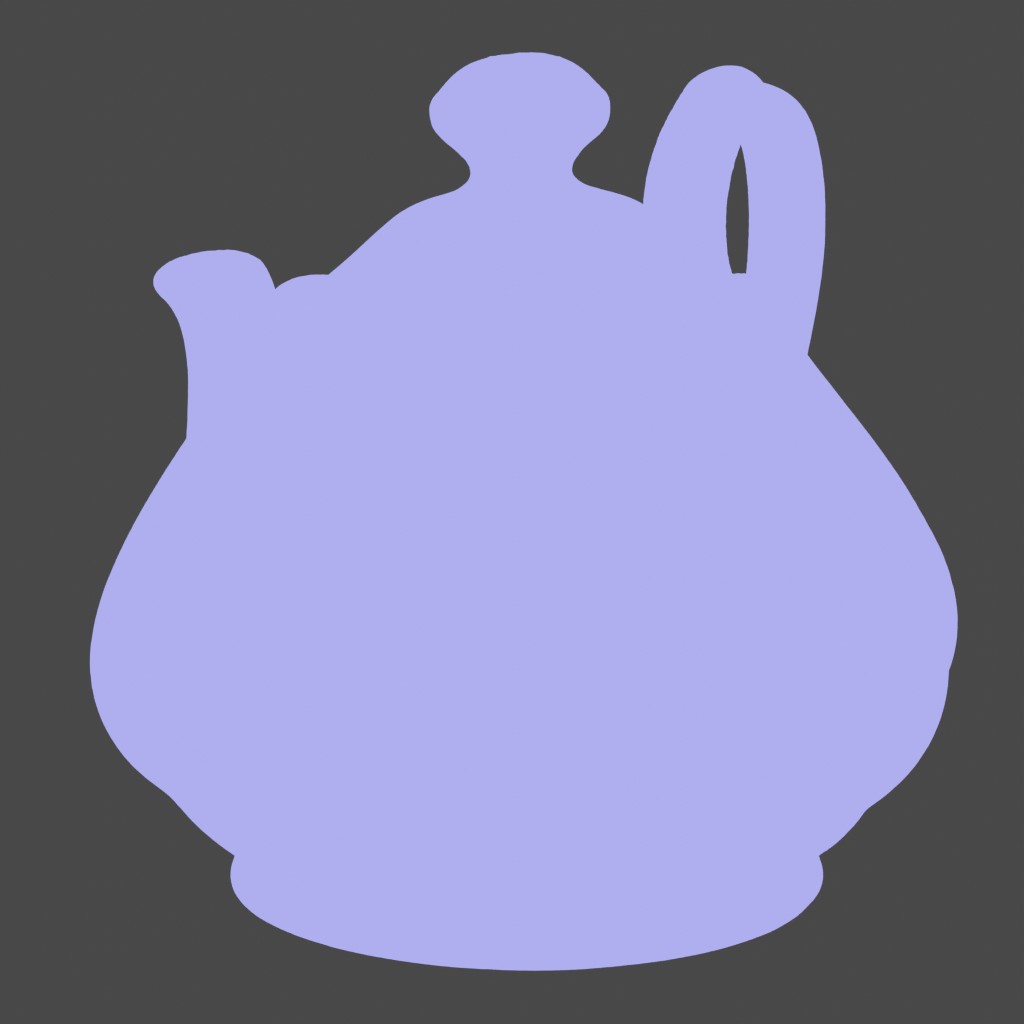}
    \includegraphics[width=0.19\linewidth]{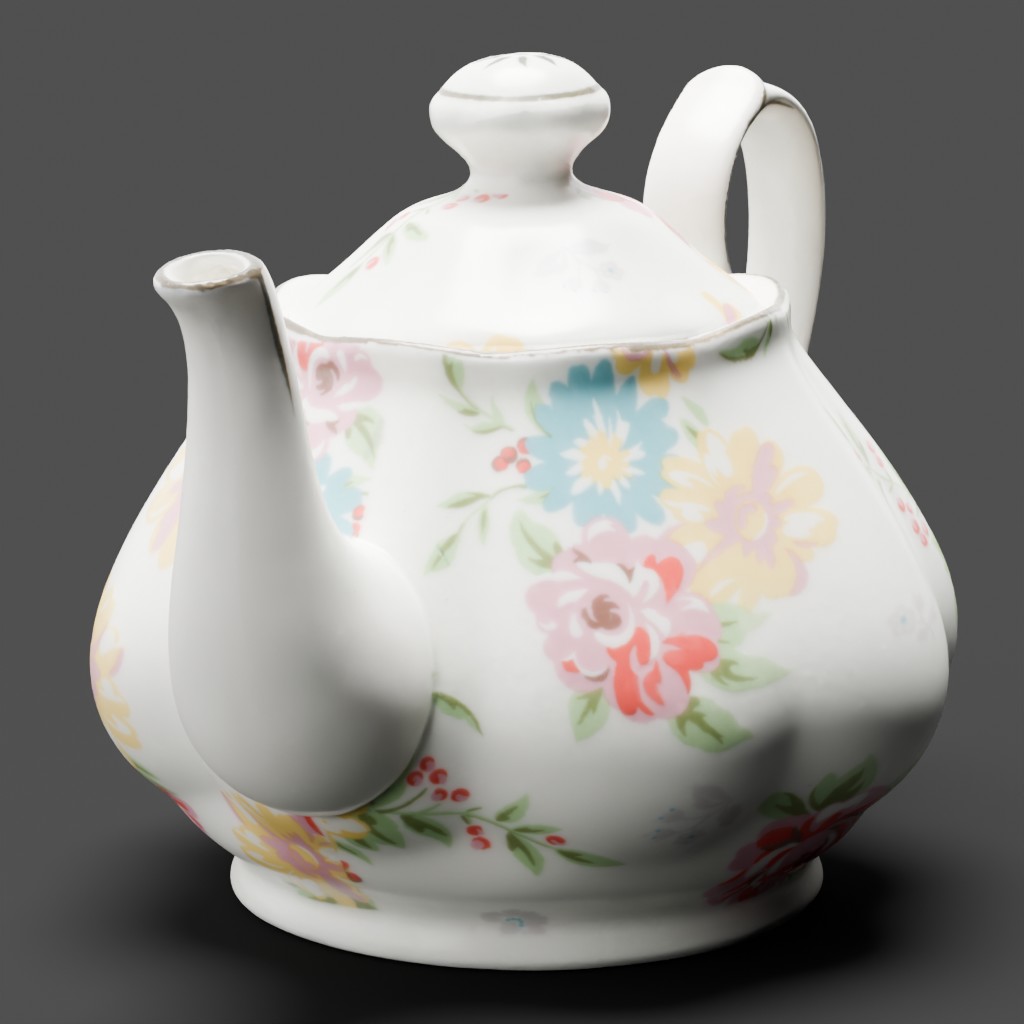}
    \caption{PBR Materials of the example DTC objects (the list of objects in Fig.9 \textbf{Row 3}). From left to right: albedo map, roughness map, metallic map, normal map, and PBR rendering.}
    \label{fig:more_PBR_maps_row3}
\end{figure*}

\begin{figure*}[t]
    \centering
    \includegraphics[width=0.19\linewidth]{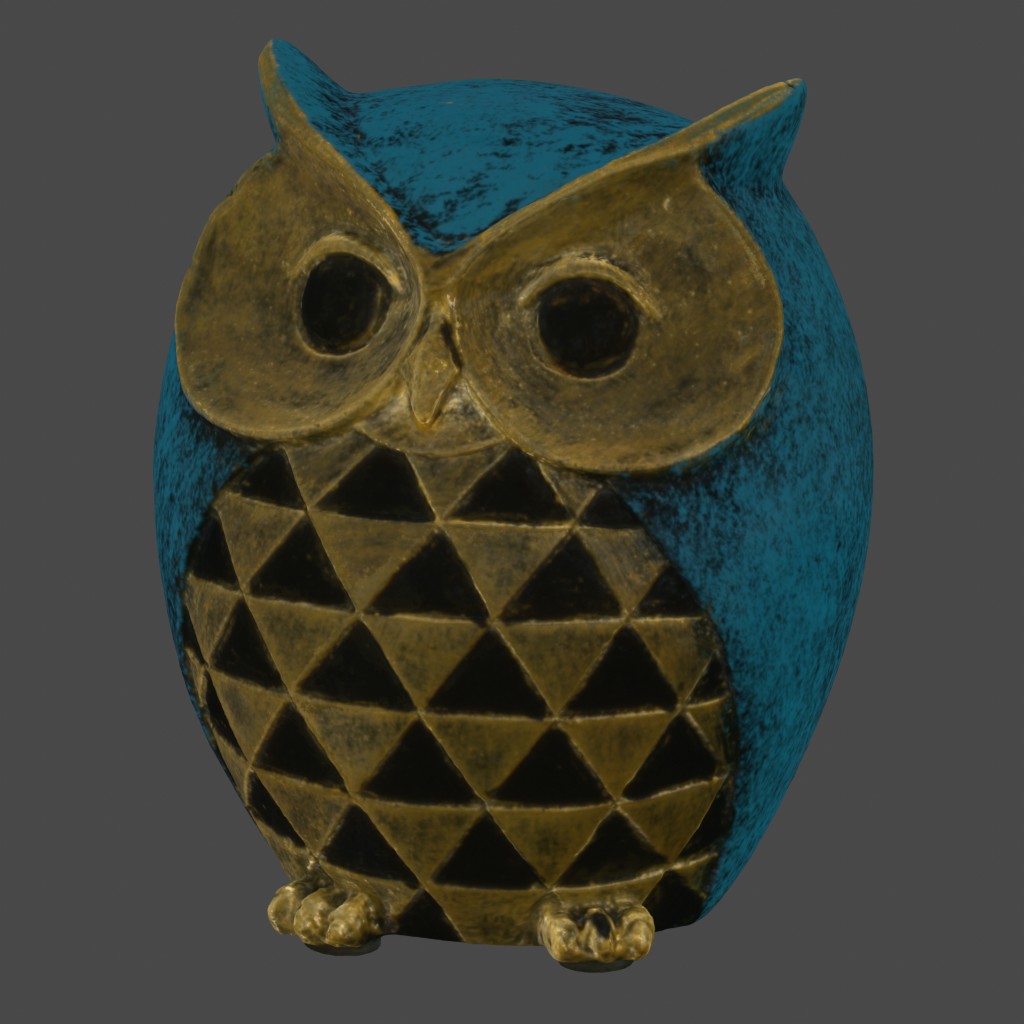}
    \includegraphics[width=0.19\linewidth]{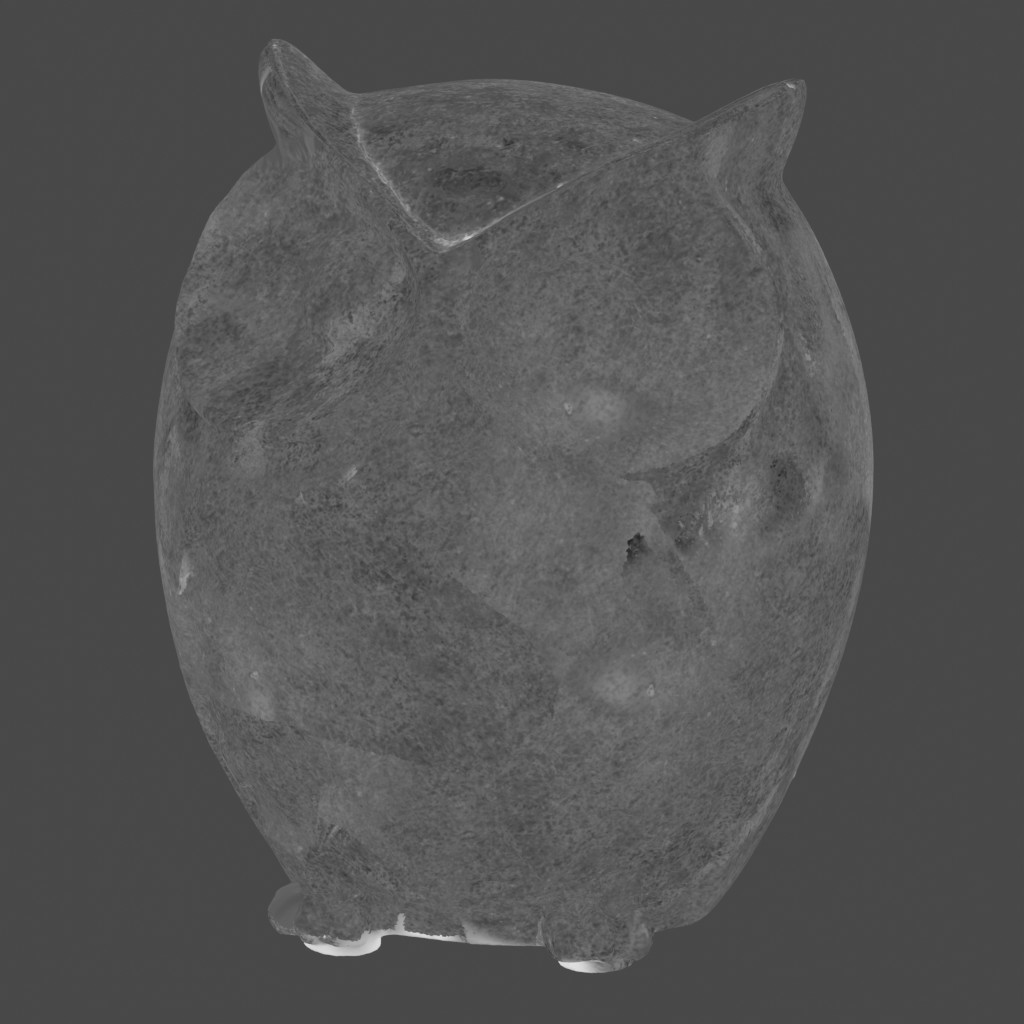}
    \includegraphics[width=0.19\linewidth]{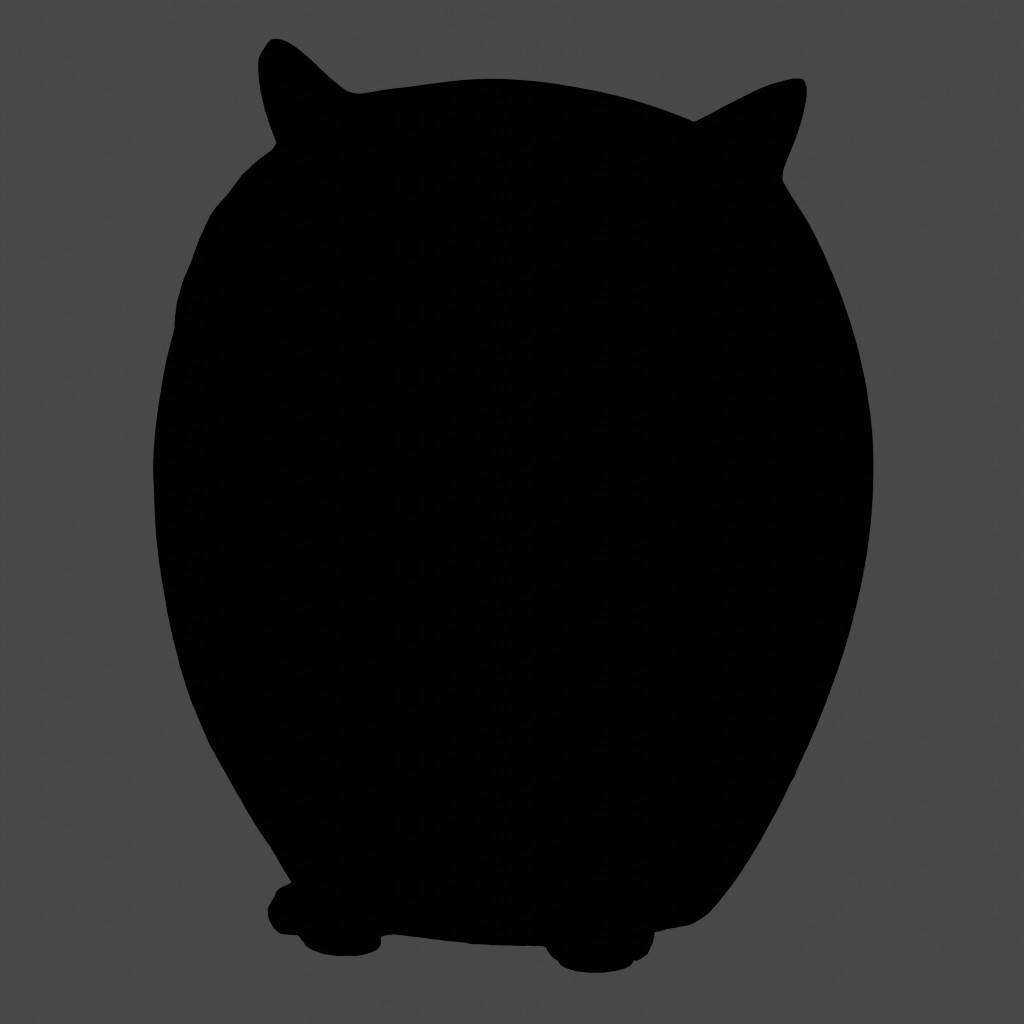}
    \includegraphics[width=0.19\linewidth]{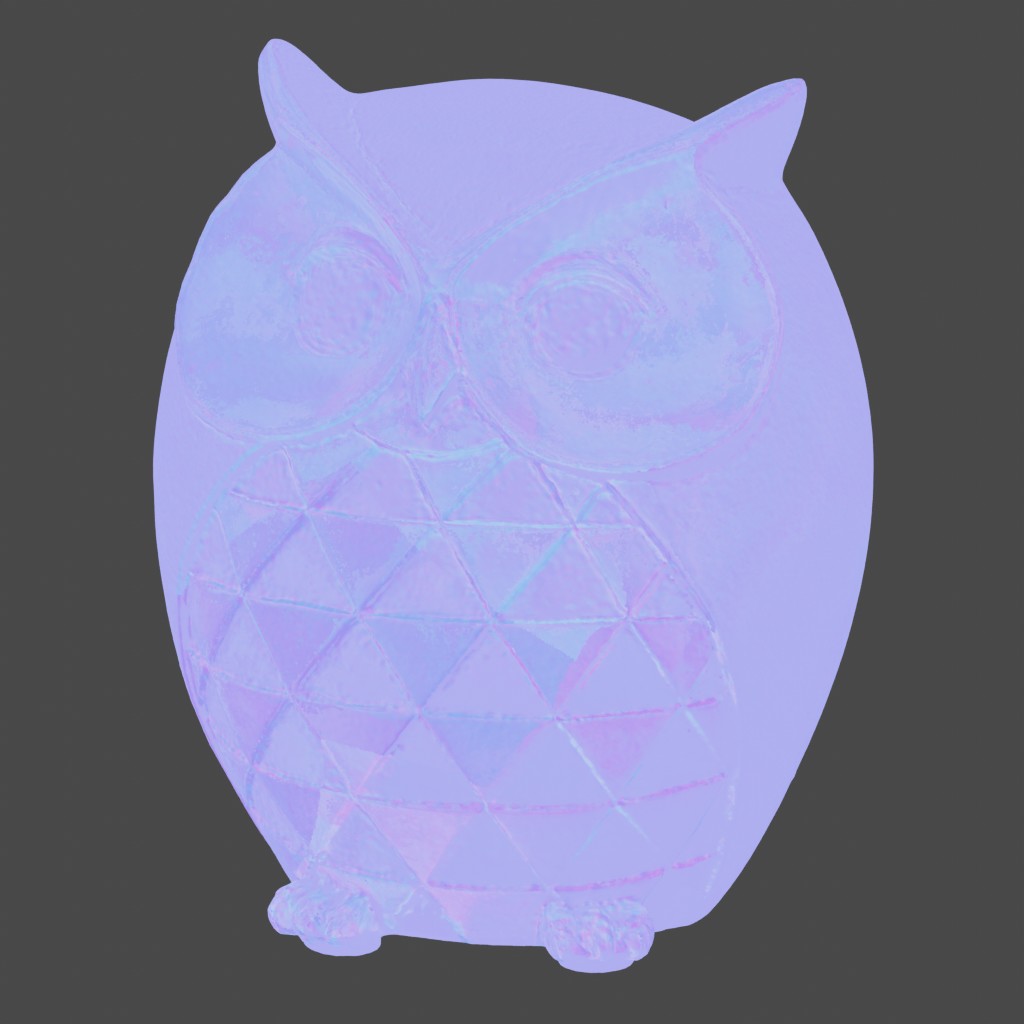}
    \includegraphics[width=0.19\linewidth]{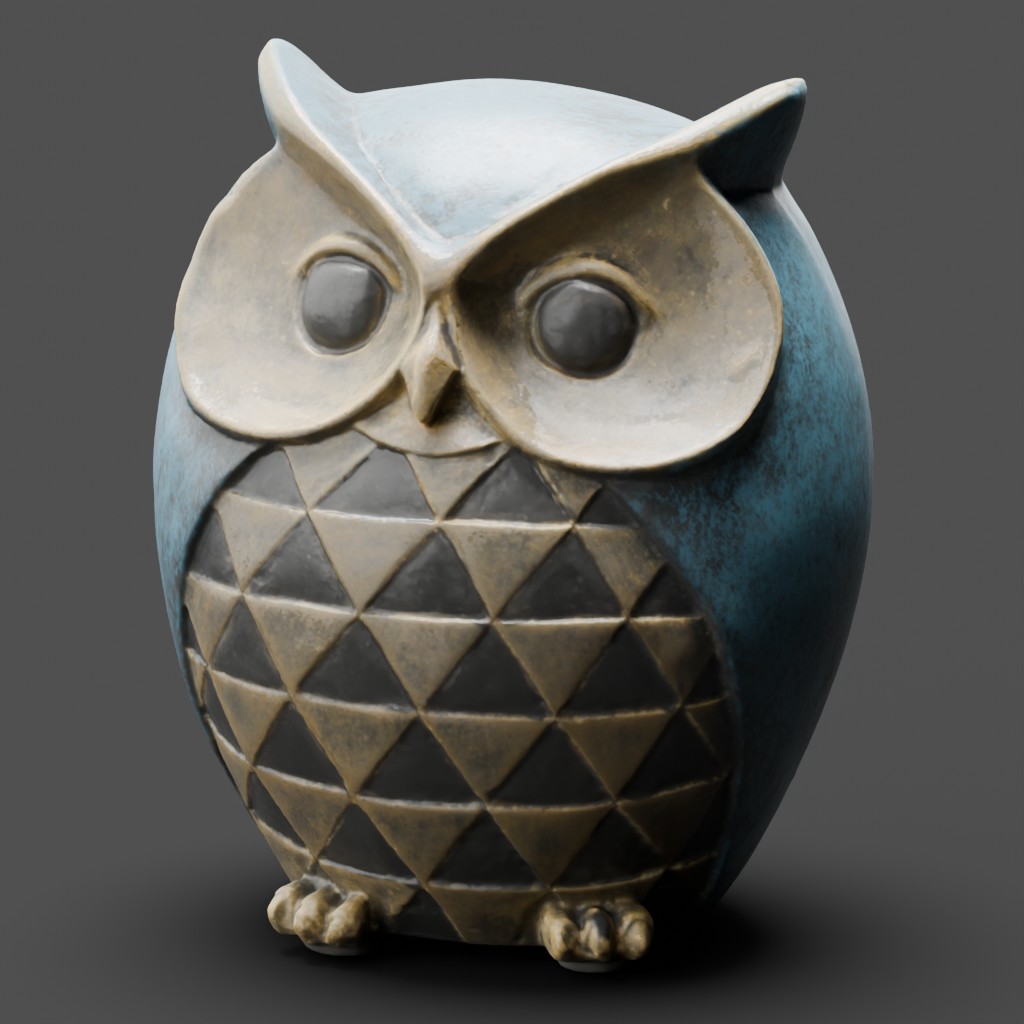} \\
    \includegraphics[width=0.19\linewidth]{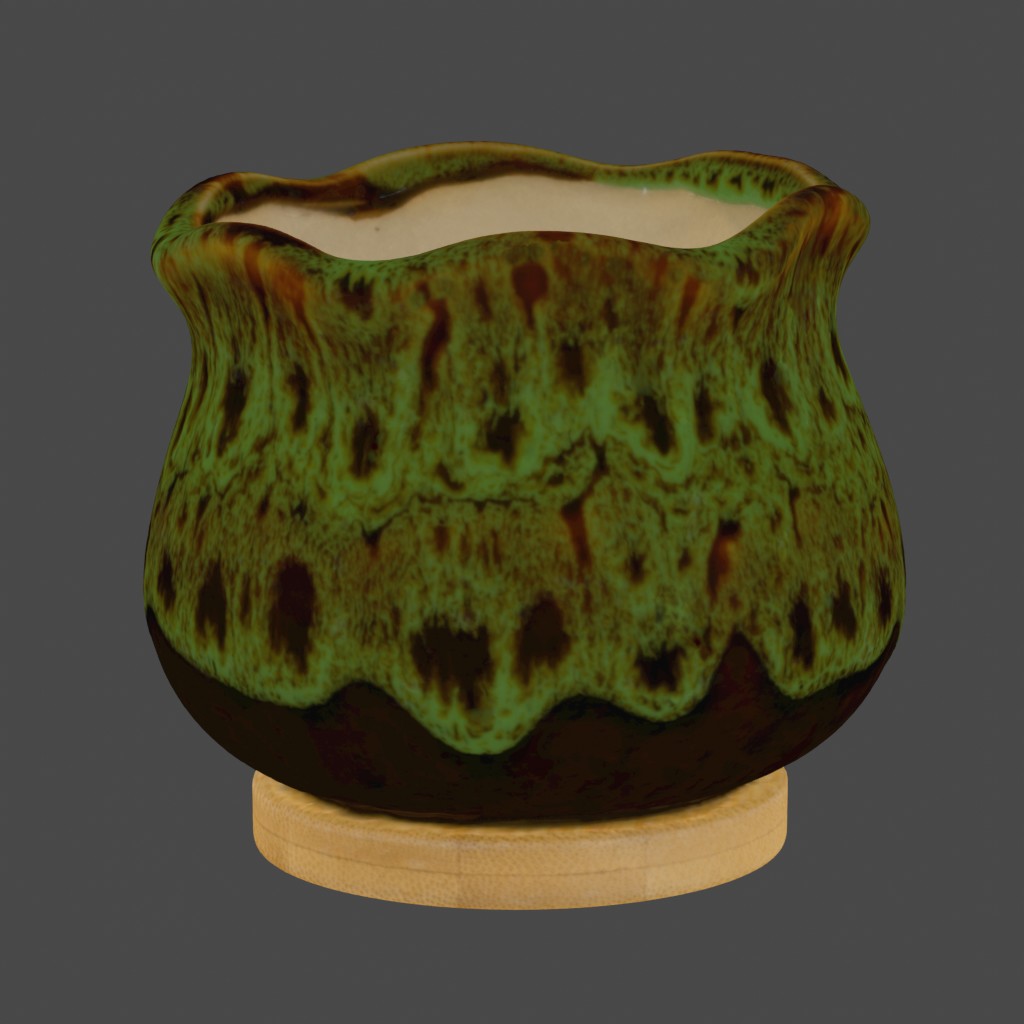}
    \includegraphics[width=0.19\linewidth]{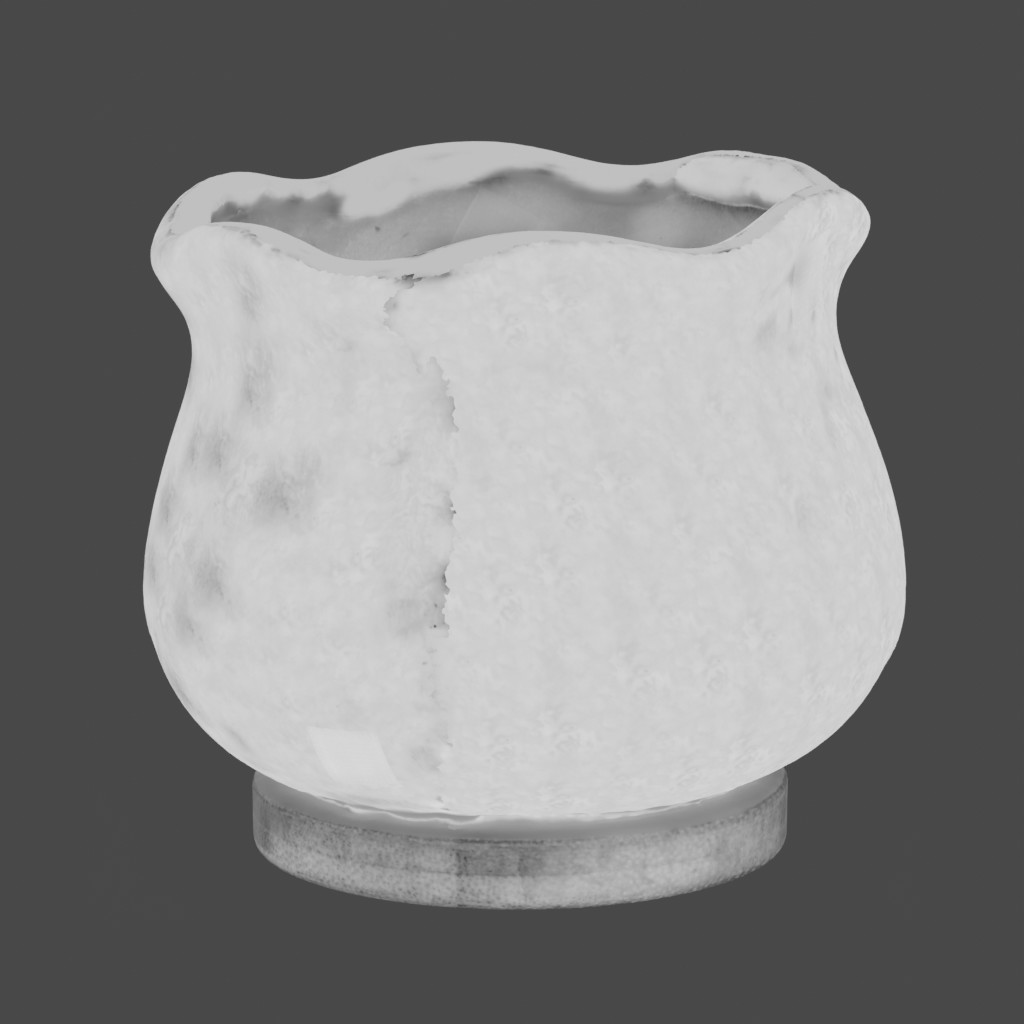}
    \includegraphics[width=0.19\linewidth]{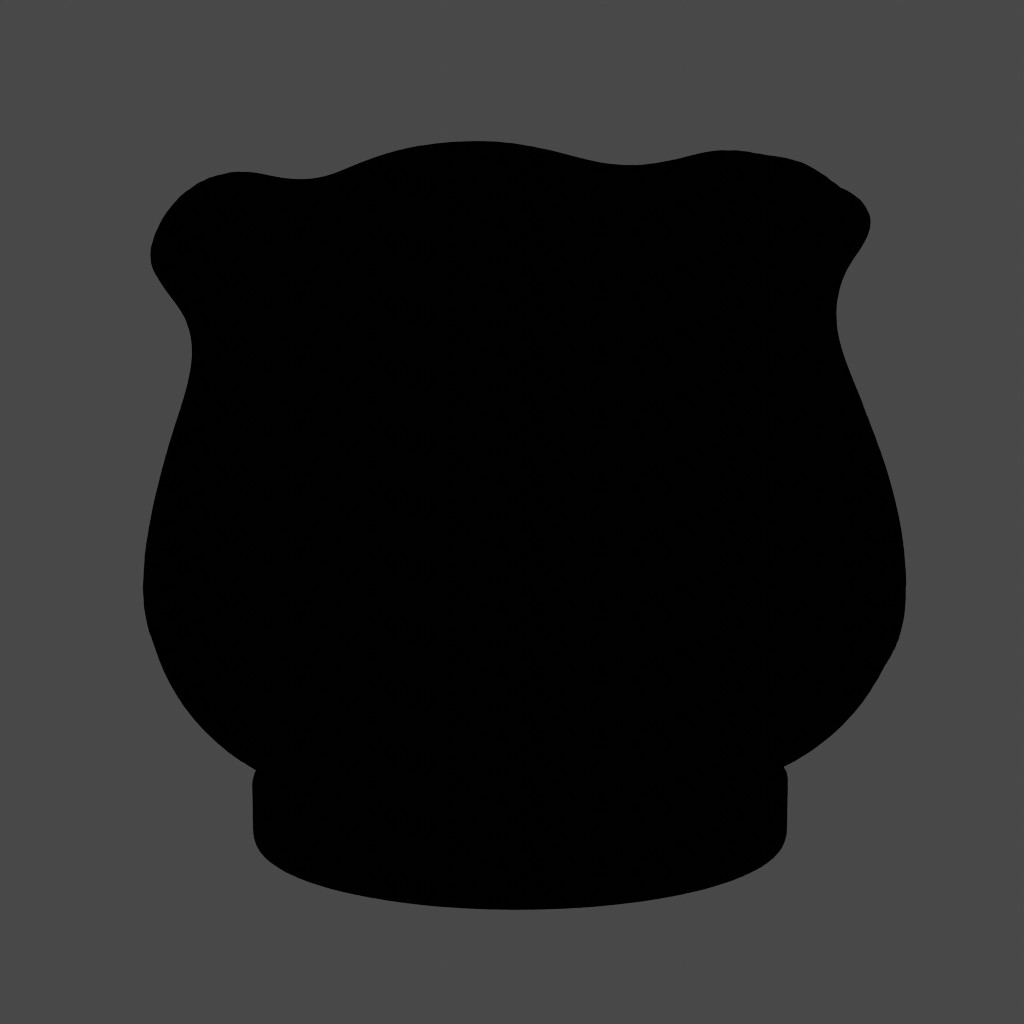}
    \includegraphics[width=0.19\linewidth]{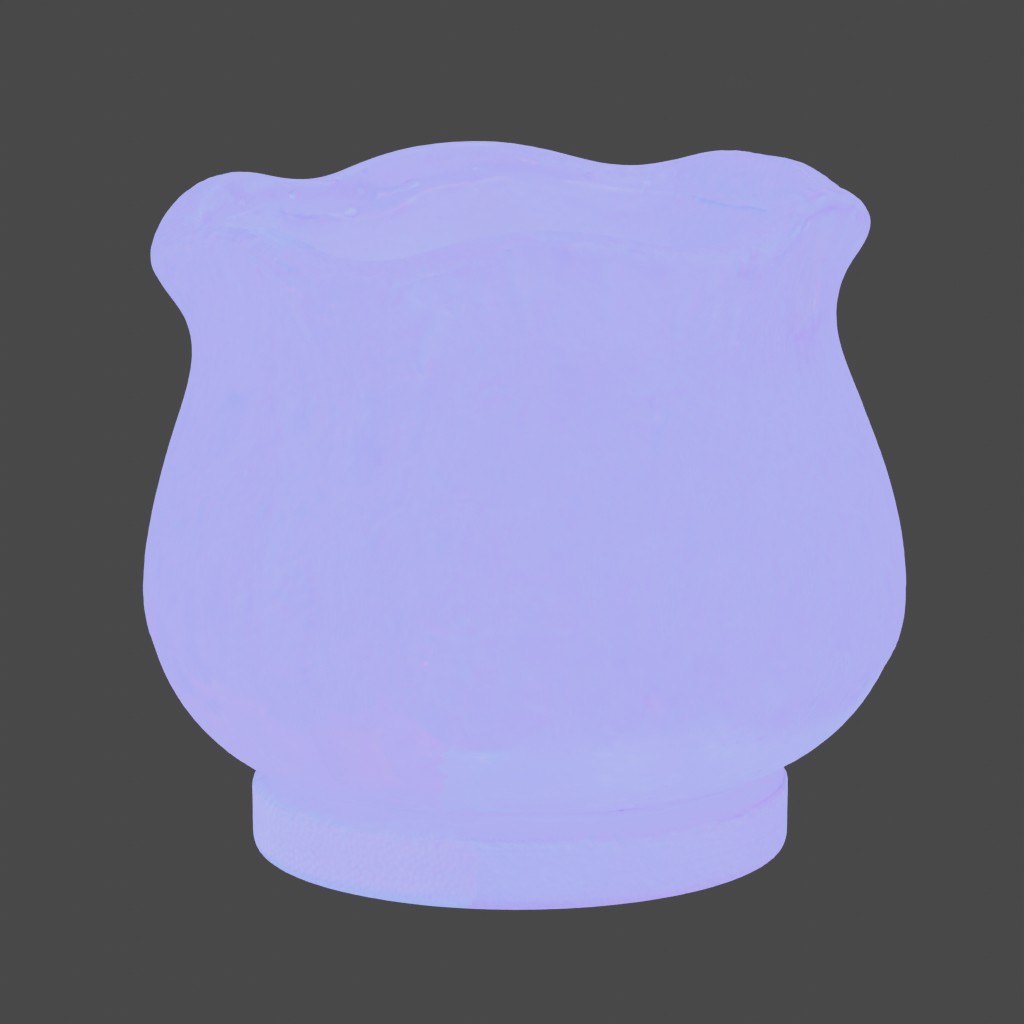}
    \includegraphics[width=0.19\linewidth]{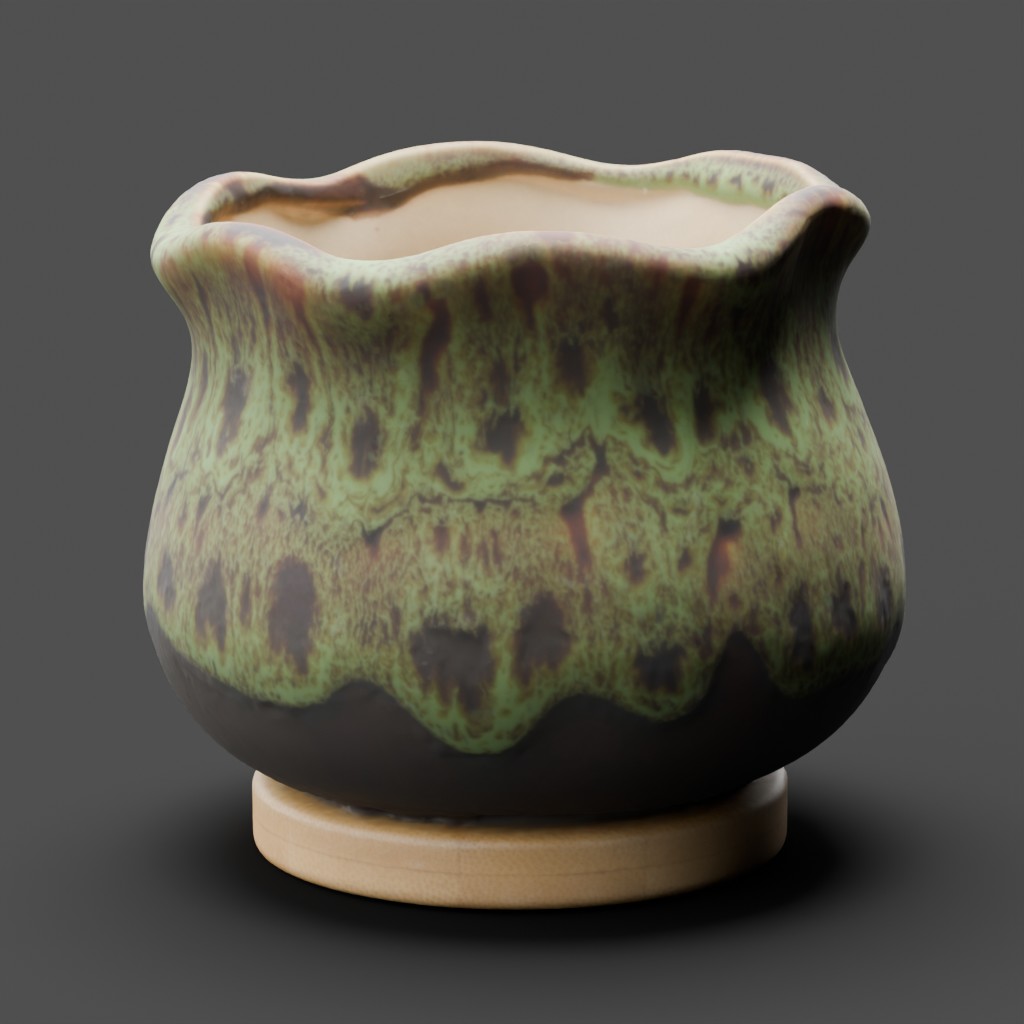} \\
    \includegraphics[width=0.19\linewidth]{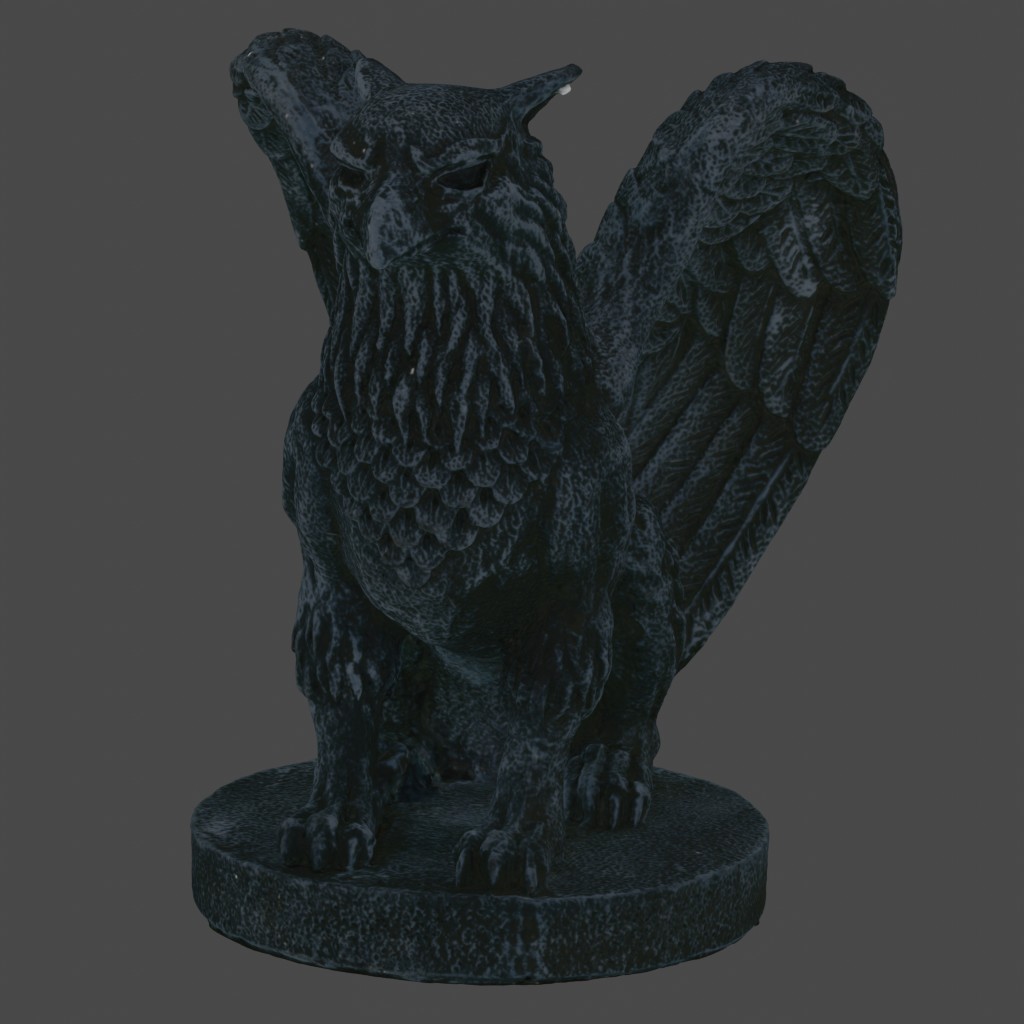}
    \includegraphics[width=0.19\linewidth]{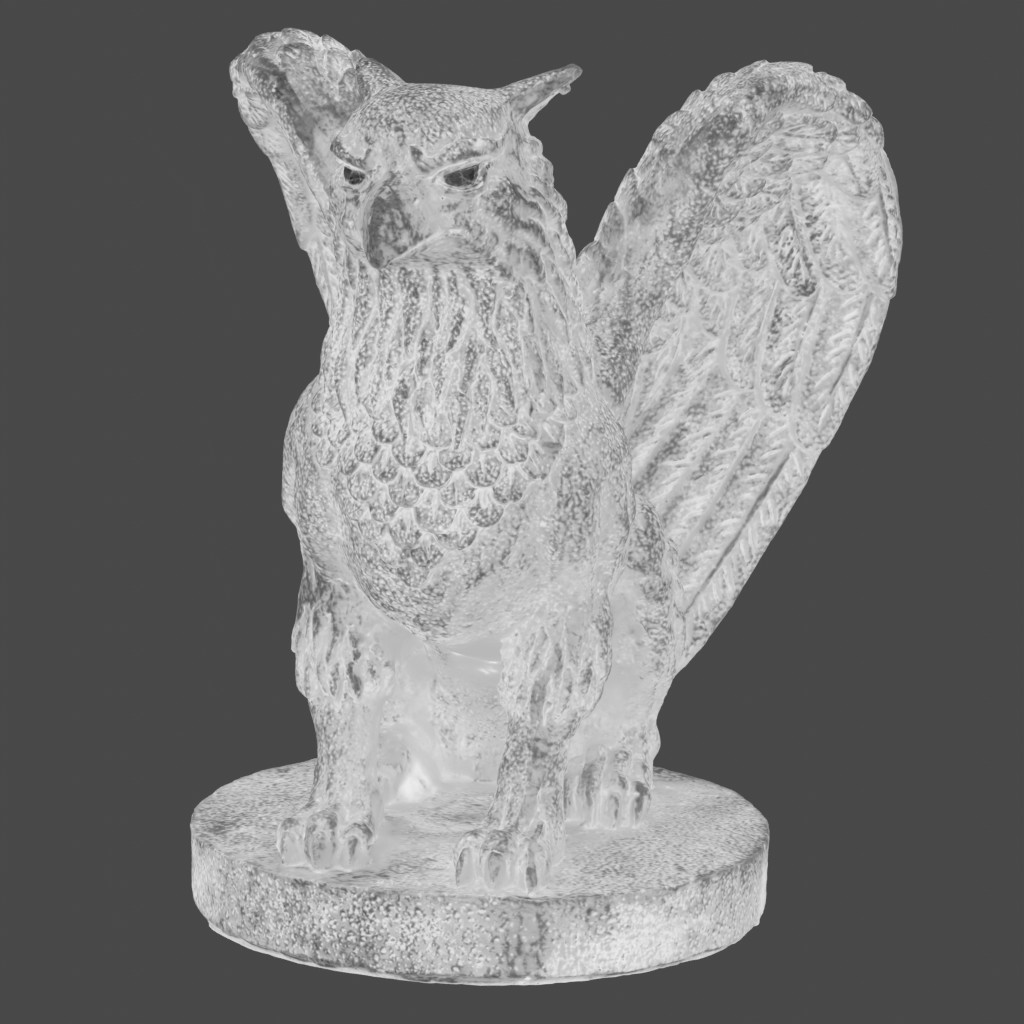}
    \includegraphics[width=0.19\linewidth]{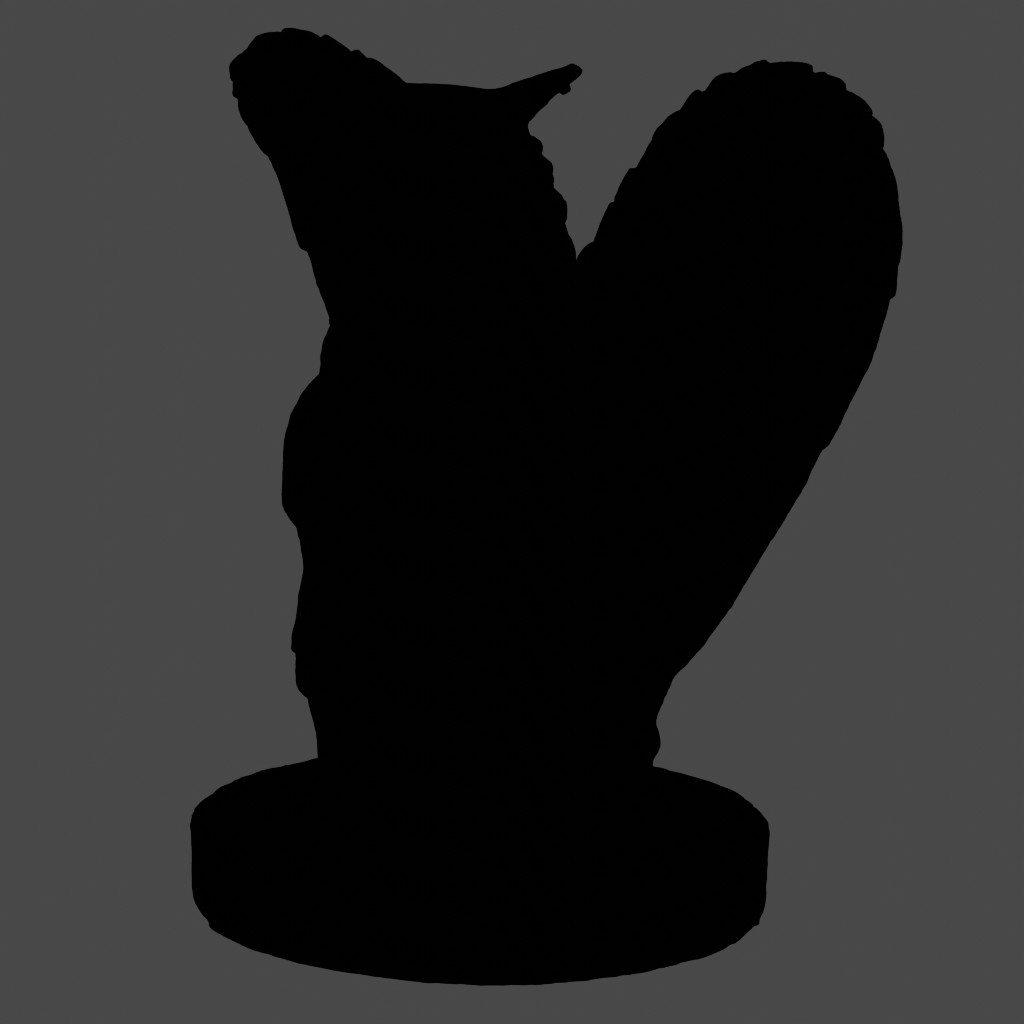}
    \includegraphics[width=0.19\linewidth]{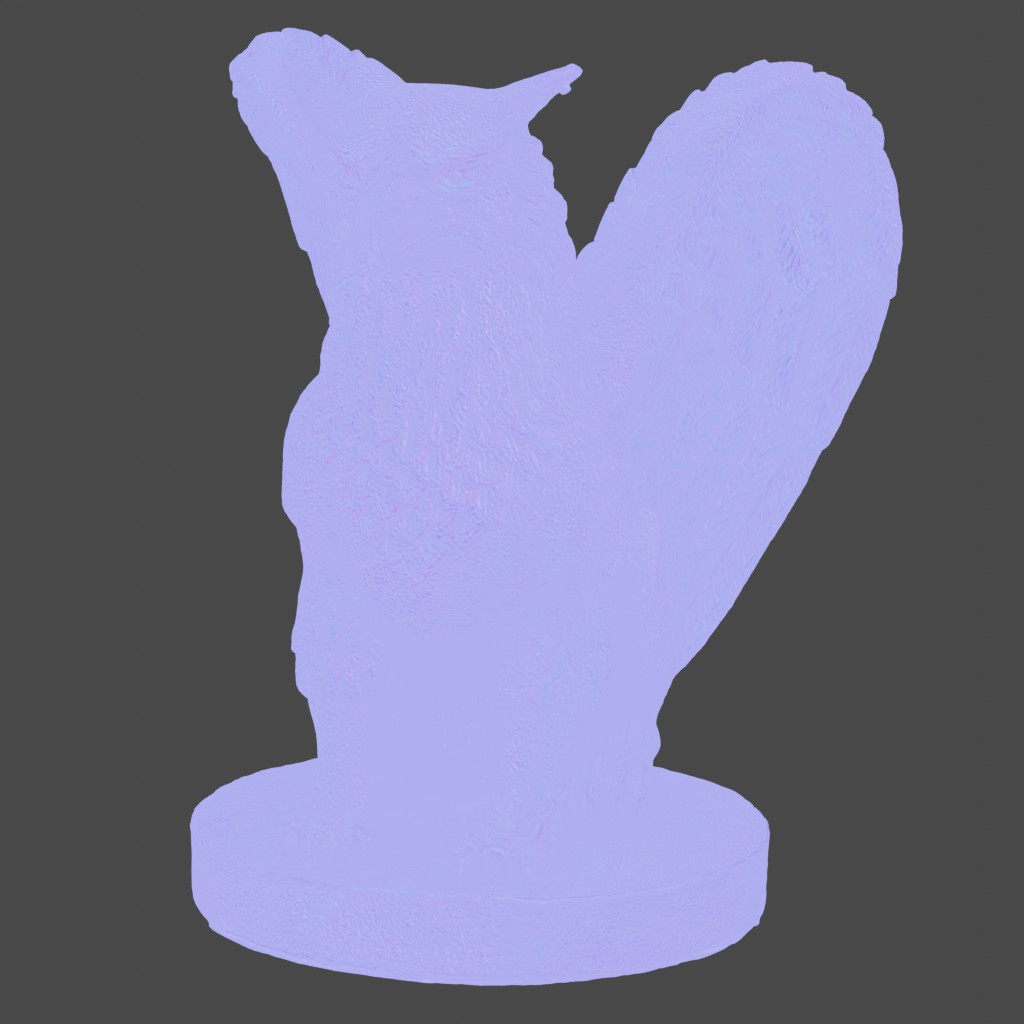}
    \includegraphics[width=0.19\linewidth]{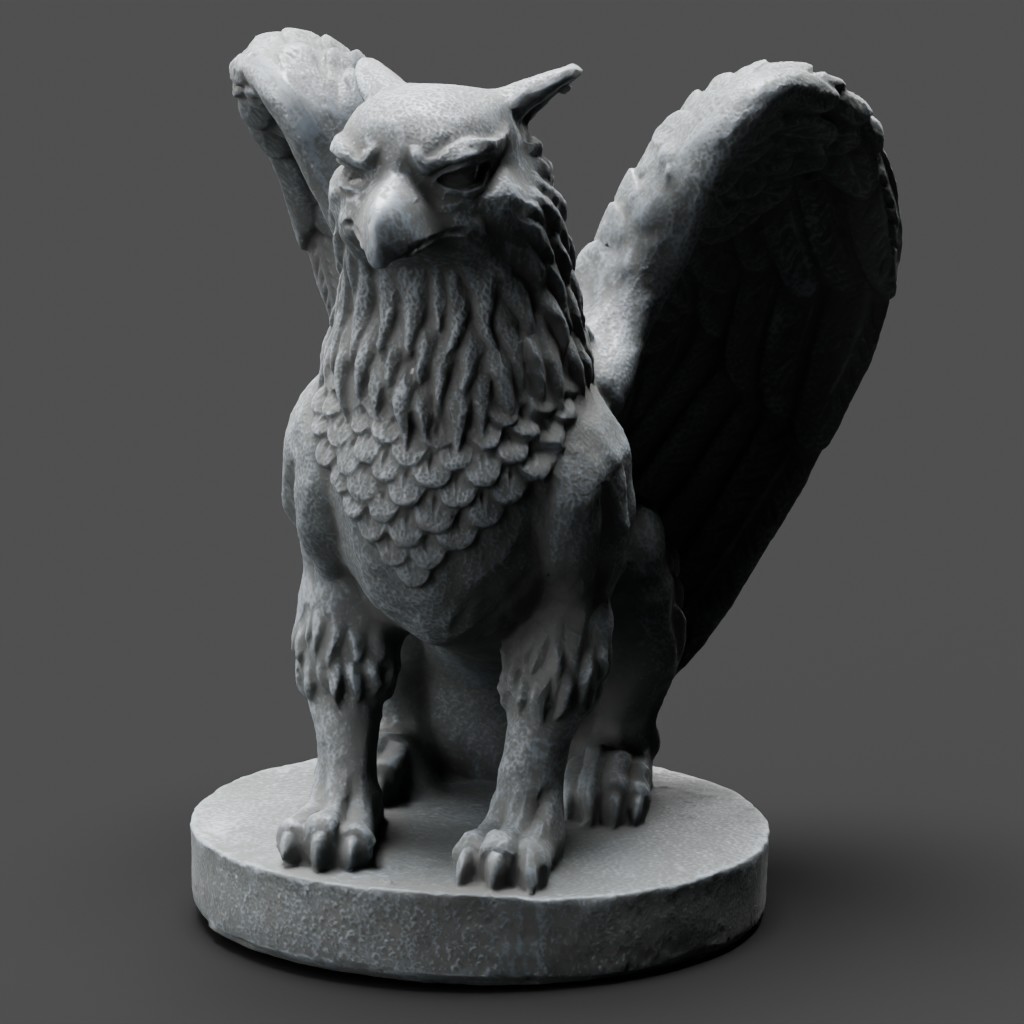} \\
    \includegraphics[width=0.19\linewidth]{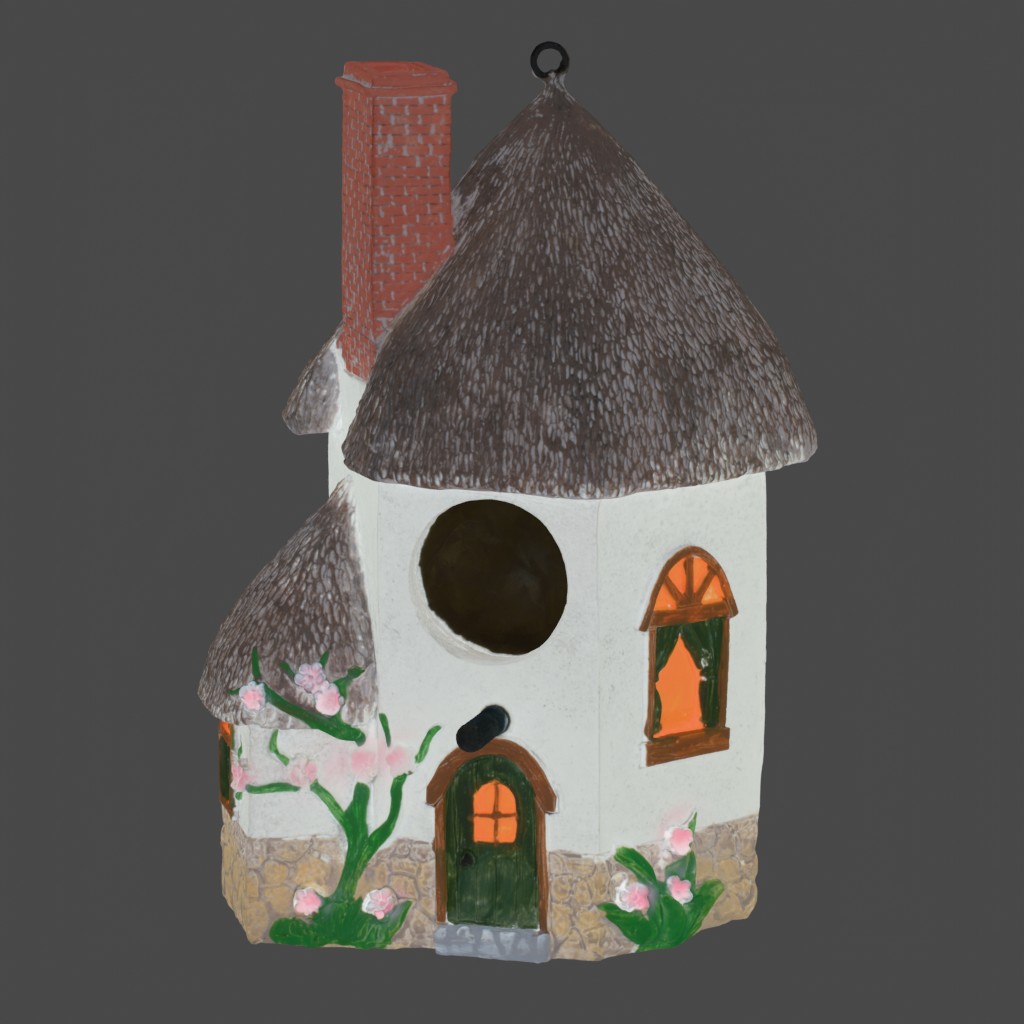}
    \includegraphics[width=0.19\linewidth]{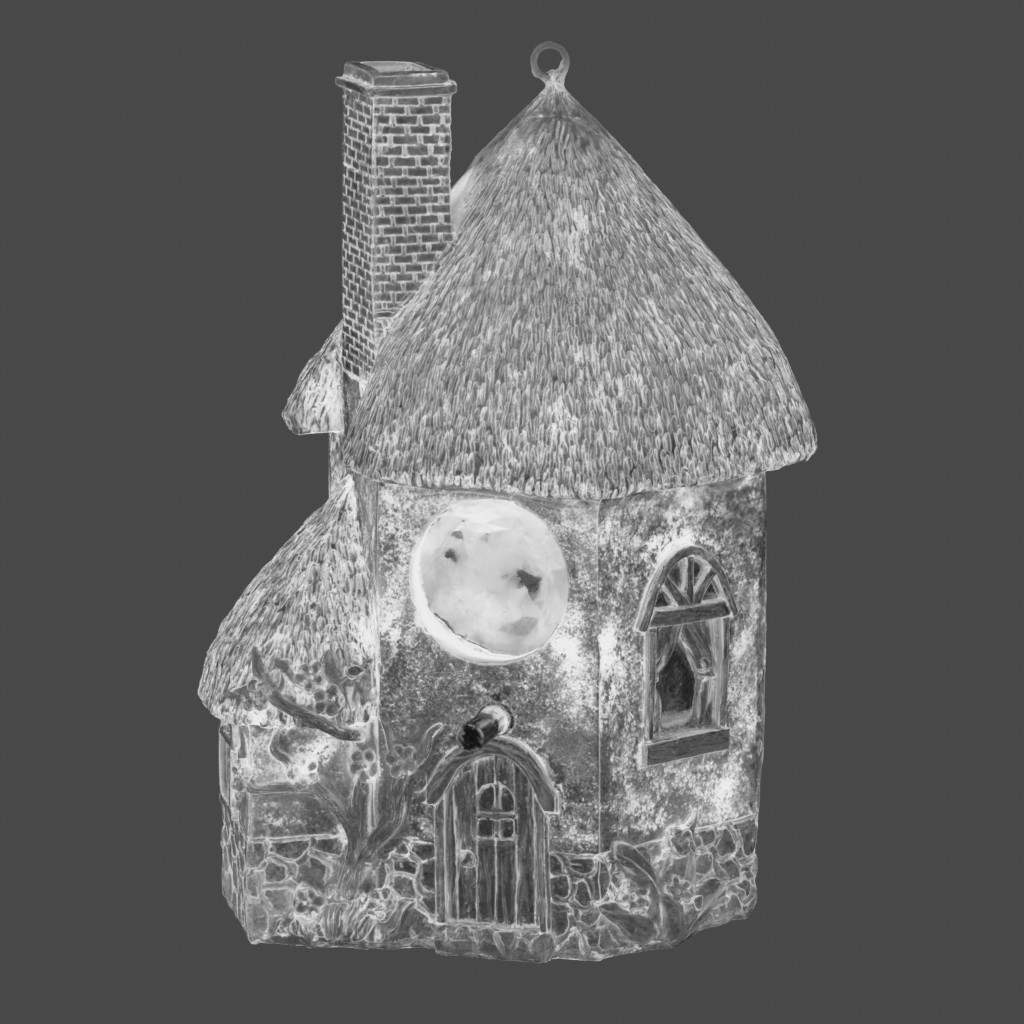}
    \includegraphics[width=0.19\linewidth]{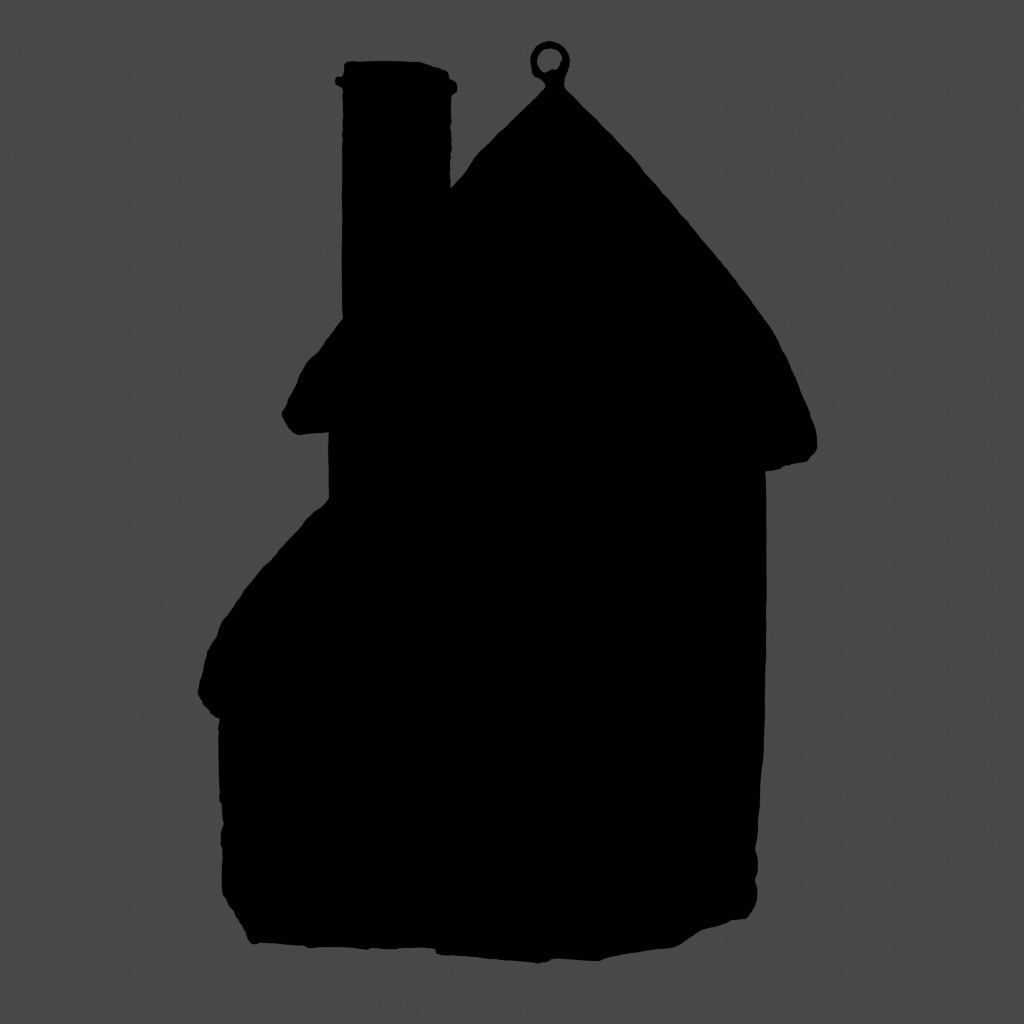}
    \includegraphics[width=0.19\linewidth]{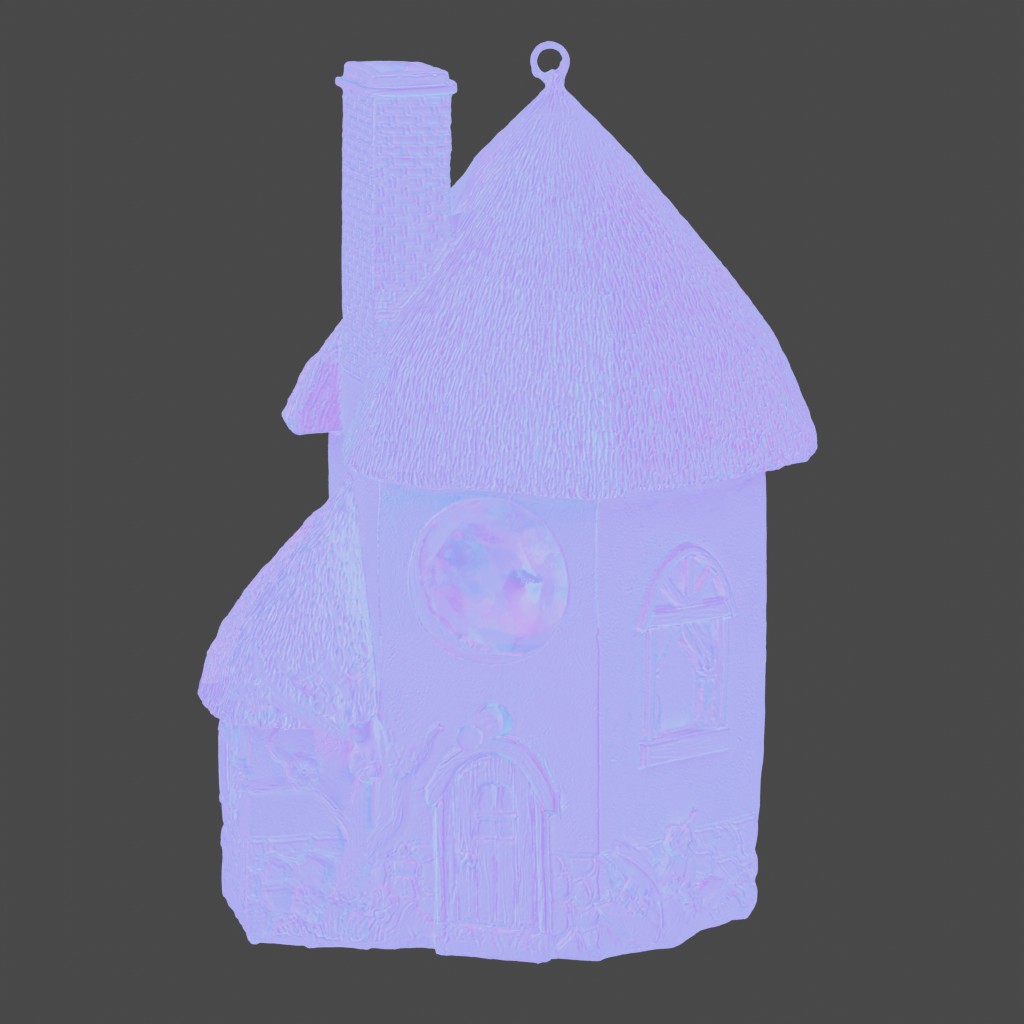}
    \includegraphics[width=0.19\linewidth]{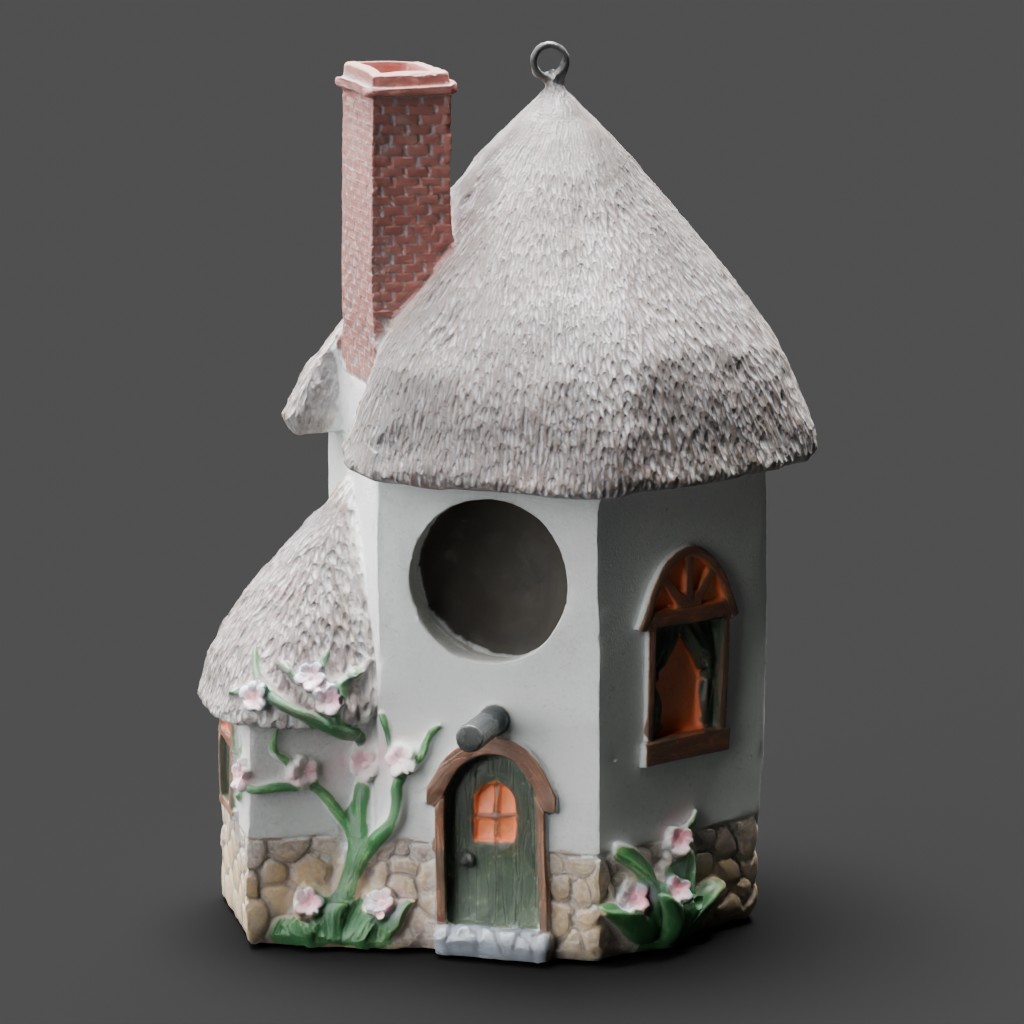} \\
    \includegraphics[width=0.19\linewidth]{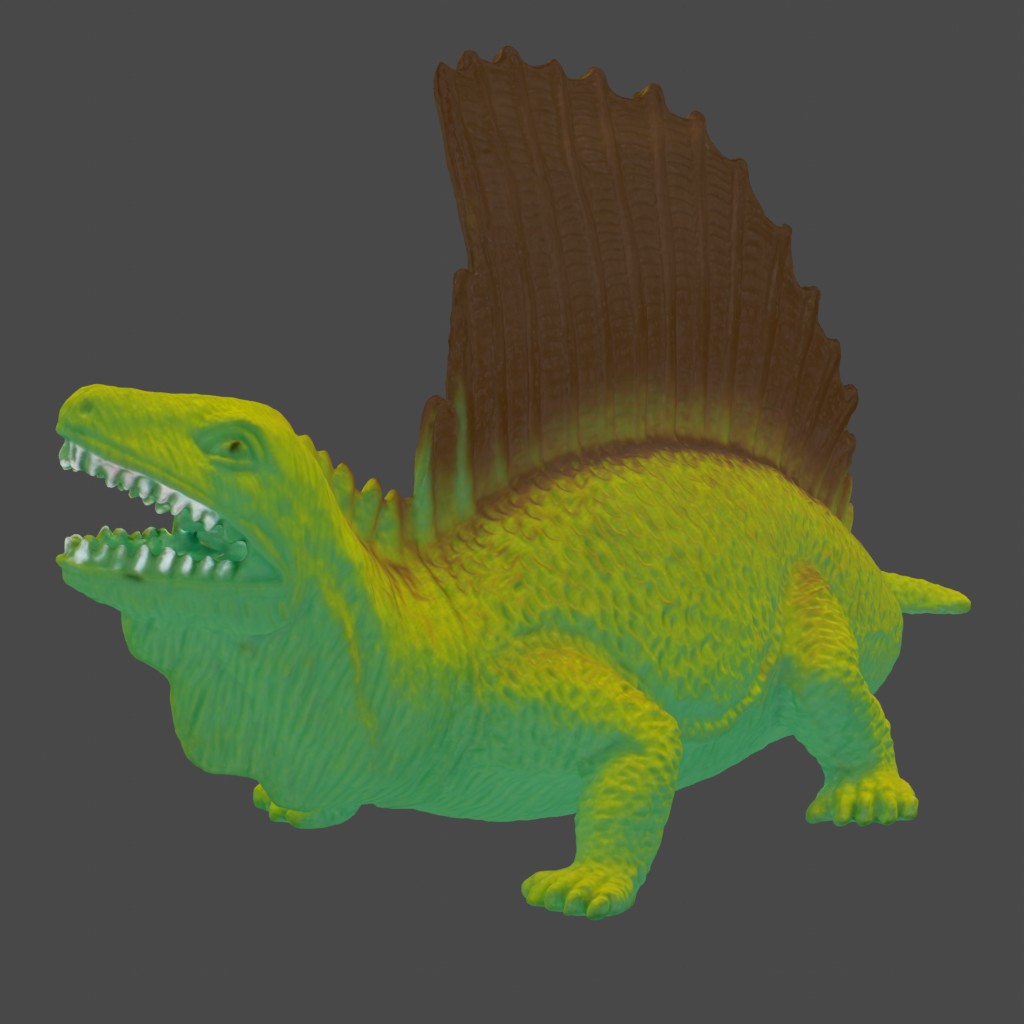}
    \includegraphics[width=0.19\linewidth]{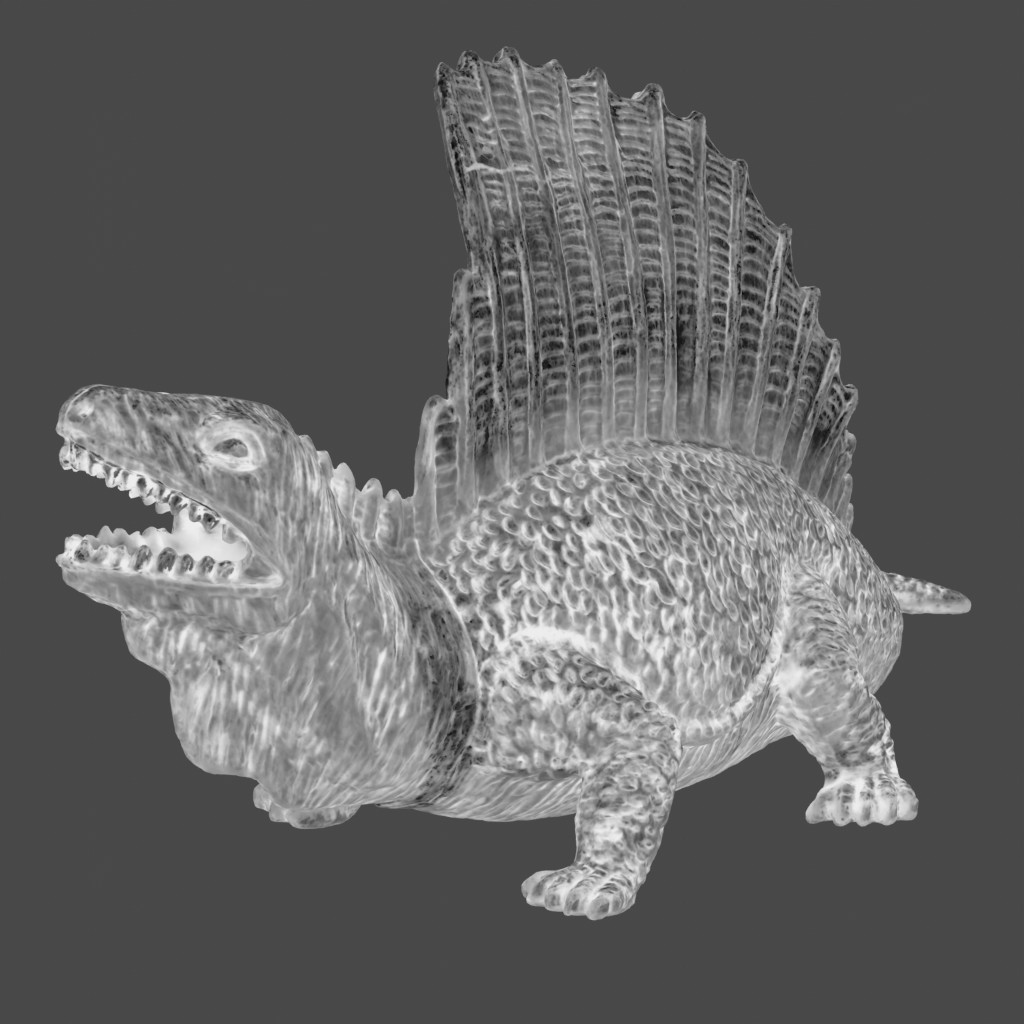}
    \includegraphics[width=0.19\linewidth]{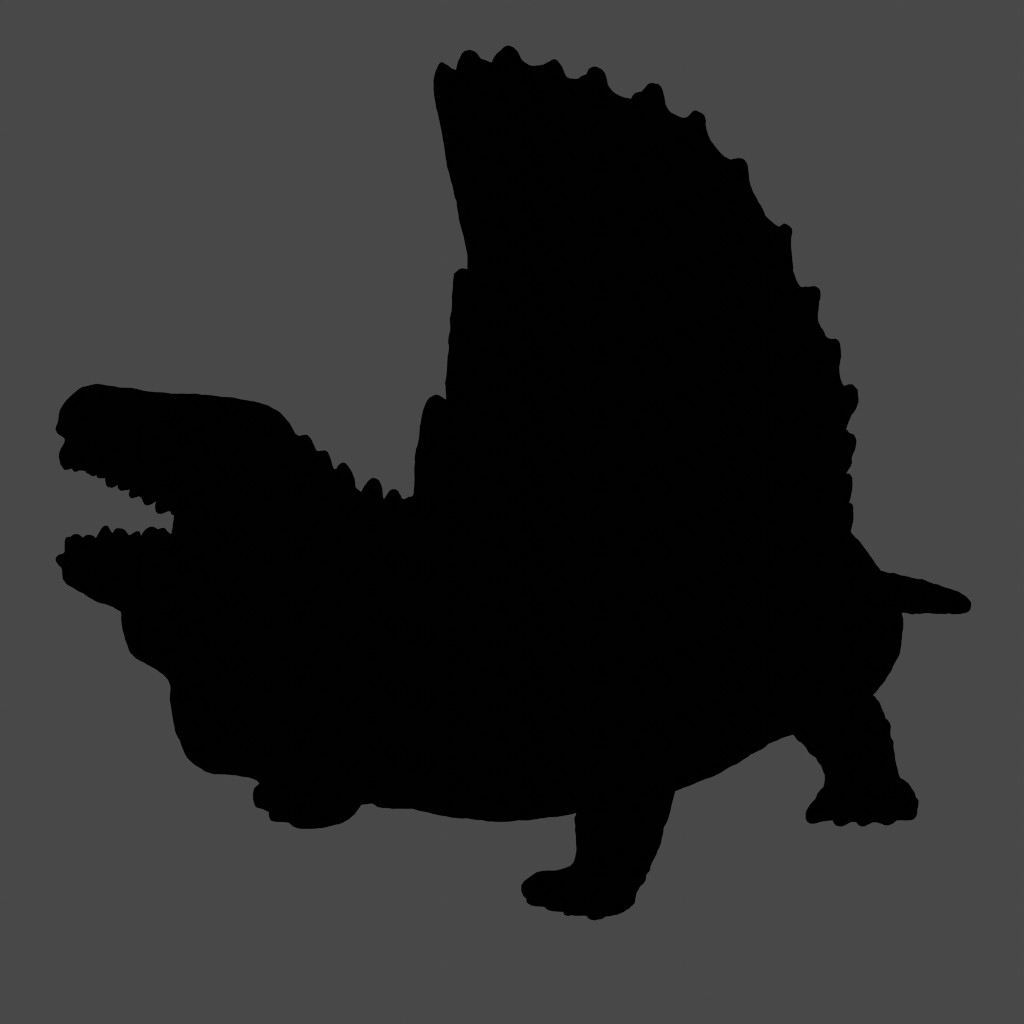}
    \includegraphics[width=0.19\linewidth]{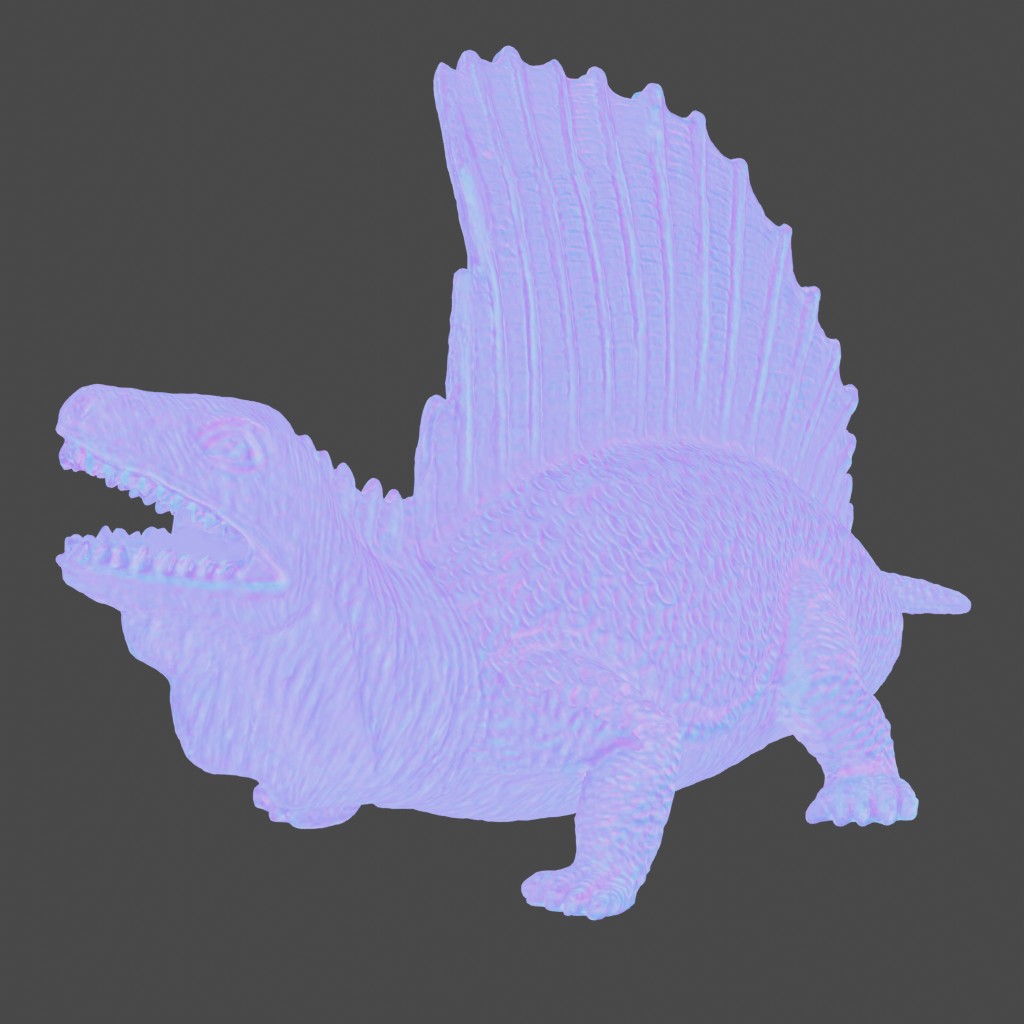}
    \includegraphics[width=0.19\linewidth]{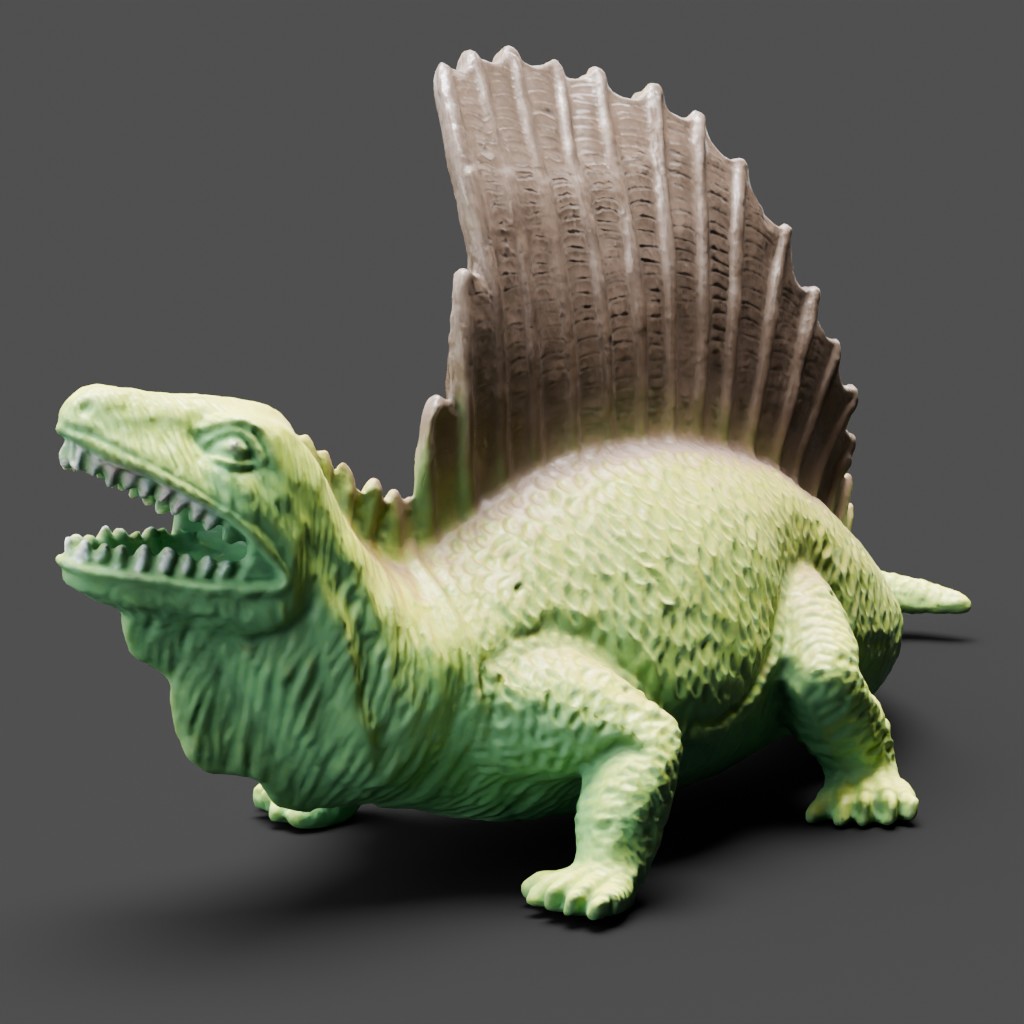}
    \caption{PBR Materials of the example DTC objects (the list of objects in Fig.9 \textbf{Row 4}). From left to right: albedo map, roughness map, metallic map, normal map, and PBR rendering.}
    \label{fig:more_PBR_maps_row4}
\end{figure*}

\begin{figure*}[t]
    \centering
    \includegraphics[width=0.19\linewidth]{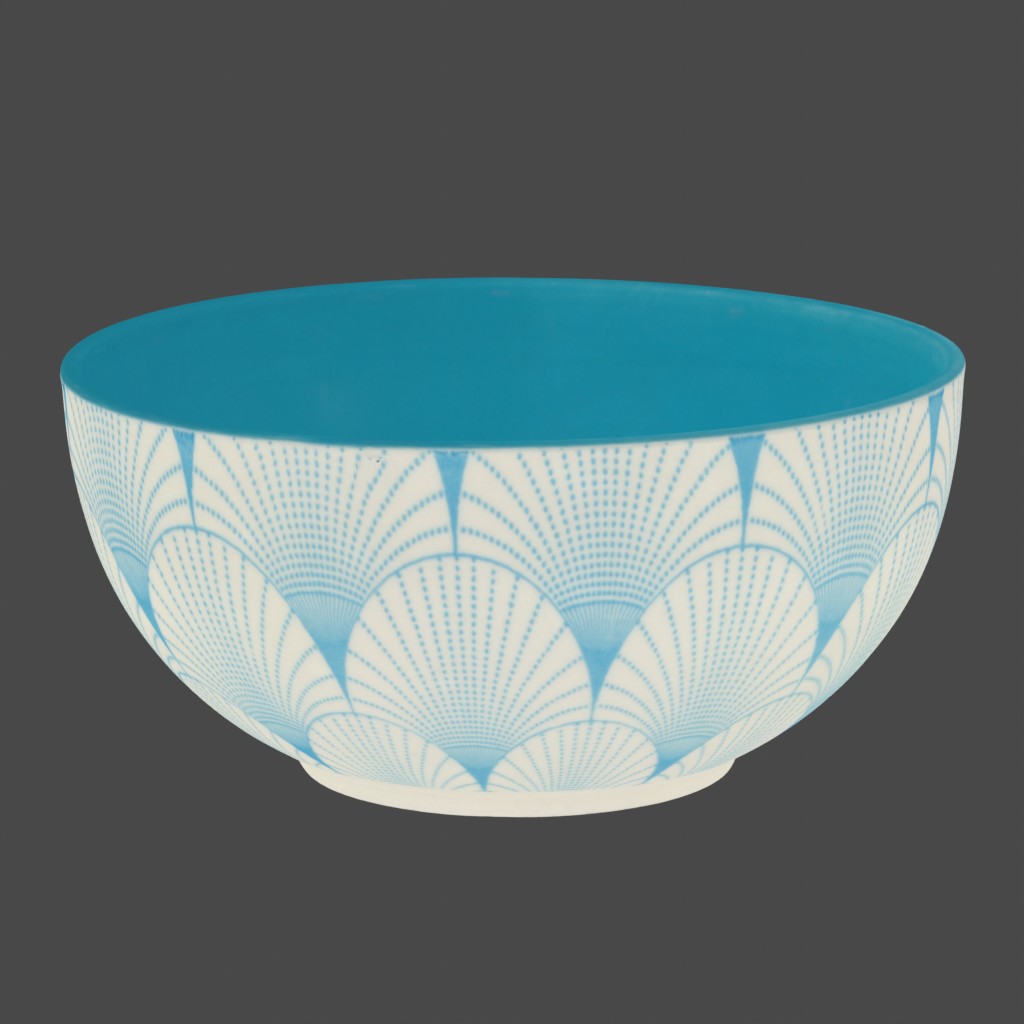}
    \includegraphics[width=0.19\linewidth]{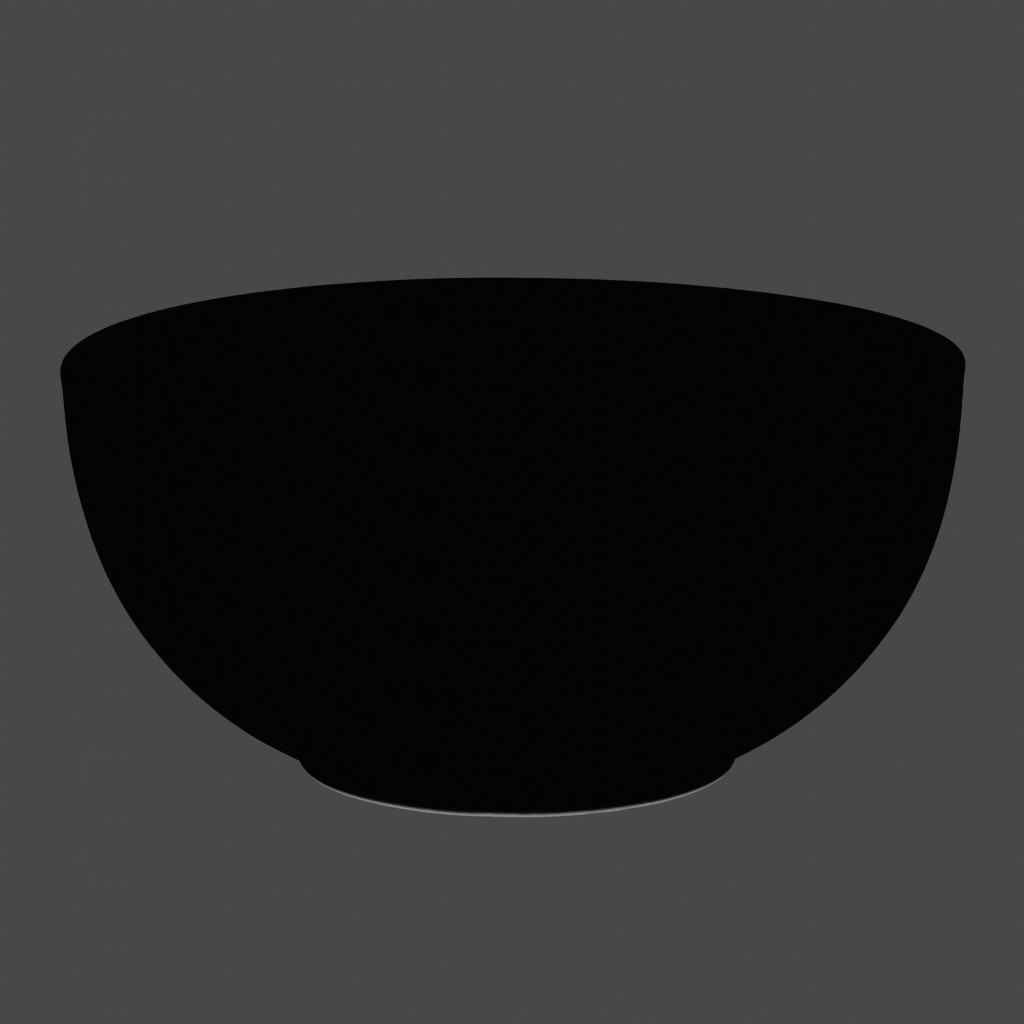}
    \includegraphics[width=0.19\linewidth]{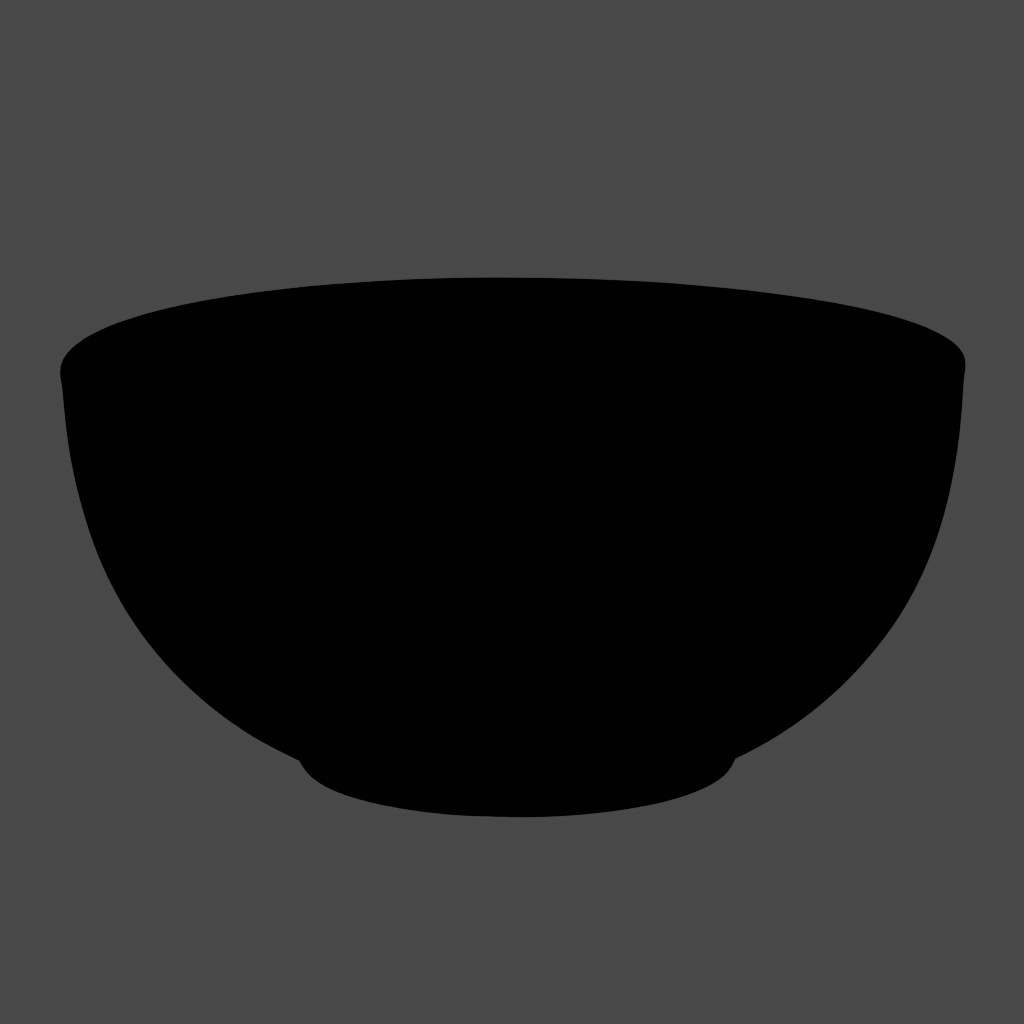}
    \includegraphics[width=0.19\linewidth]{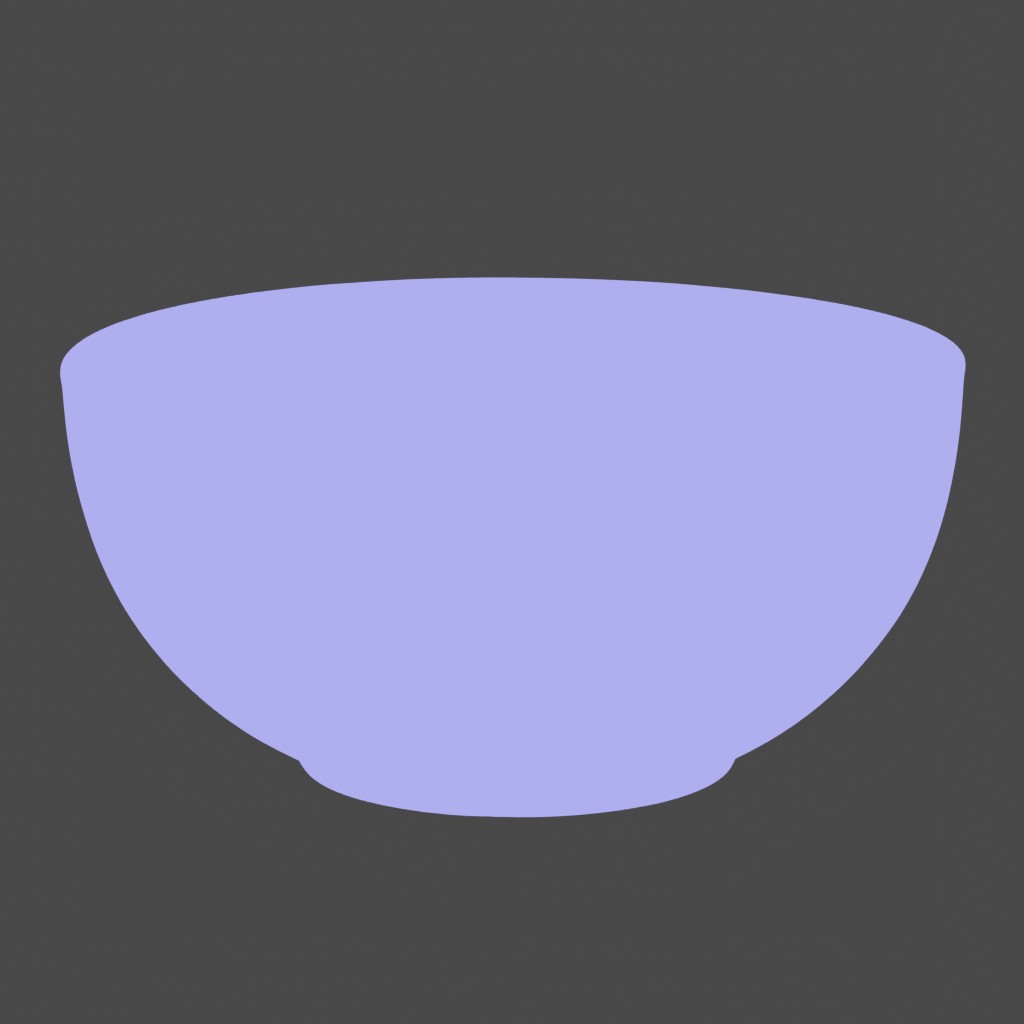}
    \includegraphics[width=0.19\linewidth]{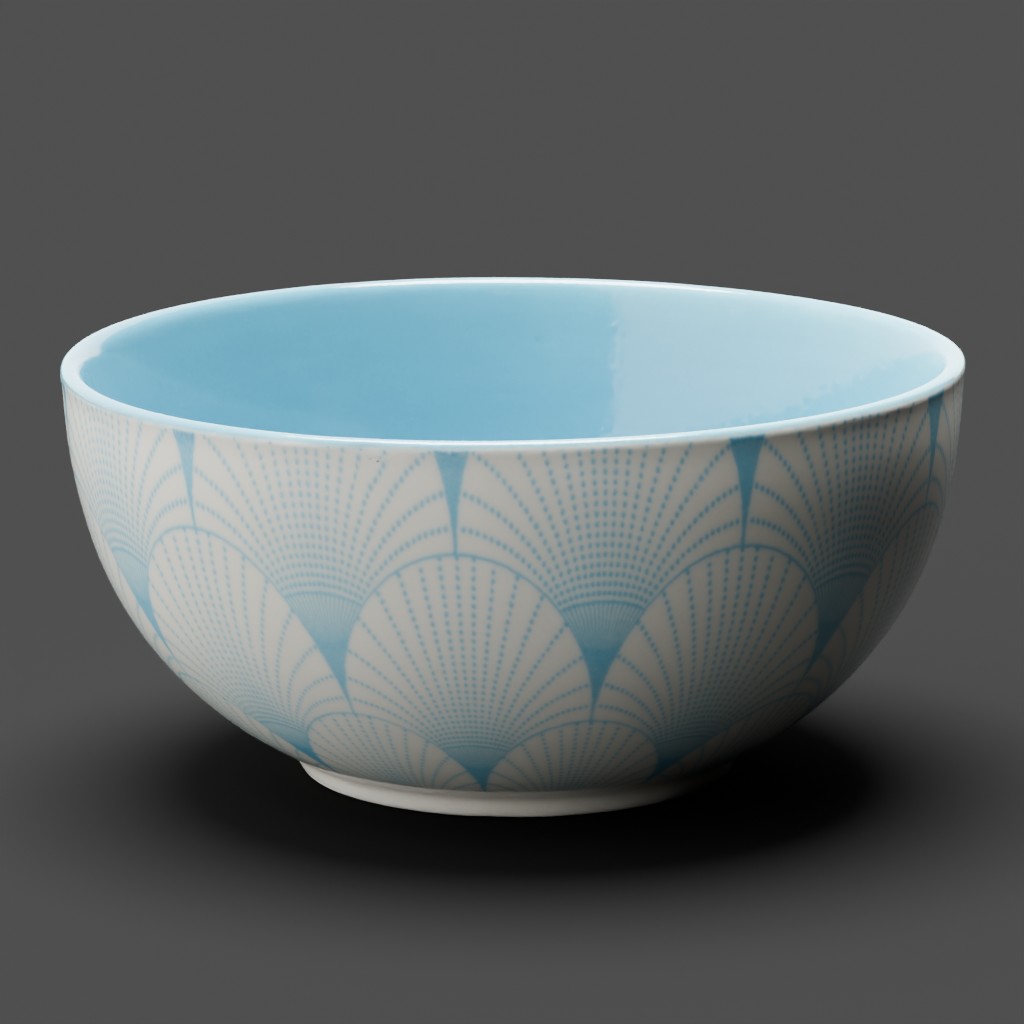} \\
    \includegraphics[width=0.19\linewidth]{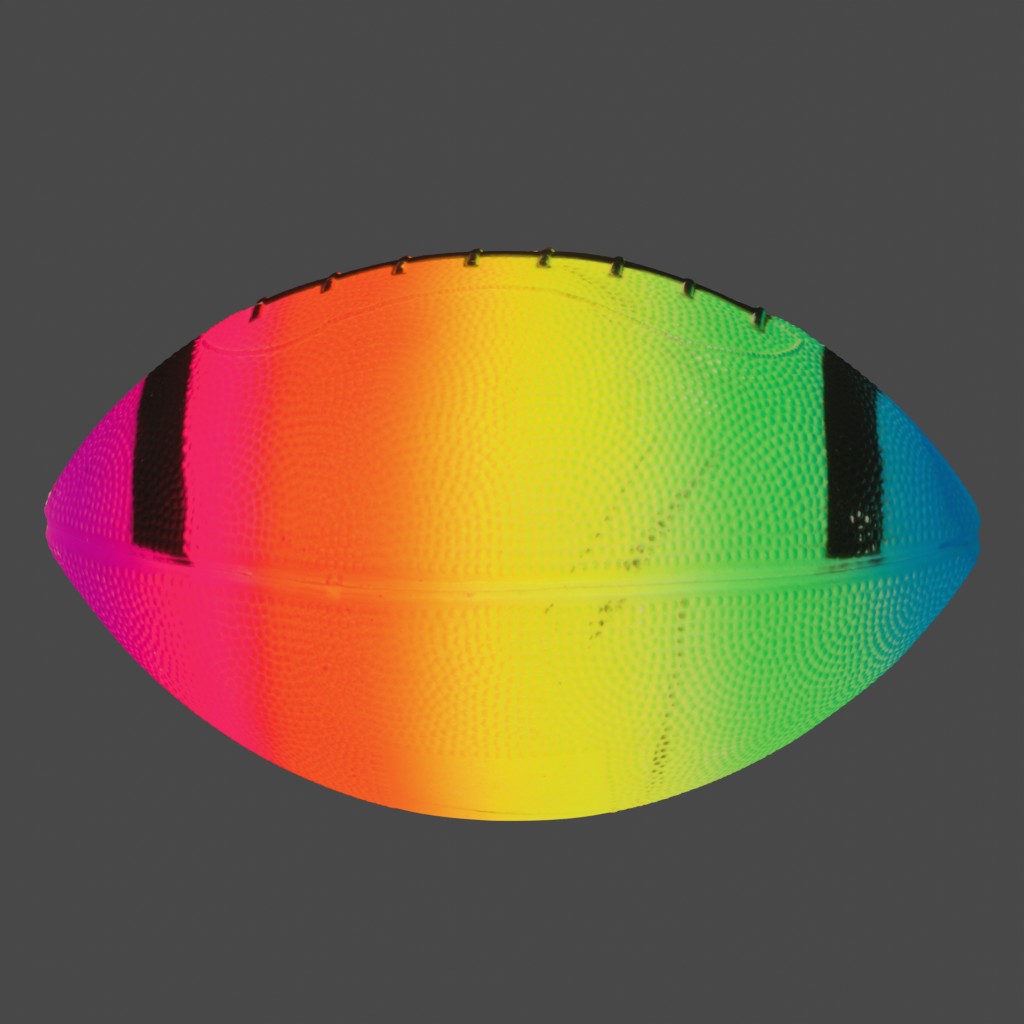}
    \includegraphics[width=0.19\linewidth]{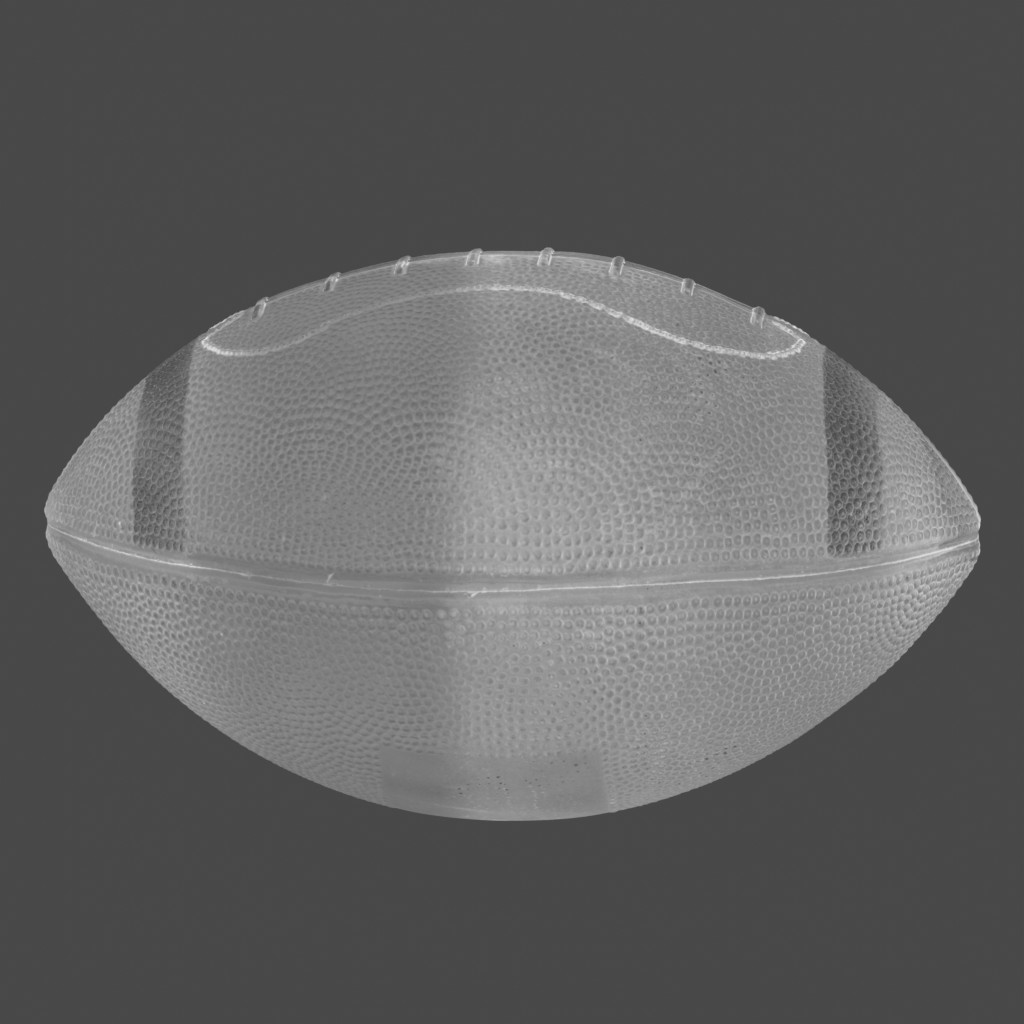}
    \includegraphics[width=0.19\linewidth]{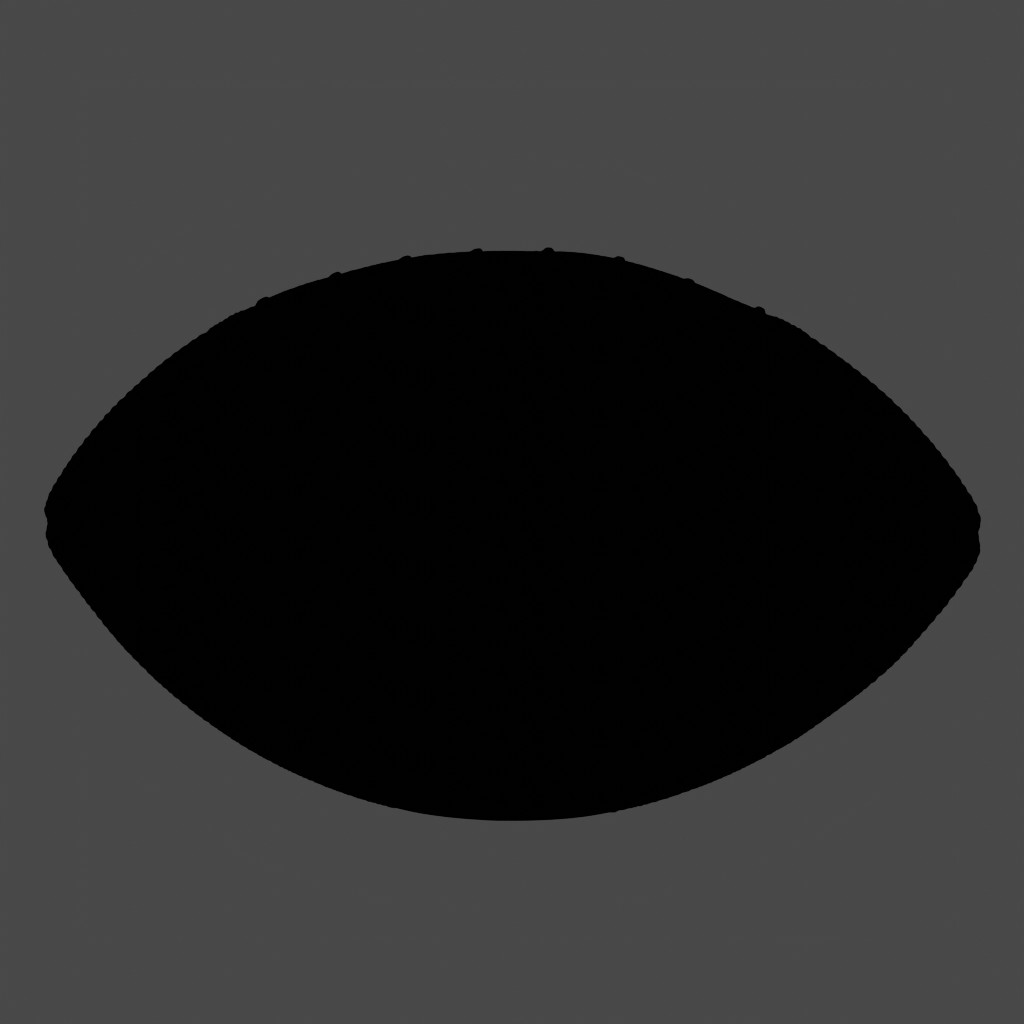}
    \includegraphics[width=0.19\linewidth]{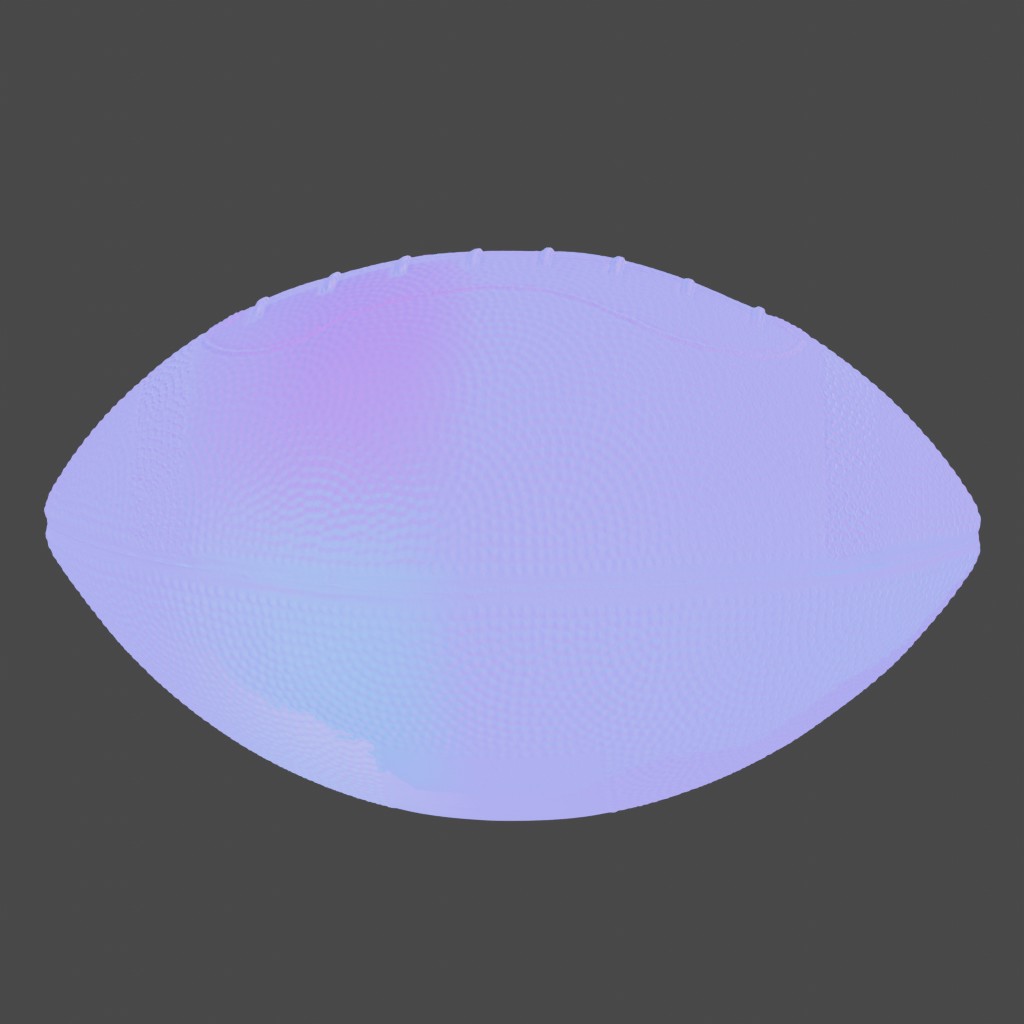}
    \includegraphics[width=0.19\linewidth]{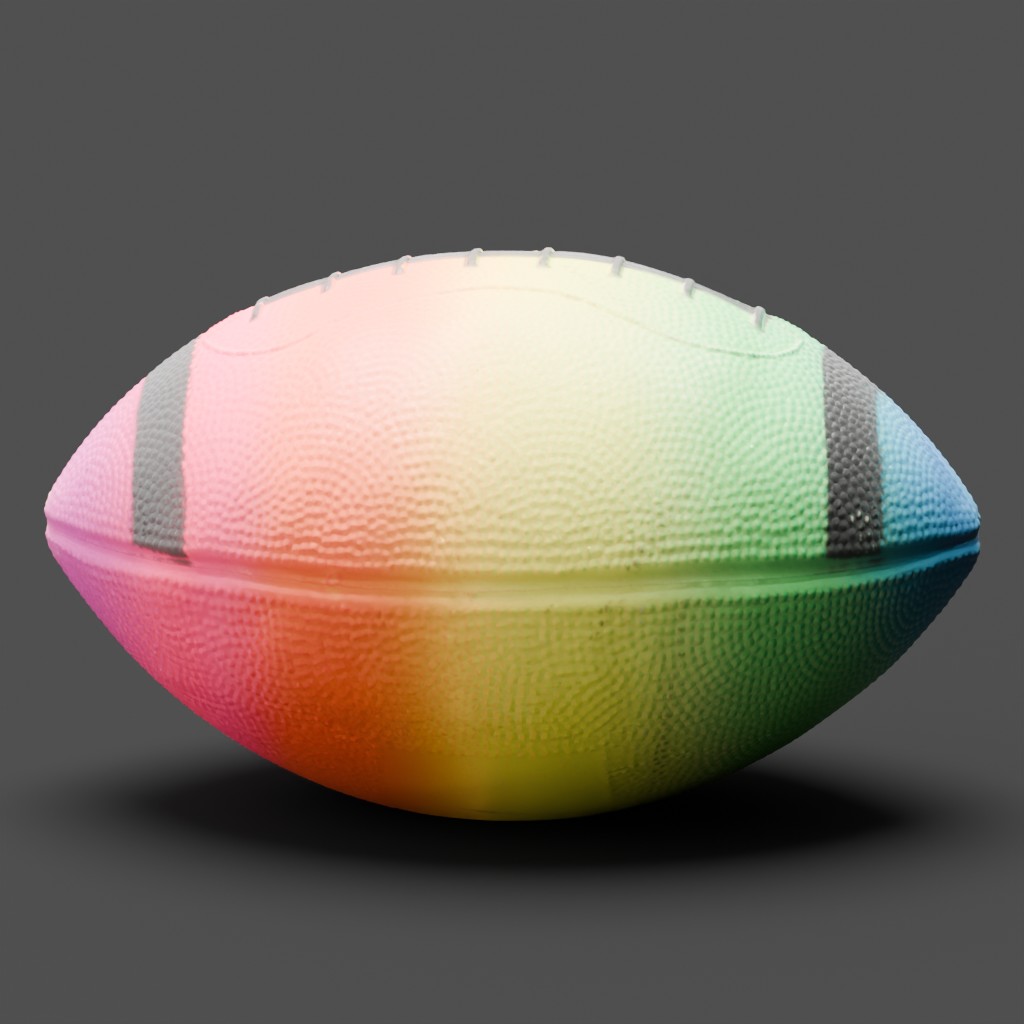} \\
    \includegraphics[width=0.19\linewidth]{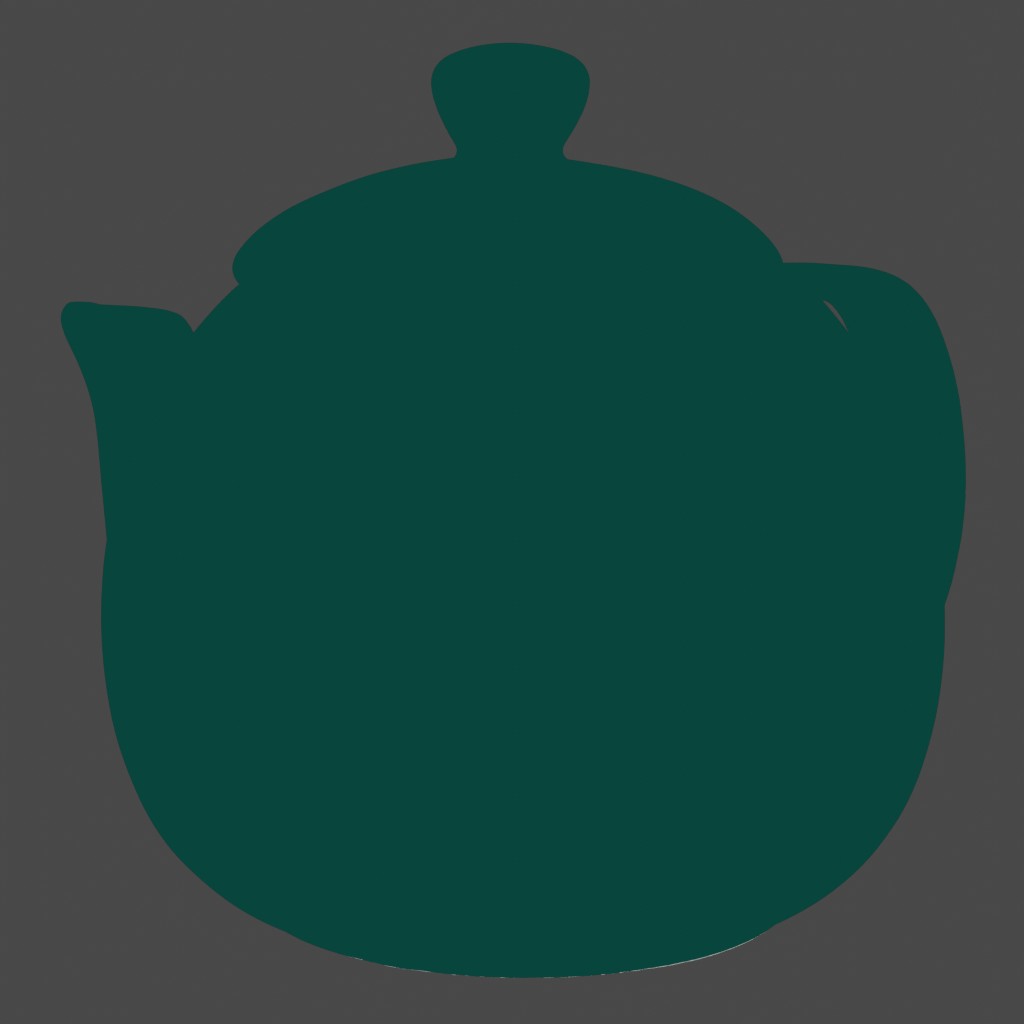}
    \includegraphics[width=0.19\linewidth]{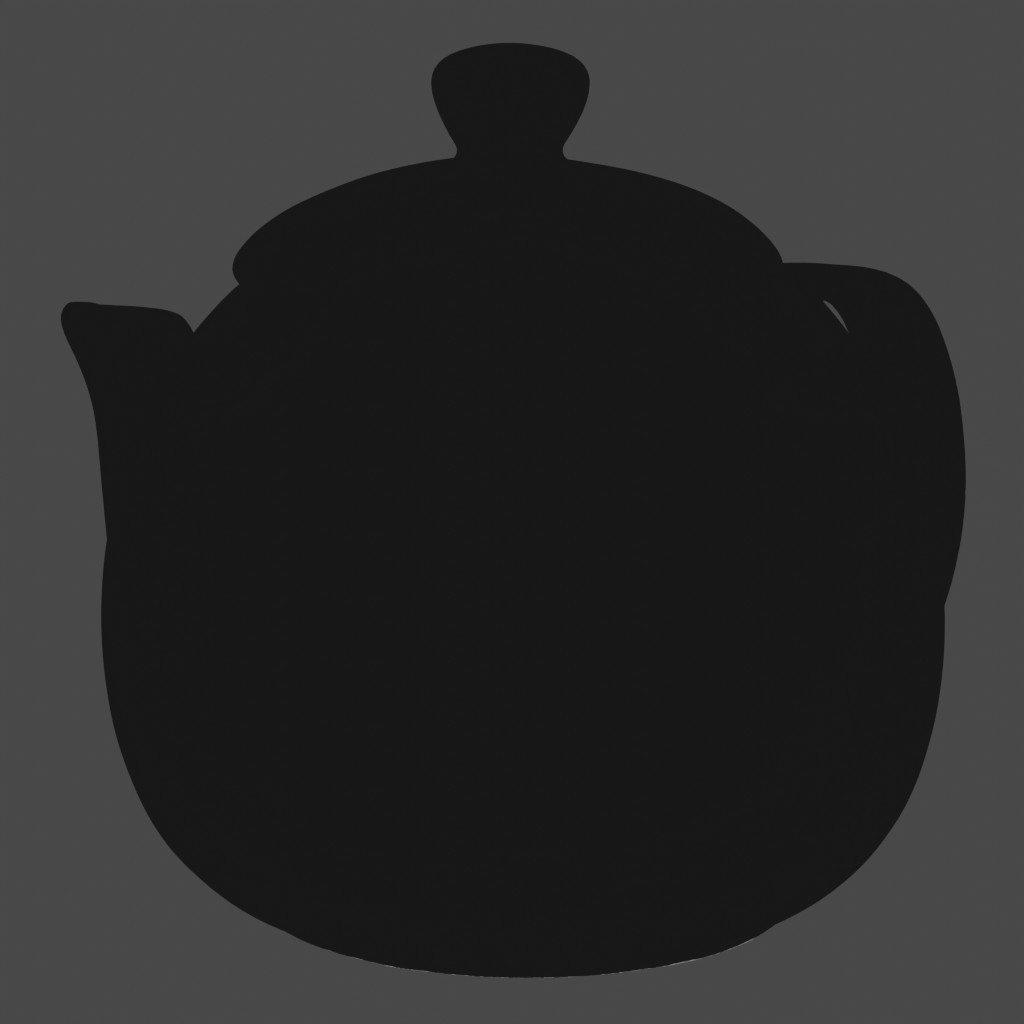}
    \includegraphics[width=0.19\linewidth]{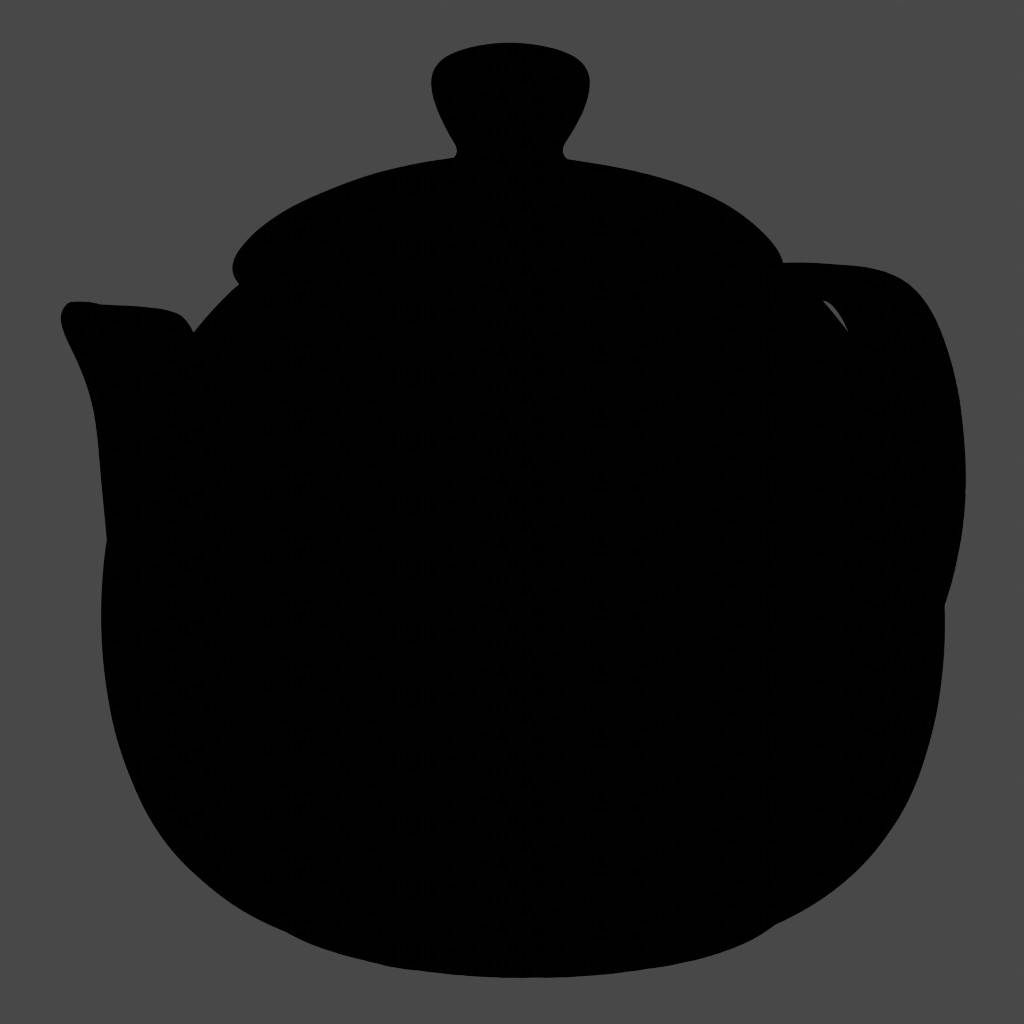}
    \includegraphics[width=0.19\linewidth]{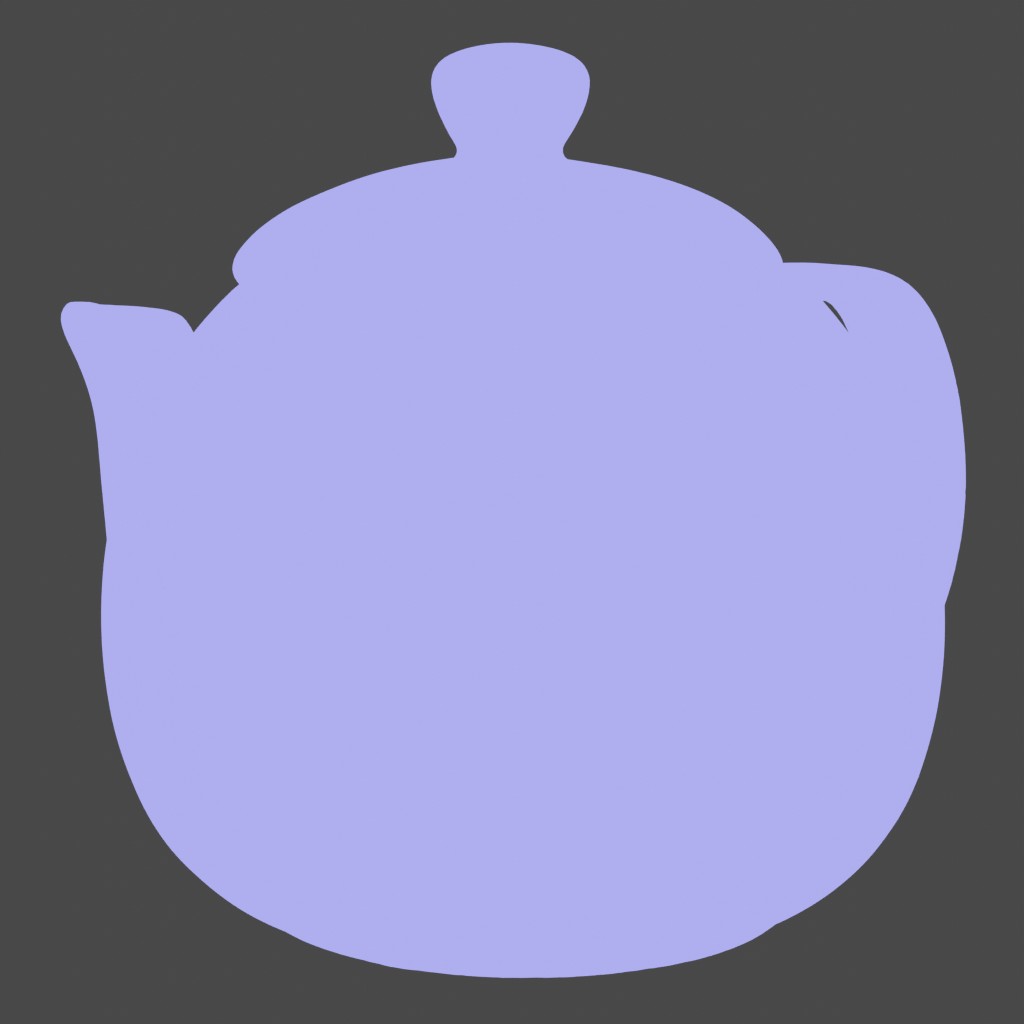}
    \includegraphics[width=0.19\linewidth]{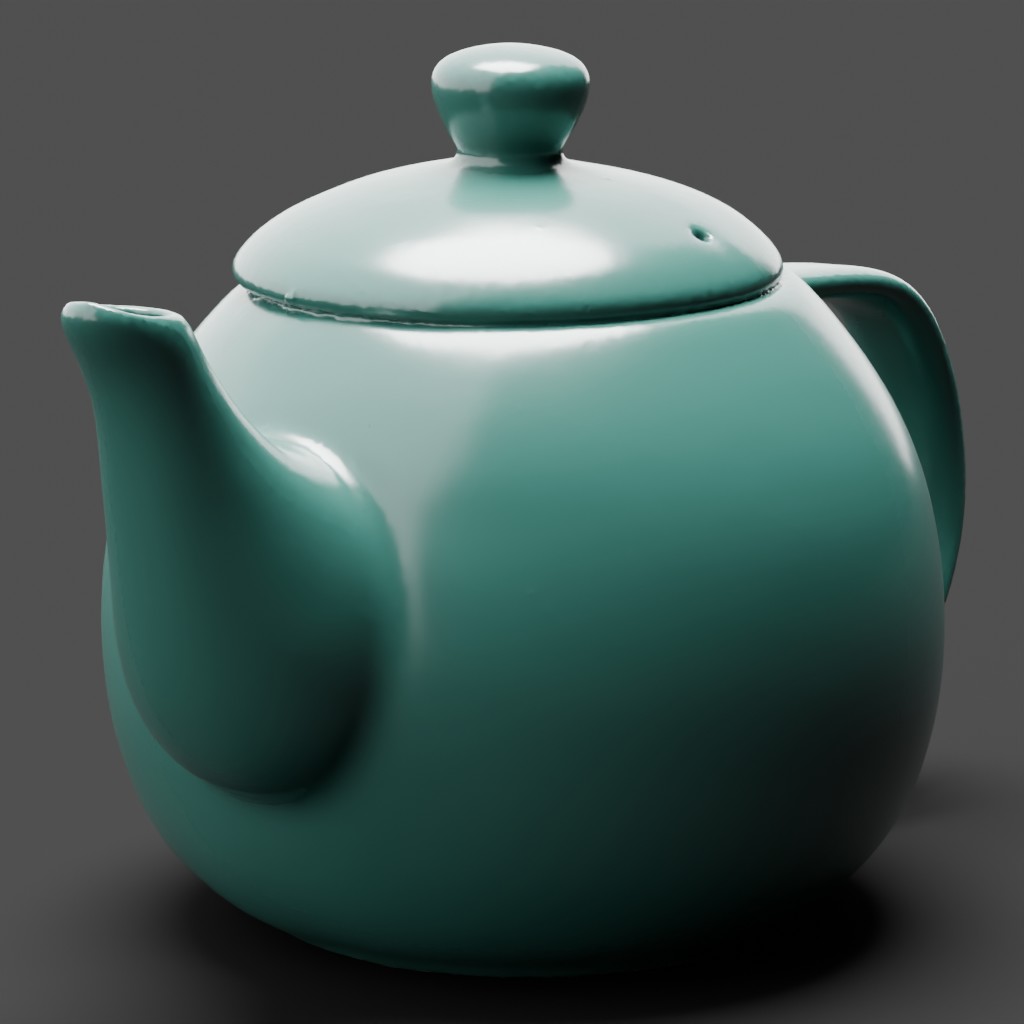} \\
    \includegraphics[width=0.19\linewidth]{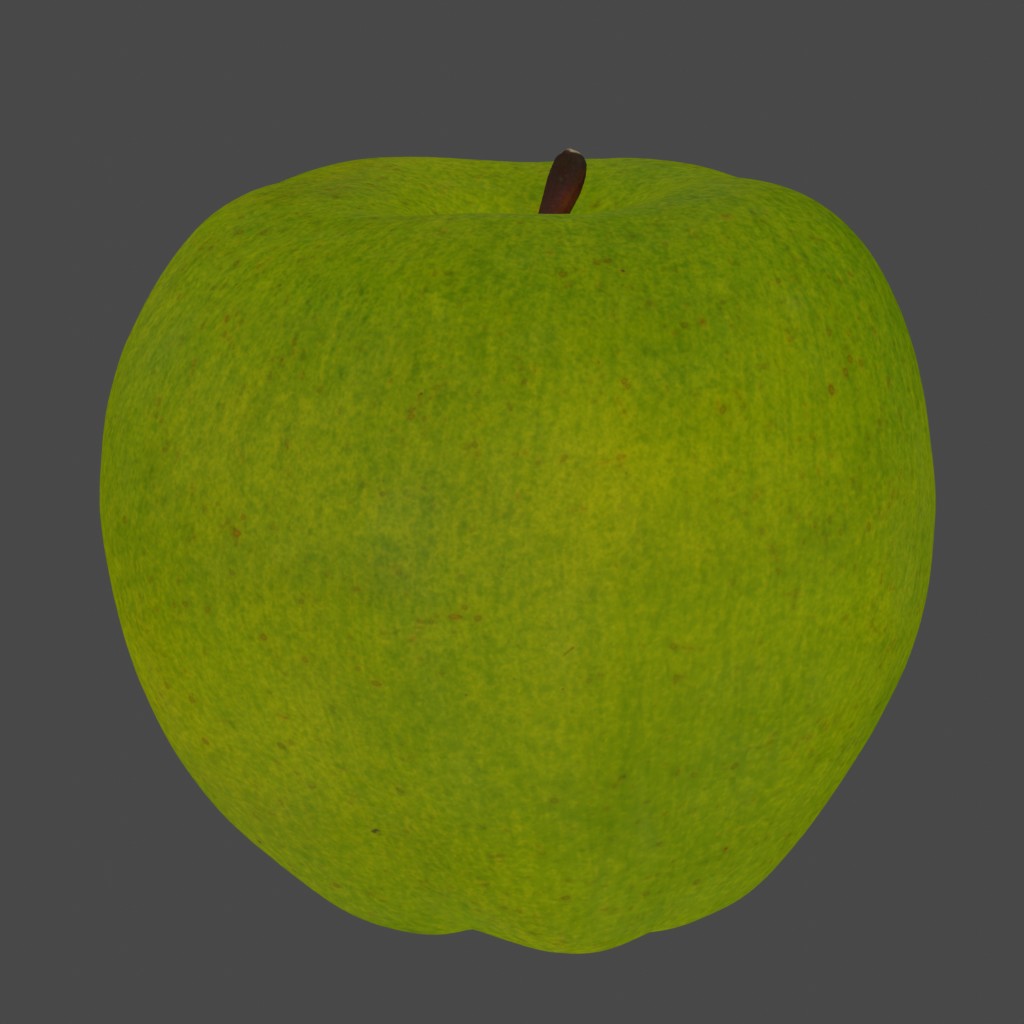}
    \includegraphics[width=0.19\linewidth]{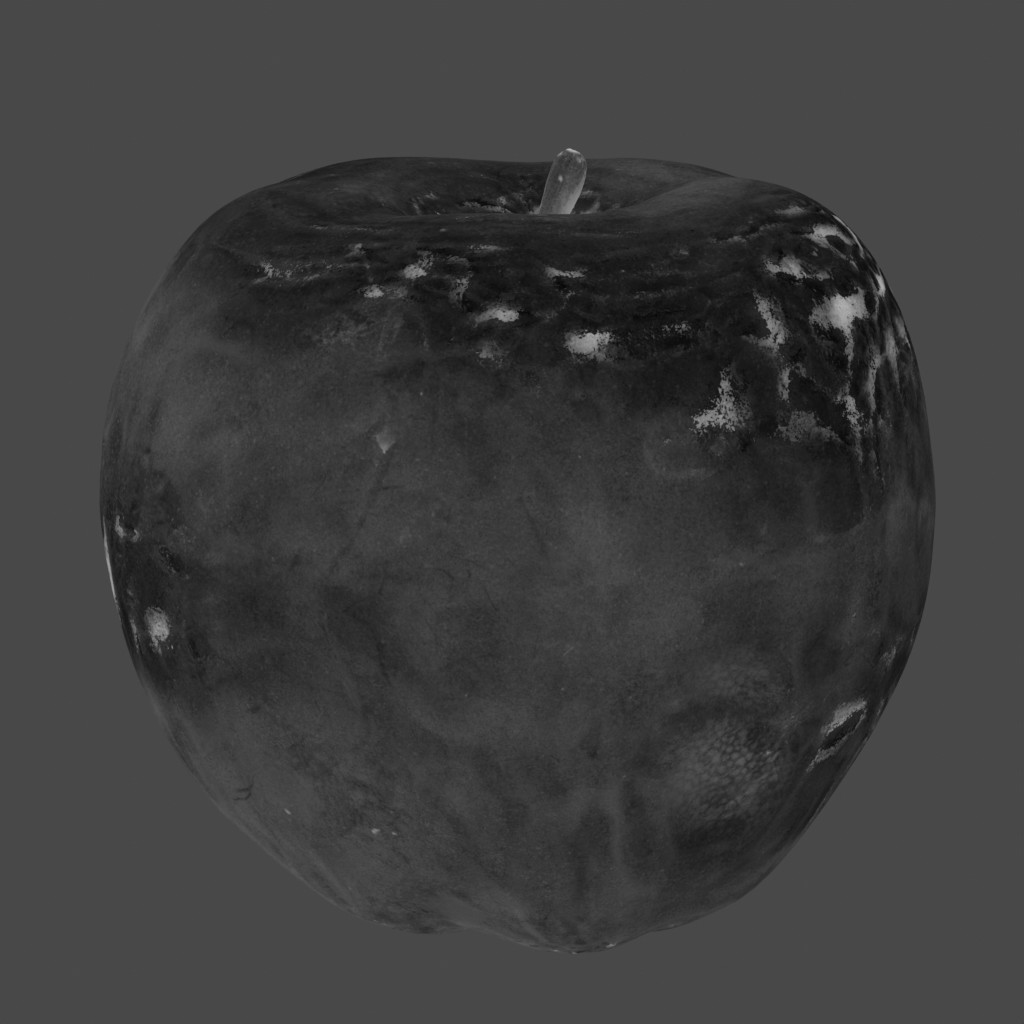}
    \includegraphics[width=0.19\linewidth]{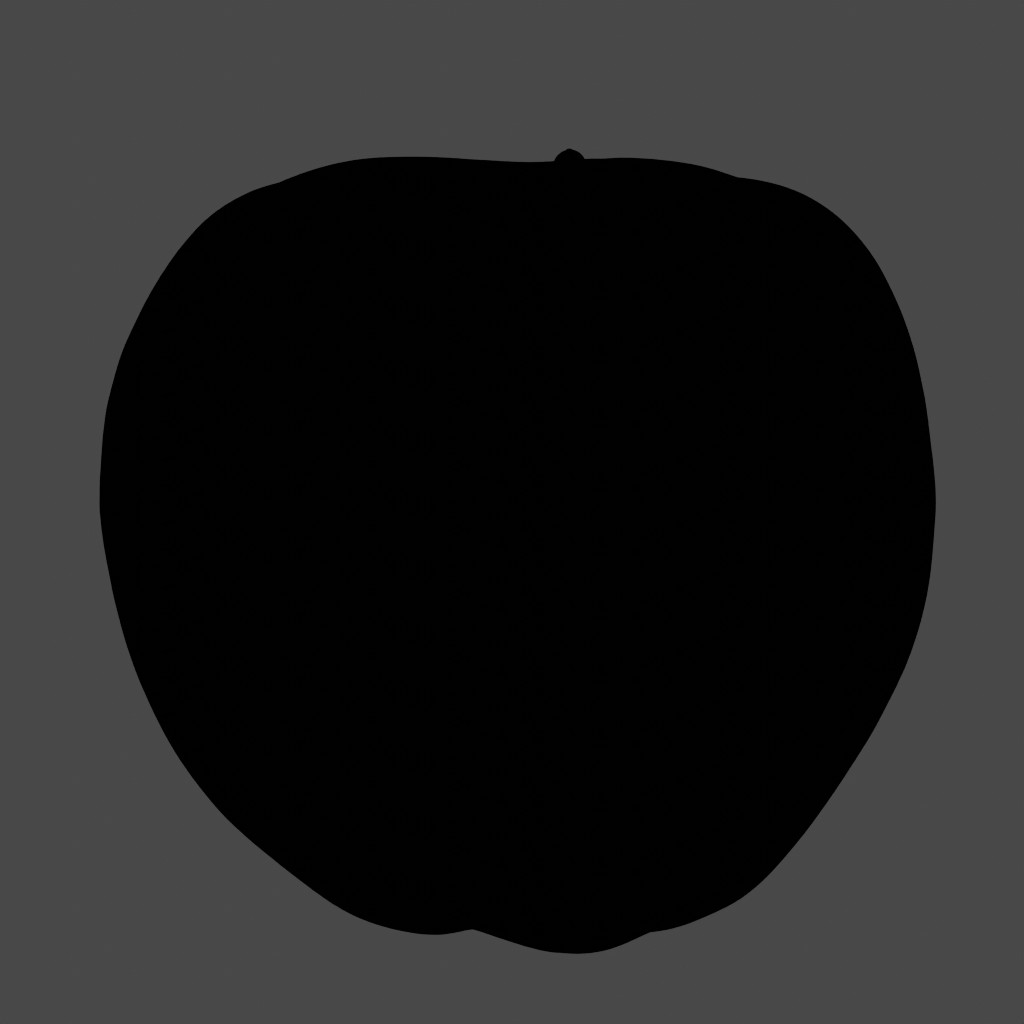}
    \includegraphics[width=0.19\linewidth]{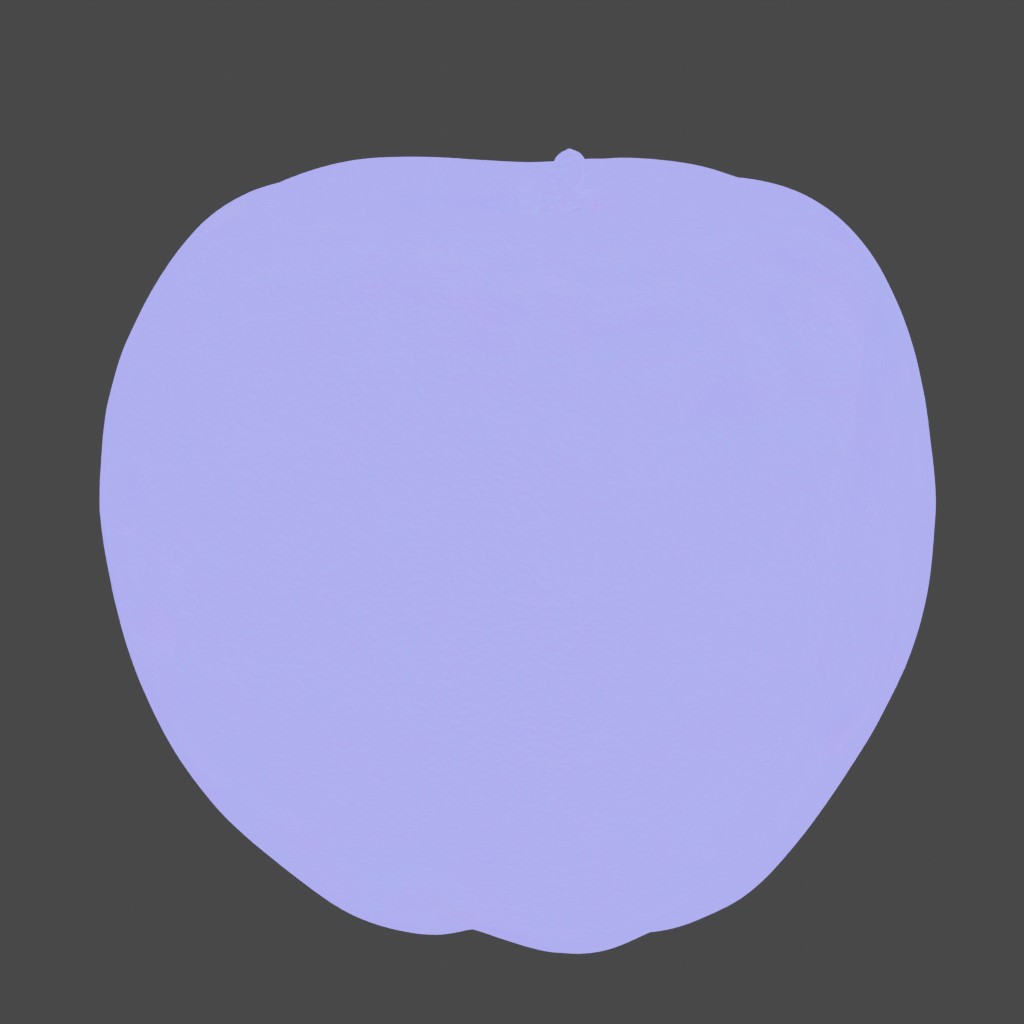}
    \includegraphics[width=0.19\linewidth]{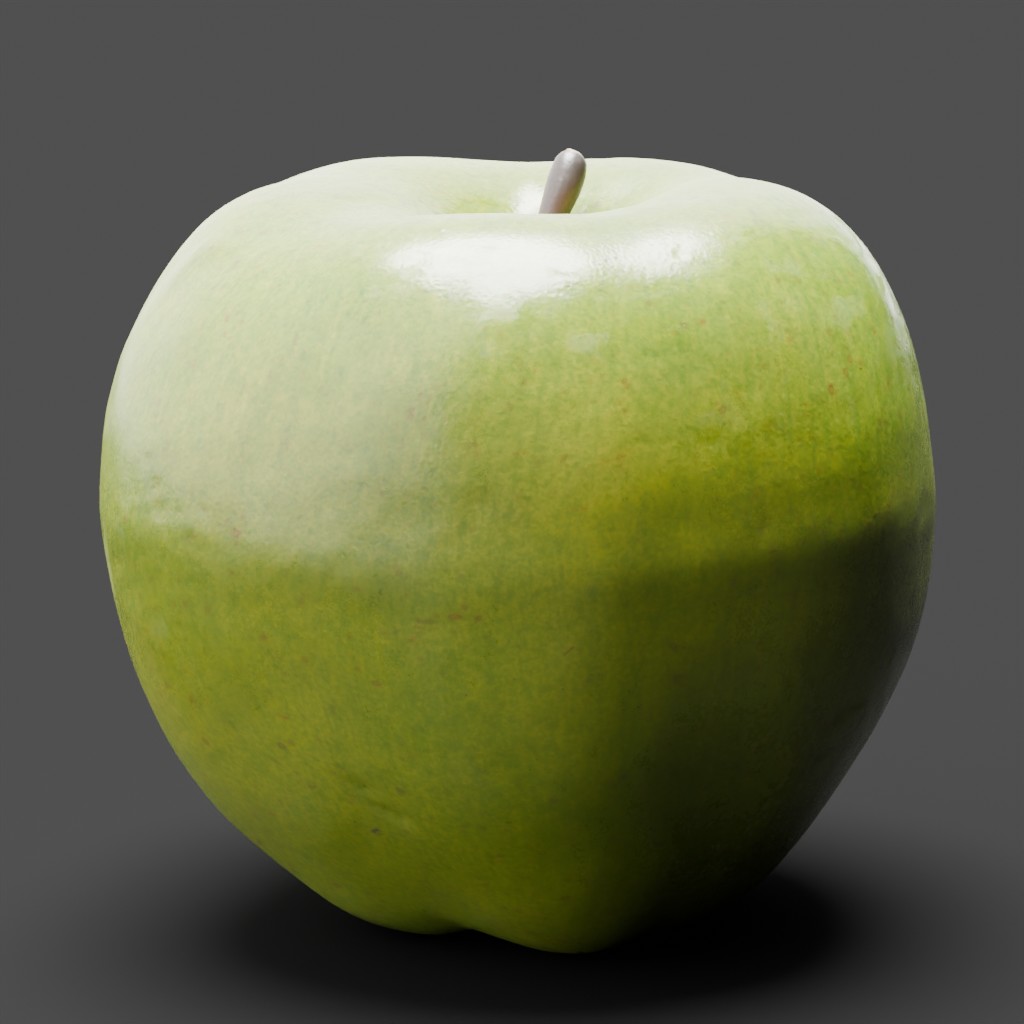} \\
    \includegraphics[width=0.19\linewidth]{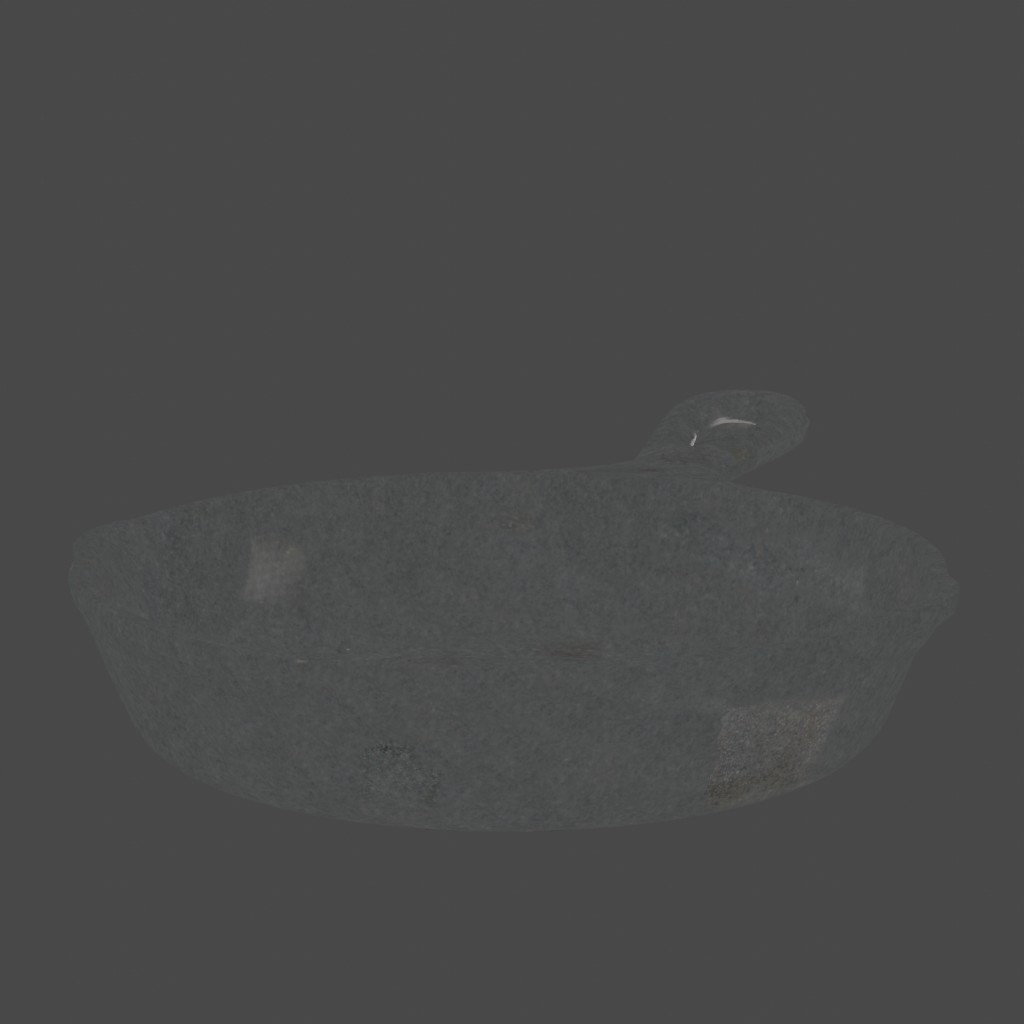}
    \includegraphics[width=0.19\linewidth]{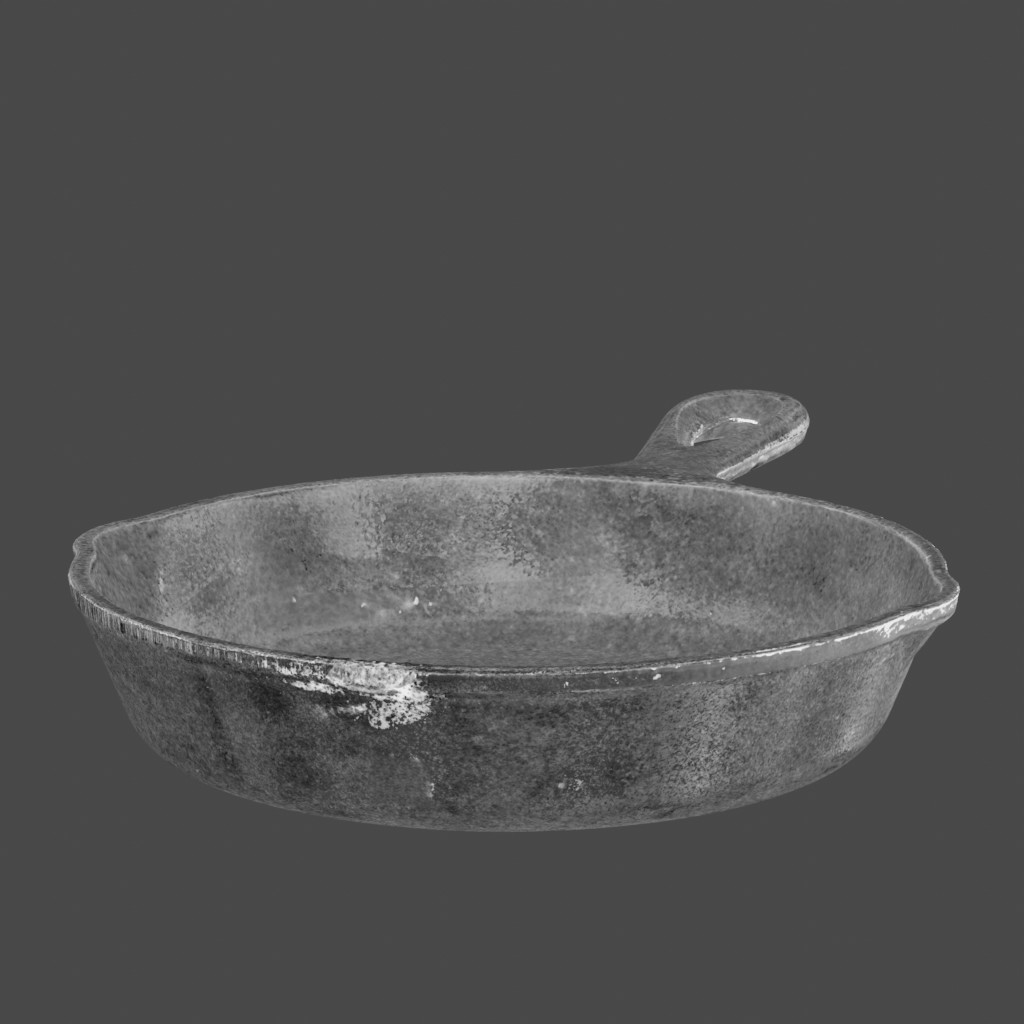}
    \includegraphics[width=0.19\linewidth]{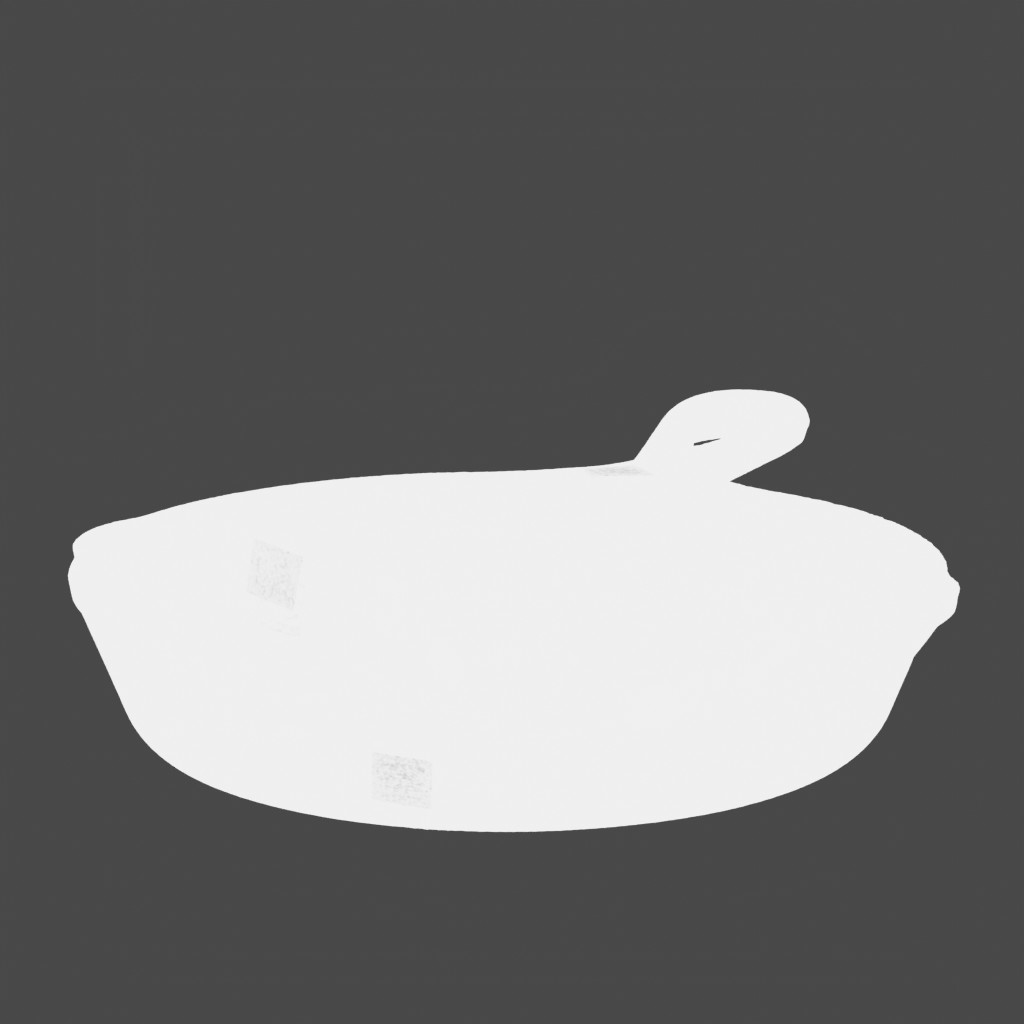}
    \includegraphics[width=0.19\linewidth]{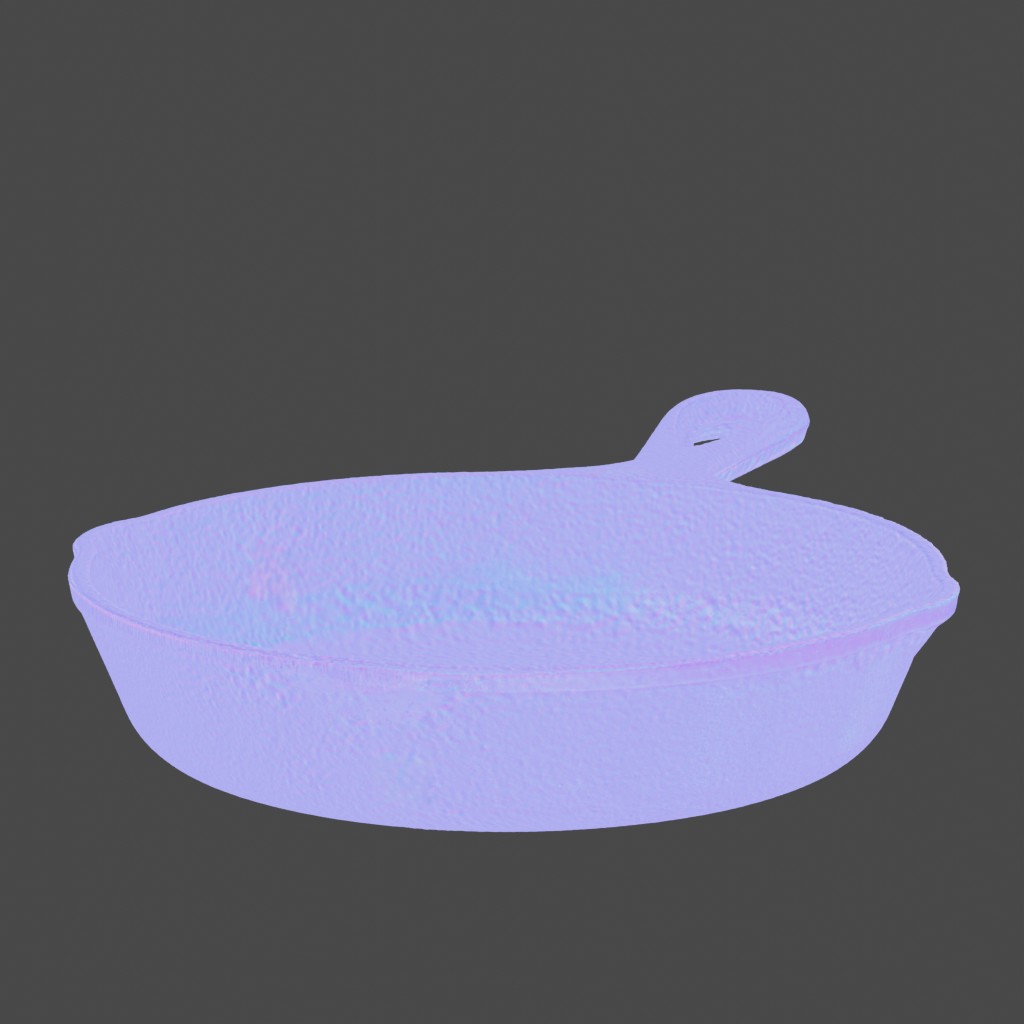}
    \includegraphics[width=0.19\linewidth]{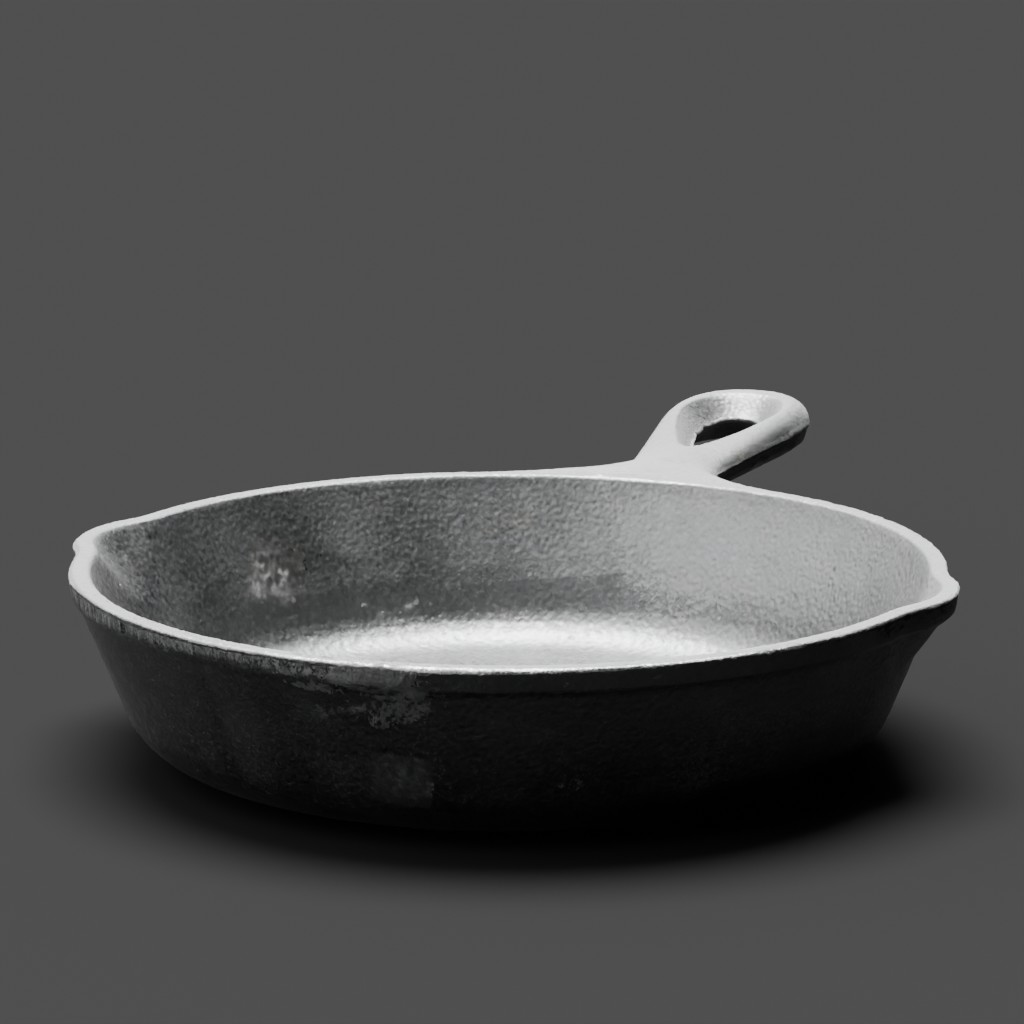}
    \caption{PBR Materials of the example DTC objects ((the list of objects in Fig.9 \textbf{Row 5})), From left to right: albedo map, roughness map, metallic map, normal map, and PBR rendering.}
    \label{fig:more_PBR_maps_row5}
\end{figure*}

For each scanned model, we generated high-quality 4K-resolution PBR material maps—including albedo, roughness, metallic, and normal maps—to achieve a photorealistic appearance. Fig.~\ref{fig:DTC_More_Models} showcases additional examples of DTC scanned digital-twin-quality 3D models, each accompanied by a full set of PBR material maps (Fig.~\ref{fig:more_PBR_maps_row1}, \ref{fig:more_PBR_maps_row2}, \ref{fig:more_PBR_maps_row3}, \ref{fig:more_PBR_maps_row4}, \ref{fig:more_PBR_maps_row5}).

\subsection{List of Models for DSLR Evaluation Data}
The full list of object models selected for DSLR evaluation data is as follows, and each model is captured under 2 different environment lighting conditions. For the 15 object recordings that are used in our benchmark evaluations, we highlight them in \textbf{bold}.
\label{subsec:model_list_DSLR}
\begin{itemize}
    \item \texttt{Airplane\_B097C7SHJH\_WhiteBlue}
    \item \texttt{Airplane\_B0B2DC5QBP\_BlueGray}
    \item \textbf{\texttt{BirdHouse}}
    \item \texttt{BirdHouse\_B0B8F27TFK\_BrownRoofYellowWalls}
    \item \texttt{BirdHouseRedRoofYellowWindows}
    \item \texttt{BirdHouseWoodenRoofGreenWall}
    \item \texttt{Bowl\_B0BQR77WRW\_LightGrey\_1\_TU}
    \item \texttt{Box\_ADTIR\_DecorativeBoxHexLarge\_Green}
    \item \texttt{Car\_38330969\_Toy}
    \item \texttt{CaramicBowlBluewithBrown}
    \item \texttt{CeramicBowlBigWhite}
    \item \textbf{\texttt{Cup\_B08TWHJ33Q\_Tan}}
    \item \texttt{Cup\_B0B3JKZW76\_Brown}
    \item \texttt{Cup\_B0CQXPND8L\_Stripes}
    \item \texttt{Dutch\_Oven\_B0B916N11D\_Black}
    \item \textbf{\texttt{Figurine\_B08FYFNYP4\_LionKing}}
    \item \texttt{Figurine\_B0983CQ2HH\_Angel}
    \item \texttt{Figurine\_B0CR3Y5T3K\_Gnome}
    \item \texttt{Gargoyle\_B005KDPAFW\_BatWings}
    \item \textbf{\texttt{Gargoyle\_B08SQMBDXY\_HandsOnKnees}}
    \item \textbf{\texttt{Gargoyle\_B0C2PNF2C1\_Meditating}}
    \item \texttt{Gravestone\_B08TBJQ5XP\_LightGrayKitty}
    \item \texttt{Hammer\_B000FK3VZ6\_Wood}
    \item \texttt{Home\_ADTIR\_A6116F6\_DraganBoxSmall\_Wood}
    \item \texttt{Home\_ADTIR\_L041OV8\_Rinnig\_PlateHolder}
    \item \texttt{Keyboard\_B07P6K5GMY\_Black}
    \item \texttt{Keyboard\_B0CL8S2DW9\_Pink}
    \item \texttt{Kitchen\_Spoon\_B008H2JLP8\_LargeWooden}
    \item \textbf{\texttt{Mallard\_B082D168CK\_MintGreen}}
    \item \textbf{\texttt{Mallard\_B09LV16HD5\_LightBrown2}}
    \item \textbf{\texttt{Mallard\_B0BPY18VHR\_White}}
    \item \texttt{Mallard\_B0C6MQWM21\_BlackWhite}
    \item \textbf{\texttt{Mouse\_B0CHNVBBLF\_Honeycomb\_1}}
    \item \texttt{Pan\_B0CFQWYJZ8\_BlackWoodHandle}
    \item \texttt{Pan\_B0CHW1KK8Z\_Black}
    \item \textbf{\texttt{Planter\_B0C4G81ZPF\_Cat}}
    \item \texttt{Pottery\_B097S319TR\_Woman}
    \item \textbf{\texttt{Pottery\_B0CJJ59SLH\_BlueHairFairy}}
    \item \textbf{\texttt{Shoe\_B000ZP6MIY\_Navy7L\_TU}}
    \item \texttt{Spoon\_B08M3XNKYR\_Slotted}
    \item \texttt{TeaPot\_B074ZQYRP7\_BrownDragonShaped}
    \item \texttt{TeaPot\_B07GL8MH3X\_PinkFlamingo}
    \item \texttt{TeaPot\_B07QP5MFQ1\_BlackCastIron}
    \item \texttt{TeaPot\_B084G3K8TD\_YellowBlackSunflowers}
    \item \textbf{\texttt{TeaPot\_B08HSDHBM4\_BlackGoldLeaves}}
    \item \textbf{\texttt{TeaPot\_B00ESU7PFG\_WhiteRoseFlowers}}
    \item \textbf{\texttt{TeaPot\_B01KFCZB2Y\_WhiteWoodHandle}}
    \item \texttt{Vase\_B09ZGXSVTT\_White\_TU}
    \item \texttt{Vase\_B0BV44B4R4\_BlueBirdsYellowBirds}
    \item \texttt{Vase\_Corrected}
\end{itemize}

\subsection{List of Models for Egocentric Evaluation Data}

The full list of models selected that contains pairs of egocentric data is as follows, and each model is captured with both \textit{active} and \textit{passive} trajectories. For the 15 object recordings that are used in our benchmark evaluations, we highlight them in \textbf{bold}.
\label{subsec:model_list_egocentric}
\begin{itemize}
    \item \texttt{Airplane\_B097C7SHJH\_WhiteBlue}
    \item \texttt{Airplane\_B09WN2RN15\_Black\_1}
    \item \texttt{BasketPlasticRectangular}
    \item \texttt{BirdHouseToy}
    \item \textbf{\texttt{BirdHouse\_B0B8F27TFK\_BrownRoofYellowWalls}}
    \item \texttt{BirdHouse\_B004HJE8AS\_WhiteWallsTwoPorches\_2}
    \item \texttt{BirdHouseRedRoofYellowWindows}
    \item \texttt{BirdHouseWoodenRoofGreenWall}
    \item 
    \texttt{BirdHouse\_B08GYBKJ8N\_RedBarn}
    \item \texttt{Birdhouse\_B09FJYJYDQ\_BoatHouse}
    \item \texttt{BirdHouseMetalRoofYellowWall\_1\_TU}
    \item \texttt{BirdHouse\_A79823645\_BearWithLogStump}
    \item \texttt{BirdHouseWoodenRoofRedWall}
    \item \texttt{BlackCeramicDishLarge}
    \item \texttt{BlackCeramicMug}
    \item \texttt{Block\_B007GE75HY\_RedBlue}
    \item \texttt{Bowl\_B0BQR77WRW\_LightGrey\_1\_TU}
    \item \texttt{Bowl\_B07ZNJ5RQV\_Orange\_TU}
    \item \texttt{CandleDishSmall}
    \item \texttt{CrateBarrelBowlRed}
    \item \textbf{\texttt{CeramicBowlBigWhite}}
    \item \texttt{CelebrateBowlPink}
    \item \texttt{Car\_38330969\_Toy}
    \item \texttt{CaramicBowlBluewithBrown}
    \item \texttt{Candle\_B0B2JQWNNQ\_White}
    \item \texttt{Candle\_B0B764F39X\_Turquoise}
    \item \texttt{Candle\_B09MP8NDML\_White}
    \item \texttt{Calculator\_B0C7GP2D5C\_Purple}
    \item \textbf{\texttt{Cup\_B08TWHJ33Q\_Tan}}
    \item \texttt{Cup\_B0B3JKZW76\_Brown}
    \item \texttt{Cup\_B0CQXPND8L\_Stripes}
    \item \texttt{Cup\_B08TWHJ33Q\_Gray}
    \item \texttt{Cup\_B09QCYR1SL\_Pitcher\_1}
    \item \texttt{Cup\_B01LYONYPB\_SkyBlue}
    \item \texttt{Dumbbell\_B00PY62Y90\_Black}
    \item \texttt{Dumbbell\_B0CN56CQPS\_PurpleCoolGray}
    \item \texttt{Dutch\_Oven\_B0B916N11D\_Black}
    \item \texttt{FakeFruit\_B076H96CS1\_Banana}
    \item \texttt{FakeFruit\_B09992T572\_WatermelonSlice}
    \item \texttt{FakeFruit\_B0815W7RKC\_StarFruit}    
    \item \texttt{Figurine\_B08FYFNYP4\_LionKing}
    \item \textbf{\texttt{Figurine\_B0983CQ2HH\_Angel}}
    \item \texttt{Figurine\_B0CR3Y5T3K\_Gnome}
    \item \textbf{\texttt{Flask}}
    \item 
    \texttt{Gargoyle\_B07GHVQ3C4\_BatCat}
    \item \textbf{\texttt{Gargoyle\_B08SQMBDXY\_HandsOnKnees}}
    \item \texttt{Gargoyle\_B0C3RQ5254\_Bronze}
    \item \texttt{Gargoyle\_B0BYTQT173\_Dragon}
    \item \texttt{Gargoyle\_B0C2PNF2C1\_Meditating}
    \item \texttt{Gargoyle\_B00N08IU24\_Dog}
    \item \texttt{Gravestone\_B08TBJQ5XP\_LightGrayKitty}
    \item \texttt{Gravestone\_B07WDGH3NR\_GrayAngel}
    \item \texttt{Hammer\_B000FK3VZ6\_Wood}
    \item \texttt{Hammer\_B0BN6FXDQ7\_BlueHandle}
    \item \texttt{Hammer\_B01N63ONKY\_DrillingSledge}
    \item \texttt{Kitchen\_Cup\_B09G2WNN61\_DarkBlue}
    \item \texttt{Kitchen\_Spoon\_B008H2JLP8\_LargeWooden}
    \item \texttt{Kitchen\_Spoon\_D146567C\_Green\_1}
    \item 
    \texttt{LargeLightGreyDish}
    \item \texttt{Mallard\_B082D168CK\_MintGreen}
    \item \textbf{\texttt{Mallard\_B09LV16HD5\_LightBrown2}}
    \item \textbf{\texttt{Mallard\_B0BPY18VHR\_White}}
    \item \texttt{Mallard\_B0C6MQWM21\_BlackWhite}
    \item \texttt{Mallard\_B00RTSJU7K\_Red}
    \item \texttt{Mouse\_B0CHNVBBLF\_Honeycomb\_1}
    \item \texttt{Pan\_B0CFQWYJZ8\_BlackWoodHandle}
    \item \texttt{Pan\_B0CHW1KK8Z\_Black}
    \item \texttt{Planter\_B0C4G81ZPF\_Cat}
    \item \texttt{PlasticBowlGreen\_1\_TU}
    \item \textbf{\texttt{Pottery\_B097S319TR\_Woman}}
    \item \texttt{Pottery\_B0CJJ59SLH\_BlueHairFairy}
    \item \texttt{Pottery\_B07TGC6TGL\_White}
    \item \texttt{Pottery\_B075SX9GVK\_White}
    \item \texttt{Pottery\_B0CF8FW987\_Man}
    \item \textbf{\texttt{Shoe\_B000ZP6MIY\_Navy7L\_TU}}
    \item \texttt{Shoe\_B094ZCQK75\_Red6HL\_TU}
    \item \textbf{\texttt{Spoon\_B08M3XNKYR\_Slotted}}
    \item \texttt{TeaPot\_B074ZQYRP7\_BrownDragonShaped}
    \item \texttt{TeaPot\_B07GL8MH3X\_PinkFlamingo}
    \item \texttt{TeaPot\_B07QP5MFQ1\_BlackCastIron}
    \item \texttt{TeaPot\_B084G3K8TD\_YellowBlackSunflowers\_TU}
    \item \textbf{\texttt{TeaPot\_B08HSDHBM4\_BlackGoldLeaves}}
    \item \textbf{\texttt{TeaPot\_B00ESU7PFG\_WhiteRoseFlowers}}
    \item \textbf{\texttt{TeaPot\_B01KFCZB2Y\_WhiteWoodHandle}}
    \item \texttt{TeaPot\_B07RT7BYXL\_WhiteSnowOwl\_TU}
    \item \textbf{\texttt{Vase\_B09ZGXSVTT\_White\_TU}}
    \item \texttt{Vase\_B0BV44B4R4\_BlueBirdsYellowBirds}
    \item \texttt{Vase\_Corrected}
    \item \texttt{Vase\_B0BNX2CWW8\_YellowTall}
    \item \texttt{Vase\_B00858OOXI\_Green\_1}
    \item \texttt{Vase\_B09FYBCM1R\_BeeSunflowers}
    \item \texttt{Vase\_B0BY8PYLLC\_PinkPineapple}
    \item \texttt{Vase\_B0BNX2CWW8\_YellowShort}
    \item 
    \texttt{WhiteContainerBox}
    \item 
    \texttt{WhiteClip\_1}
    \item \texttt{WoodBlocks\_B00FIX22YQ\_BlueArch}
    \item \texttt{WoodBlocks\_B098PHYN3P\_SmallOctagon}
    \item 
    \texttt{WoodenBoxSmall}
    \item 
    \texttt{WoodenFork}
\end{itemize}

\subsection{An Example of DSLR Data}
\label{subsec:example_dslr_data}
\begin{figure}
    \centering
    \includegraphics[width=\linewidth]{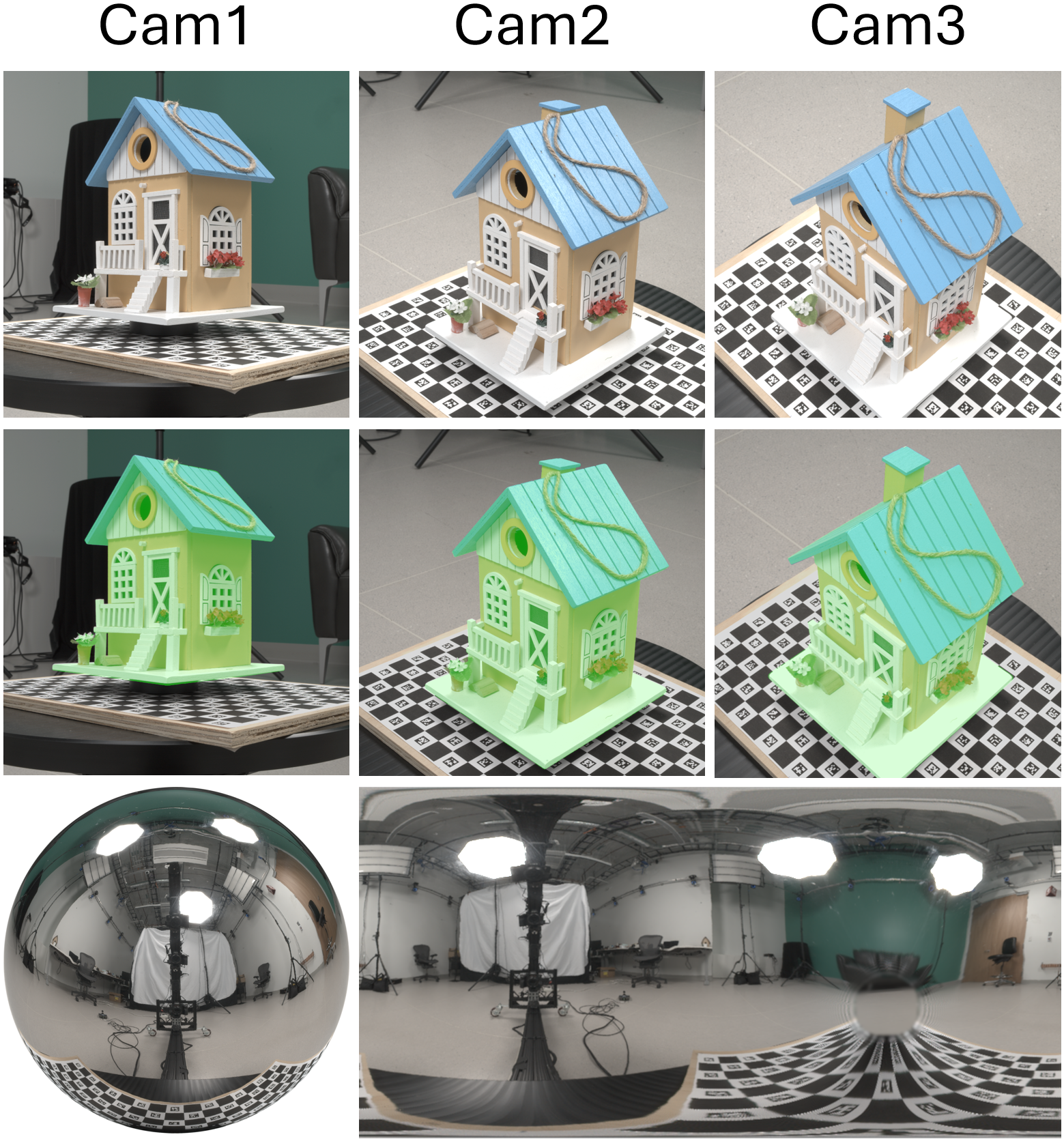}
    \caption{Example DSLR evaluation data: captured images with 3 DSLR cameras from different elevation angles (first row), the object mask overlaid on each image (second row) and the recovered HDR envmap (last row). }
    \label{fig:dslr_example_data}
\end{figure}
As described in Sec.~\ref{subsec:dslr}, we used three DSLR cameras mounted on a rig to capture 360-degree inward-facing photos of each object. As shown in Fig.~\ref{fig:dslr_example_data}, these DSLR cameras were positioned at different elevation angles and captured images circularly around the object with a 9-degree interval between shots. Each camera captured 40 photos, resulting in a total of 120 photos per object. For each photo, we provide the corresponding camera pose and an accurate foreground object mask (2nd row overlaid on the original image in Fig.~\ref{fig:dslr_example_data}) to facilitate straightforward and reliable evaluation. Additionally, to support the evaluation of relightable reconstruction, we utilized a mirror ball to capture and reconstruct an HDR environment map (bottom row in Fig.\ref{fig:dslr_example_data}).

\subsection{An Example of Egocentric Data}
\label{subsec:example_ego_data}

\begin{figure*}[t]
    \centering
    \includegraphics[width=0.19\linewidth]{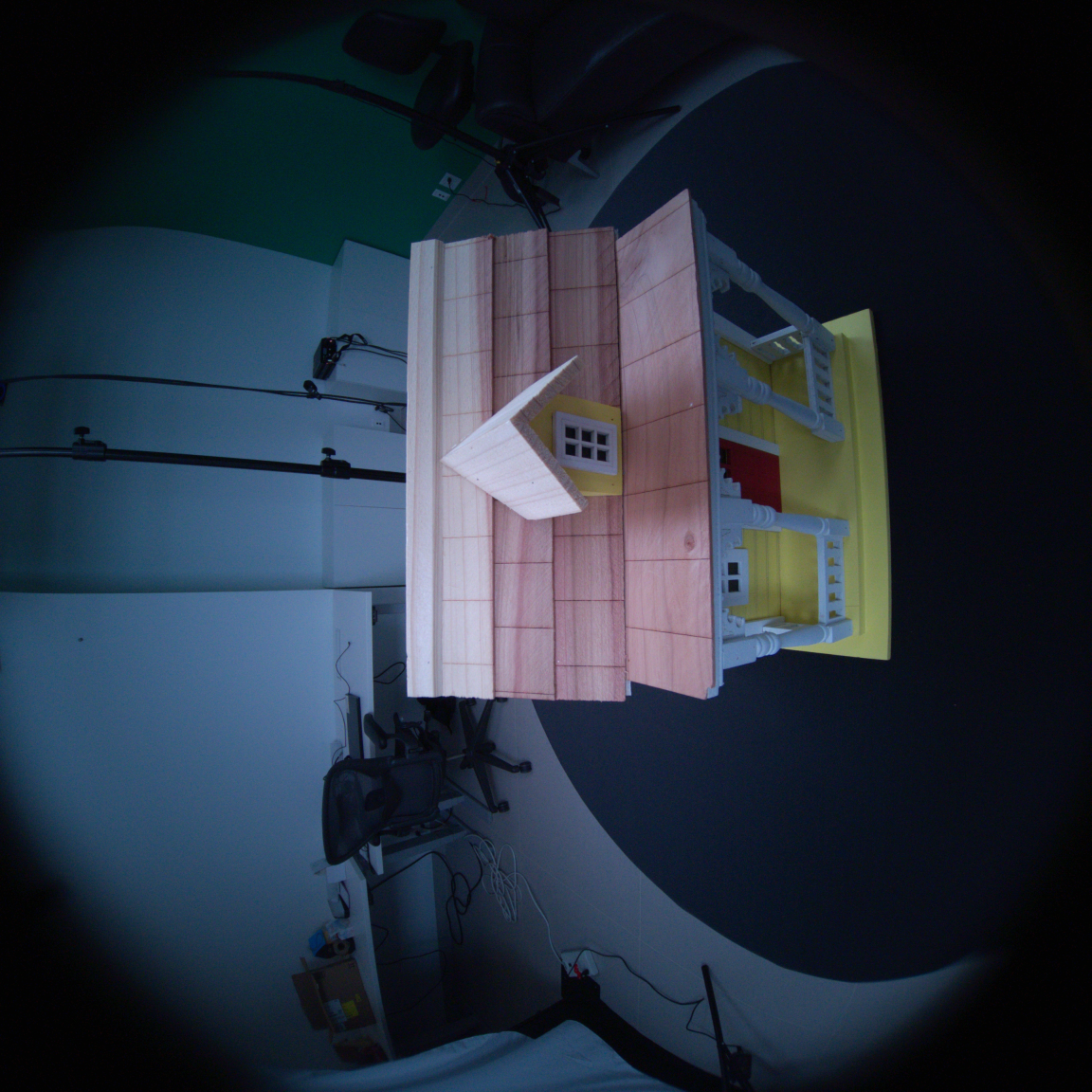}
    \includegraphics[width=0.19\linewidth]{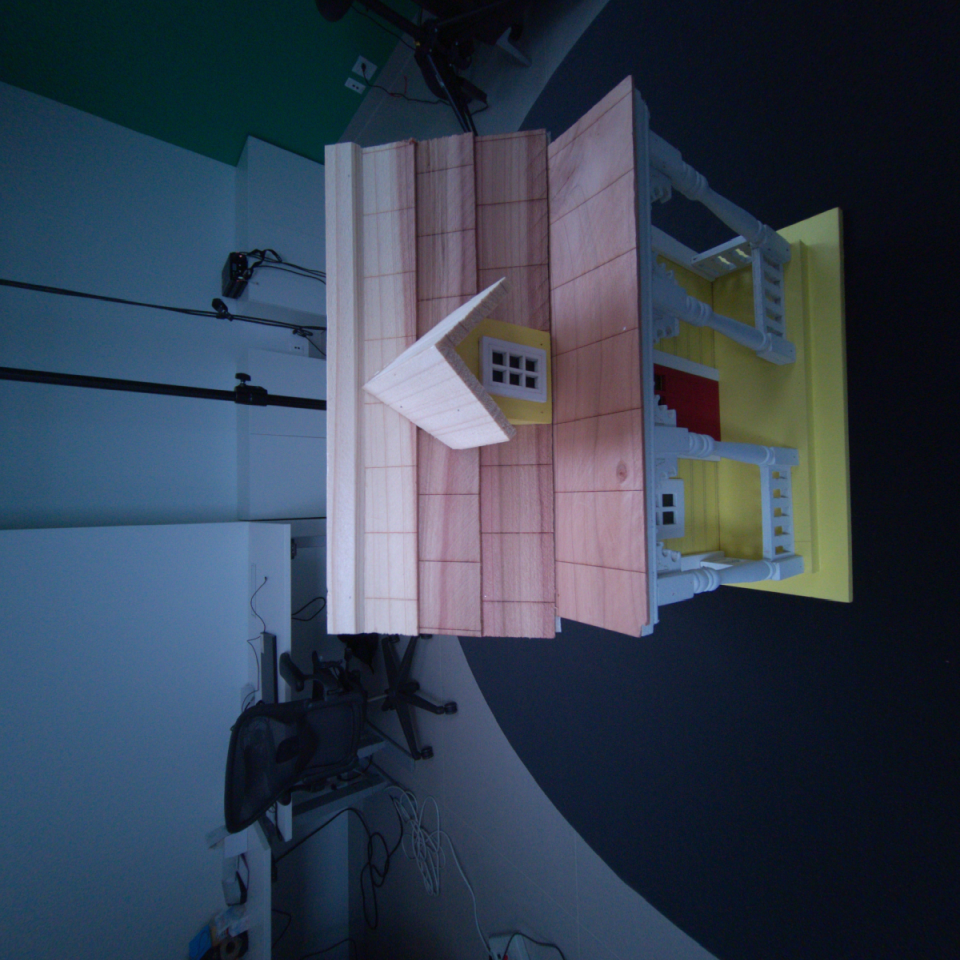}
    \includegraphics[width=0.19\linewidth]{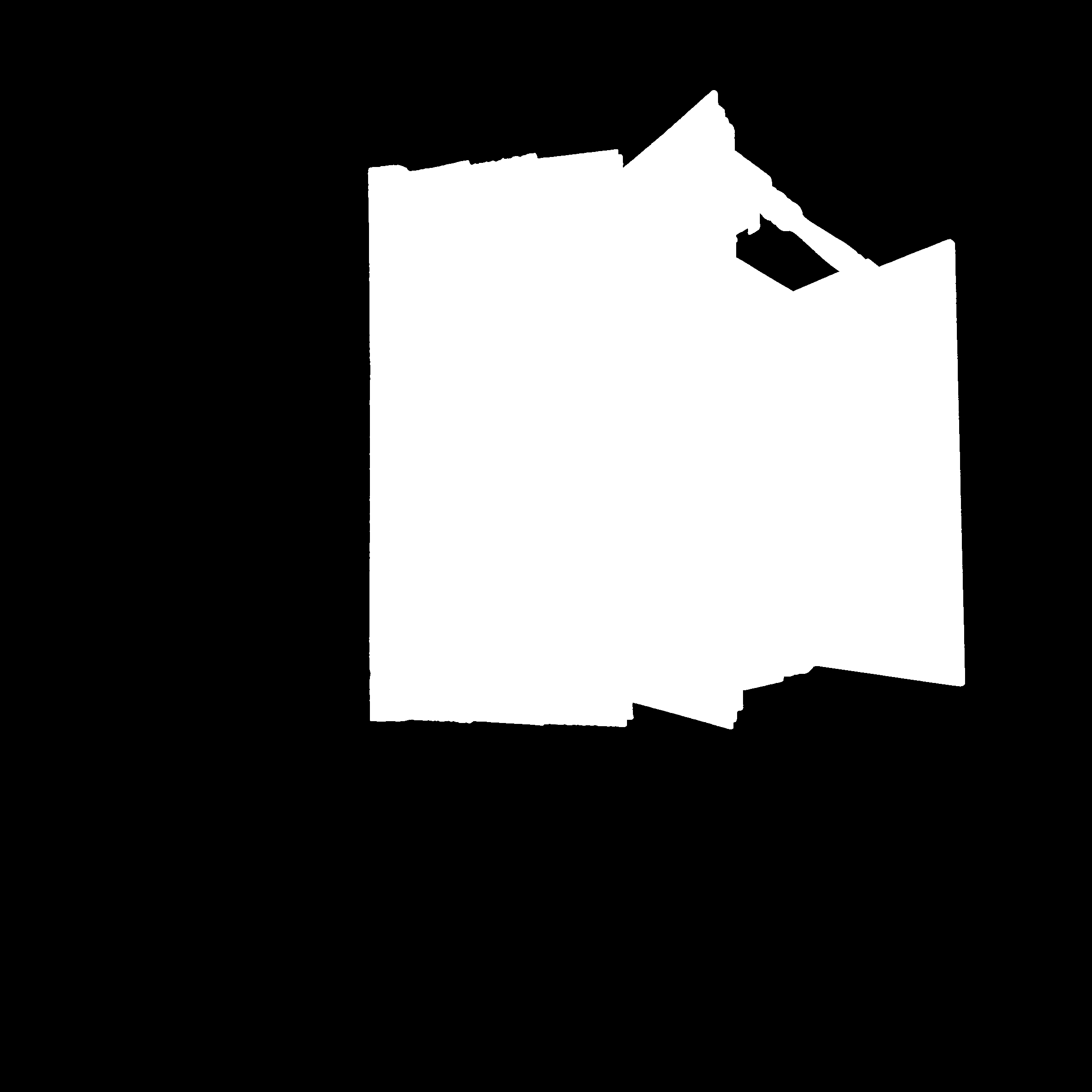}
    \includegraphics[width=0.19\linewidth]{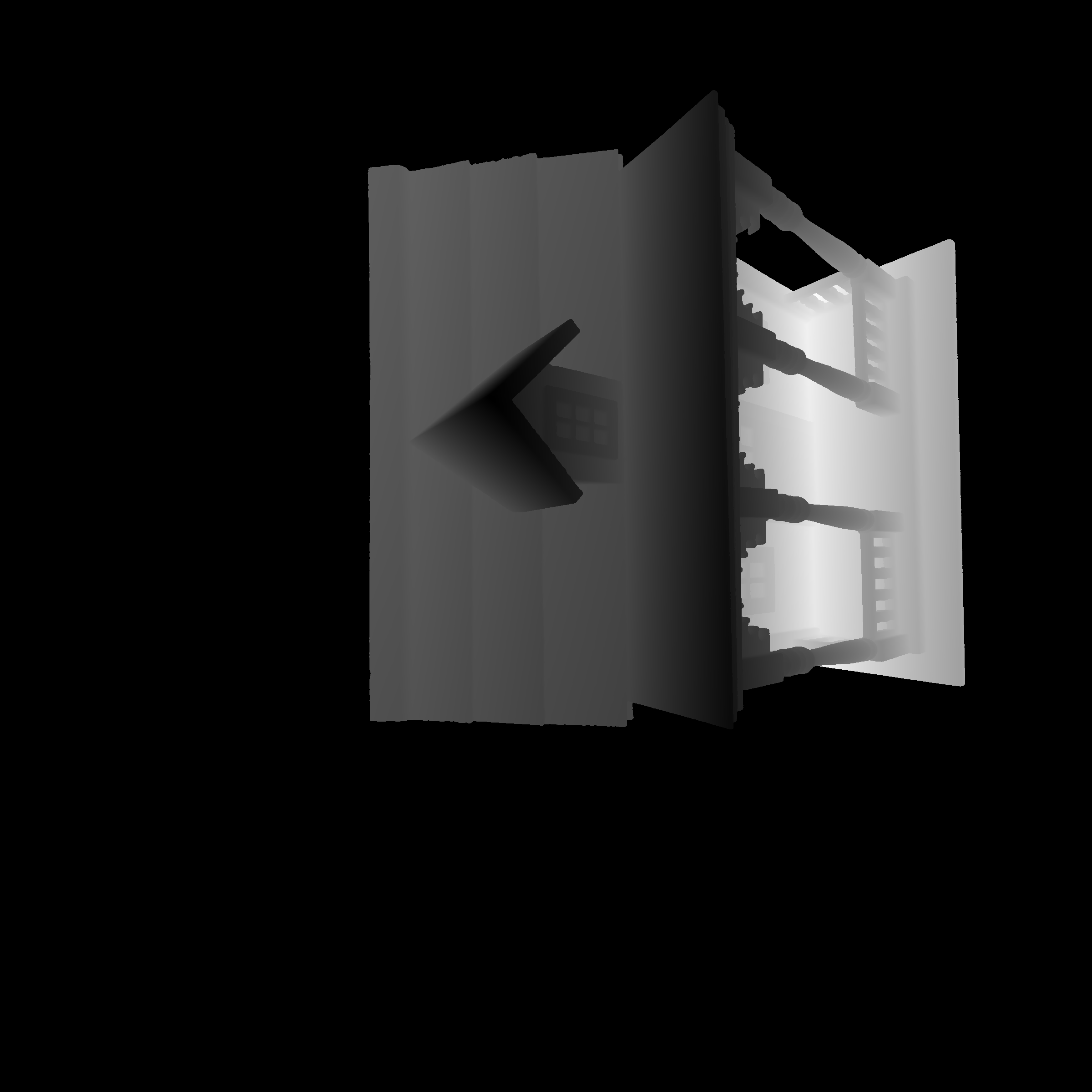}
    \includegraphics[width=0.19\linewidth]{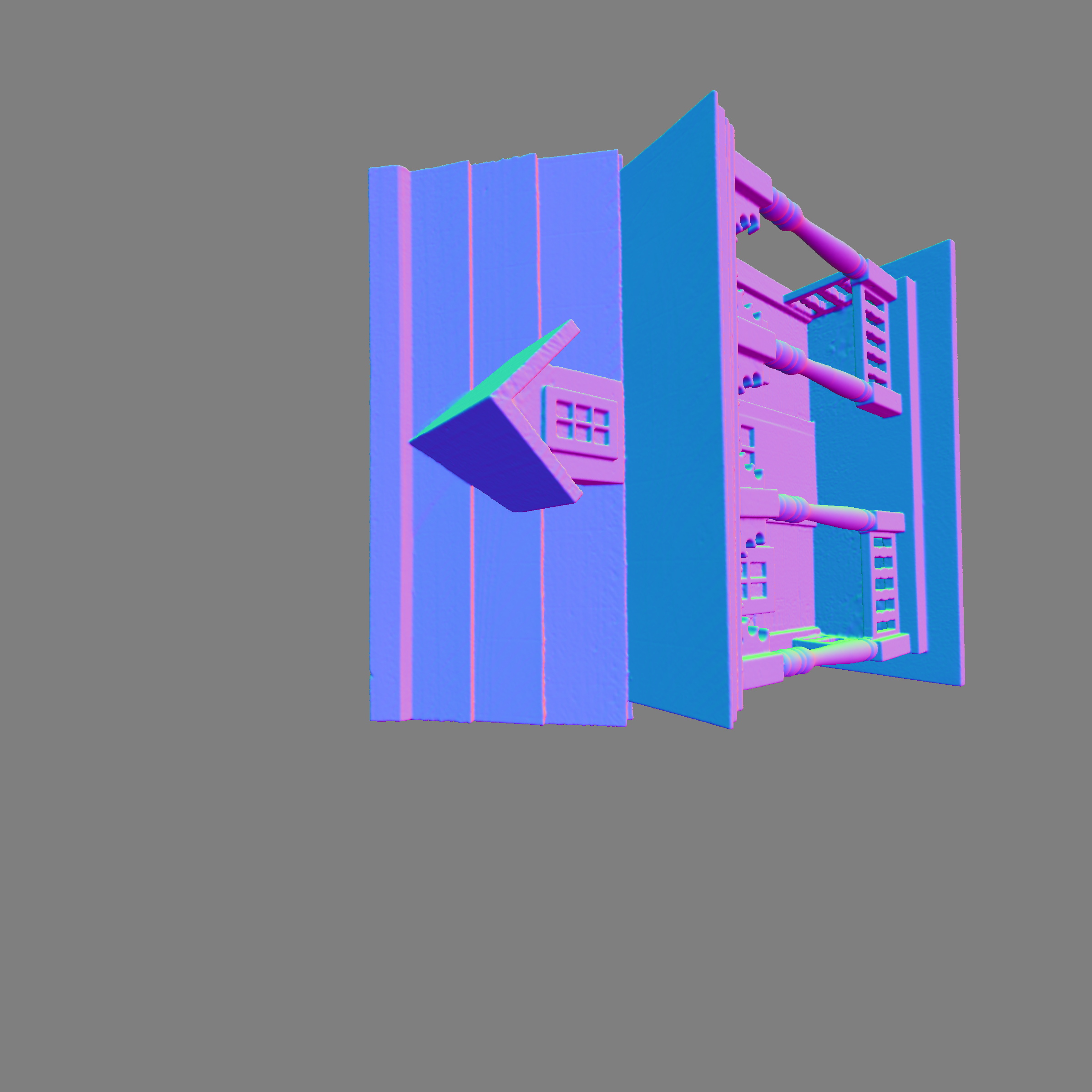}
    \caption{An example of visualization for the egocentric recordings. From left to right, we present the original egocentric view from the fisheye RGB camera input, the rectified image view, object mask, rendered depth, and rendered normal. The egocentric recording are gravity aligned respect to the global trajectory coordinate. We use the rectified image for all evaluations.}
    \label{fig:aria_examples}
\end{figure*}
\begin{figure}[t]
    \centering
    \includegraphics[width=0.95\linewidth]{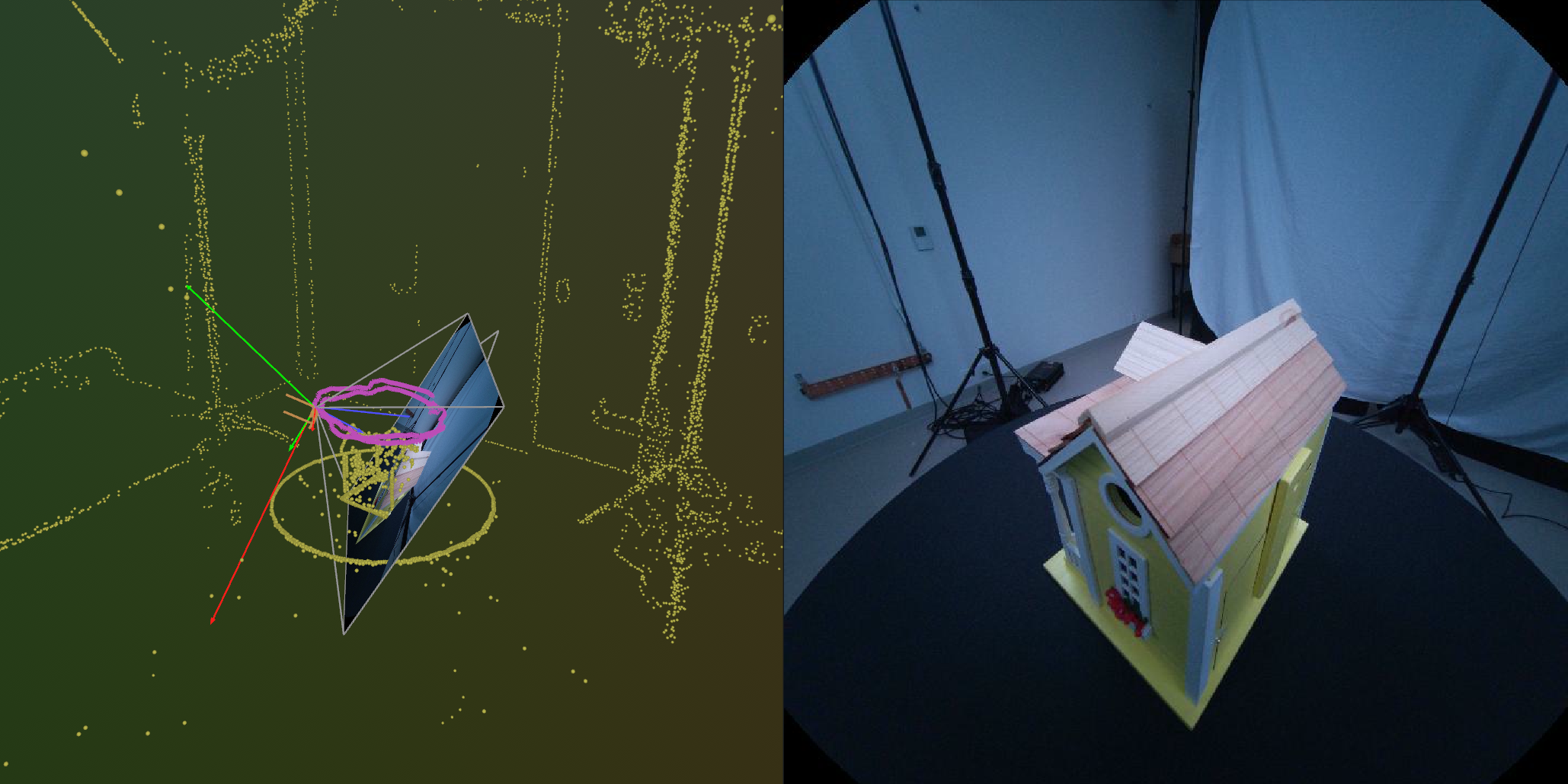}
    \includegraphics[width=0.95\linewidth]{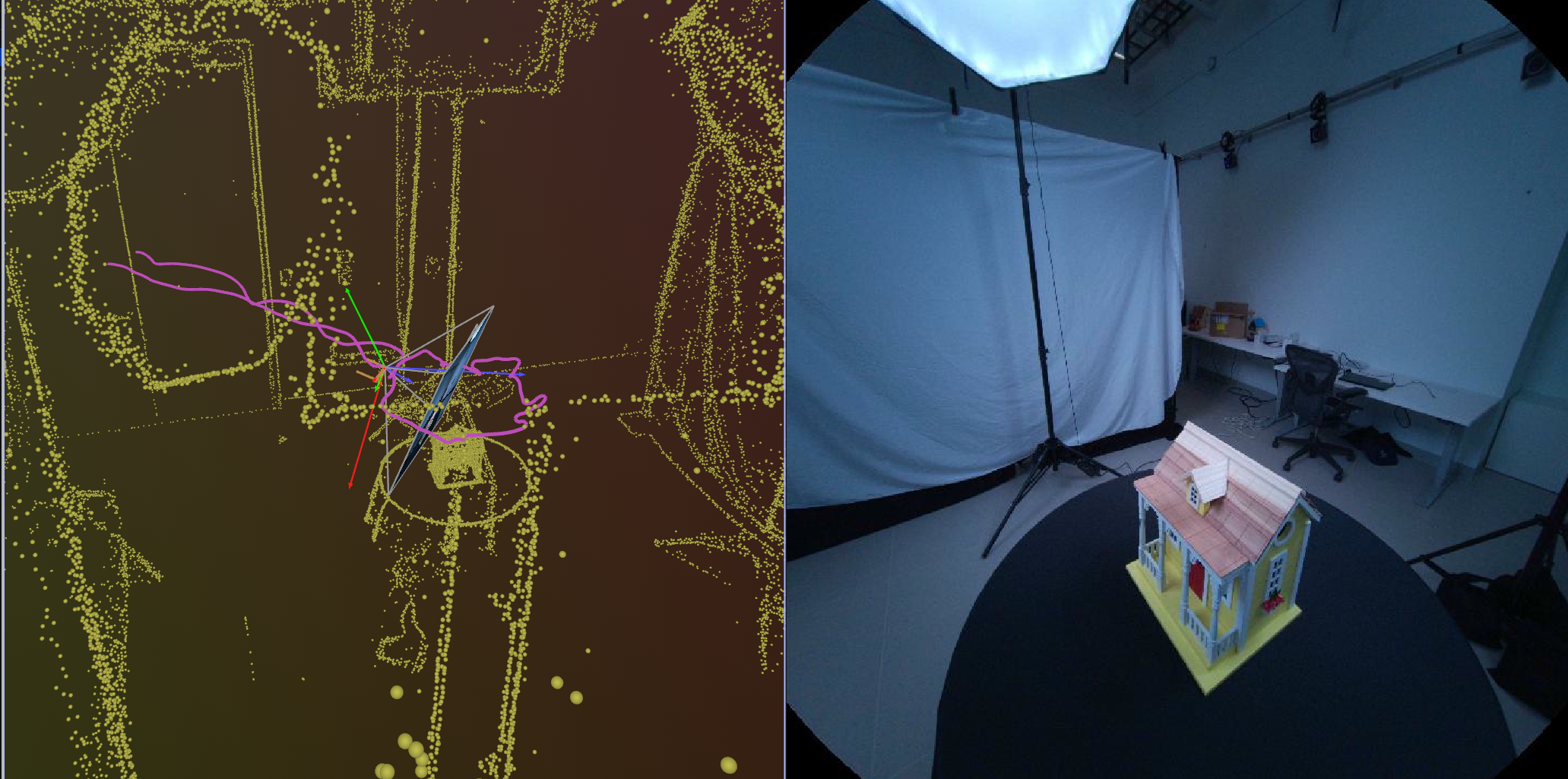}
    \caption{An snapshot visualization of the active (above) and passive (below) of egocentric recordings in the 3d space. The left of image shows the trajectory (visualized in Violet color) of the camera movement in each scenario. The yellow points are the scene point cloud.}
    \label{fig:aria_examples_passive_active}
\end{figure}

Fig.~\ref{fig:aria_examples} provides an example visualization of the egocentric recording observed from its RGB camera and the rendered ground truth using the 3D aligned digital twin model using the method described in Sec.~\ref{subsec:aria_eval_data}. The raw RGB input from Project Aria is a wide angle fisheye camera, which limits the amount of existing off-the-shelf baselines using pin-hole camera models. To resolve this limitation, we rectify all the raw images using a linear camera model, and render the corresponding ground truth of the linear camera model as well. All the evaluations on egocentric recordings are performed using the linear rectified images as input and output. We use the focal length 1200 and 2400x2400 resolution which retains the original pixel resolution and field of view from the raw image input. 

It is worth noting the image input from Project Aria is gravity aligned. Both the raw image and rectified image will appear as 90 degree counter-clock wise rotated according to the gravity direction without explicitly updating the camera calibration information. For the reconstruction and evaluation, we do not rotate the image or update the calibration. We only rotate the rendered images 90 degree clockwise for the benefit of visualization purpose (e.g. Fig.~\ref{fig:aria_baseline_comparisons})

We provided two different type of recordings to benchmark reconstruction from egocentric views. Fig.~\ref{fig:aria_examples_passive_active} shows a snapshot of the egocentric recording. We demonstrate each type of camera trajectory overlaid in the 3D space using the scene point cloud. The active trajectory represents a common object-centric 360 view of the objects while the passive trajectory represents a more casual walk around the object from different distances.

\section{Complementary Benchmarks}
\label{sec:complementary_benchmarks}

In this section, we provide complementary results on the different applications of benchmarking in Sec.~\ref{sec:benchmark_SOTA}. First, we include the qualitative comparisons of baselines in the inverse rendering application using the DTC dataset. We use the same baselines described in Sec.~\ref{subsec:inverse_rendering_dslr}. Second, as indicated in the main paper, we include a new baseline evaluation for sparse-view reconstruction in Sec.~\ref{sec:lrm_dslr} for DSLR data and Sec.~\ref{sec:lrm_passive} for egocentric data. Third, we present the qualitative comparisons of egocentric active reconstruction baselines described in Sec.~\ref{subsec:egocentric_nvs}. 

\paragraph{Explanations on Evaluation Metrics.} In Sec.~\ref{subsec:inverse_rendering_dslr}, we evaluate the quality of \textit{geometry reconstruction} by comparing the rendered depth maps, normal maps, and predicted 3D meshes to the ground truth. For depth maps, we use the Scale-Invariant Mean Squared Error (\textbf{SI-MSE}) as the evaluation metric. Normal maps are evaluated using \textbf{Cosine Distance}, and for the 3D meshes, we compute the \textbf{Bi-directional Chamfer Distance} between sampled surface points. For \textit{novel view synthesis} and \textit{novel scene relighting}, we utilize widely adopted metrics: Peak Signal-to-Noise Ratio for both HDR (\textbf{PSNR-H}) and LDR (\textbf{PSNR-L}), Structural Similarity Index Measure (\textbf{SSIM})~\cite{wang2004image}, and Learned Perceptual Image Patch Similarity (\textbf{LPIPS})~\cite{zhang2018unreasonable}. We follow PhySG~\cite{physg2021} and adapt all these metrics to be scale-invariant, addressing the ambiguity in material and lighting decomposition. We use the same metrics for novel-view synthesis in the egocentric novel-view synthesis applications as well. The same evaluation metrics are used in StanfordORB~\cite{kuang2023stanfordorb}.

\begin{figure*}[t]
    \centering
    \small
    \begin{tabular}{cccccc|ccccc}
    & {\hskip 10pt}Normal & {\hskip 5pt}Depth & {\hskip 5pt}\begin{tabular}{@{}c@{}}Novel \\ View\end{tabular} & {\hskip 8pt}Relight & Albedo & 
    {\hskip 8pt}Normal & {\hskip 5pt}Depth & {\hskip 7pt}\begin{tabular}{@{}c@{}}Novel \\ View\end{tabular}& {\hskip 6pt}Relight & Albedo\\
    \raisebox{15pt}[0pt][0pt]{Ground Truth} &
        \multicolumn{5}{c}{\includegraphics[width=0.42\linewidth]{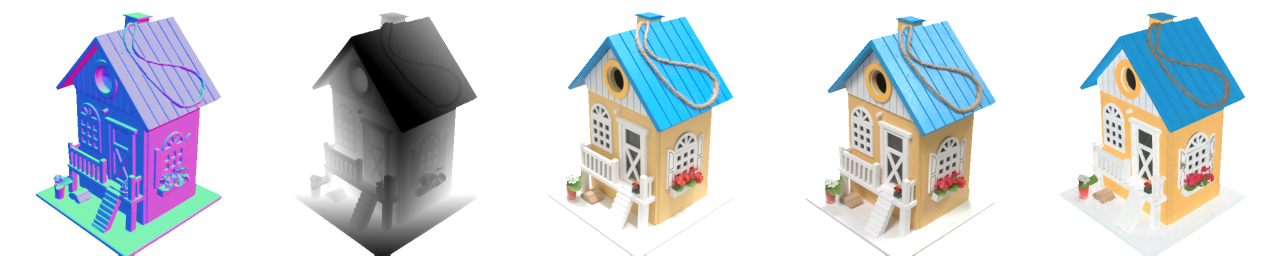}} &
        \multicolumn{5}{c}{\includegraphics[width=0.42\linewidth]{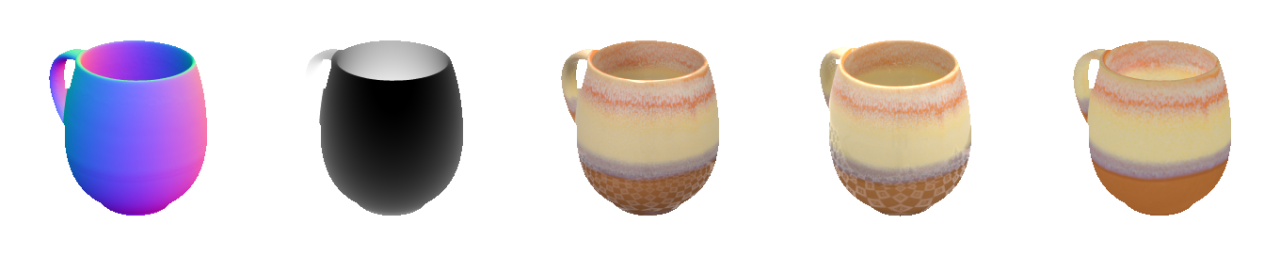}}
        \\
    \raisebox{15pt}[0pt][0pt]{Neural-PIL} &
        \multicolumn{5}{c}{\includegraphics[width=0.42\linewidth]{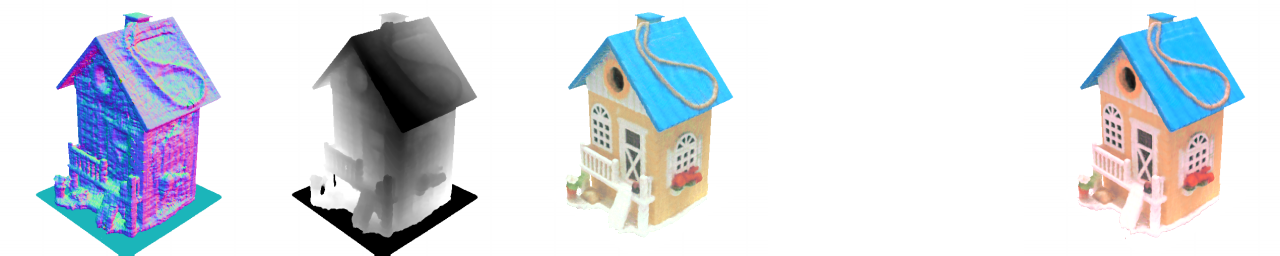}} &
        \multicolumn{5}{c}{\includegraphics[width=0.42\linewidth]{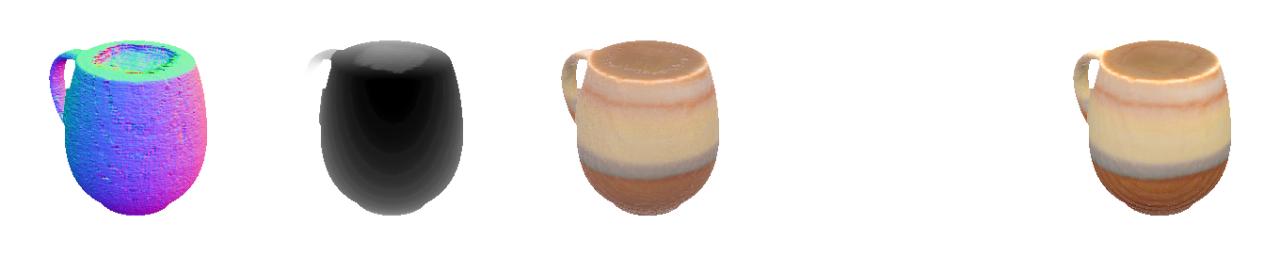}}\\     
    \raisebox{15pt}[0pt][0pt]{PhySG} &
        \multicolumn{5}{c}{\includegraphics[width=0.42\linewidth] {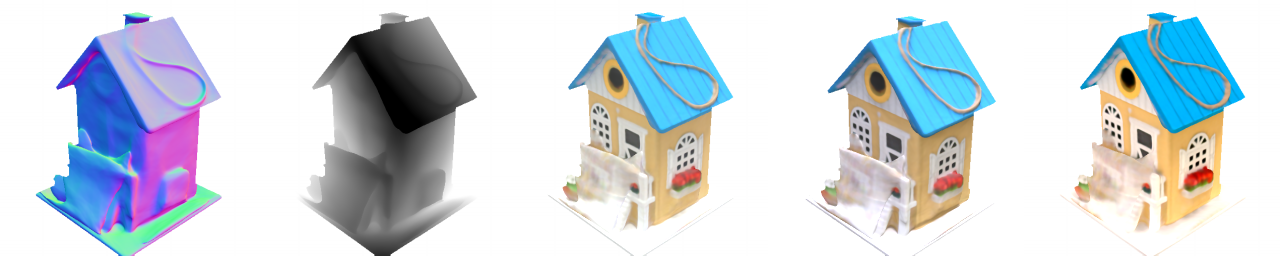}} &
        \multicolumn{5}{c}{\includegraphics[width=0.42\linewidth]{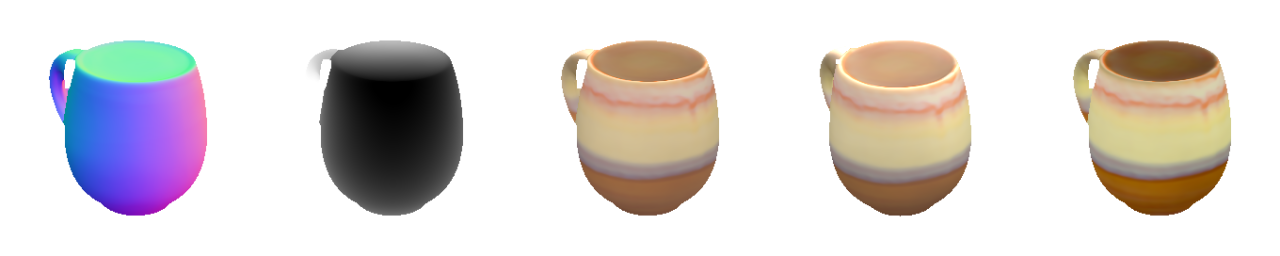}}\\
    \raisebox{15pt}[0pt][0pt]{NVDiffRec} &
        \multicolumn{5}{c}{\includegraphics[width=0.42\linewidth]{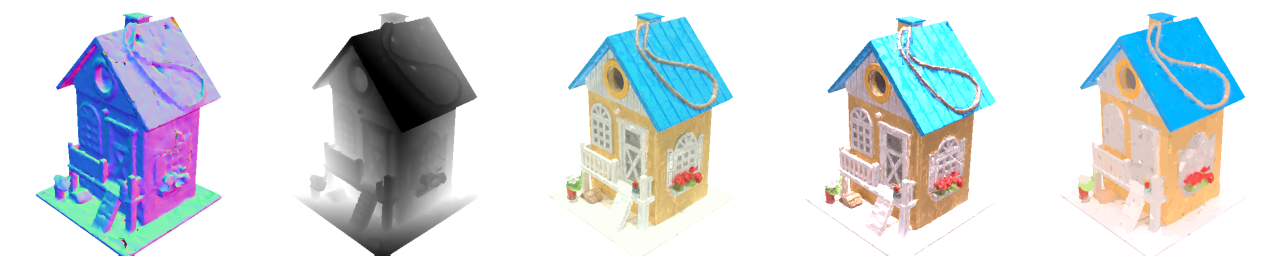}} &
        \multicolumn{5}{c}{\includegraphics[width=0.42\linewidth]{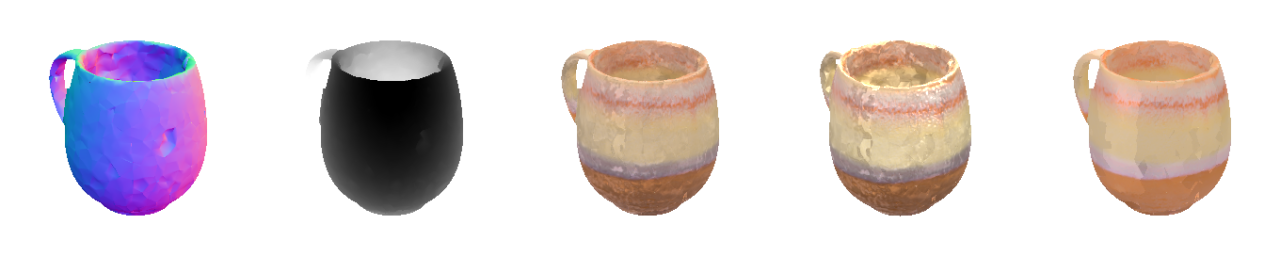}}\\
    \raisebox{15pt}[0pt][0pt]{NeRD} &
        \multicolumn{5}{c}{\includegraphics[width=0.42\linewidth]{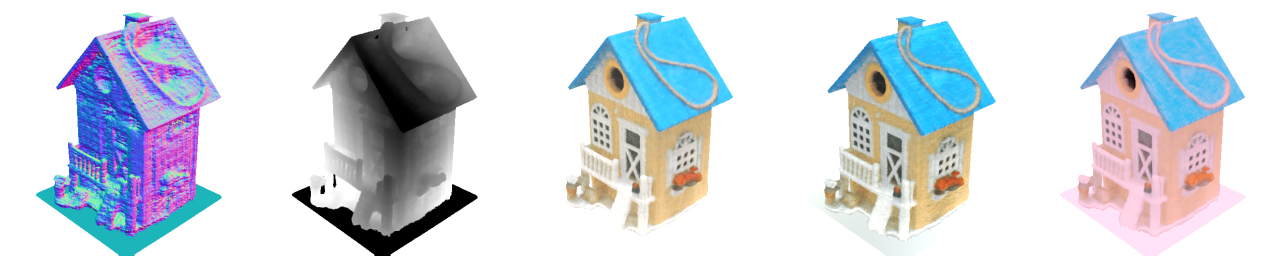}} &
        \multicolumn{5}{c}{\includegraphics[width=0.42\linewidth]{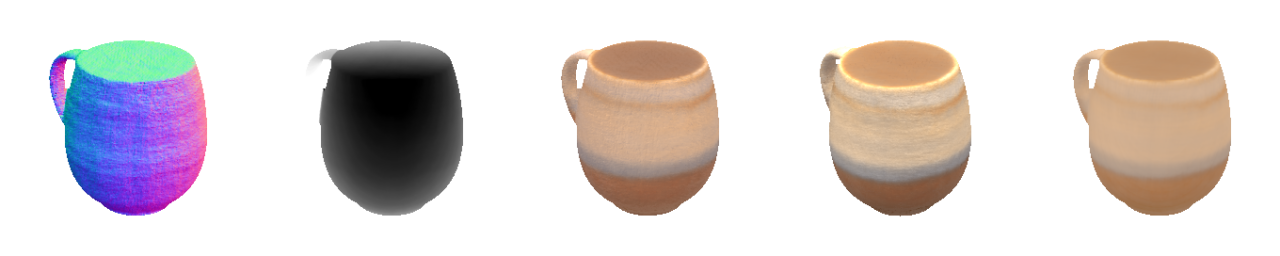}}\\
    \raisebox{15pt}[0pt][0pt]{InvRender} &
        \multicolumn{5}{c}{\includegraphics[width=0.42\linewidth]{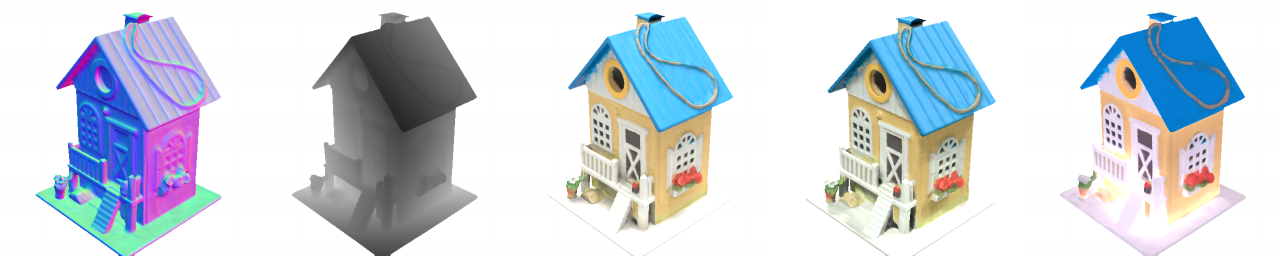}} &
        \multicolumn{5}{c}{\includegraphics[width=0.42\linewidth]{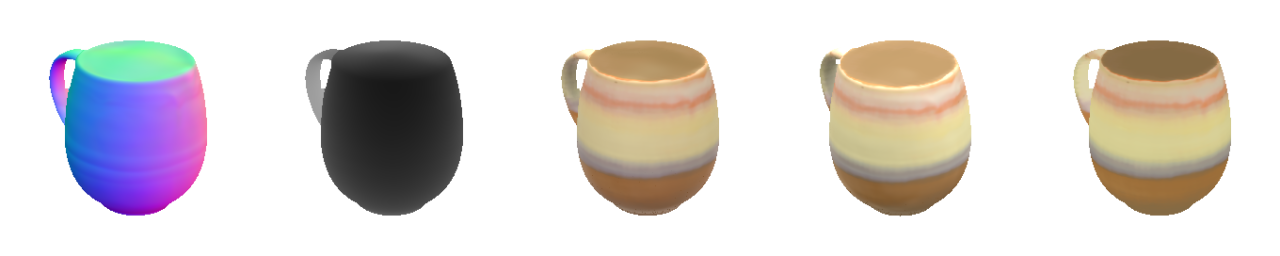}}\\
    \raisebox{15pt}[0pt][0pt]{NVDiffRecMC} &
        \multicolumn{5}{c}{\includegraphics[width=0.42\linewidth]{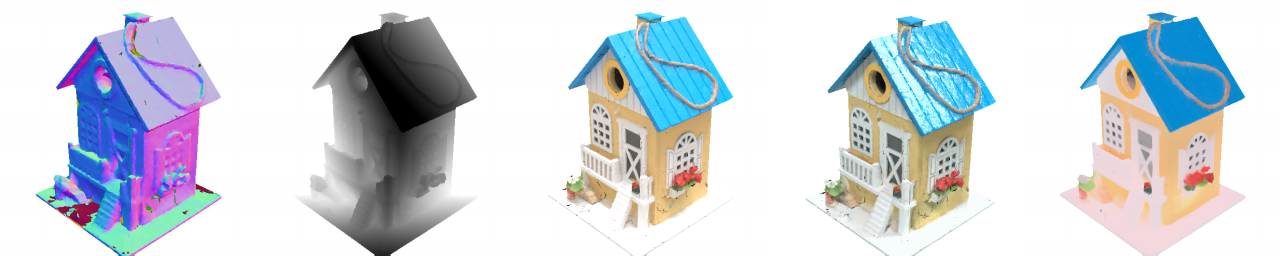}} &
        \multicolumn{5}{c}{\includegraphics[width=0.42\linewidth]{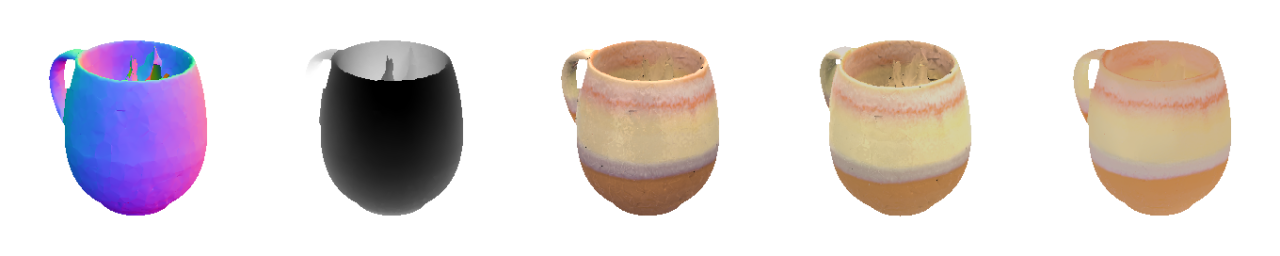}}\\
    \end{tabular}
    \caption{\label{fig:ir_qualitative}
    \textbf{Qualitative Comparisons of Baseline Methods}. We show qualitative comparisons of baseline methods, used in Table~\ref{tab:benchmark}, on two different DTC models \texttt{BirdHouseToy} (Left) and \texttt{Cup\_B08TWHJ33Q\_Tan} (Right).
    }
\end{figure*}
\label{sec:more_benchmarks}
\subsection{Inverse Rendering Baseline Comparisons}
We provide qualitative comparisons as a supplement to the inverse rendering baseline comparisons in Sec.~\ref{subsec:inverse_rendering_dslr}. Our DTC dataset contains objects with varying geometric complexity, ranging from simpler ones, \eg, cups (\cref{fig:ir_qualitative} Right), to complex ones, \eg, toy birdhouses (\cref{fig:ir_qualitative} Left). The surface material ranges from diffuse ones, \eg, toy houses, to highly reflective ones, \eg, cups. 
For the two examples in \cref{fig:ir_qualitative}, we observed that methods using surface-based representations, \ie, PhySG and InvRender, and hybrid representations, \ie, NVDiffRec and NVDiffRecMC, tend to produce smoother geometry compared to methods using NeRF~\cite{mildenhall2020nerf}-based representations, \ie, Neural-PIL and NeRD. 

\begin{table}
    \caption{Quantitative evaluation of trained LRM-VolSDF baseline for view synthesis on \textbf{GSO} dataset with different number of images. It achieves the state-of-the-art performance compared to prior work.}
    \label{tab:lrmvolsdf_baseline}
    \vspace{-0.1in}
    \setlength{\tabcolsep}{6pt}
    \renewcommand{\arraystretch}{0.4} 
    \small
    \centering
    \begin{tabular}{lccc}
    \toprule
     LRM-VolSDF & PSNR ($\uparrow$) & SSIM ($\uparrow$) & LPIPS ($\downarrow$) \\
    \midrule
    MeshLRM \cite{wei2024meshlrm} & 28.13 & 0.923 & 0.093 \\
      4 images & 28.72 & 0.940 & 0.070 \\
      8 images & 30.19 & 0.947 & 0.061 \\
     \bottomrule
    \end{tabular}
\end{table}

\begin{table}
    \caption{Quantitative results on novel view synthesis for sparse-view reconstruction using DTC DSLR data.}
    \label{tab:lrm_dslr}
    \vspace{-0.1in}
    \setlength{\tabcolsep}{4pt}
    \renewcommand{\arraystretch}{0.4} 
    \footnotesize
    \centering
    \begin{tabular}{lcccc}
    \toprule
    & PSNR-H ($\uparrow$) & PSNR-L ($\uparrow$) & SSIM ($\uparrow$) & LPIPS ($\downarrow$) \\
    \midrule
     LRM-VolSDF & 20.53 & 21.16 & 0.993 & 0.006 \\
    \bottomrule
    \end{tabular}
\end{table}

\begin{figure}
    \centering
    \includegraphics[width=\columnwidth]{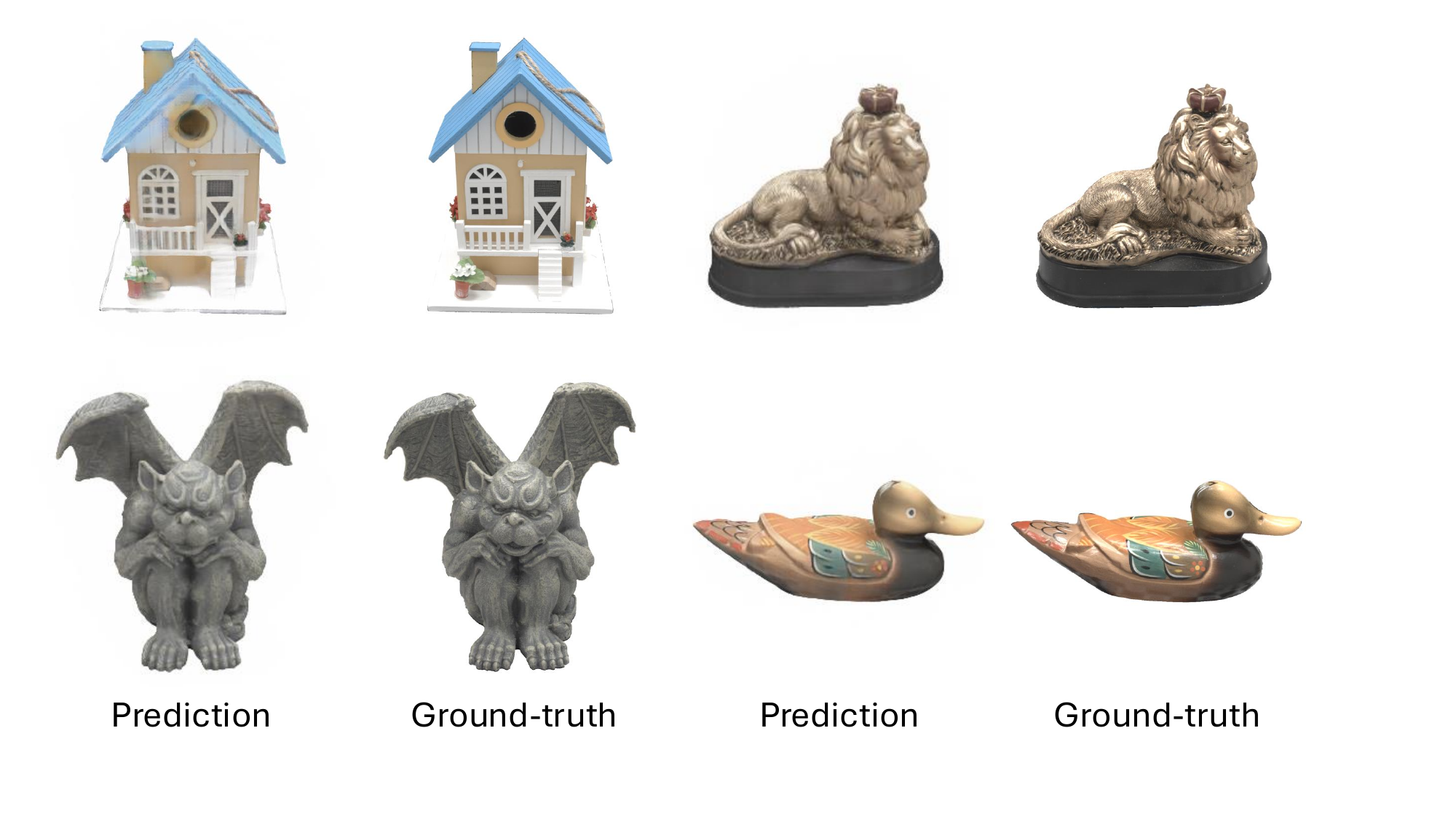}
    \caption{LRM view synthesis results from DSLR data.}
    \label{fig:lrm_dslr}
\end{figure}

\subsection{Sparse-view Reconstruction Application for DSLR data}
\label{sec:lrm_dslr}
We provide a sparse-view reconstruction baseline on our DSLR data using learning based reconstruction method. To build the baseline, we train an LRM model similar \cite{wei2024meshlrm} that achieves state-of-the-art mesh reconstruction results as demonstrated on public synthetic dataset on GSO\cite{downs2022google} dataset in Table \ref{tab:lrmvolsdf_baseline}. 

\paragraph{LRM baseline} Our LRM baseline has the same transformer architecture and tri-plane representation as \cite{wei2024meshlrm}. It consists of 24 self-attention blocks with feature dimension 1024 and 16 heads of attention. The tri-plane token number is $32\times 32$. Each token is decoded to a $8\times 8$ feature patch through a linear layer, which leads us to final tri-plane resolution of $256 \times 256$. During training, our LRM directly outputs SDF value instead of density and we use the volume ray tracing method proposed in \cite{yariv2021volume} to render images for supervision. Compared to the NeRF representation used in \cite{wei2024meshlrm}, our LRM baseline is robust to train and can achieve accurate geometry reconstruction. We evaluate our LRM-VolSDF baseline on synthetic GSO dataset. Results are summarized in Table \ref{tab:lrmvolsdf_baseline}. With 8 input views, our baseline achieves reconstruction accuracy comparable to the state-of-the-arts. 

\paragraph{Experiments and evaluation} We randomly sample 16 images from the 120 training views as inputs to our LRM-VolSDF network. The testing view is the same as the dense-view inverse rendering experiments. The quantitative numbers for view synthesis are summarized in Table~\ref{tab:lrm_dslr}. We observe that all three metrics are much worse compared to the results from synthetic GSO dataset, indicating the existence of a domain gap between synthetic and real data. Fig.~\ref{fig:lrm_dslr} visualize some of our reconstruction results. We observe that for relative simple shape, our model generalizes well and achieves highly detailed reconstruction that closely match the ground-truths. However, it struggles at reconstructing more complicated object, such as the Birdhouse. Further investigation is required to understand the gap on real world dataset. 

\subsection{Sparse-view Reconstruction Application for Egocentric data}
\label{sec:lrm_passive}

\begin{figure}
\centering
\includegraphics[width=\columnwidth]{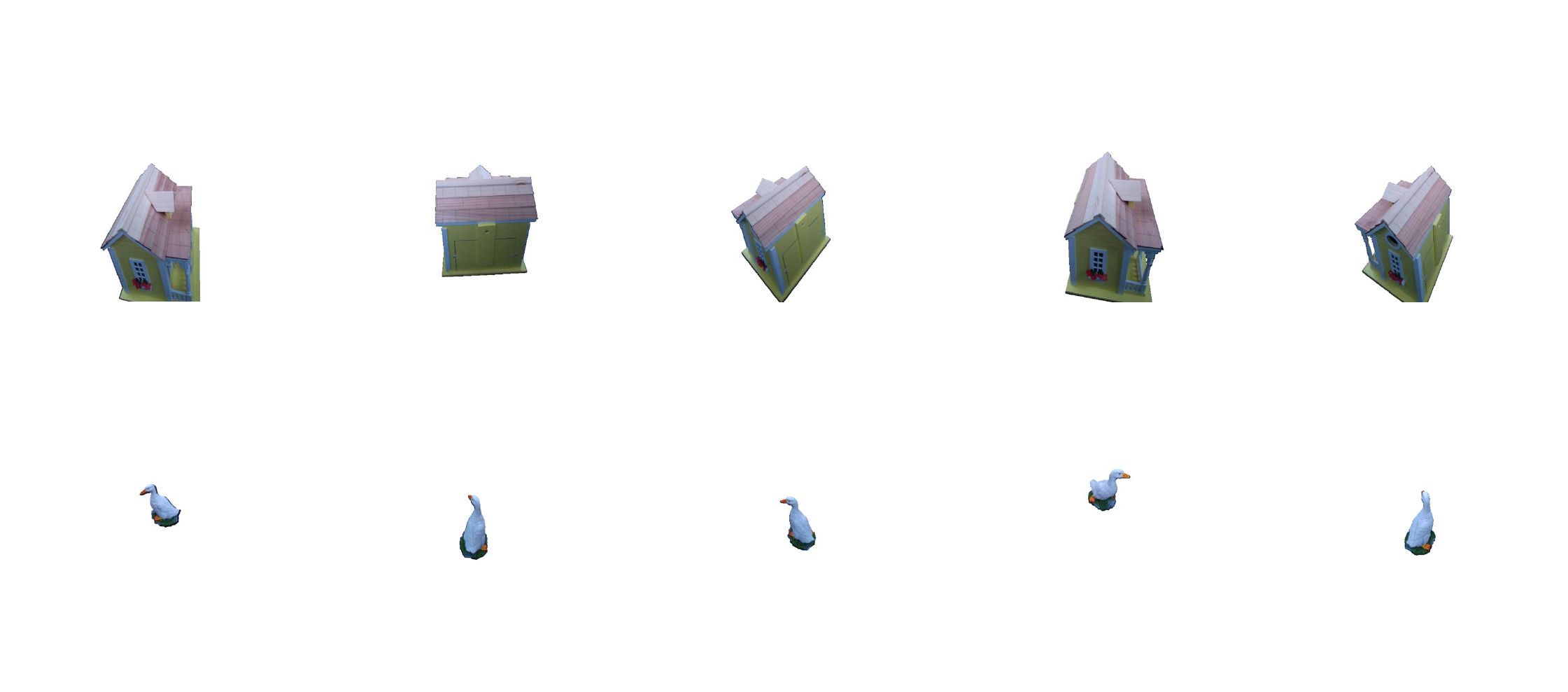}
\caption{Example input data of egocentric sparse-view reconstruction. We used the masked pixels from the rectified egocentric passive data as input. Some objects can appear very small due to the challenges in free-view movement in egocentric passive trajectory.}
\label{fig:passive_example}
\end{figure}

\begin{figure}
\centering
\includegraphics[width=\columnwidth]{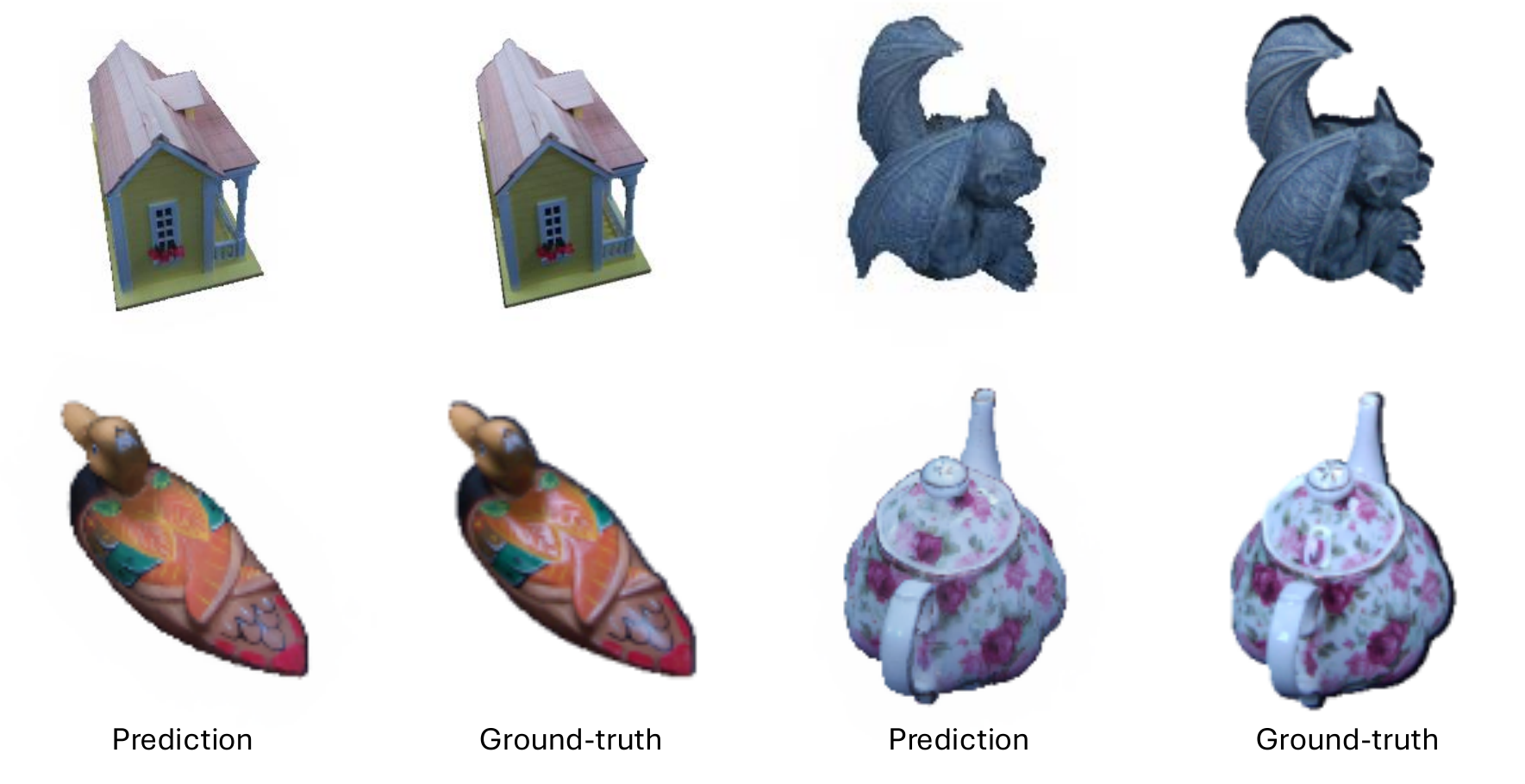}
\caption{LRM view synthesis results from egocentric data.}
\label{fig:lrm_passive}
\end{figure}

\begin{table}
    \caption{Quantitative results on novel view synthesis for sparse-view reconstruction using egocentric \emph{passive} data.}
    \label{tab:lrm_passive}
    \vspace{-0.1in}
    \setlength{\tabcolsep}{4pt}
    \renewcommand{\arraystretch}{0.4} 
    \footnotesize
    \centering
    \begin{tabular}{lcccc}
    \toprule
    & PSNR-H ($\uparrow$) & PSNR-L ($\uparrow$) & SSIM ($\uparrow$) & LPIPS ($\downarrow$) \\
    \midrule
     LRM-VolSDF & 20.03 & 21.72 & 0.971 & 0.065 \\
    \bottomrule
    \end{tabular}
\end{table}

We use the egocentric \emph{passive} recordings in this evaluation, which is a challenging sparse-view reconstruction setting where we only have casual observation of an object captured by an egocentric RGB camera. Compared to the sparse-view reconstruction for DSLR data where the object always appear in the image center and occupies a large portion of the image, the egocentric recording has natural human moving trajectory and object can appear small in many views. 

Fig.~\ref{fig:passive_example} shows several examples of the masked objects that we use as input to this evaluation. We can see the camera poses are much more diverse and object only occupies a very small portion of the image. This setting will defy any classical optimization-based 3D reconstruction method. We build a baseline using the same LRM-VolSDF model as mentioned in Sec.~\ref{sec:lrm_dslr}. We tested the model on the challenging data without any fine-tuning. 

\paragraph{Experiments and Evaluation.} For each sequence, we randomly sample 24 views from the middle 1/3 of the whole sequence as this is when our Aria glasses are relatively close to the object. We use 16 of the 24 views as inputs and 8 views as ground-truth for testing. Similar to the DSLR experiment, we use centralized cropping to place the object in the center of the image and modify the plucker rays representation accordingly. Qualitative and quantitative results are summarzed in Fig. \ref{fig:lrm_passive} and Tab. \ref{tab:lrm_passive} respectively. We observe that despite that the camera poses distribution of input views is very different from our training data, the LRM-VolSDF generalizes to the new type of input data. Our PSNR numbers for the masked region are much lower than those of synthetic data but still achieve reasonable performance. We calculate the LPIPS and SSIM based on the full image which results in higher numbers. This evaluation expose the domain gap in learning based reconstruction map for egocentric data input.
We hope our current experimental results can be a promising starting point. 

\subsection{Dense-view Reconstruction for Egocentric data}

\begin{figure*}
    \centering
    \includegraphics[width=0.49\linewidth]{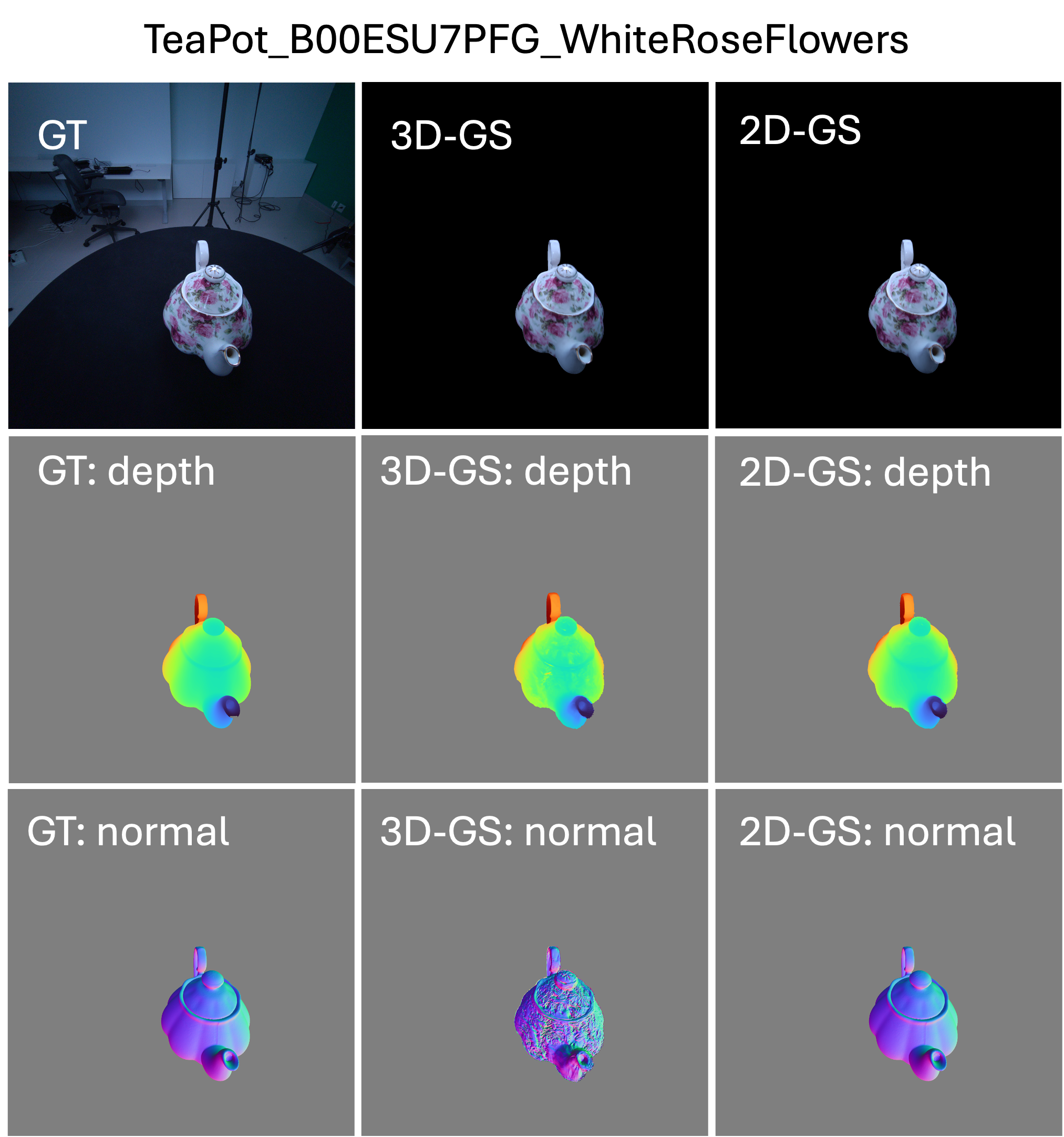}
    \includegraphics[width=0.49\linewidth]{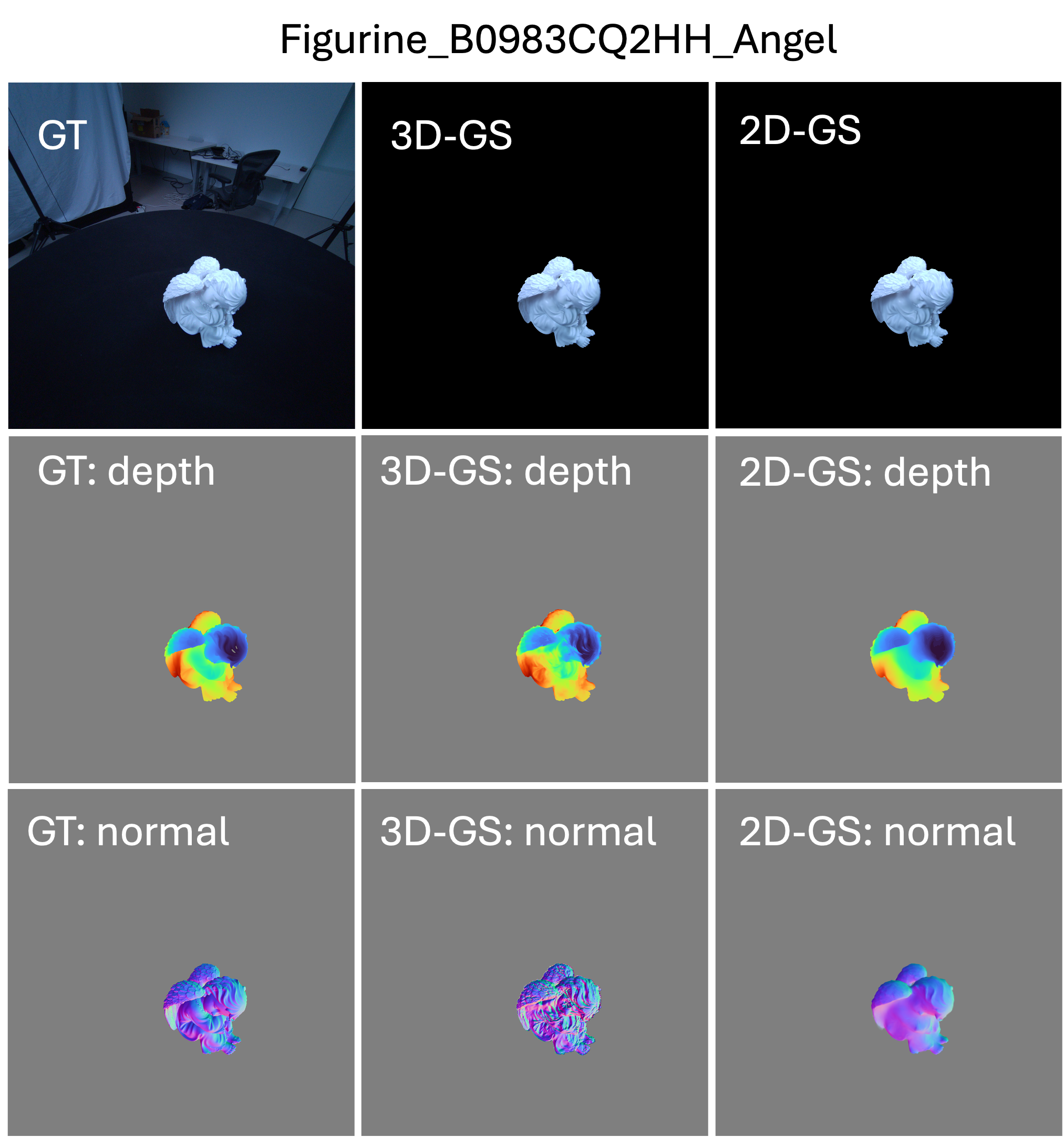}
    \includegraphics[width=0.49\linewidth]{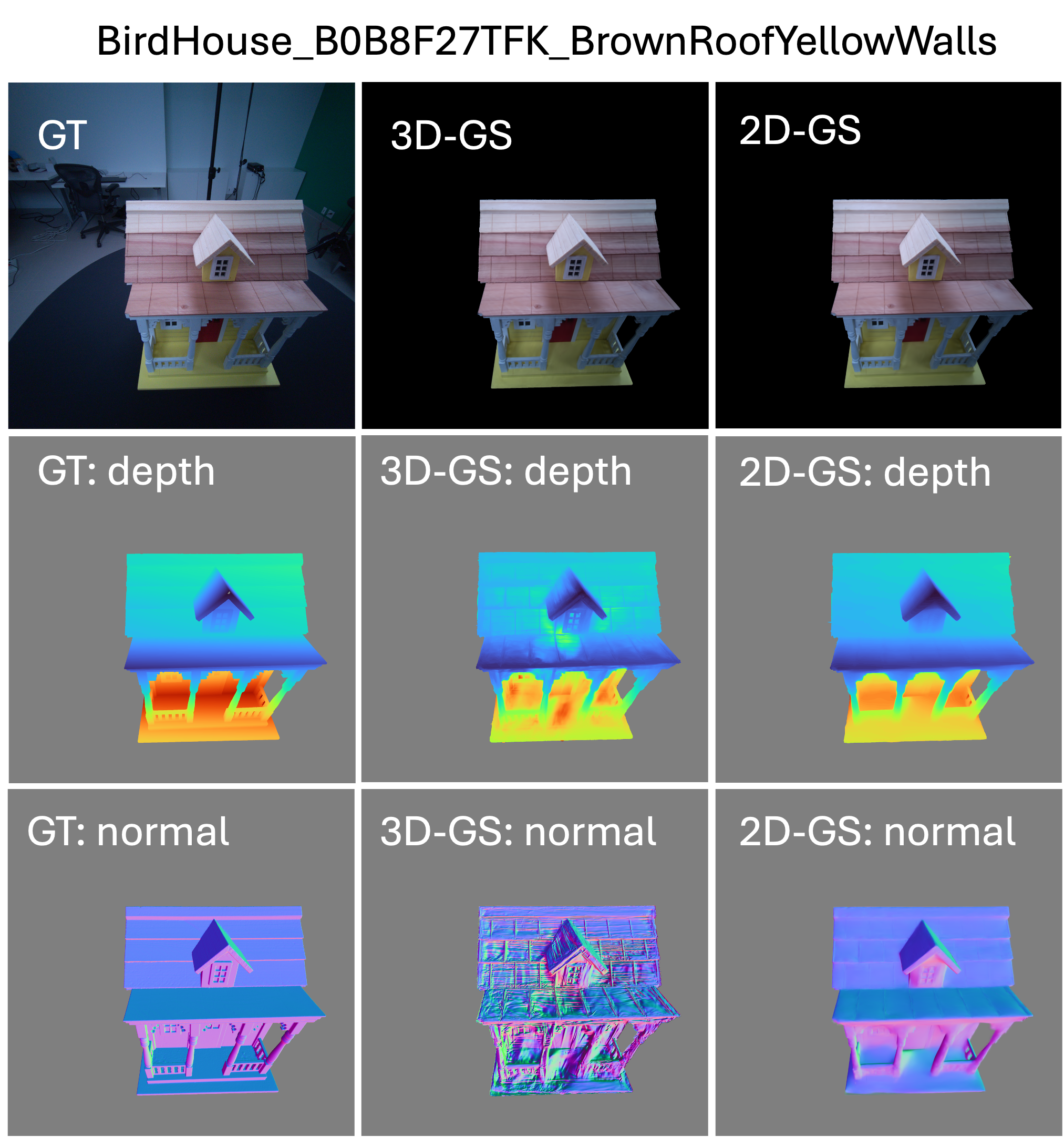}
    \includegraphics[width=0.49\linewidth]{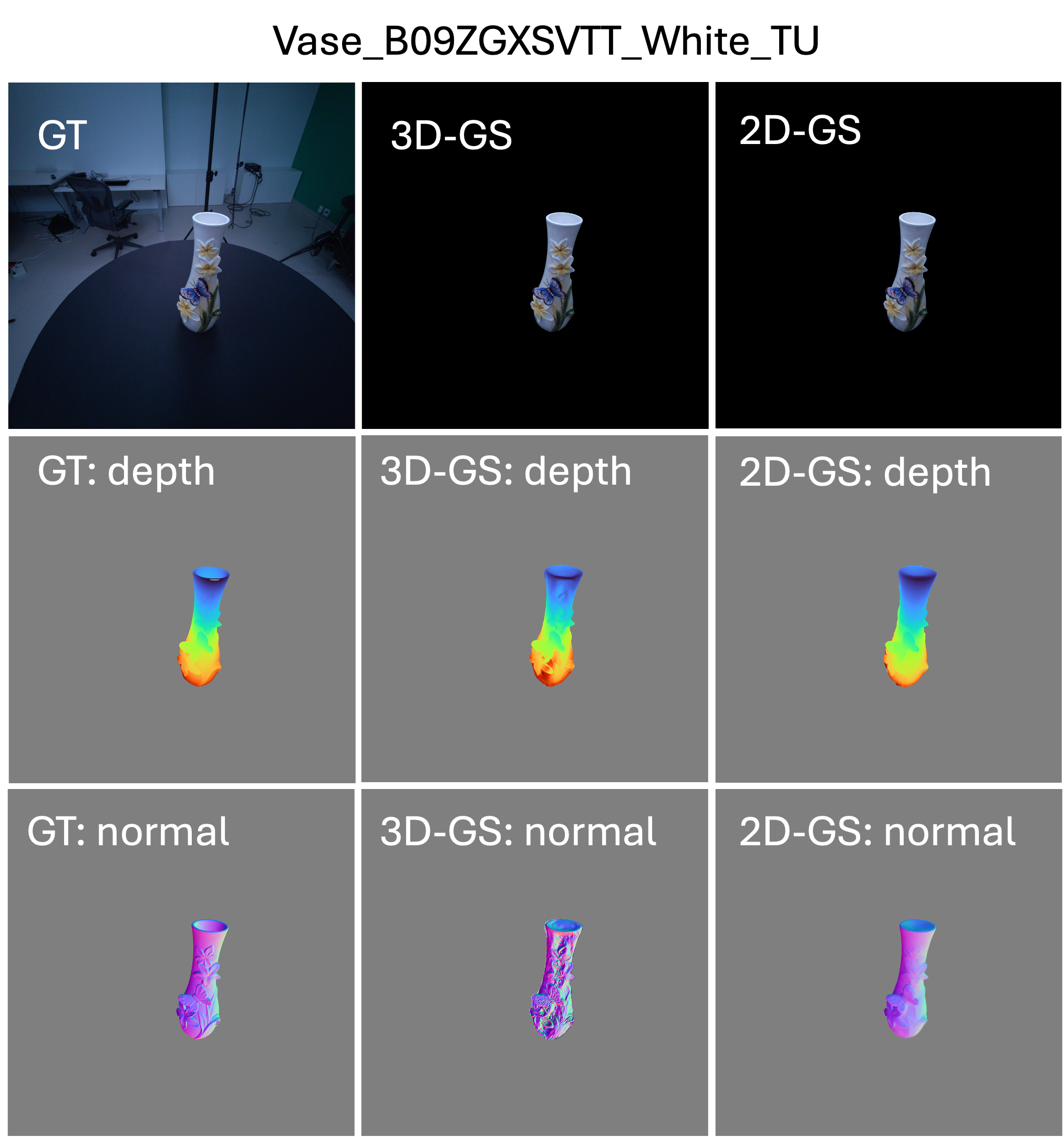}
    \caption{Qualitative comparisons of baselines reconstructions on egocentric recordings. We compare 3D-GS\cite{kerbl3Dgaussians} and 2D-GS\cite{huang20242Dgs} to the ground truth using the modalities of rendered images, depth and normal. Both baselines can provide near photoreal view synthesis on the held-out validation view compared to ground truth. However, the comparison in their geometry (depth and normal) indicate the existing methods still fall short to recover the details in digital twin reconstruction. For the benefit for visualization, we rotate all the images 90 degree clockwise.}
    \label{fig:aria_baseline_comparisons}
\end{figure*}

As a complement to Sec.~\ref{subsec:egocentric_nvs}, we provided qualitative comparisons of novel view synthesis in Fig.~\ref{fig:aria_baseline_comparisons}.
We rendered the depth from both baselines in a normalized depth range. For 3D-GS, we acquire the normal from its point cloud ray casted from each pixels. For 2D-GS, we directly use the predicted normal from the rasterizer. 




\paragraph{Results analysis:} Both 3D-GS\cite{kerbl3Dgaussians} and 2D-GS \cite{huang20242Dgs} can provide near photo-real view synthesis of the objects with high quality view-dependent effect. 3D-GS perform slightly better in PSNR metric consistent across all objects. In terms of shape reconstruction, 2D-GS performs better in particular to recover surface normals. Compared to the ground truth rendering, the estimated depth from 3D-GS is significantly noiser which also leads to visible artifacts in its normal map, while the 2D-GS tends to predict smoother depth and normal and can ignore certain details. For objects with simple shape (e.g., \texttt{Teapot\_B00ESU7PFG\_WhiteRoseFlowers}), both methods perform well. However, for objects with complex shapes, both methods fail to recover the complex geometric details in the object shape (e.g., \texttt{BirdHouse\_B0B8F27TFK\_BrownRoofYellowWalls}). The challenge in shape reconstruction indicates the direction for future research work in this area. 

\section{Experimental Details For Robotic Experiments}
\label{appendix:robotic_experiments_details}

\subsection{Experimental Objects}
The specific objects that we use from the \name and Objaverse-XL datasets~\footnote{The version of Objaverse-XL used in this work excludes all 3D models sourced from Sketchfab. Further, no Polycam assets were obtained from the Polycam source site} are as follows:

\name dataset:
\begin{itemize}
    \item \texttt{Cup\_B01LYONYPB\_SkyBlue}
    \item \texttt{Cup\_B0BR43SPKJ\_Blue}
    \item \texttt{Cup\_B0CJBZT7N5\_Black}
    \item \texttt{Cup\_B0CMPB8FNY\_MountainBluebell}
    \item \texttt{Cup\_B0CYL5PSR3\_Gray}
    \item \texttt{Cup\_B08PTSRWF8\_Green}
    \item \texttt{Cup\_B0CMD4LX4D\_DarkBlue}
    \item \texttt{Cup\_B08TWHJ33Q\_Tan}
    \item \texttt{Cup\_B0CNJP2KZF\_GreenOrange}
    \item \texttt{Cup\_B094NQH2YM\_BlackGold}
    \item \texttt{Cup\_B0BXB21T7Q\_Blue}
    \item \texttt{Cup\_B0CPX832ZP\_Floral}
    \item \texttt{Cup\_White}
    \item \texttt{Cup\_B09L8DS2ZB\_DarkBrown}
    \item \texttt{Cup\_B0C3X3WY2Q\_SageGreen}
    \item \texttt{Cup\_B0CQTF6GF1\_RedWhite}
    \item \texttt{Cup\_B09QCYR1SL\_Pitcher\_1}
    \item \texttt{Cup\_B0C81BCSXQ\_WhiteRainbow}
    \item \texttt{Cup\_B0CQXPND8L\_Stripes}
    \item \texttt{Cup\_B09YDPVRM7\_RedTeaCup}
    \item \texttt{Cup\_B0CDWXPDK1\_Zebra}
    \item \texttt{Cup\_B0CR45H24G\_Blue}
    \item \texttt{Cup\_B0B3JKZW76\_Brown}
    \item \texttt{Cup\_B0CHJYN6F1\_Black}
\end{itemize}

For Objaverse-XL, we collect a subset of $24$ objects by filtering objects by the STL file format and by the word \texttt{cup} appearing in the provided metadata tags, and then manually filtering the results by inspecting the object files to only include objects that can be considered ``cups'' by a human judge.

\subsection{Simulation and Data Collection}
When importing objects into the PyBullet simulator, we obtain separate collision meshes by performing convex decomposition of each object with CoACD~\citep{wei2022coacd} with a concavity threshold of $0.01$, and rescale each object to have a maximum bounding box side length of $0.137$ m (to match that of the test object). Since not all Objaverse-XL objects come with textures, we randomly color Objaverse-XL objects uniformly in RGB space.

For data collection, we uniformly randomly select one object from the considered object set to be placed into the scene. We uniformly randomly place the object within a $0.4 \times 0.4$ m box centered at the scene origin at a $z$ height of $0.10$m, and allow it to fall to the floor workspace. If the object rolls away (further than a $0.8 \times 0.8$ m bound), we re-sample the initial position and re-attempt the object placement.   

\textbf{Pushing:} For pushing, we generate pseudo-random trajectories by executing a scripted policy in two stages: First, the robot samples a starting position from the push by randomly sampling an angle $\theta$ and radius $r$. It then takes steps to move its end-effector to the x-y location $(x_{obj} + r\sin(\theta), y_{obj} + r\cos(\theta))$, where $(x_{obj}, y_{obj})$ is the initial object position. After reaching within $2$cm of this location, the robot begins to push the object, by moving from its current end-effector position toward the object location. All steps are normalized to have a magnitude of $0.03$m, and the action at each step is affected by uniform random noise drawn from $\mathcal{U}(-0.05, 0.05)$ in each axis. The trajectory ends after $35$ steps.

\textbf{Grasping:} For grasping, we generate successful grasps by sampling grasping candidates and filtering only the successful ones. After adding an object to the scene, we uniformly randomly sample a grasping position $(x, y, z)$ where $x = x_{obj} + \mathcal{U}(-0.05, 0.05)$,  $y = y_{obj} + \mathcal{U}(-0.05, 0.05)$, $z = 0.2 + \mathcal{U}(-0.1, 0.1)$ where $(x_{obj}, y_{obj})$ is the initial object position. The robot then performs four steps: (1) It moves in the air to the target grasp position $x, y$ coordinates, (2) it moves vertically downwards to a height of $z$, (3) it closes the gripper, and (4) it lifts to a height of $0.5$m. The grasp is considered successful if the object is at least $0.1$m above the floor after step (4).

\subsection{Policy Training}
We train convolutional neural network policies to regress actions representing $xy$ position changes and the $(x, y, z, \theta)$ grasp position for pushing and grasping respectively. For pushing, we train goal-image-conditioned policies, relabeling goals using a hindsight sampling mechanism. We sample goals uniformly randomly from future timesteps in the trajectory of a sampled initial state, and use the first action after the initial state as the label. 

We use similar architectures for pushing and grasping. The architecture consists of an encoder and an MLP head for action prediction. Below we describe the architecture for goal-conditioned pushing, with differences for the grasping policy noted. The input image size for both initial and goal images is $256 \times 256 \times 3$.

For pushing, we normalize action labels by multiplying each dimension by $100$, and for grasping, we normalize the last dimension (grasp angle) by dividing by $10$ such that it has similar magnitudes to the other action label dimensions. 

\textbf{Encoder:}
\begin{itemize}
    \item \texttt{Conv2d(3, 32, 3, stride=2, padding=1)}
    \item ReLU
    \item \texttt{Conv2d(32, 64, 3, stride=2, padding=1)}
    \item ReLU
    \item \texttt{Conv2d(64, 64, 3, stride=2, padding=1)}
    \item ReLU
    \item Flatten
    \item \texttt{Linear((image\_size/8)$^2 \times 64$, 512)}
    \item ReLU
\end{itemize}

Note that for the pushing policy, the observation and goal image share encoder weights.

\textbf{Action Prediction Head:}
\begin{itemize}
    \item \texttt{Linear($512\times 2, 256$)} (the input feature dimension is $512$ for grasping, which has a single image input as opposed to $1024$ and two image inputs for pushing, which have their features concatenated together).
    \item ReLU
    \item \texttt{Linear(256, 256)}
    \item ReLU
    \item \texttt{Linear(256, 2)} (for grasping, the final output size is $4$)
\end{itemize}

We train policies using a mean-squared-error loss using the Adam optimizer with a learning rate of $3e-4$ and a batch size of $32$, holding out $5\%$ of the data for validation. We train for $200$ epochs and take the checkpoint with lowest validation loss for evaluation. Training uses a single NVIDIA Titan RTX GPU. 

\subsection{Evaluation}

For evaluation, we generate test sets of $100$ initial conditions for each task using the ``\texttt{cup\_scene003}'' object from the StanfordORB dataset. For pushing, we generate trajectories in the same way as during data collection and use the initial state and final state as the initial condition and target goal for the policy. We roll out the policy for $35$ steps and compute the final position error of the object in the $x,y$ dimensions. For grasping, we generate initial object positions in the same way as in data collection and the policy performs a single grasp attempt, which is determined successful if the object ends at least $0.1$m above the working surface. 

In Figure~\ref{fig:detailed_robotic_pushing_results}, we present detailed results for the pushing experiments, plotting the success rates for each policy with respect to varying success thresholds based on final object position error.



\end{document}